\documentclass[twoside,12pt]{article}
\usepackage{epsfig}

\def\Journal#1#2#3#4{{#1} {#2} (#4) #3 }

\def\NPB{{\em Nucl. Phys.} B}

\def\PLB{{\em Phys. Lett.} B}

\def\PRL{\em Phys. Rev. Lett.}

\def\PREP{\em Phys. Rep.}

\def\PRD{{\em Phys. Rev.} D}

\def\ZPC{{\em Z. Phys.} C}

\newcommand{\be}{\begin{equation}}
\newcommand{\ee}{\end{equation}}
\newcommand{\bea}{\begin{eqnarray}}
\newcommand{\eea}{\end{eqnarray}}

\newcommand{\F}{$ F_{2}(x,Q^2)\:$} 
\newcommand{\Fc}{$ F_{2}\,$}

\newcommand{\as}{$\alpha_s\,$}
\newcommand{\amz}{$\alpha_s(M_Z^2)\,$} 
\newcommand{\bs}{\overline{s}}
\newcommand{\bc}{\overline{c}}
\newcommand{\bu}{\overline{u}}
\newcommand{\bd}{\overline{d}}

\newcommand{\bU}{\overline{U}}
\newcommand{\bD}{\overline{D}}

\newcommand{\xpom}{$x_{IP}\,$}

\newcommand{\ipb}{pb$^{-1}\,$} 

\newcommand{\pb}{\,pb$^{-1}$}

\newcommand{\empz}{E-p_z}
\begin{document}

\title{ \vspace{1cm} Collider Physics at HERA}
\author{M.\ Klein$^{1}$, R.\ Yoshida$^{2}$\\ 
\\
$^1$University of Liverpool, Physics Department, L69 7ZE, Liverpool,\\ 
United Kingdom\\
\\
$^2$Argonne National Laboratory, HEP Division, 9700 South Cass Avenue,\\ 
Argonne, Illinois, 60439, USA\\
}
\maketitle
\begin{abstract}
From 1992 to 2007, HERA, the first electron-proton collider, operated at
cms energies of about 320\,GeV and allowed the investigation of 
deep-inelastic and photoproduction processes at the highest energy scales accessed
thus far. This review is an introduction to, and
a summary of, the main results obtained at HERA during its operation.
\end{abstract}
%\eject
\begin{center}
\vspace{2cm}
To be published in {\it Progress in Particle and Nuclear Physics}
\end{center} 
\begin{flushright}
\vspace{-17cm}
ANL-HEP-PR-08-23
\end{flushright}

\newpage
\tableofcontents
\newpage

\section{Introduction}
HERA  was the culmination of 50 years of experimentation
with electron, later also muon and neutrino, beams to explore the
structure of the proton, first in elastic and subsequently in inelastic
scattering. The idea had been simple: use the pointlike lepton
as a probe to study the strong interaction phenomena 
of the nucleons and their internal structure.
HERA emerged from a series of electron-proton 
accelerator studies in the 70's as the most energetic version of
an $ep$ collider possible. It was built in the 80's for maximum
possible luminosity, and with the capablility to scatter polarised electrons and positrons
off protons, at a center of mass energy, $\sqrt{s}$, 
of about 320\,GeV. The first HERA data were taken in summer 1992.
It ceased operations in June 2007 after a long, 
successful data taking period of 16 years.

In deep inelastic lepton-nucleon scattering (DIS),
the proton structure is probed by a virtual photon ($\gamma$), or weak
boson ($Z_0$ or $W^{\pm}$), which carries a four momentum squared
$Q^2 \leq s$.   The momentum transferred
is inversely proportional to the  spatial distance inside the 
proton which can be 
resolved with the photon or the weak boson probe.  High energy
DIS experiments are thus microscopes
with a resolution power extending much beyond the eye, optical microscopes
or modern laser techniques. The basic experimental interest
and challenge over decades has naturally been to enlarge $Q^2$ 
by increasing~$s$.

The $ep$ collider  at the DESY laboratory in Hamburg, Germany,
had its first phase of operation, ``HERA I'', from 
1992 through 2000. In this period, the collider experiments
H1 and ZEUS each recorded data corresponding to integrated luminosities
of approximately 120\,pb$^{-1}$ of $e^+p$ and 15\,pb$^{-1}$ of $e^-p$
collisions. The energy of the electron (positron) beam was about 
27.5\,GeV.  The energy of the proton beam was initially set 
to 820\,GeV until it was increased to 920\,GeV, in 1998, and
kept at that value.

The HERA collider was then upgraded to increase the specific luminosity by
a factor of about four, as well as to provide longitudinally polarised
lepton beams to the collider experiments.  
The second data taking phase, ``HERA II'', began in 2003,
after completion of the machine and detector upgrades and when
unexpected large background problems were finally overcome.
It ended in 2007.  The H1 and ZEUS experiments each recorded approximately
200\,pb$^{-1}$ of $e^+p$ and 200\,pb$^{-1}$ of $e^-p$ data with 
electron (positron) energy of approximately 27.5 GeV and proton energy of
920 GeV.  The lepton beams had an average polarisation
of approximately $\pm$ 30\% with roughly equal samples of
opposite polarities recorded.

In the last three months of HERA operation, data with lowered proton
beam energies of 460\,GeV and 575\,GeV were taken, each experiment recording
approximately 13\,pb$^{-1}$ and 7\,pb$^{-1}$, respectively.  The primary
purpose of this data is the measurement of the longitudinal  proton 
structure function $F_L$.

The physics at an $ep$ collider, which was to follow
the fixed target experiments performed in the 70's, was summarised
in a well-known article by C.\,Llewellyn Smith and B.\,Wiik\,\cite{wiiksmith}
written 30 years ago.  
HERA has fulfilled those early expectations but delivered much more
insight. While some predictions, such as the unification
of electromagnetic and weak interactions, at $Q^2$ values
as large as the squared masses of the weak bosons, were confirmed, 
new, original and unforeseen, views on the structure of the proton and 
the dynamics of quark-gluon interactions were gained.  The observations
at HERA led to an enormous development in the understanding of
the underlying field theory of the strong interactions, Quantum 
Chromodynamics (QCD).
New fields of study emerged and existing fields entered a new precision
phase; among these are the physics
of high parton densities, the physics of high scale
photoproduction and photon structure,  the chromodynamic theory of
heavy quark production and of hard diffraction. 
The limits set at HERA in the search for new heavy particles, as
predicted, for example, in supersymmetric theories, are competitive
with those set at the $e^+e^-$ collider LEP and the $p \overline{p}$ collider
Tevatron.  As of this writing, no compelling indications of physics beyond the
Standard Model have been found at these three colliders searching in the
few hundred GeV energy range.   The Large Hadron Collider (LHC)
is now set to expand the search to the TeV scale. The proton 
structure determined with precision 
by H1 and ZEUS at the $ep$ collider HERA is of crucial
importance~\cite{heralhcws} for
the physics of $pp$ collisions at the LHC.

The current review is an overview of, and an introduction to,
the physics results as obtained during the years of operation 
of HERA, with emphasis on the HERA~I results
since much of the more recent data is still
being analysed. During the years 1992-2007,
H1 and ZEUS have each published
more than one hundred and fifty journal publications.  
Given the space limitations, 
not all subjects can be covered in the present review. The reader
is referred to the web pages of the experiments \cite{webh1zeus}
and to the series of international workshops on ``Deep Inelastic
Scattering and QCD'' \cite{disworkshops} at
which, since 1994,
the physics at HERA and related results have been discussed annually, and 
in depth. This paper
focuses on the results of ZEUS and H1 and therefore cannot
do justice to the impressive amount of theoretical work, on
which the interpretations of the HERA data often rely. The readers
are referred to the publications of the collaborations
and papers cited therein. A review on early HERA results has
been given in \cite{allenhal}. A recent general introduction to the
physics of deep inelastic scattering is \cite{robinmandy}.

The paper is organised as follows: in Section 2  HERA accelerator
issues are discussed, the $ep$ kinematics and the two collider
detector concepts are presented. Section 3 is devoted to the
basic inclusive cross section measurements and the structure
functions are derived. Section 4 presents the approaches of ZEUS
and H1 to extract the quark and gluon momentum distributions
in the proton from QCD fits to the inclusive data. Section 5 discusses 
measurements of jet production and their importance to the determinations
of the gluon distribution and the strong coupling constant,
$\alpha_s$. Section 6 is devoted to the results of the production
of charm and beauty quarks and their understanding
in QCD. Section~7 presents results on diffraction, including inclusive
scattering, vector meson production and the new field of
deeply virtual Compton scattering (DVCS).
Section 8 summarises the results on tests of the electroweak
theory, in neutral and in charged current $ep$  interactions. Finally, in 
Section 9, the main results on searches for new physics at high
scales are presented. 

The paper closes with a short summary 
and outlook. As this paper is being written there is still much 
analysis activity devoted to the publication of
new and more accurate results, mainly based on the
HERA II period of operation and upgraded detectors. In some areas,
work on  the combination
of H1 and ZEUS results has begun, which is expected
to lead to the ultimate accuracy of HERA data.
Some combined H1-ZEUS data, such as those of inclusive DIS cross sections,
have already been presented in preliminary form; however, as there 
are no combined published results available yet, their
presentation is deferred essentially  to a later review publication.

\section{Accelerator and Detectors}
\subsection{Introduction}
HERA, the ``Hoch Energie Ring Anlage"
at DESY, Hamburg, was the first, and so-far the only, accelerator 
complex in which
electrons and protons were collided. In the 60's and 
70's several lepton-proton experiments were performed. Among these were the
pioneering SLAC electron  experiments\,\cite{slac} and the CERN muon 
experiments\,\cite{bcdms,emc}, in which 
a lepton beam of energy $E_l$ between 20 and 280\,GeV was scattered off a
stationary nucleon target. The maximum energy squared in the
center of mass system for fixed target lepton-proton scattering is $s=2M_pE_l$
with the proton mass $M_p$.  The available cms 
energy sets the basic kinematic limits in any
deep inelastic lepton-nucleon scattering (DIS) experiment.

The SLAC experiment, in 1968, resolved the proton's structure down to
a distance of $10^{-15}$\,m at which it was discovered that the proton 
has point-like constituents. Since
the energy available to lepton beams was limited to a few hundred
GeV, several proposals were put forward to build an electron-proton
collider as a two-ring structure with different types of particles
being stored. In such a configuration the energy $s$ becomes equal
to $4 E_e E_p$ which is about $2E_p$ times larger than the fixed target
energy, for a given lepton beam energy.  Since the late 60's,
Bjoern Wiik and colleagues had considered such machines and proposed
to probe proton's structure deeper with an $ep$ collider
at DORIS\,\cite{dorisep}, later at
the $e^+e^-$ colliders PEP\,\cite{pepep}, TRISTAN\,\cite{tristanep} and  
PETRA (PROPER)\,\cite{proper} and subsequently
at the proton  ring accelerator SPS at CERN
(CHEEP)\,\cite{cheep}.  Another $ep$ collider proposal, to use the
Tevatron accelerator, was
put forward by a Canadian group
(CHEER)\,\cite{cheer}. When the decision
to build a large electron-positron collider, LEP, at CERN was made, 
DESY endorsed the HERA proposal\,\cite{hera84}, i.e. the plan to
build an electron-proton collider at highest attainable energy, i.e.
$s \simeq 10^5$\,GeV$^2$ with
$E_e \simeq 30$\,GeV and $E_p \simeq 800$\,GeV. 
This design was challenging since it was to use a chain of two
times four pre-accelerators (see Table\,\ref{tab-herachain})
only some of which already existed at DESY, and enormous efforts 
were required to
ensure its reliable operation and high luminosity. The design
was made for positron- as well as electron-proton collisions
and to achieve a large degree of longitudinal lepton beam polarisation,
using the Sokolov-Ternov effect\,\cite{Sokolov} 
and spin-rotators around the interaction regions.

Two general purpose collider detectors with nearly $4 \pi$ acceptance
were proposed in 1985, H1\,\cite{H1prop} and ZEUS\,\cite{ZEUSprop},
and were built in the 80's. They were  
operated over the 16 years of HERA operation. Two further experiments
at HERA were built and run in the fixed target mode. 
The HERMES experiment\,\cite{HERMES} (1994-2007) 
used the polarised $e^{\pm}$ beam
to study spin effects in lepton-nucleon interactions using a
polarised nuclear target.
The HERA B experiment\,\cite{HERAB} (1998-2003) was designed to
investigate B meson physics and nuclear effects
in the interactions of the proton beam halo
with a nuclear wire target. The results of the fixed target experiments are
not discussed in this review.

The collider detectors H1 and ZEUS were designed primarily for
deep inelastic $ep$ scattering at highest momentum transfers, $Q^2$, and
large final state energies. Thus, much attention was paid to the
electromagnetic and hadron calorimeters. The H1 collaboration 
chose liquid argon as active material for their main calorimeter 
to maximize long term reliability.   The ZEUS collaboration chose
scintillator active media and
uranium as the absorber material  for the 
desired equalization of the calorimeter ``$e/\pi$'' response to electrons and
hadrons. The large calorimeters were complemented by large area wire
chamber systems for muon momentum and hadron shower energy tail 
measurements. Because the $e$ and $p$ beam energies were very
different, the detectors were asymmetric with extended 
coverage of the forward (proton beam) direction.
Drift chambers inside the calorimeters,
both in H1 and in ZEUS, were segmented into a forward and a central 
part. Later, in H1 starting in 1996 and in ZEUS from 2003 onwards, 
silicon detectors near the beam pipe were installed for precision vertexing
and tracking. Both apparatus were complemented with detector systems
positioned near the beam axis in the accelerator tunnel, to measure 
backward photons and electrons, mainly for the determination of the 
interaction luminosity, and to tag leading protons and neutrons in 
the forward direction.
Thanks to enormous efforts of several hundreds of engineers and physicists,
both experiments took data for the entire time of HERA's operation
at high efficiency. The only exception was a period of two years after 
the modification of the interaction region in 2001, which had caused 
unexpected problems to the $ep$ operation.

\subsection{Accelerator}
%
%HERA was the only $ep$ collider built so far. 
Most of the data at HERA were taken
with electron or positron energies of $E_e \simeq 27.6$\,GeV 
and $E_p=920$\,GeV.
In the first years, 1992-1997, $E_p$ was set to 820\,GeV. In the last months
of operation, March-May (June) 2007, $E_p$ was lowered to 460\,(575)\,GeV.
Typical beam currents were $I_p \simeq 100$\,mA and
$I_e$ ranged from about 35\,mA at
injection to $10-15$\,mA when the beam was dumped for refilling.
Initially the use of PETRA as a pre-accelerator
caused some problems due to eddy-current effects in HERA proton magnets
at low energies and for tight injection aperture. With 174 colliding
bunches the design luminosity of $1.4 \cdot 10^{31}$\,cm$^{-2}$s$^{-1}$
was reached in 1997.   The accelerator performance then stabilised and could
be improved to give a total delivered integrated luminosity, between
1992 and mid 2000, of close to 200\pb\, to each collider experiment.  
In the summer 2000,
HERA operation was stopped for modifications of the
interaction regions devoted to a further significant
enhancement of the luminosity.

For HERA, with the matched proton and electron beam sizes
and head-on collisions, the luminosity is essentially given 
by the product of the brightness, $N_p/\epsilon_p$, with 
the electron bunch current divided by the square root of the $\beta$
functions in $x$ and $y$.  Methods considered to increase the luminosity
included  improvements in the
injector chain which would overcome space charge effects limiting the
brightness and an increase of currents with added power.
Eventually, the luminosity increase was achieved by reducing the 
$\beta$ functions at the interaction point (IP), from $\beta_{xp}=7$\,m
and $\beta_{yp}=0.5$\,m to  $\beta_{xp}=2.46$\,m and $\beta_{yp}=0.28$\,m.
This required the separation of the $e$ and $p$ beams 
much closer to the IP  than in HERA\,I, 
i.e. at 11\,m instead of 24\,m. Moreover, combined function
(bending and focusing) magnets were introduced as close as
2\,m to the H1 and ZEUS interaction points. This forced H1 and ZEUS
to modify the inner detector configurations, and thus in 2000 the
physics program for $ep$ scattering with $Q^2$ between about 
0.1 and 2\,GeV$^2$ was terminated.
It took some time until the effect of the superconducting
magnets on the vacuum near the IR was finally understood and
regular, approximately bi-monthly,
warming up and cooling down cycles of these magnets
were introduced which improved the vacuum.

The period 2001-2003, following the modification of the IR, was
extremely demanding. Beam induced background caused the
central drift chambers of H1 and ZEUS to trip at very low currents. 
Identification of the background sources and their suppression
became of vital importance for the experiments and the luminosity
upgrade to be successful.  In an intense period of 
particularly strong collaboration between the machine and the detector
experts, a series of beam based experiments and background
simulations was performed over a period of more than one year
\cite{h1bgd,zeusbgd}.

The different components of the background, those from the $e$ and
$p$ beams, were investigated by varying the beam currents; special runs
with one type of beam only were also made.
% DONT DELETE THIS FROM TEX
%The drawn drift chamber current $I_c$, and similarly the counting rate
%of scintillator and silicon detectors monitoring the beam,
%for given beam currents $I_e$ and $I_p$,  has the following
%components
%\begin{equation} \label{idrift}                                               
%I_c = I_0 + a_{sy} I_e + b_e I_e p_L + b_p I_p p_R.
%\end{equation}  
%Here $I_0$ is the drift chamber current offset. The coefficient
%$a_{sy}$ measures the chamber current induced by synchrotron radiation
%which is proportional to the electron beam current. The finite vacuum
%pressure on  the left (incoming electron beam) side, $p_L$,  contributes
%a current $\propto I_e$ to $I_c$. Similarly the pressure on the
%right (incoming proton beam) side gives rise to a current proportional
%to the proton beam current.  The pressure $p_L$ as $p_R$ 
%has a static component $p^0_{L,R}$ and a dynamic part which arises
%from synchrotron radiation causing emission of ions from the beam pipe
%into the evacuated beam pipe region.  Putting these current dependencies
%together one finds the following quadratic form for $I_c$
%\begin{equation} \label{idbgd}
%I_c = I_0 + A I_e + B I_e^2 + C I_p + D I_e I_p.
%\end{equation}
%Measurements with varying beam currents, including runs with only
%electrons or only protons stored in HERA, allowed the four current
%coefficients, Equation~\ref{idbgd},  to be determined.  
Experimentally, it turned out that the dominant contribution 
% chamber current was related  to D,
was due to the dynamic interplay of synchroton induced effects
with the proton beam: the new beam-line components combined
with the bending of the electron beam 
closer to the IR than in HERA\,I caused ions to be released, which were 
scattered by the
proton beam and its halo. 
Further  problems arose
due to the narrow aperture, from back-scattered synchrotron radiation and
with leaks at flanges when the temperature load had risen too high due to
mechanical or beam steering problems. The three successful counter
measures were modifications of the beam absorber mask system,
improved pumping  
and steady operation at slowly increasing currents to clean the surfaces
using the synchrotron radiation itself. 
%Further evolved studies regarded 
%the chemical composition of the ions, as digested from the measured
%multiplicity distributions in the central chambers, and detailed
%simulations of the beam line which lead to an understanding of the
%location of the background sources, extending up to $15$\,m 
%in $e$ beam direction. 
From the end of 2003 onwards, routine
operation could be resumed with a specific luminosity increase
as planned by a factor of about four and high positron/electron and proton 
currents. The annual delivered luminosity in 2006 was about 200\,\ipb,
as much as had been delivered in the whole phase of HERA\,I.
The machine was running very reliably and with high luminosity when its operation
was ended.
\begin{figure}[htbp]
   \centering
   \includegraphics[height=9cm]{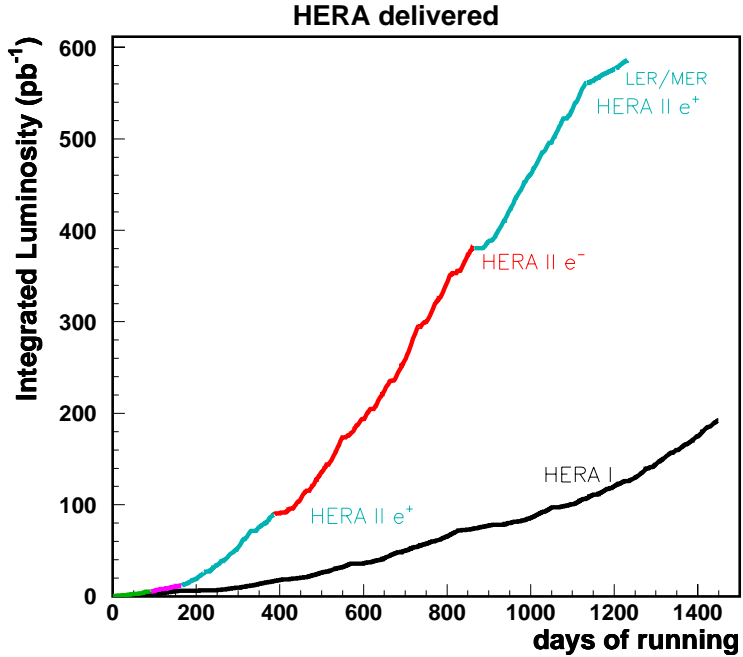}
\begin{minipage}{16.5 cm} 
   \caption{Luminosity delivered by HERA to each of
 the collider experiments, H1 and ZEUS, as a function of the
 number of days of operation. HERA was operated in an initial 
 phase (HERA I), 1992-2000, with unpolarised lepton beams, mainly with $e^+$.
 In the upgraded phase (HERA II), 2003-2007, the luminosity
 was increased and polarised $e^{\pm}$ data were taken with about
 equal amounts in terms of charge and polarisation states. The operation
 of HERA ended with two runs at 450\,GeV (LER) and 575\,GeV (MER)
 proton beam energy, in 2007.}
   \label{fig:lumi}
\end{minipage}
\vspace{-1cm}
\end{figure}

\begin{table}[htdp]
\begin{center}
\begin{minipage}{16.5 cm} 
\caption{Chain of (pre)accelerators at HERA and their energies.
The main ring is 6223\,m in circumference.
High field (5\,T) superconducting dipole magnets
bend a proton ring of up to 1\,TeV of energy. The electron ring requires
high power, provided by a system of  normal and superconducting 500\,MHz cavities, 
to compensate for the synchrotron radiation losses. 
\vspace{0.2cm}
} 
\label{tab-herachain}
\end{minipage}
\begin{tabular}{|c|c|c|c|}
\hline 
\bf{Protons} &  $\bf{E_p}$ & \bf{Electrons} & $\bf{E_e}$ \\
\hline
Source & 20 keV & Source & 150 keV  \\
RFQ & 750 keV & LINAC II & 450 MeV \\
LINAC III & 50 MeV & PIA & 450 MeV  \\
DESY III & 8 GeV & DESY II & 7 GeV  \\
PETRA & 40 GeV & PETRA & 12 GeV  \\
HERA & 920 GeV & HERA & 27.5 GeV  \\
\hline
\end{tabular}
\end{center}
\end{table}

Synchrotron emission leads to a build-up of 
transverse polarisation\,\cite{Sokolov} with a characteristic 
time of about 30 minutes at HERA. Since 1994 this was used for the
experimental programme of the HERMES experiment which was surrounded
by spin rotators, a pair of dipole magnets flipping the spin from transverse
to longitudinal orientation and back. In the first phase of HERA
the collider experiments were not equipped with spin rotators. 
Polarisation effects, for unpolarised protons,
result only from weak boson exchange at high $Q^2$ as the interference of
$Z_0$-photon  exchange in neutral current scattering occurs at 
at a size of order $10^{-4}Q^2$/GeV$^2$. 
%It so requires high luminosity for efficient exploitation and 
Thus, the spin rotators around H1 and ZEUS  were installed only for
the luminosity upgrade phase, HERA\,II.
For HERMES, depolarising effects from the 
coils of H1 and ZEUS were compensated 
during HERA\,I with special magnets near the beam axis. 
Although the compensating magnets
had to be removed when the focusing magnets were installed
for the luminosity upgrade,
HERA still achieved longitudinal $e^{\pm}$ beam
polarisations with luminosity weighted
means of typically 30\% and maximum  values of up to $40-50$\,\%.

Eventually
both H1 and ZEUS achieved total data collection efficiencies of 70-80\%
and collected large $e^{\pm}p$ data
samples corresponding to integrated luminosities of nearly 500\pb \,each, 
as is illustrated in Figure~\ref{fig:lumi}. The HERA operations ended
with efficient low proton beam
energy runs, in which   13\pb\, at $E_p=460$\,GeV
and 7\pb\, at an intermediate energy of 575\,GeV  were collected
in only three months. This would have taken about a year in the old, HERA\,I, 
configuration.

Physics at HERA would have profited from further running. 
Also one would have wished
to accelerate deuterons at HERA in order to study the structure
of the neutron in the new kinematic range,
as was proposed in 2003\,\cite{ed}. Nevertheless,
the 16 years of data taking and analysis so far have already had a great
impact on the understanding of the partonic structure of the
proton and on the development of the theory of strong interaction
dynamics. After the final publications, expected in a few years' time, the
HERA programme will have spanned nearly three decades of exciting physics
and experimentation with the participation of perhaps a thousand experts.   

\subsection{Deep Inelastic Scattering Kinematics}
\label{diskine}
The deep inelastic $ep$ scattering cross section, 
$d^2\sigma/dxdQ^2$, of the inclusive reaction $ep \rightarrow eX$  
depends on the energy $s = 4 E_e E_p$ and on two kinematic variables,
usually taken to be the four-momentum transfer squared, $Q^2$, and
Bjorken $x$. At HERA, $x$ is obtained from the measurement
of the inelasticity $y$, $Q^2$ and $s$ as $x=Q^2/(sy)$. The 
salient feature of the HERA collider experiments is the possibility
to determine the event kinematics in neutral current (NC) scattering from the
electron $e$, or from the hadronic final state $h$, defined as all final state
particles except the scattered lepton (and radiated photons associated
with the lepton), or using a combination of the
two. This leads to a maximum exploitation of the available kinematic
range and to a redundant control of the measurement which was
absent in the fixed target DIS measurements. Exploiting the NC
calibration for the measurement of the charged current (CC) processes, 
in which there
is no scattered electron, is an important means to reliably reconstruct
the inclusive scattering kinematics for the reaction $ep \rightarrow \nu X$.
The choice of the most appropriate kinematic reconstruction method
for a given phase space region is a non-unique decision based
on resolution, measurement accuracy and radiative correction effects.
H1 most frequently uses the ``electron'' and ``sigma methods'' while
ZEUS most often uses the ``PT'' and ``double angle methods'' to 
reconstruct $x$ and $Q^2$. These methods are introduced below.

The four-momentum transfer squared, $Q^2$, can be calculated from the
incoming ($k$) and outgoing ($k'$) electron four-momenta
 as follows
\begin{equation}\label{q2e}
Q_e^2 = -q^2 = -(k-k')^2 = 2 k k' = 2 E_e E'_e (1-\cos{\alpha_e}) = 4 E_eE_e' \cos^2{\theta_e/2}
\end{equation}
neglecting the electron mass and
defining $\theta_e = \pi -\alpha_e$ 
to be the angle between the scattered electron direction
and the proton beam axis \footnote{
The angles are defined between the directions of the 
outgoing particles and the proton beam. This
somewhat unfortunate choice can lead to some confusion since 
Rutherford `back-scattering' at HERA is the scattering of
the electron at high $Q^2$ at small `forward' angles.}.  The angle $\theta_e$
is measured with the tracking detectors, or by the determination of the
impact point at the face of the calorimeters and the event-wise 
reconstructed position of the interaction vertex.
Here $E_e'$ is the scattered electron energy, measured by the calorimeters.
The inelasticity is given by $k$, $k'$ and $P$, the four momentum
of the incoming proton, and similarly to $Q_e^2$, it can be calculated
using the electron kinematics as
\begin{equation} \label{ye}
y_e = \frac{Pq}{Pk}=1-\frac{Pk'}{Pk}=1 - \frac{E_e'}{2E_e} (1- \cos{\theta_e}).
\end{equation}
Often these relations are expressed as
\begin{equation}
 y_e = 1-\frac{\Sigma_e}{2 E_e} ~~~~~~~~~ 
Q^2_e = \frac{p_{t,e}^2} {1 - y_e}, 
 \label{eq:emeth}
\end{equation}
in which $\Sigma_e$ is the difference between the energy, $E_e'$,
and the longitudinal momentum, $p_{z,e}$, of the scattered
electron, and $p_{t,e}$ is its transverse momentum.

Similar relations are obtained from the hadronic final state
reconstruction~\cite{jb}
\begin{equation}
 y_h  = \frac{\Sigma_h}{2 E_e} ~~~~~~~~~ 
Q^2_h = \frac{p_{t,h}^2} {1 - y_h},
 \label{yjb}
\end{equation}
where
\begin{equation} \label{e-pz}
\Sigma_h = \sum_i{(E_i-p_{z,i})}
\end{equation}
is the hadronic $E-p_z$ variable and the sum, as for the
transverse momentum, extends over the calorimeter cells
(and tracks in some cases)
of the hadronic final state. In the reconstruction of the
energy, the momentum measurements
from the tracking detectors are also used
and the corresponding energy deposits are removed from the sum.
The hadronic energy is  given by the relation
\begin{equation}
E_h=\frac{p_{t,h}}{\sin{\theta_h}}.
\end{equation} \label{Eh}
The angle of the hadronic system, $\theta_h$, can  be obtained
from measurements of the energy and momentum components
with the calorimeters as
\begin{equation} \label{thetanh}
\cos{\theta_h} = \frac{p_{t,h}^2 -\Sigma_h^2}{p_{t,h}^2 + \Sigma_h^2}
%\end{equation} \label{thetah}
%
~~~~or~~~~ 
%\begin{equation}
\tan{\frac{\theta_h}{2}} = \frac{\Sigma_h}{p_{t,h}}.
\end{equation} 
In the naive quark parton model (QPM), $\theta_h$ represents the angle
of the struck quark.

A straightforward calculation using Equations~\ref{q2e},\,\ref{ye} 
and $x_e = Q^2_e/sy_e$ relates the uncertainties on
the electron energy and angle to the uncertainties  
on $Q^2$ and  $x$  as follows
\begin{eqnarray} \label{qxele}                                                     
 \delta Q_e^2/Q_e^2 &=& \delta E_e'/E_e' +  \tan{(\theta_e /2)} \cdot
  \delta\theta_e \nonumber \\                                                                           
 \delta x_e/x_e &=& 1/y_e \cdot \delta E_e'/E_e' \nonumber\\                           
            & & + [(1-y_e)/y_e \cdot \cot{(\theta_e /2)}                                
 + \tan{(\theta_e /2)}] \cdot \delta\theta_e .                                    
\end{eqnarray}                                                                  
The uncertainty of Bjorken $x_e$ thus appears to
diverge at small $y$, in the presence of a possible miscalibration
of the electron energy measurement or a misalignment affecting
the polar angle measurement. On the contrary, the uncertainty of
$Q^2$ has no divergent behaviour, and indeed the electron method
allows $Q^2$ to be reconstructed optimally over the full kinematic range.
There is also a term, $\propto \tan{(\theta_e/2)} \cdot \delta \theta_e$,
which becomes sizable at very large angles, when the electron
is scattered into the backward region, at small $Q^2$. This was
an important motivation for building  dedicated Backward 
Detectors, BST (H1) and BPT (ZEUS), as supplements to the
original apparatus, to extend the acceptance and   
to reduce the influence of the resulting cross section
uncertainty to a tolerable level by measuring $\theta_e$
extremely precisely at $\theta_e$ approaching $180^{\circ}$
(see Sections 2.4 and 3.4.2).

Relations similar to Equation~\ref{qxele} can be obtained 
for the  $Q^2$ and $x$ uncertainties in the hadronic final state
reconstruction:                                                                             
\begin{eqnarray} \label{qxjb}  
 \delta Q^2_h/Q^2_h &=& (2-y_h)/(1-y_h) \cdot \delta E_h/E_h   \nonumber\\             
& & +[ 2 \cot{\theta_h} + y_h/(1-y_h) \cdot \cot{(\theta_h /2)}] 
\cdot \delta\theta_h   \nonumber\\                                                                 
 \delta x_h/x_h &=& 1/(1-y_h) \cdot \delta E_h/E_h \nonumber\\                       
            & & + [-2 \cot{\theta_h}                                
            + (1-2y_h)/(1-y_h) \cdot \cot{(\theta_h/2)}]                          
            \cdot \delta\theta_h .                                         
\end{eqnarray}                                                                  
Unlike the case of electron-based reconstruction,
the uncertainties, both for $Q^2$ and $x$, become large when
$y$ approaches 1. Thus the hadron reconstruction method
and the electron method are complementary.
It was proposed early in\,\cite{bkf} to
combine the variables $Q^2_e$ and
$y_h$ in the ``mixed method'' reconstruction of the event kinematics in order
to avoid the divergent feature of $x_e$.                                                                              
                                                                               
%\begin{equation} \label{KINTRANS}                                               
%\begin{array}{ccc}                                                              
%\hat{x}_e   = \frac{x}{1-\epsilon_e/(1+\epsilon_e)y} & &                       
% \hat{x}_j   = x\: \frac{(1+\epsilon_j)(1-y)}{1-y\: (1+\epsilon_j)} & \\        
% \hat{Q}^2_e = Q^2(1+\epsilon_e)                      & &                       
% \hat{Q}^2_j = Q^2 \frac{(1+\epsilon_j)^2(1-y)}                                 
%                          {1-y\: (1+\epsilon_j)} & \\                           
% \hat{y}_e   = y - \epsilon_e(1-y)                    & &                       
% \hat{y}_j   =  y(1+\epsilon_j)                                                 
%\end{array}                                                                     
%\end{equation}                                                                  
%                                                                               
%
\begin{figure}[h]
\begin{center}
%\epsfile{file=isoele.eps,scale=0.8}
\includegraphics[height=8cm]{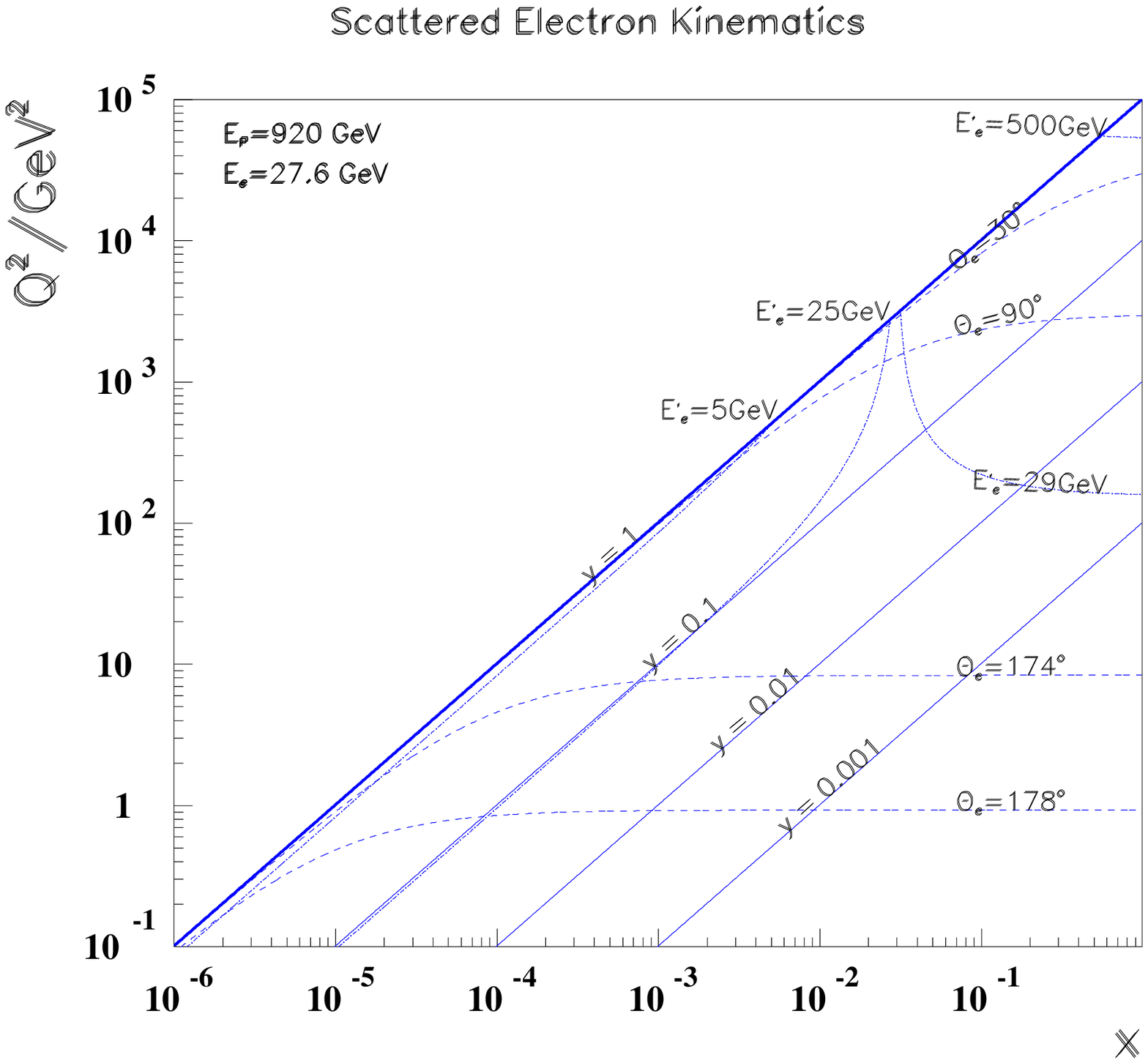}
\includegraphics[height=8cm]{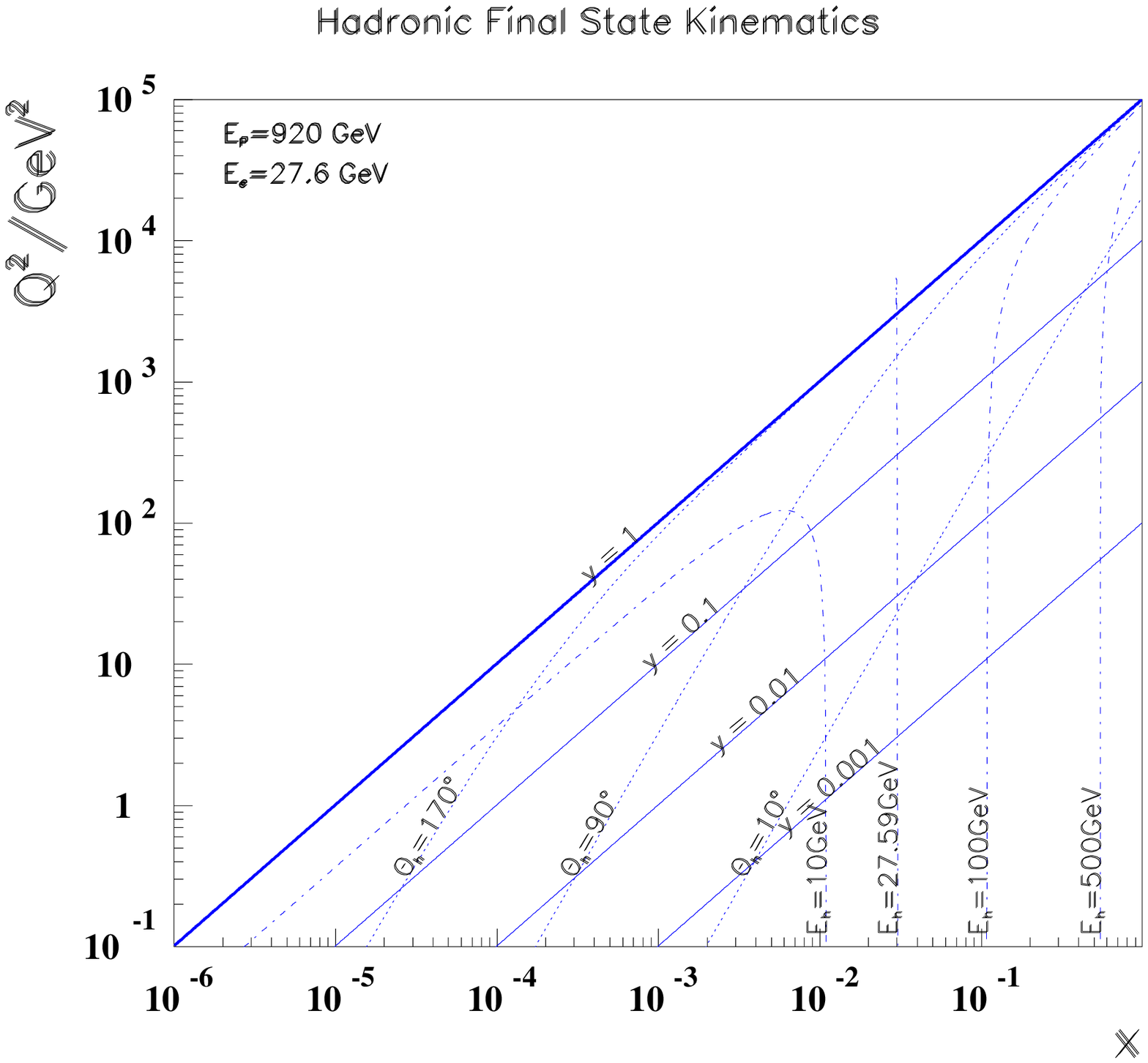}
\begin{minipage}{16.5 cm} 
\caption{Electron scattering kinematics (left) and kinematics of the
hadronic final state (right).  Dashed: lines of constant scattering angle,
$\theta_e$ (left) and $\theta_h$ (right), in the $Q^2,x$ plane. Dashed
dotted: lines of constant energy, $E_e'$ (left) and $E_h$ (right). 
For the electron energy a large %$Q^2, x$ 
region (left), in which $E_e' \simeq E_e$ is observed, because of the
relation $E_e' = Q^2/4E_e + E_e (1-y) \simeq E_e$, for
small $y$ and $Q^2$. The $E_e'$ distribution
thus exhibits a characteristic ``kinematic  peak'' at $E_e$ which
is often used to cross check or calibrate the scattered electron energy
scale.  One also recognises in the left plot that in forward
direction, at very high $Q^2$, the scattered electron carries very high
energy of a few hundred GeV.
The energy of the hadronic final state, $E_h$, varies rapidly with  Bjorken $x$
at intermediate $x$ values.
At high $x$, the hadronic final state becomes 
very energetic and much of the energy is emitted in forward direction.
At $x=E_e/E_p = 0.03$ the iso-$E_h$ line is independent of $Q^2$
and $E_h=E_e$.}
\label{kineplanes}
\end{minipage}
\end{center}
\end{figure} 
  
A further refinement in the kinematics reconstruction
was introduced  with the ``sigma method''\,\cite{sigma}. 
Various versions of this method exist; the basic idea relies on the
exploitation of the total $E-p_z$ variable                                                            
\begin{equation} \label{eq:sigma}
 \empz = E'_e (1-\cos{\theta_e}) + 
 \sum_i \left(E_i - p_{z,i}\right) =  \Sigma_e + \Sigma_h,
\end{equation}
which for non-radiative events equals $2E_e$. Thus $y_h$,
Equation~\ref{yjb}, can be redefined and one obtains
\begin{equation}\label{eq:yh}
 y_{\Sigma}  = \frac{\Sigma_h} {E-p_z} ~~~~~~~~~~~~~ Q^2_{\Sigma}=\frac{p^2_{t,e}}{1-y_{\Sigma}}.
\end{equation} 
An important advantage of this construction is its reduced sensitivity
to miscalibrations by using the final state energies and momenta
instead of the incoming electron beam energy. The energy-momentum
conservation requirement $E-p_z \simeq 2E_e$ has often been used to 
remove radiative events from the analysis. 

A further method, the ``double angle method'',
 developed independently in \cite{standa} and 
\cite{hoegerda}, uses the possibility to express $Q^2$ and $x$
by the electron and hadronic scattering angles which again have reduced
sensitivity to energy miscalibrations
\begin{equation}\label{qxda}
y_{DA} = \frac{\tan{(\theta_h/2)}}{\tan{(\theta_e/2)} + \tan{(\theta_h/2)}}
~~~~~~~~~Q^2_{DA}= 4 E_e^2 \cdot  
\frac{\cot{(\theta_e/2)}}{\tan{(\theta_e/2)} + \tan{(\theta_h/2)}}.
\end{equation}
An application of this idea is the double angle energy calibration method,
in which the electron energy scale is determined from the measurements
of $\theta_e$ and of $\theta_h$, for regions of phase space in which
both the electron and the hadrons are well contained in the apparatus.

H1 developed various versions of kinematic reconstruction methods.
Diffractive analyses (see Section~7) 
often used a specific definition of the inelasticity as
\begin{equation}
y = y_e^2 + y_{DA} (1-y_{DA}).
\end{equation}
A further idea has been to replace the electron beam energy 
in the calculation of $x$, similarly
as it was replaced in $y_{\Sigma}$.
This defines an extension of the $\Sigma$ method which takes
radiative effects at the lepton vertex completely into account giving
\begin{equation}
x_{\Sigma} = \frac{Q^2_{\Sigma}} {2 E_p (E-p_z) y_{\Sigma}} = \frac{Q_{\Sigma}^2} {2 E_p \Sigma_h}.
\end{equation}

The hadronic final state, due to the particles lost in
the beampipe as well as energy lost in the inactive material in the detector, 
is, in general, less well-determined than the scattered electron.
In the ``PT method'' of reconstruction 
\cite{pt_method} used by the ZEUS collaboration, the fact that 
owing to the well-measured 
electron, $\delta_{PT}=P_{T,h}/P_{T,e}$ gives a good
event-by-event estimate of the hadronic energy loss is used to improve
both the resolution and uncertainties of the reconstructed $y$ and $Q^2$.
The PT method uses all measured variable, in a way specific
to the ZEUS detector, to optimise the resolution over the entire 
kinematic range measured, namely,
\begin{equation}
 \tan{\frac{\theta_{PT}}{2}} = \frac{\Sigma_{PT}}{P_{T,e}},\;\;
 \Sigma_{PT} = 2E_e\frac{{C(\theta_h,P_{T,h},\delta_{PT})}\cdot\Sigma_h}
                       {\Sigma_e+{C(\theta_h,P_{T,h},\delta_{PT})}\cdot\Sigma_h}.
\end{equation}
The variable $\theta_{PT}$ is then substituted for $\theta_h$ in the formulae
for the double angle method to determine $y$ and $Q^2$. The 
detector-specific function
$C$ is $\Sigma_{true,h}/\Sigma_{h}$ as a function of $\theta_h$, $P_{T,h}$ and
$\delta_{PT}$ and is determined from Monte Carlo simulations.

The double differential cross section has to be 
independent of how $Q^2$ and $x$ are reconstructed.
Many cross-checks of the results employ the underlying differences in the
reconstruction algorithms and detector inputs.
The measurement redundancies can also be expoited, as demonstrated 
in the case of the recent study of methods to average the H1 and ZEUS
measurements\,\cite{sasha}, to reduce the final measurement
uncertainty to, eventually, about 1\,\%.

% \cite{h1zeus}.

%
\subsection{Detectors}
The H1 \cite{h1det} and ZEUS \cite{zeusdet} detectors are nearly
$4\pi$ hermetic apparatus with a solenoidal field of
1.2\,T and 1.43\,T, respectively, built to investigate high energy
$ep$ interactions at HERA. Their main design features are determined by
the need to measure energies and momenta up to a few hundred GeV,
as is shown in the kinematic plane plots (Figure~\ref{kineplanes}).

\begin{figure}[htbp]
\begin{center}
\includegraphics[height=11.2cm]{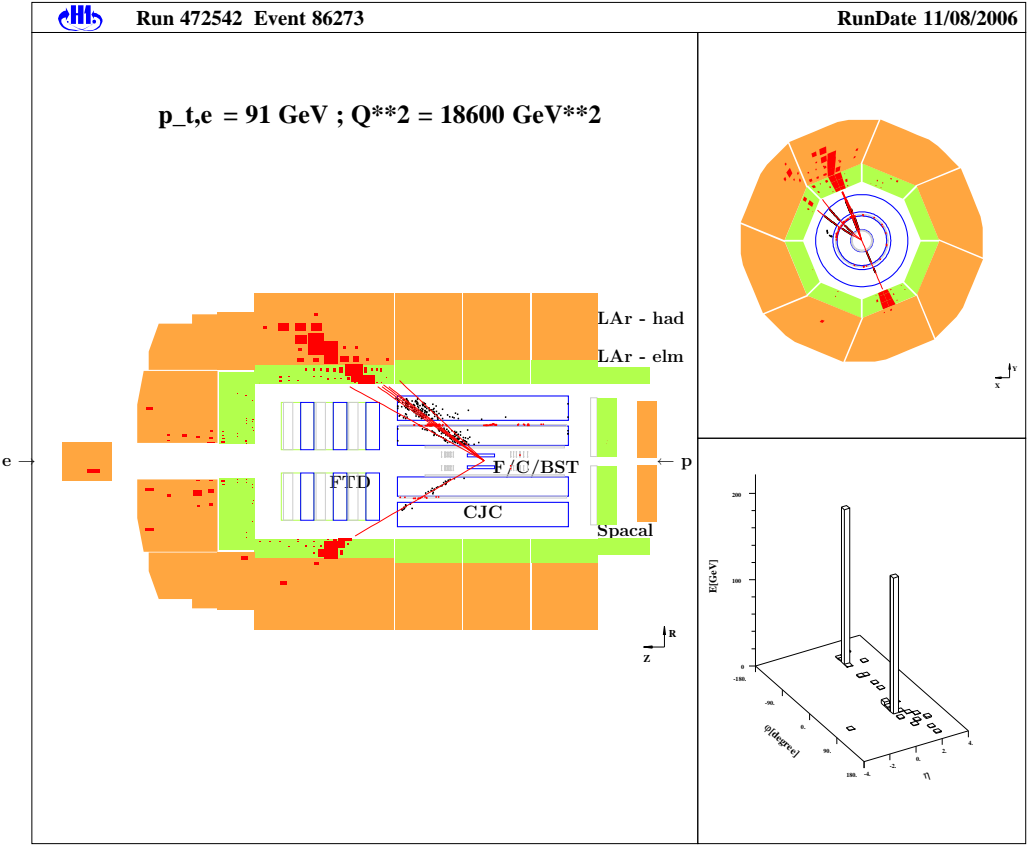}
%\epsfile{file=h1detev.eps,scale=0.8}
\begin{minipage}{16.5 cm} 
\caption{ Deep inelastic event from neutral current  $ep$ scattering
as registered in the H1 apparatus in its upgraded, HERA II, configuration.
The electron beam enters from the left and the proton beam from the right. 
On the left, an $r z$ cross section shows the main
components of the H1 detector as denoted, see text.  The
event has a transverse momentum of the scattered electron of 91\,GeV
and a $Q^2$ of 18600\,GeV$^2$ which correspond to $y=0.55$
and $x=0.33$. Thus the electron scattering angle 
is about $\theta_e = 30^{\circ}$
and $E_e' = 180$\,GeV, i.e. at high $Q^2$ and large $x$ the
incoming electron is scattered in the forward direction
and carries a rather large energy (cf. Figure\,\ref{kineplanes}). 
The right top figure shows the event
in the $x,y$ projection exhibiting transverse momentum balance
between the scattered electron and the hadronic final state as is
characteristic for NC events. The lego plot
visualizes the energy deposition in the LAr cells exhibiting the
 narrow jet structure of the hadronic final state 
emerging from the struck quark.}
\label{h1event}
\end{minipage}
\end{center}
\end{figure}

\begin{figure}[htbp]
\begin{center}
\includegraphics[height=9.2cm]{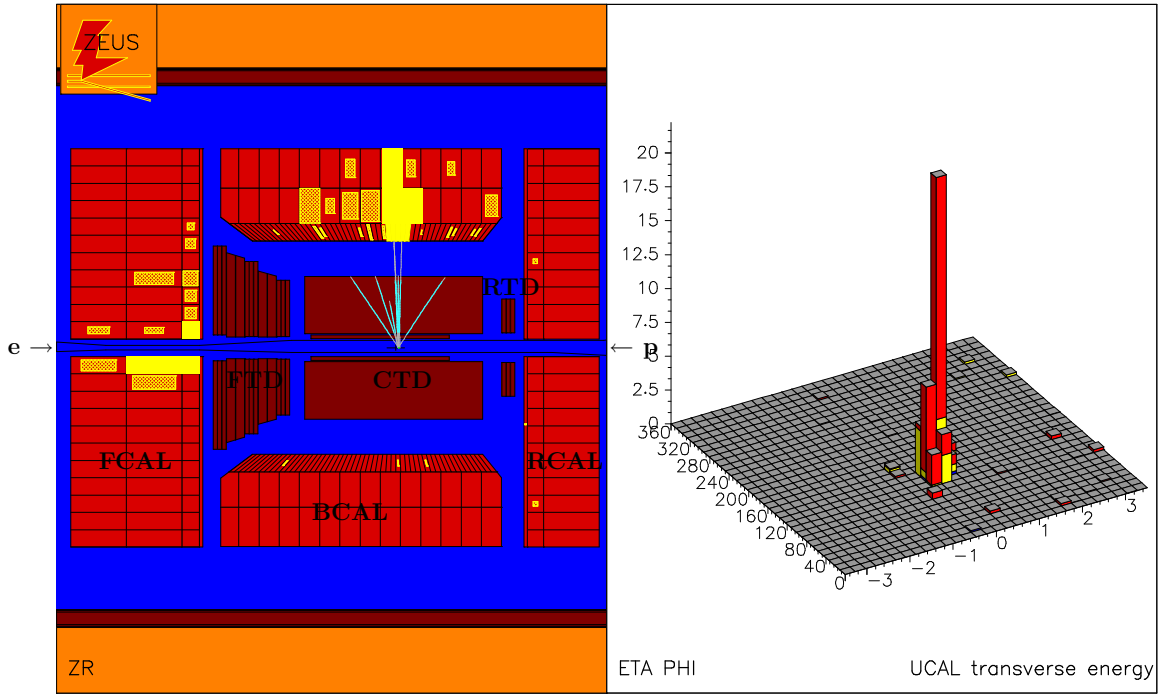}
%\epsfile{file=,scale=0.8}
\begin{minipage}{16.5 cm} 
\caption{Deep inelastic event from $ep$ charged current scattering
as registered in the ZEUS apparatus, in its HERA\,I configuration.
The electron beam enters from the left and the proton beam from the right.
On the left, an $r z$ cross section shows the main
components of the ZEUS detector as denoted, see text. The right figure
shows the measured transverse energy in the calorimeter.
The event is a typical charged current scattering event with
an energetic jet, unbalanced in transverse momentum,
and some energy produced from the proton remnant in
forward direction. For this event one finds  
$\theta_h \simeq 90^{\circ}$ and $p_{t,h} \simeq 33$\,GeV.
From Equations~\ref{yjb}-\ref{thetanh} this corresponds
to  $Q^2 \simeq 2700$\,GeV$^2$,
$y \simeq 0.6$ and thus $x \simeq 0.05$, since at this time, 
HERA was operated with $E_p=820$\,GeV. 
 }
\label{zeusevent}
\end{minipage}
\end{center}
\end{figure}

Due to the large difference between $E_p$ and $E_e$, the HERA apparatus,
unlike similar devices at LEP or the Tevatron, are asymmetric. Large energy
deposits, up to energies close to $E_p$,
occur in the proton beam (forward) direction, and the
particle multiplicity, due to the hard interaction and the proton remnant,
is often very large. On the other hand, the hemisphere in the
electron (backward) direction is less occupied,
the energies deposited are below or comparable to $E_e$
and the multiplicities are low. Such considerations have determined the final
layouts of the H1 and the ZEUS detectors, as illustrated with the
event displays showing an NC event in H1, Figure~\ref{h1event}, and
a CC event in ZEUS, Figure~\ref{zeusevent}.  

The main component of the H1 detector is  the Liquid Argon Calorimeter (LAr).
It has an inner electromagnetic part, which in the event of
Figure~\ref{h1event}  is seen to
fully contain the scattered electron energy, and an outer hadronic part,
which is seen to contain the fraction of jet energy 
leaking out of the electromagnetic LAr.  The LAr is surrounded by
a superconducting coil producing a solenoidal magnetic field of
1.2 T and an instrumented iron structure acting as a shower
tail catcher and a muon detector. Tracks are measured in the
central tracking detector which contains three silicon detectors,
the forward, central and backward trackers, FST, CST and BST. These are
followed towards outer radii by the Central Inner Proportional Chamber,
CIP, and the Central Jet Chamber, CJC. The CJC has two concentric parts,
separated by the Central Outer-$z$ Chamber, COZ. Tracks in forward
direction are measured in the Forward Tracking Detector, FTD, a set of
planar drift chamber modules of different azimuthal orientation.

The main component of the ZEUS detector is
the Uranium-Scintillator calorimeter. Shown in Figure~\ref{zeusevent} 
are the forward, barrel and rear components (F/B/RCAL) which  
are surrounded by a muon detector.
The calorimeter is segmented into inner electromagnetic parts,
of $5 \times 20$\,cm$^2$ tower sizes in the barrel, followed by hadronic
parts of coarser granularity, $20 \times 20$\,cm$^2$.
The space visible inside the BCAL is occupied by the coil giving a field
of 1.43\,T and a preshower detector.
The tracking detector consists of a central cylindrical jet type drift
chamber (CTD) and sets of planar drift chambers 
in  forward (FTD) and rear (RTD) directions. During the upgrade phase,
for HERA\,II, the FTD was modified by introducing straw tubes and
inside the CTD a micro vertex detector (MVD) was added, comprising 
forward and central silicon strip detectors.

Besides the main parts of the H1 and ZEUS detectors,
there were two instrumented areas further away from the
interaction region, both in H1 and ZEUS:
\begin{itemize}
\item 
In the electron beam (backward) direction, several `electron taggers'
were placed, at 8\,m, 33\,m and 103\,m for H1 and at 6\,m, 8\,m,
35\,m and
44\,m for ZEUS, in order to measure the scattered electron energy
in a small range of $\theta_e$ close to 180$^{\circ}$. Some of these
taggers were installed for only a part of
the data taking period.  The electron 
tagger system enabled, in different acceptance 
regions of $y_e$ corresponding to the different positions,
inelastic $ep$ scattering processes 
to be tagged at very small $Q^2$.
The taggers thus allowed an efficient control and analyses
of photoproduction processes. A process of particular interest is
Bethe-Heitler scattering, $ep \rightarrow ep \gamma$, whose cross-section is known
theoretically to about 0.5\% accuracy and is thus suitable for the
determination of the interaction luminosity. Therefore both H1 and ZEUS 
were also equipped with photon taggers
positioned at $\simeq 100$\,m down the $e$ beam line, 
at 180 degrees. The measurement accuracy of the luminosity
thus far is about 1.5\% for both experiments.
\item In the proton beam
(forward) direction, protons and neutrons may be scattered at a small angle,
via soft processes such as vacuum or pion exchange, respectively.  
Leading protons are a characteristic 
signature for diffractive processes such as hard
diffraction or elastic vector meson production.
Both H1 and ZEUS installed so-called forward or leading-proton
spectrometers for HERA I, FPS and LPS, at about 100\,m
distance from the interaction point. During the upgrade,
H1 kept its FPS and added another, very forward proton spectrometer,
VFPS, positioned at 220\,m away. ZEUS decided to remove the
LPS during the upgrade phase. Similarly, H1 operated a forward
neutron calorimeter (FNC) throughout the HERA lifetime while ZEUS
confined the FNC operation to the first phase of HERA. 
\end{itemize}
The detectors are complemented by systems of veto counters
to reject beam background with time-of-flight measurements.
In the forward direction, H1 had a muon detector system comprising
a toroid magnet inside a system of drift chambers. Similarly 
ZEUS had a toroid magnet with drift tubes for muon track
reconstruction and limited-streamer tubes for triggering.

The H1 and ZEUS detectors functioned very reliably although 
intense and challenging efforts in the operation, calibration
and maintenance were required. The main components, the central drift chambers,
the big calorimeters and the large area muon detectors ran 
over 15 years with high efficiency and without a need for replacements of the
basic hardware. Parts of the readout and trigger electronics
were replaced or updated. In 1997 the H1 central chamber had to be rewired
when ageing effects were observed. Despite much higher backgrounds
in HERA II, no ageing effects were seen in either of the 
H1 and ZEUS tracking chambers, apart from a small reduction of pulse-heights 
with time. The purity of the argon liquid in H1 was monitored and was found 
to worsen at a very small, tolerable rate of 0.05\,ppm per year.
The argon was exchanged only once, in 1995.
Similarly, the scintillator material in the ZEUS Uranium calorimeter
survived all the operation well.

The installation of focusing magnets near the beam pipe
in the course of the luminosity upgrade, in 2001, required modifications
of the H1 and ZEUS interaction regions. It also provided the
only significant time period during which large detector upgrades,
as mentioned above, could be
made. The H1 detector upgrade program involved a large number
of components; similarly ZEUS installed new hardware
at various places of their apparatus.
The original detector configurations are described in \cite{h1det} 
and \cite{zeusdet}. 
%Characteristics of the components are summarised in 
%Table\,\ref{tabdeth1z}. The upgraded and new components are
%listed in Table\,\ref{tabupgrade}.
%%
%\begin{table}[htdp] \label{tabdeth1z}
%\begin{center}
%\begin{tabular}{|l||l|}
%\hline
%H1 & ZEUS \\   
%\hline
%\hline
%LAr  & U-Sc \\
%\hline
%CJC $\delta p/p = 0.001 p $ & CTD  $\delta p/p = 0.002 p $\\
%~~~~~~$ 7 - 173^{\circ}$ & ~~~~~~~ $ 7 - 173^{\circ}$ \\
%\hline
%\end{tabular}
%\caption{Detector configuration of H1 (left) and ZEUS (right)
%in HERA I. EXPAND OR DROP?}
%\end{center}
%\end{table}
%%
%%
%\begin{table}[htdp]
%\begin{center}
%\begin{tabular}{|l||l|}
%\hline
%H1 & ZEUS \\ 
%\hline
%\hline
%FST BST  & MVD forward and central\\
%\hline
%FTD  planar drift chambers & FTD straw tubes \\
%\hline
%Lumi & Lumi \\
%\hline
%VFPS & -- \\
%\hline
%\end{tabular}
%\caption{Detector upgrade components of H1 (left) and ZEUS (right)
%in HERA II}
%\end{center}
%\label{tabupgrade}
%\end{table}
%

%\section{Proton Structure Functions and QCD Fits}
%\input{hch3}
\section{Proton Structure Functions}
\subsection{Introduction}
The scattering amplitude for electron-proton scattering is a product
of lepton and hadron currents times the propagator characteristic
of the exchanged particle, a photon or $Z_0$ in neutral current scattering,
a $W^{\pm}$ in charged current scattering. The inclusive scattering
cross section therefore is given by the product of two tensors,
\begin{equation}
\frac{d^2\sigma}{dxdQ^2} = \frac{2 \pi \alpha^2}{Q^4 x} 
\sum_j{\eta_j L_j^{\mu \nu} W_j^{\mu \nu}},
\label{siglw}
\end{equation} 
where $j$ denotes the summation over $\gamma$, $Z_0$ exchange
and their interference for NC, and $j=W^+$ or $W^-$ for CC. 
The leptonic tensor $L_j^{\mu \nu}$ is related to the coupling 
of the electron with the
exchanged boson and contains the electromagnetic or the weak couplings,
such as the vector and axial-vector electron-$Z_0$ couplings, $v_e$ and $a_e$,
in the NC case.  This leptonic part of the cross
section can be calculated exactly in the standard electroweak $U_1 \times SU_2$
theory.  The hadronic tensor, however, describing the interaction of
the exchanged boson with the proton, can only be reduced to a sum of
structure functions, $F_i(x,Q^2)$, but not be fully calculated.
Conservation laws reduce the number of basic structure functions
in unpolarised $ep$ scattering to $i=1-3$.
In perturbative QCD the structure functions are related to 
parton distributions $f$ $via$ coefficient functions $C$
\begin{equation} \label{factoreq}
 [F_{1,3},F_2] = \sum_i{\int_0^1{[1,z]\frac{dz}{z} C_{1,2,3}(\frac{x}{z},\frac{Q^2}{\mu_r^2},
 \frac{\mu_f^2}{\mu_r^2},\alpha_s(\mu_r^2)) \cdot f_{i}(z,\mu_f^2,\mu_r^2)}},
\end{equation} 
where $i$ sums the quark $q$, anti-quark $\overline{q}$ and gluon $g$ contributions
and $f_i(x)$ is the probability distribution
of the parton of type $i$ to carry a fraction $x$
of the proton's longitudinal momentum. 
The coefficient functions are  exactly calculable
but depend on the factorisation and renormalisation scales $\mu_f$ and 
$\mu_r$. The parton distributions are not calculable but have to be
determined by experiment. Their $Q^2$ dependence obeys evolution
equations (see Section 4).  A general factorisation
theorem\,\cite{factor}, however, has proven the parton distributions 
to be universal, i.e. to be independent of 
the type of hard scattering process. This makes 
deep inelastic lepton-nucleon scattering a
most fundamental process: the parton distributions in the proton
are measured best with a lepton probe and may be used
to predict hard scattering cross sections at, for example, the LHC.
The parton distributions  are derived 
from measurements of the structure functions in NC and CC scattering,
as is discussed below.

The following sections describe the structure function formalism,
and present important NC and CC measurements.
Both H1 and ZEUS have collected data samples of roughly $50-100$\,pb$^{-1}$ 
of unpolarised
$e^+ p$ data and of similar luminosity in the two charge and opposite
helicity states (see Table\,\ref{tabpoldata}).
\begin{table}[htdp]
\begin{center}
\begin{minipage}{16.5 cm} 
\caption{Integrated luminosity (L), in pb$^{-1}$, collected by H1 and ZEUS,
for specific data selections, distinguished according to 
the beam charge and helicity (P) state. The unpolarised data
were taken in HERA I. The H1-ZEUS differences in polarisation
and luminosity are due to somewhat different analysis criteria,
data taking efficiencies and an asymmetry in rotator switches
between H1 and ZEUS during the $e^-p$ run,
in the spring of 2006. While this review is being written a significant
part of the polarised data, taken in the years 2004-2007,
is still being analysed.
\vspace{0.2cm}
}
\label{tabpoldata}
\end{minipage}
\begin{tabular}{|l||l|l||l|l|l|l||l|}
\hline
exp & $e^+$ (0) & $e^-$ (0) & $e^+$ (P) & $e^+$ (-P) & $e^-$ (P) & $e^-$ (-P) & sum\\ 
\hline
H1 (P)  &  - & -     & 33 & 37 & 36 & 26 & -- \\
\hline
H1 (L)  &  100 & 15& 95 & 75 & 46 & 103 & 432 \\
\hline
ZEUS (P)  &  - & -      & 32 & 37 & 30 & 27 & -- \\
\hline
ZEUS (L)  &  100 & 15&  98 & 76&  81 & 105 & 475\\
\hline
\end{tabular}
\end{center}
\end{table}

\subsection{Structure Functions and Parton Distributions}
%HERA, the highest energy DIS laboratory, has covered a
%wide range in $x$ and $Q^2$ unaccessible hitherto.
%With accurate measurements of
%NC and CC processes using electron and positron beams, a 
%unique set of cross sections could be measured which allows a rather
%complete set of parton distributions to be derived.
%
The neutral  current deep inelastic ep scattering cross section, at tree level,
is given by a sum of generalised structure functions
according to 
\begin{eqnarray} \label{ncsi}     
 \sigma_{r,NC} =\frac{d^2\sigma_{NC}}{dxdQ^2} \cdot \frac{Q^4 x}{2\pi \alpha^2 Y_+}                                                     
% \sigma_{r,NC} = \sigma_{r,0} + \sigma_{r,\gamma Z} + \sigma_{r,ZZ} &=&                   
  =            {\bf F_2} + \frac{Y_-}{Y_+} {\bf xF_3} -\frac{y^2}{Y_-} {\bf F_L},
% \sigma_{r,CC} = \sigma_{r,WW} &=&                                                  
%              {\bf W_2} + \frac{Y_-}{Y_+} {\bf xW_3}
\end{eqnarray}                                                                  
where the electromagnetic coupling, the photon              
propagator and a helicity factor are absorbed 
in the definition of a reduced cross section $\sigma_r$, and $Y_{\pm}=1 \pm (1-y)^2$.                                                                              
The functions ${\bf F_2}$ and  ${\bf xF_3}$ 
depend on the lepton beam charge and                  
polarisation ($P$) and on the electroweak parameters,                     
e.g. on the weak boson masses  ($M_Z,~M_W$) and
the fine-structure constant  $\alpha$, as~\cite{klri};
\begin{eqnarray} \label{strf}                                                   
 {\bf F_2^ \pm} &=& F_2 + \kappa_Z(-v_e \mp P a_e) \cdot F_2^{\gamma Z} +                      
  \kappa_Z^2 (v_e^2 + a_e^2 \pm 2 P v_e a_e) \cdot F_2^Z \nonumber \\                     
 {\bf xF_3^ \pm} &=&  \kappa_Z( \pm a_e + P v_e) \cdot xF_3^{\gamma Z} +                       
  \kappa_Z^2( \mp 2 v_e a_e - P (v_e^2+a_e^2)) \cdot xF_3^Z,                                   
\end{eqnarray} 
  with 
  $\kappa_Z(Q^2) =   Q^2 /[(Q^2+M_Z^2)(4\sin^2 \Theta \cos^2 \Theta)]$ 
and the weak mixing angle     $\sin^2  \Theta=1 -M^2_W /M^2_Z$.
In the hadronic tensor  decomposition~\cite{derman} the structure
functions  are well defined                          
quantities, independently to the parton model. In the Quark Parton Model (QPM)  
the longitudinal structure function is zero\,\cite{cgross} and the 
two other functions are given by the sums and differences of
quark ($q$) and anti-quark ($\overline{q}$) distributions as
\begin{eqnarray} \label{ncfu}                                                   
  (F_2, F_2^{\gamma Z}, F_2^Z) &=&x \sum (e_q^2, 2e_qv_q, v_q^2+a_q^2)(q+\bar{q})            
                                 \nonumber \\                                   
  (xF_3^{\gamma Z}, xF_3^Z) &=& 2x \sum (e_qa_q, v_qa_q) (q-\bar{q}),                         
\end{eqnarray} 
where the sum extends over all up and down type quarks
and $e_q =e_u,e_d$ denotes the electric charge of up- or
down-type quarks. The vector and axial-vector weak couplings of the fermions
($f=e,u,d$) to the $Z_0$ boson in the standard electroweak model
are given by
\begin{equation} \label{va}
 v_f= i_f - e_f 2 \sin^2 \Theta ~~~~~ a_f=i_f
 \end{equation}
 where   $e_f = -1,2/3,-1/3$ 
 and $i_f=I(f)_{3,L}=-1/2,1/2,-1/2$ denotes the
 left-handed weak isospin charges, respectively. Thus the vector coupling of the
 electron, for example, is very small, $v_e=-1/2 + 2 \sin^2 \Theta \simeq 0$,
 since the weak mixing angle is roughly equal to 1/4. In a famous
 experiment~\cite{prescott}
  in polarised electron-nucleon scattering at very low $Q^2$,
 it was observed that $a_e \simeq -1/2$ which proved the absence of
 right-handed weak currents in the available energy range, and thus the
validity of the most simple weak doublet structure, now called
the standard electroweak theory~\cite{GWS}.

Out of the eight structure functions entering in Equation~\ref{ncsi} four have been
accessed at HERA.  At low $Q^2$, the cross section is determined 
by $F_2$ and $F_L$. The longitudinal structure function
$F_L$ is measured at low $x$. The large gluon density in this region leads to a rather
large value of $F_L$ compared to those from the fixed target measurements
which were measured at a much larger $x$.

At low $Q^2$ and low $y$ the reduced cross section 
to a very good approximation is given by $\sigma_r = F_2(x,Q^2)$. Indeed
the accurate knowledge of the proton structure, at low $x$
from HERA and at large $x$ from fixed target experiments,
is mostly due to precision measurements of $F_2$. At $y > 0.5$, $F_L$ makes a sizeable
contribution to $\sigma_{r,NC}$. In the DGLAP
approximation of perturbative QCD, to
lowest order, the longitudinal structure function is given by \cite{am}
\begin{equation}
        F_L(x) = \frac{\alpha_s}{4 \pi} x^2
        \int_x^1 \frac{dz}{z^3} \cdot \left[ \frac{16}{3}F_2(z) + 8
        \sum e_q^2 \left(1-\frac{x}{z} \right) zg(z)
        \right],
\label{altmar}
\end{equation}
which at low $x$ is dominated by the gluon contribution (see below). 
A measurement of $F_L$ requires a variation of the beam energy
%in order to remove the dominant $F_2$ contribution to 
since
$\sigma_{r,NC}(x,Q^2;s) = F_2(x,Q^2) -y^2 F_L(x,Q^2)/Y_+$;
$F_L$ is obtained as the slope in $y^2/Y_+$  at  fixed $x$ and $Q^2$  
($y=Q^2/sx$).
Two further structure functions can be accessed with 
cross section asymmetry measurements
in which the charge and/or the polarisation of the lepton beam are varied.

Defining a 
reduced cross section, similar to $\sigma_{r,NC}$, for 
the inclusive polarised charged current $e^{\pm} p$ 
scattering  as
\begin{equation}
 \label{Rnc}
 \sigma_{r,CC} =  
  \frac{2 \pi  x}{Y_+ G_F^2}
 \left[ \frac {M_W^2+Q^2} {M_W^2} \right]^2
          \frac{{\rm d}^2 \sigma_{CC}}{{\rm d}x{\rm d}Q^2},
\end{equation}
one obtains a sum of structure functions, analogous to Equation~\ref{ncsi}, as
\begin{eqnarray}
 \label{ccsi}
 \sigma_{r,CC}^{\pm}=
 \frac{1\pm P}{2}( W_2^\pm   \mp \frac{Y_-}{Y_+} xW_3^\pm - \frac{y^2}{Y_+} W_L^\pm)\,.
\end{eqnarray}
Here the Fermi constant 
$G_F$ is defined~\cite{hector} using the weak boson
masses, in very good agreement with $G_F$ determined
from the measurement of the muon lifetime~\cite{pdg}.  
In the QPM (where $W_L^\pm = 0$),
the structure functions represent
%lepton beam-charge dependent 
sums and differences,
that depend on the charge of the lepton beam,
of quark and anti-quark distributions given by
\begin{eqnarray}
 \label{ccstf}
    W_2^{+}  =  x (\bU+D)\hspace{0.05cm}\mbox{,}\hspace{0.1cm}
  xW_3^{+}  =  x (D-\bU)\hspace{0.05cm}\mbox{,}\hspace{0.1cm} 
    W_2^{-}  =  x (U+\bD)\hspace{0.05cm}\mbox{,}\hspace{0.1cm}
 xW_3^{-}  =  x (U-\bD)\,.
\end{eqnarray}

Using these equations one finds
\begin{eqnarray}
\label{ccupdo}
 \sigma_{r,CC}^+ \sim x\bU+ (1-y)^2xD, ~~\\
 \sigma_{r,CC}^- \sim xU +(1-y)^2 x\bD . ~~
\end{eqnarray}
Combined with Equation~\ref{strf}, which approximately reduces to
\begin{eqnarray} \label{f23ud}
\sigma^{\pm}_{r,NC} \simeq [c_u (U+\bU) +c_d(D+\bD)] + \kappa_Z [d_u(U-\bU) + d_d (D-\bD)] \nonumber \\
%F_2 = x (e_u^2 (U+\bU) + e_d^2 (D + \bD)] \hspace{0.05cm}\mbox{,}\hspace{0.1cm}
%xG_3 = x [e_u a_u (U -\bU) + e_d a_d (D -\bD)]
\mbox{with} \hspace{0.2cm} c_{u,d} = e^2_{u,d} + \kappa_Z (-v_e \mp P a_e) e_{u,d}v_{u,d} 
\hspace{0.2cm} \mbox{and} \hspace{0.2cm} d_{u,d} = \pm a_e a_{u,d} e_{u,d},
\end{eqnarray}
one finds that the NC and CC cross section measurements at HERA 
determine the complete set   $U$, $D$, $\bU$ and $\bD$,
i.e. the sum of up-type, of down-type and of their 
anti-quark-type distributions. 
Below the $b$ quark mass threshold,
these are related to the individual quark distributions as follows
\begin{equation}  \label{ud}
  U  = u + c    ~~~~~~~~
 \bU = \bu + \bc ~~~~~~~~
  D  = d + s    ~~~~~~~~
 \bD = \bd + \bs\,. 
\end{equation}
Assuming symmetry between sea quarks and anti-quarks, 
the valence quark distributions result from 
\begin{equation} \label{valq}
u_v = U -\bU ~~~~~~~~~~~~~ d_v = D -\bD.
\end{equation}

Equation~\ref{f23ud} also shows that the vector and axial vector couplings
of up and down type quarks can be determined at HERA. As will be discussed
below, recent QCD fits have given quite remarkable results on joint
determinations of parton distributions and of the weak 
neutral current coupling constants of the up and down type quarks.
One notices that the underlying assumption here is that
all up as all down type quarks have the same couplings, e.g. $a_u=a_c$,
which holds in the Standard Model.
\subsection{Measurement Techniques}
The inclusive neutral current scattering process is identified by
the unique signature of an isolated electron cluster above some 
minimum energy, of a few GeV, in the LAr/Spacal (H1) or the U/Sc (ZEUS)
calorimeters. In charged current scattering, a large transverse momentum, summed  
over all hadronic final state particles, is required (e.g. $p_T(h) > 12$\,GeV).
The NC and CC scattering events are required to have an
interaction vertex within the nominal interaction region, $\pm 35$\,cm 
using the central trackers or, at the trigger level, of $\pm 2$\,ns from the
nominal interaction time. Unlike H1, the time resolution of the ZEUS calorimeter
is suitable for distinguishing genuine $ep$ scattering from beam background outside
the interaction region. 

NC events, unlike CC events, are balanced in the transverse
plane. Therefore the hadronic calorimeter can be calibrated requiring the ratio
$p_T(h)/p_T(e)$ to be close to one.  The calibration of the
electromagnetic calorimeters is based on several types of events,
in which the electron
energy scale can be established, such as the QED Compton process,
elastic $\rho$ production in DIS and events within the ``kinematic peak'' 
region of phase space,  in which
the scattered electron has an energy very close to the initial beam energy. 
Also, comparisons with the
tracking detector measurements and the electron energy determined with 
the double angle method (see Section\,\ref{diskine}) are used. 
%, and, for lower $Q^2 < 500$\,GeV$^2$ it uses also the kinematic peak method. 
Further adjustments to the linearity of the energy scale determination
are made, for example in the case of the H1 Spacal calorimeter,
with the reconstruction of the $\pi_0$ and the $J/\psi$ mass. 
The calorimeters are aligned by matching the energy deposits with tracks found by the
tracking system; also elastic Compton events, with 
characteristic back-to-back $e-\gamma$ energy deposits in the
transverse plane, are exploited for this purpose.

The basic measurement technique is to compare the reconstructed
event distributions with complete Monte Carlo simulation samples with 
much larger statistics
than the data.
Backgrounds are subtracted from the data; most important of these are the
events due to photoproduction where
the scattered electron escapes near the beam axis and an energy deposit 
in the calorimeter is wrongly attributed to be
the scattered electron. At small scattered
electron energies, $E_e' \geq 2$\,GeV, this background may well exceed
the genuine DIS contribution. Therefore, in some analyses the calorimetric 
electron identification is complemented by a track associated with 
the calorimeter cluster. Additionally, efficiencies, such as those from
the trigger condition
used to select the CC or NC events, are determined from the data
and are used to correct the simulation.  Radiative corrections to the
tree level cross section\,\cite{hector} in NC are much reduced by requiring
the longitudinal momentum in the event to be conserved, i.e.
$E-p_z \simeq 2E_e$ (see Section \,\ref{diskine}). 
The simulation is normalised to
the measured luminosity, which is known to within $1-2$\,\% at HERA 
using Bethe-Heitler scattering, $ep \rightarrow ep \gamma$. 
The cross section, binned according to detector resolution, is then 
determined as the ratio of the background subtracted number of
DIS events in data to the simulated events, each corrected for
the luminosity of the data sample and the simulation, respectively,
and multiplied by the DIS cross section used in the simulation.
With such techniques, refined in several aspects,
both ZEUS~\cite{allzeus} and H1~\cite{allh1} have published, since 1993,
about 15 papers each on the measurements of  NC and
CC cross sections at HERA.
\subsection{Low $\bf{Q^2}$  and  $\bf{x}$ Results}
\subsubsection{The Discovery of the Rise of $\bf{F_2(x,Q^2)}$}
\begin{figure}[h] 
\centering 
\includegraphics[height=7.6cm]{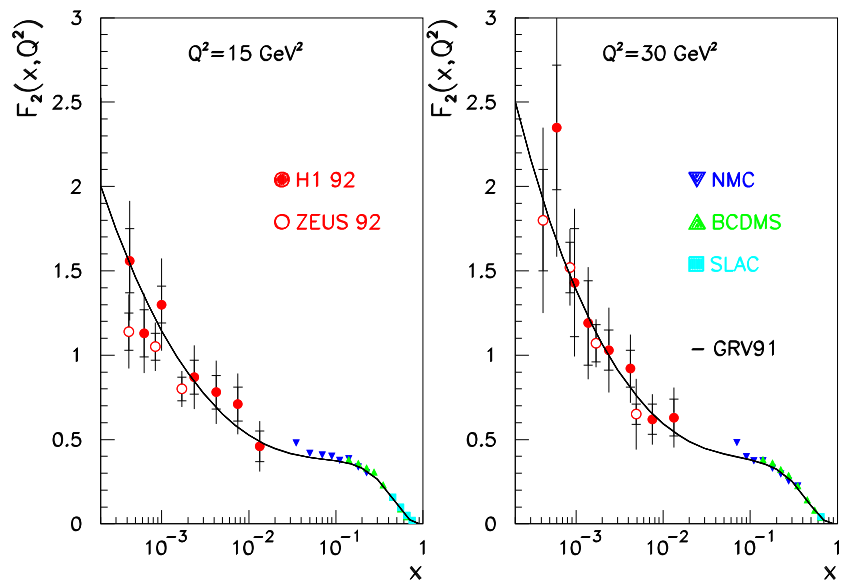}
\begin{minipage}{16.5 cm} 
\caption{The first measurements of H1 (solid points) and ZEUS (open points)
of the proton structure function $F_2(x,Q^2)$ based on the data
taken in 1992 shown as a function of Bjorken $x$.  The H1 data
at low (larger) $x$ were obtained with the electron (mixed) method.
The ZEUS data were obtained with the double angle method (see text).
The HERA collider
is equivalent to a fixed target lepton proton beam experiment
of about 50\,TeV lepton beam energy. The HERA experiments were thus able to
extend the kinematic range of the $F_2$ data provided by the
fixed target electron (SLAC) and muon (BCDMS, NMC) proton experiments
by two orders of magnitude into the then-unknown domain of low $x$.
The curve shows a result  of the so-called dynamic parton distribution
approach. }
\label{fig:f92}
\end{minipage}
\end{figure}

 The first measurements of the proton structure
function \F, shown in Figure~\ref{fig:f92},
were based on only 0.03\,pb$^{-1}$ of data, taken in 1992.
With these first measurements the rise of \Fc  towards low $x$ was discovered.
This rise is in agreement with general expectations on the
low $x$ (large $\omega=1/x$) asymptotic limit of QCD~\cite{classics74};  
however, QCD could not predict the actual scale ($Q^2$) at which the limit
would be applicable. The
fixed target data available up to 1992 led to many extrapolations towards 
low $x$; for example, classic Regge theory~\cite{donland} expected $F_2$
to continue to be flat at low $x$. The curve shown 
in Figure~\ref{fig:f92} was the expectation of
a model\,\cite{grv91}, in which the parton distributions 
were generated dynamically
and evolved from a very low scale $Q^2 \simeq 0.3$\,GeV$^2$
towards higher $Q^2$. In this approach a rise towards low $x$ was
indeed expected.
%
%\subsubsection{More Accurate Data at Low $Q^2$}

\begin{figure}[h]
\begin{center}
\includegraphics[height=8.8cm]{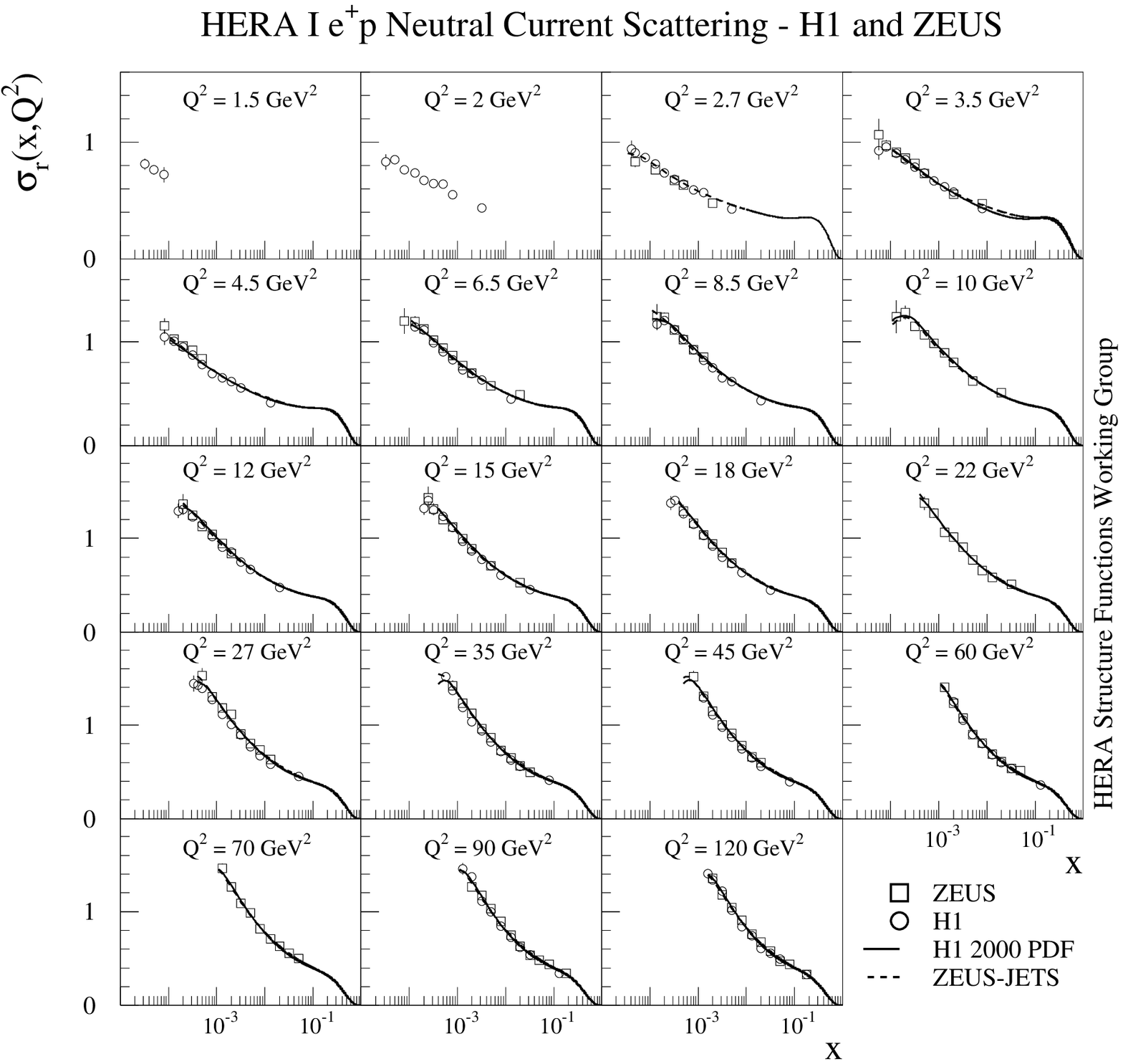}
\includegraphics[height=8.8cm]{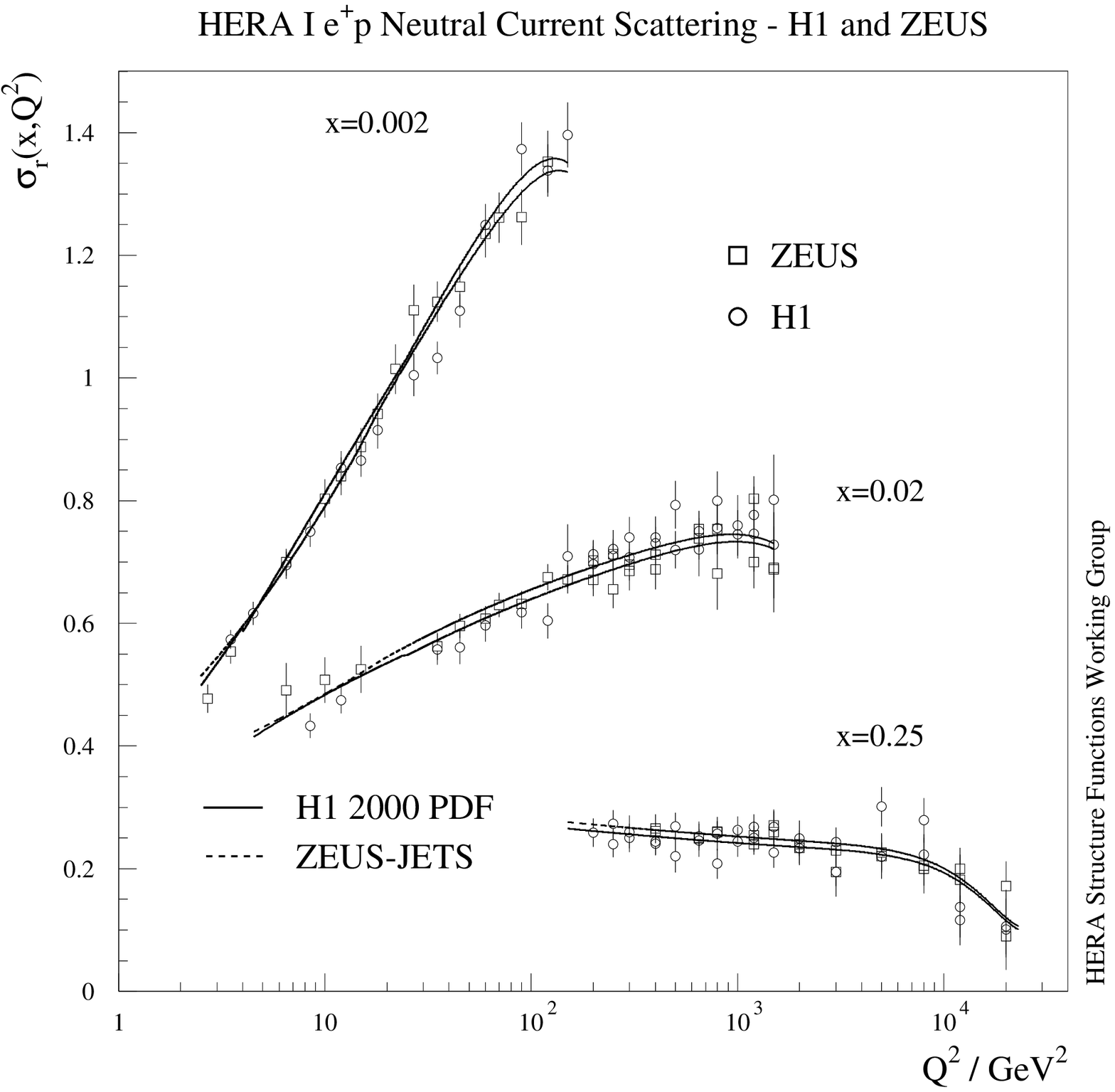}
\begin{minipage}{16.5 cm} 
\caption{Deep inelastic neutral current $e^+p$ scattering cross
 section measurements, from the HERA~I data taking period,
  as a function of $x$ in different $Q^2$
 intervals (left) and  
 for  three selected $x$ bins as a function of $Q^2$ (right).
  Open points: H1, open squares: ZEUS.
    The error bars show the total uncertainty.
    The curves are NLO QCD fits as performed 
   by H1 and ZEUS to their own data. The flattening of the curves
   for the two lower $x$ values at large $Q^2$ is due to the 
   effect of $F_L$ (right plot).
    At $x = 0.25$ the data and the curves decrease visibly for
     $Q^2$ above a few 1000\,GeV$^2$. This effect results from
      the $\gamma Z$ interference which reduces
       the cross section in positron-proton scattering at high $Q^2$,
        cf. Equation~\ref{strf}
        , $\propto a_e xF_3^{\gamma Z}$ with $a_e=-1/2$.}
\label{figbins}
\end{minipage}
\end{center}
\end{figure}

With increasing luminosity and improving understanding of such
measurements, the accuracy increased from about 20\% in the
first data to 2\% for the most recent measurements.  The data of H1 and
ZEUS have been rather consistent and mostly in good agreement for the
various NC and CC cross section data. The best published
measurements, thus far, of the reduced cross section 
at low $Q^2$ ($< 200$\,GeV$^2$) are
displayed in Figure~\ref{figbins}.   The left plot demonstrates the
rise of the cross section 
from the rise of $F_2$ towards low $x$ in all bins of $Q^2$.
In the intermediate $Q^2$ 
range the cross section, $\sigma_{r} = F_2 -y^2F_L/Y_+$,
is observed to flatten or even turning over under the influence of the 
$F_L$ term at high $y$. Dedicated measurements and analyses
are being performed to extract $F_L$ as is discussed below.

Another dramatic observation on the behaviour of $F_2(x,Q^2)$
was its strong dependence at fixed $x$ on $Q^2$, (see Figure~\ref{figbins} right). 
Thus the very first observations of scaling,
the independence of $F_2(x,Q^2)$ on $Q^2$, has been confirmed
only for $x \simeq 0.1$ where it was first discovered at SLAC\,\footnote{
For rather accidental reasons the energy and angular range of the
SLAC $ep$ spectrometer was such that the measurements correspond to
$x \simeq 0.1$. Therefore the first observations of inelastic DIS\,\cite{slac}
occurred in  a region where the behaviour of $F_2$ could
rather directly be related to the quark interpretation of the proton structure.
Departures from this behaviour, which were expected in 
field theory, were first observed at Fermilab \cite{fnal74},
and with more accurate data, they could be shown subsequently
to be logarithmic in $Q^2$. This was a major success in the
process of establishing Quantum Chromodynamics as the 
appropriate field theory for  quark-gluon dynamics.}.
 The H1 and ZEUS measurements illustrate the important
differences between the scaling violations at high $x$, where $F_2$
decreases as a function of $Q^2$ due to gluon bremsstrahlung, and 
at low $x$, where $F_2$ increases as a function of $Q^2$
mainly due to $q \overline{q}$ pair production
from photon-gluon fusion. The strong rise with $Q^2$ at low $x$
is a major discovery of HERA. 
\subsubsection{Remarks on  Low $\bf{x}$ Physics}
Confrontation of the measurements
with the DGLAP evolution equations (see Section\,4) leads to the result
that at low $x$ the quark contribution to the $\ln Q^2$ derivatives
is nearly negligible. This has two consequences: the measurement
as presented in Figure~\ref{figbins} (right) can be interpreted as
the observation of a very large gluon distribution $xg(x,Q^2)$
at low $x$ which at fixed $Q^2$ rises with $x$. Furthermore,
neglecting the quark part (see \cite{prytz}),
one can infer from the second DGLAP equation that 
 $ xg(x,Q^2)  \propto  \exp{\sqrt{c~\ln T~\ln (1/x)}}$ with
 $T=\alpha_s(Q^2)/\alpha_s(Q^2_0)$   \cite{muel}
and get a valid estimate of $F_2$ and its behaviour in $Q^2$ and $x$.
Indeed one is thus able to obtain that \F rises towards low $x$
as $ x^{-\lambda}$  with $\lambda \propto \sqrt{\ln T/\ln (1/x)}$.
The rise is thus expected to be stronger as $Q^2$ increases
as indeed is observed (see Figure~\ref{figbins} left).
A numerical study of this behaviour was performed in \cite{knd}.
Much deeper theoretical evaluations can be found in \cite{ball}.
Apart from the numerical success of this and further studies
(see for example \cite{haidt}) one finds that the proton structure
at low $x$ differs qualitatively from the large $x$ limit:
while the latter is given by the valence quark behaviour, the
low $x$ region is dominated by the dynamics of the QCD 
vacuum, the gluons.

Studies quoted above, and QCD fits discussed below, appear to validate
the application of DGLAP equations over most of 
the HERA kinematic
range.  However, questions remain; because $x$ is 
as small as 10$^{-5}$
for $Q^2 > 1$\,GeV$^2$,
there are expectations that terms neglected in DGLAP evolution, 
those proportional
to $\ln{(1/x)}$, should become sizable and important~\cite{bfkl}.  Also,
as $Q^2$ becomes small, the applicability of perturbative QCD becomes 
suspect and impossible when $\alpha_s$ approaches 1.

\begin{figure}[h]
\centering 
\includegraphics[height=7cm]{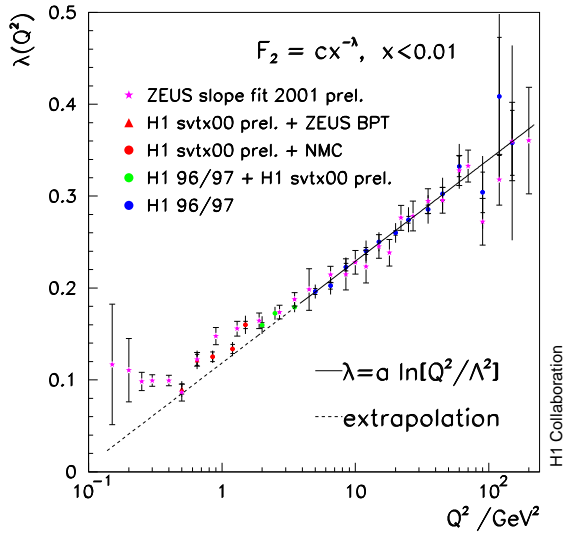}
\begin{minipage}{16.5 cm} 
\caption{Rise of $F_2$ versus $x$ as a function of $Q^2$, see
text. Note that with $Q^2$ also $x$ varies, roughly as
$x \sim Q^2/0.1 s$. The figure is from \cite{paully}.
% GET COMBINED H1-ZEUS PLOT. 
%Right:
%Simulated measurement of the reduced cross
%section $\sigma_{r,NC} = F_2 -f(y) F_L$ at $Q^2=10$\,GeV$^2$
%as is anticipated with H1. The inner error bars are the statistical 
%uncertainty and the full error bars are the expected
%systematic and statistical uncertainties added in quadrature.
%The two structure function contributions
% can be measured  as slope ($F_L$) and intercept ($F_2$)
%of $\sigma_r$ versus $f(y)=y^2/[1+(1-y)^2]$. 
}
\label{rose}
\end{minipage}
\end{figure}

The rise of $F_2$ towards low $x$ has been examined quantitatively
\cite{zeusrise,h1rise} (see Figure~\ref{rose}).
For $x < 0.01$, where the valence quarks have
a negligible role, the rise at fixed $Q^2$
may be expressed as $F_2 = C x^{-\lambda}$.
It has been measured that for $Q^2 > 3.5$\,GeV$^2$, i.e. in the
DIS region, the parameter $C$ is approximately constant and
$\lambda = 0.048\cdot(\ln Q^2/\Lambda^2)$
with $\Lambda = 0.29$\,GeV. At lower $Q^2$, the rise becomes weaker
and $C$ dependent on $Q^2$. This may be interpreted
as a change from the partonic behaviour, at larger $Q^2$,
to a hadronic behaviour \cite{dolal}
at lower $Q^2$, the transition taking place
near  $Q^2 \sim 1$\,GeV$^2$ corresponding to a resolution
distance of about 0.3\,fm. 
The so-called BPT data \cite{bpt} below 1 GeV$^2$
which were 
obtained by ZEUS with a dedicated detector placed 
near to the beampipe, together with the precision
measurements based on the central apparatus of H1 and ZEUS,
play a  crucial role in this observation.

The transition region is often studied in terms of
the ``colour dipole model'' (CDM),
which will also be discussed  in connection
with diffractive DIS in Section 7.
%
%At low $x$ 
%it is useful to view the $ep$ interaction 
%in an unconventional frame of reference:
%instead of using the parton picture and the infinite momentum
%frame\,\cite{feynman} one may consider the virtual 
%photon-proton interaction in the proton restframe. The virtual 
%photon disintegrates long before interacting with the proton,
%at a coherence length $\propto 1/x$ apart, into a dipole
%of quark and antiquark which hits the proton. This picture, 
%originally considered in \cite{gribov} and turned 
%into the colour diploe model (CDM) in \cite{cdm}, has received much
%attention over the years of HERA's existence \cite{cdmrev} because of its
%simplicity and universality.
The CDM is able to provide predictions,
using a phenomenological ansatz for a hard scattering cross 
section, for a variety of cases, as for the structure functions
$F_2$, $F_L$ but as well for the diffractive and heavy quark
structure functions, which are introduced below.  The strict
relations between these predictions lead to crucial tests
of the validity of this approach \cite{nachtcdm}, results of
which can still be expected at HERA. The CDM model
fundamentally does not distinguish between the partonic or hadronic
behaviour of the proton and thus is applicable for large as well
as very small values of $Q^2$ in DIS. It is to be also noted that
the transverse size of the dipole is $2/\sqrt{Q^2}$; thus by varying
the momentum transfer, one is able to probe the transverse
extension of the proton. Most notably at HERA, the transverse
size of gluons was measured for the first time \cite{gluontran}.

The HERA 
physics of low $x$ has important relations
to other fields. The knowledge of the rising parton densities is
of importance for investigations of the quark-gluon plasma
phase in nucleus-nucleus collisions, as are studied at RHIC \cite{rhicreview}
and at the forthcoming LHC, as well as for cosmic super-high-energy 
neutrino physics \cite{suneu}. Furthermore there is a wide range 
of forward scattering at the LHC, in which low $x$ values
are involved and it is, so far, not clear whether one may use the
DGLAP equations, evolved to the LHC range, in order to understand
large rapidity particle production characteristics. It is possible that
modifications from the large
contribution of $\ln (1/x)$ terms in this domain are needed 
for a full description.
%
%\item
%
\subsubsection{The Longitudinal Structure Function}

There is a strong interest
in $F_L$ since the longitudinal structure function
is a measure of the gluon distribution 
which is independent and complementary to  the derivative
$\partial F_2(x,Q^2) /\partial \ln Q^2$.
A measurement  of $F_L(x,Q^2)$  represents
a crucial test of QCD to high orders as had long been studied
experimentally~\cite{flexp} and theoretically~\cite{flthy}.
The importance of this test at HERA rests  also on the observation
that the determinations of the gluon density at
low $x$ and low $Q^2$ ($< 10$~GeV$^2$)
are not constrained well because of the lack of range and the
uncertainties of the measurements 
of  $\partial F_2(x,Q^2) / \partial \ln Q^2$.

\begin{figure}[h]
\centering
\includegraphics[height=8.3cm]{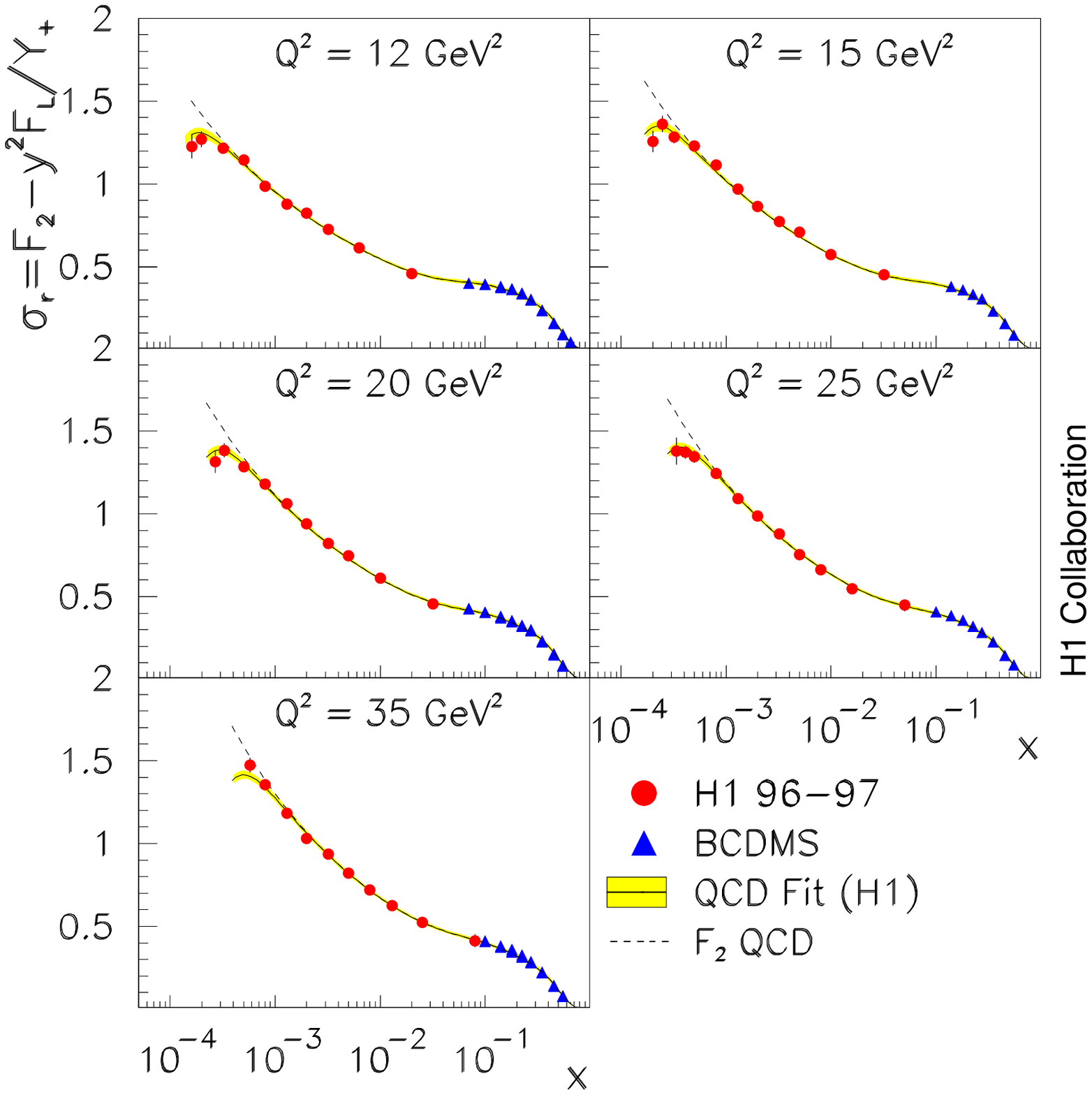}
\includegraphics[height=8.3cm]{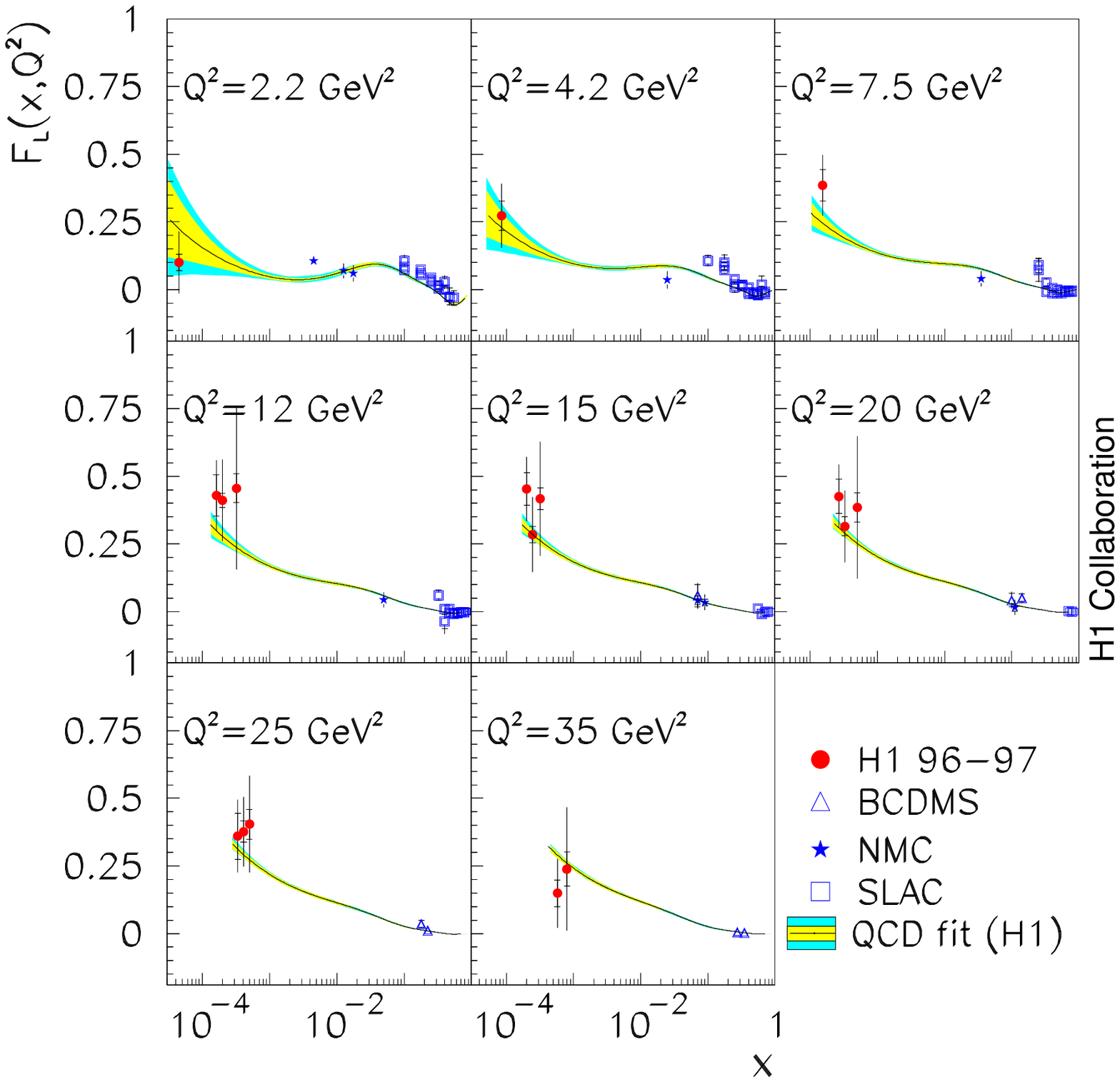}
\begin{minipage}{16.5 cm} 
\caption{Left: Measured reduced cross section as a function of $x$
in the mid $Q^2$ region. The data at lowest $x$ are seen to depart from
the extrapolated behaviour of $F_2$. This departure, for each bin, occurs at 
about $y \simeq 0.5$ and is attributed to the effect of the
longitudinal structure function $F_L$ which at low $x$ is large.
The curves are the NLO QCD fit result on $\sigma_r$ for
$F_L=0$ (dashed) and for $F_L$ (solid)
obtained in the fit. Right: The longitudinal
structure function $F_L(x,Q^2)$ for different $Q^2$ bins
as obtained by H1 at low $x$, and by charged lepton-proton
scattering experiments at large $x$ (see \cite{h1alfas}).
The curves are from the H1 $\alpha_s$ NLO QCD fit,
described below, and its uncertainties. 
}
\label{fsiy}
\end{minipage}
\end{figure}

%\item
At lowest $x$, corresponding to large $y$, the reduced cross section
departs from $F_2$ because of $F_L$.  This had been seen for the
first time 
at HERA in \cite{flh1}.  A subsequent accurate measurement of the
cross section up to high values of $y$ is shown in Figure~\ref{fsiy}.
This measurement is particularly difficult
because at high $y \simeq 1 -E_e'/E_e$, 
the scattered electron energy is rather small; $E_e' = 2.7$\,GeV
for $y=0.9$.  At such low energies the 
DIS electron signal is  mimicked by hadronic energy depositions
in photoproduction processes which poses 
severe difficulties in extracting the genuine DIS contribution.
%triggering and electron identification.
%

The measurements of the high $y$ cross section have been used
by H1 to determine the longitudinal structure function at low $x$.
This is enabled by the large kinematic coverage of HERA: 
the range in $y$ extends over two orders of magnitude, thus
at lower $y$ ($< 0.35$) one is able to determine $F_2$ which then may be
extrapolated to the high $y$ region. At larger $Q^2$, in the DIS
region, as is illustrated in Figure~\ref{fsiy}, one  may use
the NLO QCD fit to predict $F_2$ in the region 
where only $\sigma_r$ is actually
measured and subtract the $F_2$ part from the cross section.
At lower $Q^2$ pQCD cannot be trusted and different means have
been developed by H1 \cite{h1alfas,flshape}
to simultaneously determine $F_L$ with $F_2$.
The published H1 data on $F_L$ are shown in  Figure~\ref{fsiy},
further results were presented subsequently in \cite{flshape}.
The principal problem of such determinations is of course
their unavoidable dependence on the knowledge of $F_2$.

During the last 4 months of its operation, HERA was operated at
reduced proton beam energies. Data samples of 13\,pb$^{-1}$
at 460\,GeV and of 7\,pb$^{-1}$ at 565\,GeV were collected
by both ZEUS and H1.
This data will allow a 
model-independent separation of the contributions of $F_2$
and $F_L$. 
%as is illustrated in Figure~\ref{rose}. 
Since the luminosity behaves as $L \propto E_p^2$, a reduction
to half the energy requires 4 times longer running. A more precise 
measurement for statistical reasons will thus be restricted
to the lower $Q^2$ region. As this paper is being written, both the
ZEUS and H1 Collaborations are analysing the low energy run data and
results are  much anticipated. These will complement and  
perhaps validate previous attempts by
the H1 Collaboration to extract $F_L$ as were summarised briefly above.
An attempt will also be devoted to the extraction of the diffractive
longitudinal structure function, $F^D_L$, which is of particular
theoretical interest and about which nothing is known experimentally
so far\,\cite{h1floi}.

%
%unintegrated, f2c, final states ...
%\end{itemize}
 
\subsection{High $\bf{Q^2}$ Results}
\begin{figure}[htbp]
\begin{center}
\includegraphics[height=8.8cm]{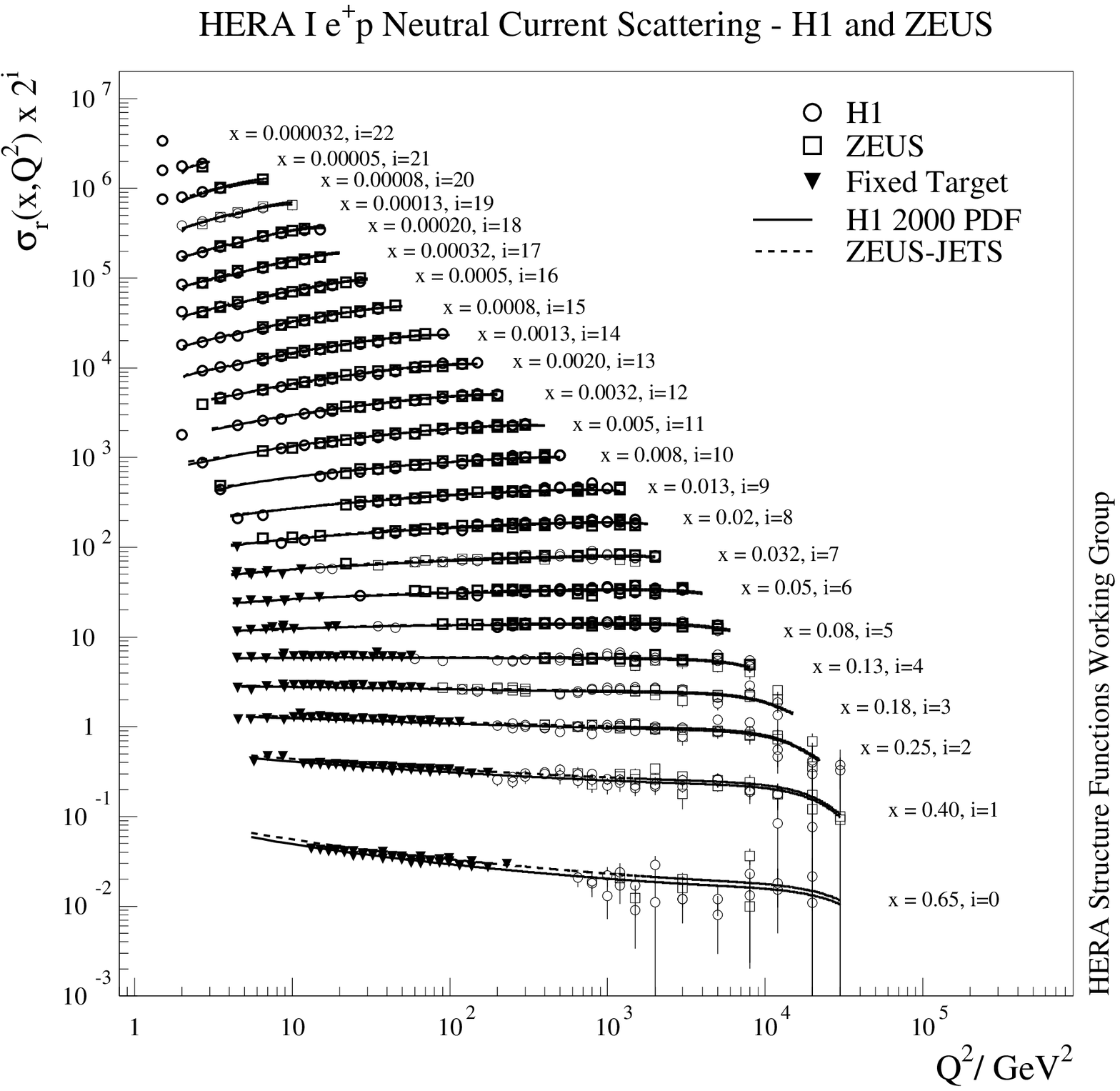}
\includegraphics[height=8.8cm]{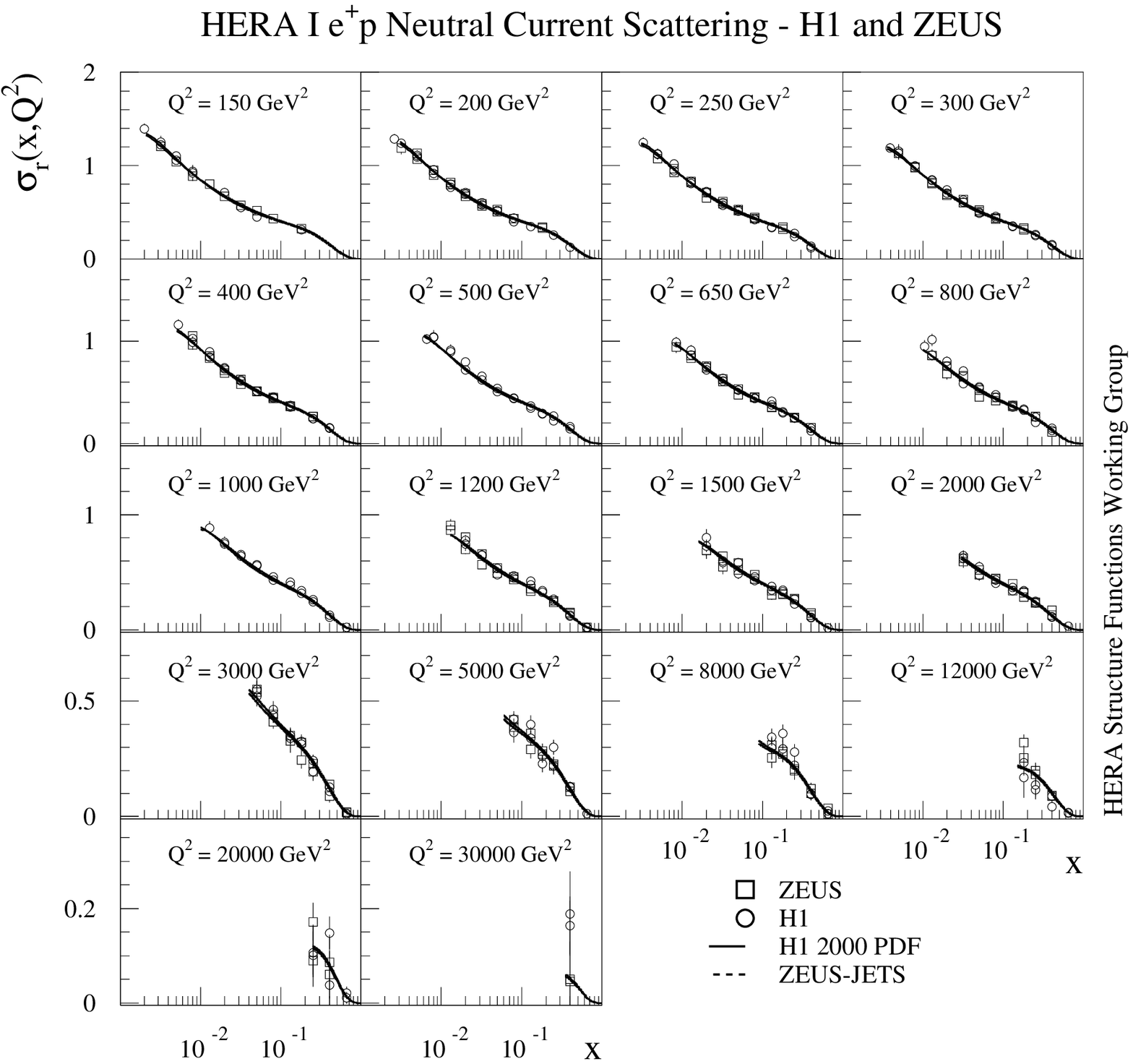}
\begin{minipage}{16.5 cm} 
\caption{ Deep inelastic neutral current $e^+p$ scattering cross
 section  data from the HERA I data taking period. Left:
 NC reduced cross section data versus $Q^2$ for different $x$. The 
 measurements are consistent with earlier DIS fixed target
 measurements but considerably extend the kinematic range
 in $Q^2$ and $x$. Right: NC cross section data
 versus $x$ in different intervals of $Q^2$.  
 The curves are NLO QCD fits as performed by H1 and ZEUS
 to their own data (see Sections 4 and 5.4).}
\label{figscal}
\end{minipage}
\end{center}
\vspace{-1cm}
\end{figure}

The large cms energy of HERA has permitted the extension of the
measurements of DIS lepton-proton scattering cross
sections, as compared to previous fixed target experiments
by about two orders of magnitude towards high momentum
transfers squared, $Q^2$. This can be seen in Figure~\ref{figscal}
(left) which presents the cross section measurements as 
obtained previously in fixed target experiments and those
from   H1 and ZEUS as a function of $Q^2$ for the
full range of Bjorken $x$ covered.  One observes scaling to hold
for the complete $Q^2$ range at $x \simeq 0.1$. At larger
$x$, negative scaling violations, due to quark bremsstrahlung,
continue to exist up to high $Q^2$. One also observes in this 
range the tendency of the cross section to decrease stronger
at highest $Q^2$ which marks the onset of the effects
of the $\gamma Z$ interference contribution which for
$e^+p$ scattering is destructive, (see Equation~\ref{strf}).
Figure~\ref{figscal} (right) shows the H1 and ZEUS data
in the high $Q^2$ region as a function of $x$. The data
agree rather well and are described by NLO QCD fits. 
HERA has thus established perturbative QCD to be valid
up to $Q^2$ values beyond $10^4$\,GeV$^2$ which 
was one of the questions posed initially~\cite{wiiksmith} when one 
envisaged a high energy $ep$ collider to be built.

\begin{figure}[htbp]
\begin{center}
\includegraphics[height=8.8cm]{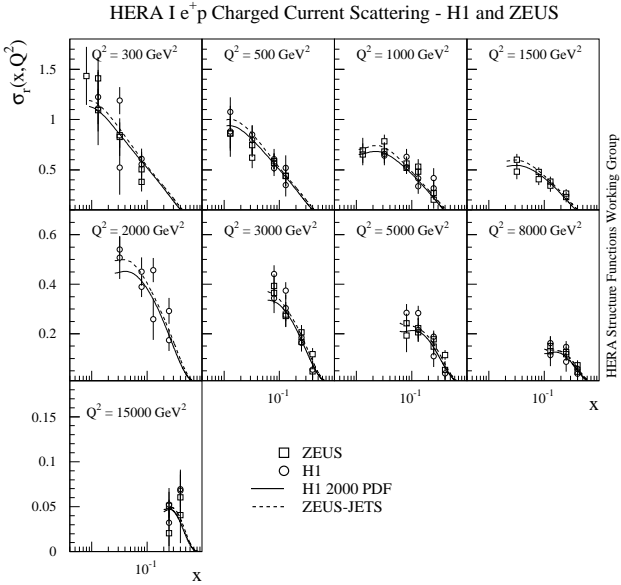}
\includegraphics[height=8.8cm]{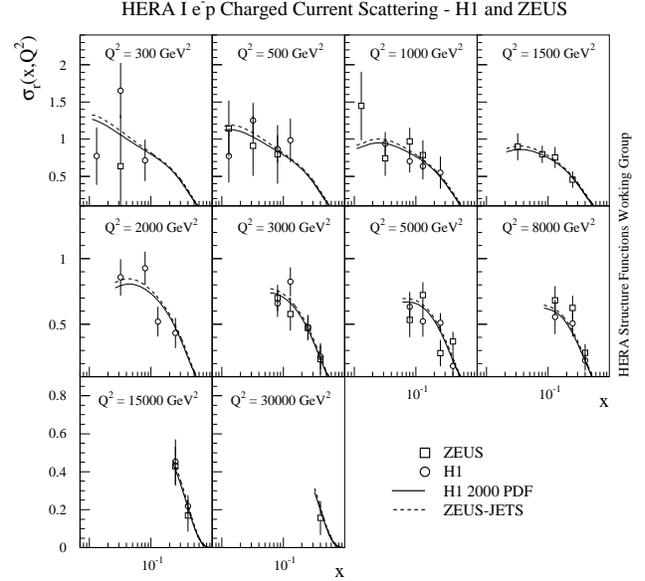}
\begin{minipage}{16.5 cm} 
\vspace{-0.3cm}
\caption{ Deep inelastic charged current $e^{\pm}p$ scattering cross
 section  for  $Q^2$ between $300$ and $30000$\,GeV$^2$,
 measured by  H1 and ZEUS using data taken in HERA\,I.
The errors shown are the total errors, which 
are dominated by the statistical uncertainty. 
 The curves are NLO QCD fits as performed by H1 and ZEUS
 to their own data (see Sections 4 and 5.4).}
\label{figccep}
\end{minipage}
\end{center}
\end{figure}

Unlike fixed target experiments, which required separate
electron/muon- and neutrino-nucleon scattering apparatus,
neutral and charged current data could be
taken simultaneously at HERA because at larger $Q^2 \geq 100$\,GeV$^2$
the weak interaction process $ep \rightarrow \nu X$ becomes
measurable.
% as its cross section 
%becomes comparable to the NC process. 
Figure~\ref{figccep} shows the published 
results from ZEUS and H1
on $e^{\pm}p$ CC scattering. These data are of unique nature:
they further support QCD, allow flavour contributions
to proton's structure to be disentangled (Equation~\ref{ccstf})
and, as is discussed below, they demonstrate clearly 
that the electromagnetic and the weak interactions
become of similar strength for $Q^2 \simeq M^2_{W,Z}$.
%which had not been shown directly prior to HERA~\cite{wiiksmith}.

The data presented here belong to the initial (HERA I) period of 
data taking. The luminosity increase and the polarisation
in HERA\,II have much enhanced the importance of the
high $Q^2$ NC and CC data, which will be published
in the future. Yet, the data from the HERA\,I
period have already  had a major impact on 
the determinations of the quark and the gluon
distributions in the proton,
and also on the determination of the strong coupling 
constant as is discussed in the following.

%\begin{figure}[htbp]
%\begin{center}
%\epsfig{file=ncepa1.eps,scale=0.8}
%\includegraphics[height=9cm]{ncepa1.eps}
%\caption{ Deep inelastic neutral current $e^=p$ scattering cross
% section  at low $Q^2$ and $x$
% for lower $Q^2$, $1.5 - 120$\,GeV$^2$,
% as obtained by
% H1 and ZEUS  from data taken in HERA I. 
% The curves are NLO QCD fits as performed by H1 and ZEUS
% to their own data, shown for $Q^2 \geq Q^2_{min}$ as chosen
% in the fit procedure. Note that there are data available for
% $Q^2$ below 1\,GeV$^2$, both from shifted vertex operation and from
% ZEUS also using a detector, $BPC$, near the beam pipe. 
% WOULD BE GOOD TO SHOW THEM PERHAPS VS W}
%\label{ncepa}
%\end{center}
%\end{figure}

\section{QCD Fits}
\subsection{Introduction}

In perturbative QCD  the quark and gluon
distributions $D(x,Q^2_0) = q,g$,
determined at  an initial value $Q^2_0$, can be 
evolved  further in $Q^2$ using 
a system of coupled integro-differential equations (DGLAP equations)
\begin{equation} \label{dglpeq}
  \frac{dq_i(x, Q^2)}{d\ln Q^2} =
  \frac{\alpha_s(Q^2)}{2\pi}
  \int_x^1\frac{dz}{z}
  \left[
    \sum_j q_j(z, Q^2) P_{ij}\left(\frac{x}{z}\right)
    + g(z, Q^2) P_{ig}\left(\frac{x}{z}\right)
  \right] 
\end{equation}
\begin{equation}\label{dglpeg}
\frac{dg(x, Q^2)}{d\ln Q^2} =
  \frac{\alpha_s(Q^2)}{2\pi}
  \int_x^1\frac{dz}{z}
  \left[
    \sum_j q_j(z, Q^2) P_{gj}\left(\frac{x}{z}\right)
    + g(z, Q^2) P_{gg}\left(\frac{x}{z}\right)
  \right], 
\end{equation}
which was 
introduced independently by Dokshitzer~\cite{yuri}, Gribov and 
Lipatov~\cite{griblip} and by Altarelli and Parisi~\cite{altpar}. 
The splitting  functions $P_{\alpha\beta}(x/z)$ in leading order
represent the probability for a parton $\beta$ with a momentum
fraction $z$ of the proton to emit a parton $\alpha$ with
momentum fraction $x$ of the parent parton.
From $\chi^2$ comparisons
of these QCD predictions with data, 
the quark and gluon momentum  distributions and
the strong coupling constant \amz can be determined.
The equations  can be applied in 3rd order (NNLO) 
perturbation 
theory due to a major progress, achieved recently, 
in pQCD calculations~\cite{moch}. 
Most of the parton distribution analyses, as those presented subsequently,
are still done in NLO. The next order will be important to match the
still increasing accuracy of further DIS measurements from H1 and ZEUS.
In the upcoming NNLO analyses, the dependences of the fit results,
for example, such as that for \amz, on the chosen scales $\mu$
are expected to be reduced significantly. 

The HERA data are at sufficiently high $Q^2$ at
large $x$ such that  higher twists, power corrections $\propto Q^{-2}$
can be neglected. They extend to such low values of $x$, however,
that issues as to whether the evolution equations need to be modified
to include $\ln 1/x$ terms have been under investigation for
many years \cite{lowxcoll}.
%, see section~\ref{sec:lowx}. 
Besides attempts to replace the DGLAP equations by different
equations~\cite{bfkl,ccfm}, corresponding to modified rules of multi-gluon
emission at low $x$, the effects have also been investigated of
their modification with
a resummation of small $x$ logarithms \cite{ball}.  These studies, so far,
have not lead to definite conclusions.
For the
resolution of these questions,
an increase in energy beyond HERA, as is considered with the LHeC\,\cite{lhec},
is very desirable in order to reach even lower $x$ and to perform
such investigations in a region of large enough $Q^2$ where \as is small.

The following sections present a brief
overview on the most important pQCD analyses performed so far
by ZEUS and H1 to examine the partonic content of the proton.

\subsection{Determinations of Parton Distributions} 
The quark and gluon distributions in the proton define the
various structure functions and their $x,~Q^2$ dependence
as given in Equation~\ref{factoreq}, \ref{dglpeq} and \ref{dglpeg}.  
The principal method of
extracting a complete set of quark distributions, $q(x,Q^2)$,
and the gluon distribution, $xg(x,Q^2)$, requires a parameterisation
of their $x$ dependence at some initial value of $Q^2=Q_0^2$, of 
a few GeV$^2$, as input to the DGLAP evolution equations 
and to the factorisation equation (Equation \ref{factoreq}) leading to the
structure functions. This procedure defines a pQCD prediction
of the cross-sections at all $Q^2$ and $x$,
 which can be confronted in a $\chi^2$ minimisation
procedure to the data under study. 
%In the HERA QCD analyses
%basically two programs were used, QCDNUM \cite{qcdnum},
%and QCDFIT \cite{qcdfit}, which were independently developed
%but from time to time cross checked against each other and
%against Mellin transformation code \cite{benchqcd}.
Two DGLAP evolution programs QCDNUM \cite{qcdnum}
and QCDFIT \cite{qcdfit}, which were independently developed,
are used most often in the HERA QCD analyses.  They have been 
cross-checked against each other and
against a Mellin transformation code at both the HERA Physics Workshop in
1995 and subsequently at the HERA-LHC Workshop in 2005.
%\cite{benchqcd}.

 The ZEUS and the H1 Collaborations
so far focused mainly on analyses of their own data, as are described
subsequently. There have been different attempts for global
determinations of the parton distributions, such as those 
by the MRST~\cite{mrsto} group,
by the CTEQ~\cite{cteqo} group and by Alekhine~\cite{alekhino}, in which 
much wider sets of data, in particular those from  lepton-nucleon
fixed target experiments and the Drell-Yan measurements,
are considered. Unlike the fixed-target data, the HERA data
constitute a complete set of NC and CC cross sections over
a very wide range in $x$ and $Q^2$. The additional
constraint obtained from having both $e^+ p$ and $e^- p$ data
allows the extraction
of sets of parton distributions that approach the 
accuracy of global determinations. There are specific regions
for which HERA data do not have sufficient accuracy; examples are
the high $x$ behaviour of the $u/d$ valence quark ratio, which is
better derived from $lp$ and $lD$ data, and
the $\bu -\bd$ difference, which is clearly observed
at medium $x \simeq 0.01$ \cite{towell}
in Drell Yan data
but was only barely discernable in the initial H1
data\,\cite{mkbr}. 
This could have been accessed at low $x$
with electron-deuteron data at HERA.

%Besides in the usage of input data sets and the way the pdf's are
%parameterised, an important difference is in the treatment of
%systematic errors. This regards two aspects: i) in the QCD analyses
%allowance can be made, or not, to determine optimum values
%of systematic cross section shifts, as e.g. due to a chosen 
%energy scale. This distinguishes H1 and ZEUS fits. ii) focusing
%on HERA data the parameter uncertainties can be determined
%from the conventional $\chi^2 +1$ requirement. This is practically
%impossible in global fits due to inconsistencies in the
%world's data. This distinguishes the HERA Collaboration fits
%from some of the global approaches. In comparing the uncertainties
%of pdf's, directly or  in their extrapolation to the LHC, one thus 
%has to be extremely cautious and the error definitions cannot be
%ignored. This is of possibly high practical relevance as
%their exists the principal hope to be able to determine the
%luminosity at the LHC to better than 5\% accuracy from 
%high statistics SM processes as $Z$ or $W^{\pm}$ 
%production~\cite{lumiLHC}.

An important consideration in doing pQCD pdf fits is the treatment
of systematic errors.  In the H1 fits, the correlated systematic uncertainties
are treated as fit parameters; this means the optimum value of 
systematic shifts of the cross-sections are determined through the fit.
After the initial fit, the model assumptions (the shape of the
pdf parameterisation, for example) in the fit are varied to determine
the ``model uncertainties'' which are separate from the systematic
uncertainties of the data.  In the ZEUS fits, on the other hand,
the initial fit that determines central values of the parameters is
made with the systematic offsets set to zero.  Then an ``envelope''
of $n$ systematics errors is determined by repeating the fits $2n$ times
after off-setting the cross-sections to the +1 standard deviation 
and -1 standard deviation values
allowed by the systematic uncertainties
and then adding the changes in quadrature \footnote{In practice this
is done through a matrix inversion procedure\,\cite{qcdfit}.}.
The model uncertainties are then found by repeating the central fit with
the variation on model uncertainties.  The uncertainties of the H1 pdf
results are dominated by the model uncertainties, whereas the systematic
uncertainties dominate the ZEUS data.  It is remarkable that the uncertainties
of H1 and ZEUS fits are, in the end, about the same\,\cite{uncheral}.

A further challenge for the QCD pdf determinations consists
in the treatment of the heavy flavours. At HERA, the charm and beauty
contributions to the cross section are significant (see below) and
the data are in $Q^2$ regions which include the $c$ and
$b$ thresholds.  A quark which is heavy, compared to the
QCD parameter $\Lambda \sim 0.3$\,GeV, becomes light at
very high $Q^2$ compared to  $M_{c,b}^2$. Correspondingly
there exist various
theoretical prescriptions on  the treatment of $c$ and $b$,
as light quarks, 
% from $g \rightarrow c \bc$,
as variable in their appearance or as
heavy quarks, from $\gamma g$ fusion
(see \cite{heavyqcd} for reviews). Fits have 
been presented with different heavy flavour treatments 
or versions of the variable flavour scheme (VFNS).
ZEUS (Section\,\ref{seczeusfit}),
has specifically used the VFNS scheme of \cite{mrstvfns}.
H1 has presented a fit to data at lower $Q^2$ 
using the fixed flavour scheme,
in the determination of $\alpha_s$ (Section\,\ref{secalfinc}),
or assumed the heavy quarks to be light in the 
analysis, described in sect.\,\ref{sech1fit},
which focused on their high $Q^2$ data. 
These and further differences lead to somewhat different
results. Work is in progress, both by the global fit groups
and by ZEUS and H1, to further pin down such differences and
arrive at fits, desirably at NNLO, based on the most plausible
assumptions as well as on new and possibly combined data sets.

The following sections describe briefly the different fits
of ZEUS and H1 in the determination of the parton distributions.
In Section~\ref{chjet}, the influence of di-jet data is
illustrated on the determination of the parton distributions,
in particular on $xg$ at medium $x$. Important insight
to the behaviour of the gluon distribution at low $x $
is expected from the forthcoming direct measurement of
the longitudinal structure function.
\subsubsection{The ZEUS Approach} \label{seczeusfit}
The ZEUS Collaboration performed, in recent years,
three major NLO QCD analyses, a ZEUS-S (standard) fit \cite{zeuss}, 
which includes the fixed-target DIS data in the fit,
a ZEUS-O (ZEUS data only) fit \cite{zeuso} and 
a fit to the inclusive data together
with di-jet data 
in DIS and photoproduction, ZEUS-Jets fit. The ZEUS-O fit 
exhibited the power of the HERA data in coming quite close in accuracy
and detail to the ZEUS-S fit. Here only the ZEUS-O fit
is briefly presented. The effect of adding jet cross section information,
(the ZEUS-jets fit) in particular on the gluon determination, is described 
in the next Section.

\begin{figure}[h]
   \centering
    \epsfig{file=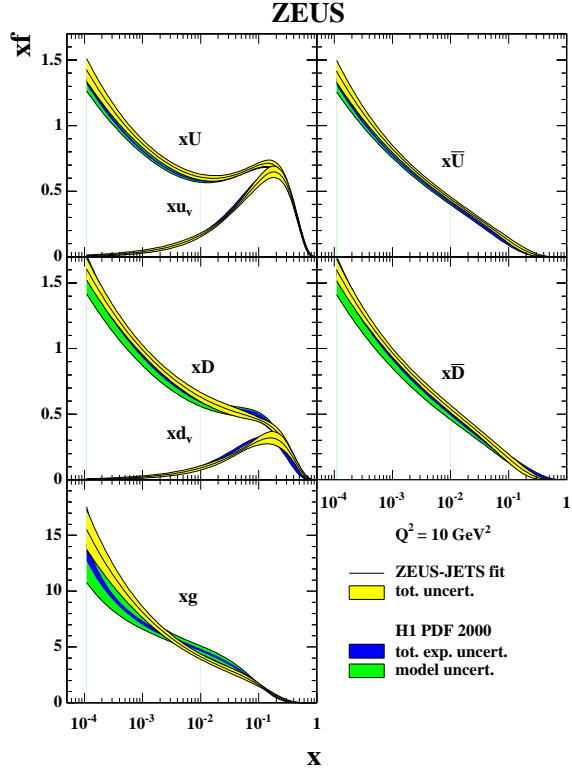,width=8cm}
\begin{minipage}{16.5 cm} 
   \caption{
Quark and gluon distributions determined by the ZEUS-Jets and the
H1PDF2000 fits vs. $x$ at $Q^2=10$\,GeV$^2$. Shown are the 
total error bands which include the experimental and the model uncertainties
(not separated for ZEUS). One notices the gluon dominance at low $x$
at which $xg$ is about 10 times larger than $xU$, for example. }
   \label{figpdfh1zeus}
\end{minipage}
\end{figure}
The main idea of the ZEUS-O pdf parameterisation, following a QCD fit
tradition, is to decompose the
quark contributions into sea quarks and valence quarks.
Four quark distribution combinations were  used:
\begin{eqnarray} \label{eqzeuso}
S=2(\bd + \bu +\bs + \bc), ~~ \Delta = \bd -\bu, ~~ d_v = d -\bd, ~~ u_v = u -\bu.
\end{eqnarray}
These distributions (times $x$) and $xg$ are parameterised as
 four-parameter polynomials of the type
\begin{equation}
xf(x) = p_1 x^p_2 (1-x)^{p_3} ( 1  +p_5 x).
\end{equation}
The strange sea was coupled to the light sea, $x\bs = 0.1 xS$,
and the charm contribution was dynamically generated \cite{zeuso}.
The most detailed information in the HERA data is on the sea, $xS$,
for which all 4 parameters were determined, using in addition the
momentum sum rule, which fixes the integral over $xg + xS + xu_v + xd_v$
to be one. There is a weak constraint coming from the charged current
data, on the difference between $\bu$ and $\bd$. Such a difference
has been established
in the muon-nucleon data as a violation of the Gottfried sum-rule
and rather accurately measured in Drell-Yan scattering \cite{towell}.
For $x \Delta$ a choice was taken to let only the normalisation
parameter $p_1$ float and set $p_2=0.5$, $p_3 = p_3(S)+2$ and $p_5 =0$.
Both valence quark distributions were used fixing $p_2 =0.5$ and requiring
the integral over $u_v$ to be 2 and over $d_v$ to be 1 (quark counting rule).
Finally $p_5=0$ was used for $xg$ which constrains the high $x$ gluon to 
be positive. Results from these fits are discussed in detail 
in \cite{zeuso}. Subsequently
fits were performed with additional information on jets. This lead generally
to a consistent set of pdf's, both with the purely inclusive ZEUS-O fit, and
with the H1 fit which is  briefly described in the following. 

A comparison of the ZEUS and H1 pdf's is presented in Figure~\ref{figpdfh1zeus}.
The agreement can be called indeed remarkable for it uses different data,
different flavour decompositions,  different parameterisations,
different $\chi^2$ definitions and independent programs imposing 
differing heavy flavour treatments. 
More detailed inspection reveals that there are differences to be understood,
as on the behaviour of $xg$ at low $x$ and $Q^2$. 
%There are discussions ongoing
%between H1 and ZEUS to choose the most appropriate way to determine the pdf's.
 %
\subsubsection{The H1 Approach} \label{sech1fit}
\begin{figure}[h]
   \centering
    \epsfig{file=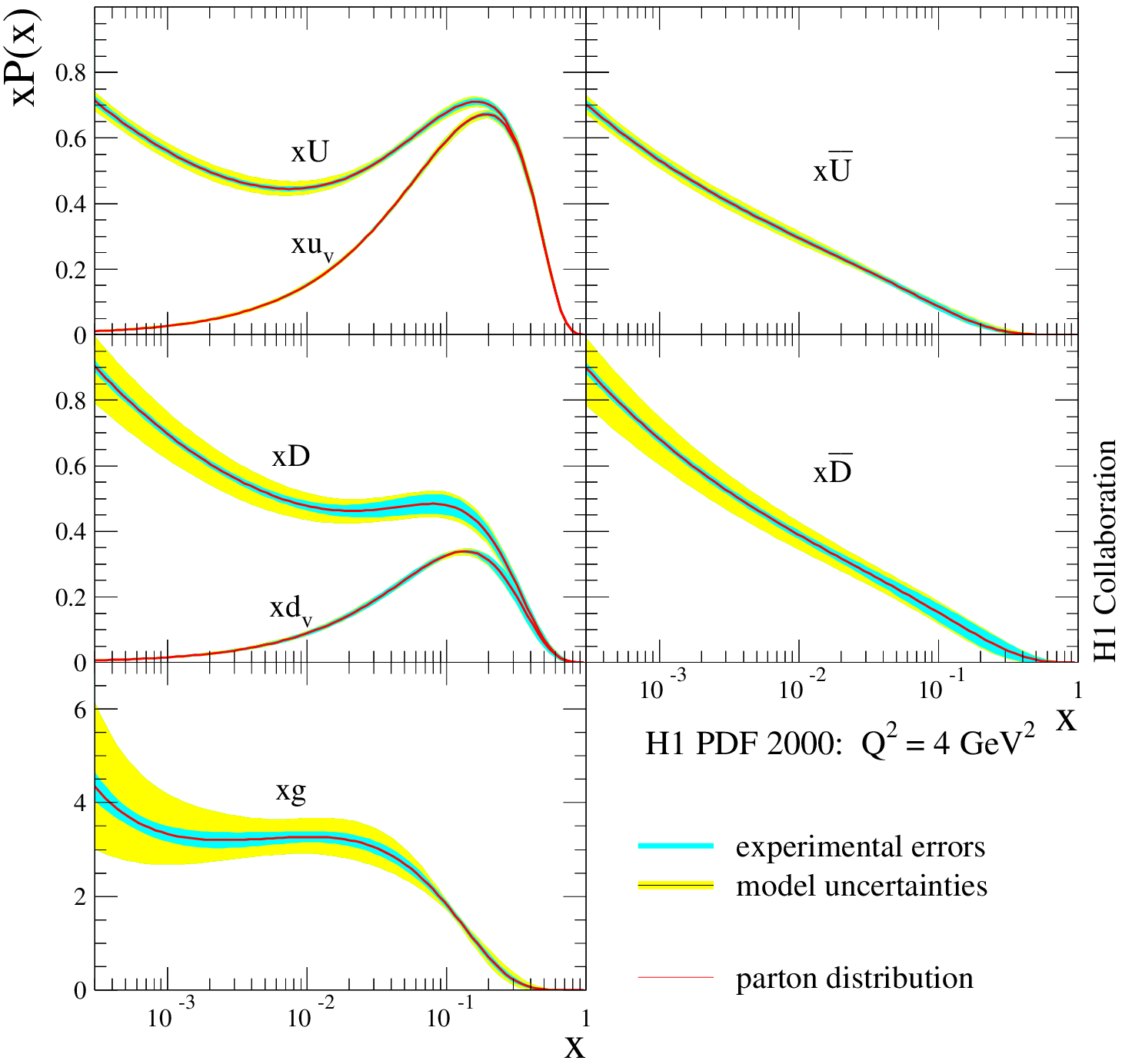,width=8cm}
      \epsfig{file=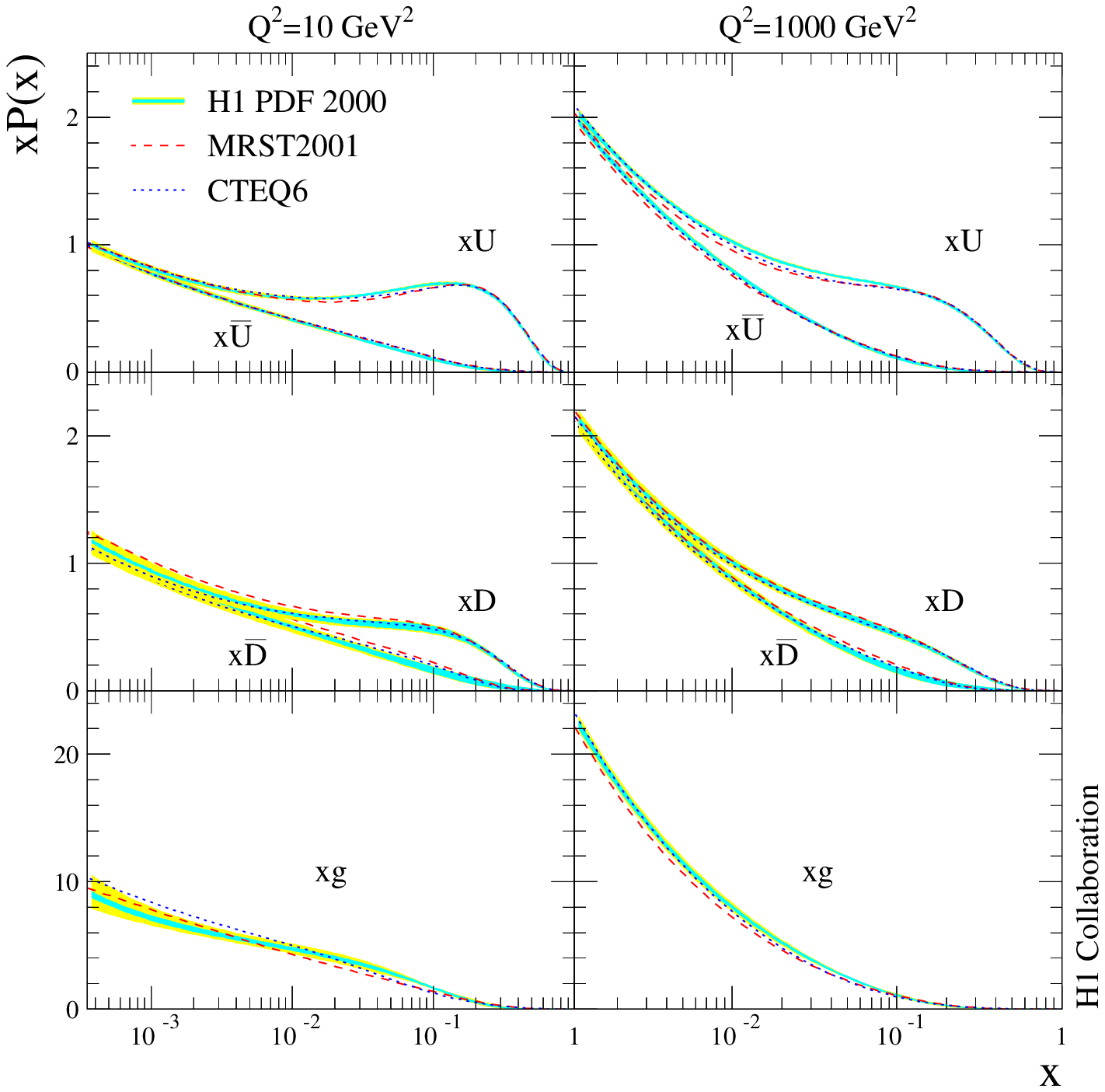,width=8cm}
\begin{minipage}{16.5 cm}  
   \caption{Determination of the sum of up, anti-up, down and anti-down quark
    distributions and of the gluon distribution in the proton based on the
    H1 neutral and charged current cross section data. Left: the parton
    distributions with their experimental and model
    uncertainties as determined by H1 at the starting scale $Q^2_0=4$\,GeV$^2$;
    Right:  for $Q^2$ of
    10  and 1000 GeV$^2$, the H1 fit is compared with results
    from MRST and CTEQ. 
     }
   \label{figpdfall}
\end{minipage}
\end{figure}

The NLO QCD analysis (H1PDF2000 fit) \cite{h1pdf2000}
of the H1 Collaboration of the CC and NC cross sections
exploits Equations \ref{ccupdo} and \ref{f23ud} which suggest a decomposition
of the various structure functions into the four basic
combinations of up ($U, \bU$) and down ($D,\bD$)
quarks and antiquarks (Equation~\ref{ud}).  The data used as
input are the fixed-target 
BCDMS data  as well as the H1 data.
The fit was developed in conjunction with the publication
of the HERA\,I high $Q^2$ data. It was thus decided to use
the simplest heavy flavour treatment and consider all quarks
to be light.  As is illustrated  in  Figure~\ref{figpdfall}  the
four effective quark distributions and the gluon distribution $xg$
are rather accurately determined.
The accuracy achieved so far, at $Q^2=Q_0^2$, for $x=0.01, 0.4$ and $0.65$, is
$1\%,\,3\%, 7\%$ for the sum of up quark distributions and
$2\%,\,10\%, 30\%$ for the sum of down quark distributions, respectively.
It is to be noted that in this particular fit the gluon distribution
comes out to be rather steeply rising towards 
lowest $x$ at $Q^2_0 = 4$ GeV$^2$, 
$\propto x^{-0.9}$, a behaviour which in most other analyses
is not confirmed. 
%This behaviour may have to do with the
%neglect of heavy quark mass effects \cite{emppriv}.
While this
low $x$ peculiarity had been noted, the fit was still
accepted because of a large uncertainty of $xg$ in this region.
Also, in the higher $Q^2$ region 
for which this analysis was optimised, the fit is
in good agreement both with previous H1 results    \cite{h1alfas}
and with the results obtained in global fits 
by the MRST \cite{MRST02} and CTEQ \cite{CTEQ02} collaborations
(Figure~\ref{figpdfall}).

The larger $x$ domain is dominated by the valence quarks.
 In this analysis, these are extracted  from the differences
$u_v=U-\overline{U}$ and $d_v=D-\overline{D}$. Note 
that this implies the assumption that sea and anti-quarks are equal
which is contradicted by some non-perturbative QCD models.

In the H1 fit  the parton distributions  at the initial
scale $Q_0^2=4$\,GeV$^2$  are parameterised
as $xP =A_p x^{B_P} (1-x)^{C_P} \cdot f_P(x)$. The function $f_P$ is a polynomial
in $x$ which is determined by requiring ``$\chi^2$ saturation" of the fits,
i.e. starting from $f_P=1$ additional terms $D_Px$, $E_Px^2$ etc. are added
and only considered if they cause a significant improvement in $\chi^2$,
half integer powers were considered in \cite{benjamin}. 
The result for fitting the H1 data has been as follows: $f_g=(1+D_gx)$, 
$f_U=(1+D_Ux+F_Ux^3)$, $f_D=(1+D_Dx)$ and $f_{\bU}=f_{\bD}=1$.
The parton distributions at low $x$ are thus parameterised as
$xP \rightarrow A_P x^{B_P}$.
The strange (charm) anti-quark distribution is coupled to the total amount of
down (up) anti-quarks as $\bs=f_c\bD$ ($\bc=f_c\bU$). 
In the absence of more detailed information, two assumptions
have been made on the behaviour of the quark and anti-quark distributions
at low $x$. It has been assumed that quark and anti-quark distributions
are equal and, moreover, that the sea is flavour symmetric.
This implies that the slopes $B$ of all four quark distributions
are set equal, $B_U=B_D=B_{\bU}=B_{\bD}$. Moreover, the normalisations of
up and down quarks are the same, i.e. $A_{\bU}(1-f_c)=A_{\bD}(1-f_s)$,
which ensures that   $\bd/\bu \rightarrow 1$ as $x$ tends to zero.
If the normalisation and slope conditions are relaxed, the
uncertainties of the pdf's at lower $x$ becomes large \cite{mkbr}.
Some constraints can be inferred
from fixed target deuteron DIS data and will also be derived from
$W^{\pm}$ asymmetry data at the LHC. A high energy
collider $eD$ experiment would deliver the ultimate answer
as to how the sea at low $x$ is composed.

\subsection{Measurements of  $\bf {\alpha_s} $ in Inclusive DIS} 
\label{secalfinc}
The scaling violations at large Bjorken $x$ are due to gluon
bremsstrahlung. In the
valence quark approximation the DGLAP equation for $F_2$ takes a non-singlet
form, i.e. there is no contribution from the gluon to the scaling violations
which are solely regulated with $\alpha_s$. At low $x$, on the contrary,
the dominant process is boson-gluon fusion and the scaling violations
are governed by the product of $\alpha_s$ and $xg$. The HERA data, unlike
previous fixed target data at lower energy, therefore have the
potential of resolving the correlation between $\alpha_s$ and the 
gluon distribution for they span such a large range in $x$ and $Q^2$.
This has been exploited by ZEUS  and H1 to determine both
the strong coupling constant and $xg$ from the inclusive DIS
data using NLO perturbative QCD.

%\subsubsection{ZEUS Procedure and Results}
The ZEUS Collaboration has used its standard (S) fit and parameterisations
as described above. In addition to the ZEUS data, a number of fixed
target experiment data was used in addition (see \cite{zeuso}),
in order to have better accuracy for the high $x$ region and constraints
on the up-down quark asymmetry at medium $x$. The heavy quarks were
treated in a variable flavour scheme. 
%The $\chi^2$ definition used a Hessian matrix technique 
%which allowed the systematic errors to vary in the error analysis. 
Standard cuts were made on the  
input data with $Q^2 > 3$\,GeV$^2$ and $W^2 > 20$\,GeV$^2$
in order to be in the DIS region and avoid higher twist contributions,
from low $Q^2$ and large $x$, respectively. The input scale at which the
pdfs are parameterised was chosen to be $Q^2_0 =7$\,GeV$^2$.
The result obtained is 
\begin{equation} \label{alfaszeus}
\alpha_s(M_Z^2) = 0.1166 \pm 0.0049~(exp) \pm 0.0018~(model)
\end{equation}
where the experimental error comprises three uncertainties, $0.0008$
from uncorrelated error sources, $0.0032$ from correlated error
sources and $0.0036$ from normalisation uncertainties. The model
uncertainties stem from variations of the fit parameters,
$Q^2_{min},~x_{min}$ and $W^2_{min}$, and of variations of
quark and gluon distribution function parameterisations. The result
changes by 0.001 if the heavy flavours are treated in the fixed flavour
scheme. Special care was devoted to the effect of higher twists as
the fit included fixed target data at rather small $Q^2$ but large $x$.
With an $x$ dependent parameterisation of power corrections
$\propto Q^{-2}$, it was observed that \amz may be lowered by 0.0032.  
%

%\subsubsection{H1 Procedure and Results}
%
The H1 Collaboration developed a dedicated procedure \cite{h1alfas}
for the determination
of $\alpha_s$ which differed in several respects from their pdf fit described
above. The input data were restricted to H1 and BCDMS lepton-proton scattering 
data when it was found that the NMC data, extending to low $Q^2$ of a few GeV$^2$
introduced a significant dependence of the \amz result on $Q^2_{min}$
(see also \cite{vogtrom}). Since this requires the movement of 
the minimum $Q^2$ 
for the fixed target data up to about 10\,GeV$^2$, the accuracy of the
BCDMS data is so overwhelming that no further fixed target data has to be
included, apart from consistency checks \cite{h1alfas}.  The analysis also
excluded the low $y$ data of BCDMS from the analysis for systematic
reasons. The second feature
of the H1 fit was the attempt to reduce the number of quark distributions
to be fitted from the usual three to two. This was possible 
as the fit was essentially on $F_2$ only, which can be written 
as a sum over singlet and non-singlet contributions as
\begin{equation} \label{f2h1deco}
F_2 = \frac{2}{9} \cdot x \Sigma + \frac{1}{3} \cdot x \Delta = 
\frac{1}{3} \cdot xV + \frac{11}{9} \cdot xA.
\end{equation}
The $V,~A$ functions are given as
\begin{equation}
V = \frac{3}{4} [3 (U +\bU) - 2 (D +\bD)] = \frac{9}{4} u_v - \frac{3}{2}d_v
+ \frac{9}{2} \bu -3(\bd + \bs) 
\end{equation}
\begin{equation}
A = \frac{1}{4} [2 (D +\bD) - (U +\bU)] =  - \frac{1}{4} u_v + \frac{1}{2}d_v
- \frac{1}{2} \bu +(\bd + \bs)
\end{equation}
for 3 flavours as heavy flavours were generated according 
to boson-gluon fusion.
In the simplified approximation, $u_v=2d_v$ and $\bu=\bd=2\bs$, one finds
that $V$ and $A$ are valence and sea quark dominated functions, respectively,
as in this approximation $V=3u_v/2$ and $A =\bu$. Despite
the reduction in the pdf space, the fit could impose the usual
momentum and quark counting rules and also a non-zero $\bu -\bd$
difference. The result obtained is 
\begin{equation} \label{alfash1}
\alpha_s(M_Z^2) = 0.1150~~\pm~~0.0017~(exp)~~^{+~~0.0009}_{-~~0.0005}~(model).
\end{equation}
The first error represents the experimental uncertainty of the data sets. 
 The second
error includes all uncertainties associated with the construction of
the QCD model for the measured cross section. A
number to be noted may be that a 100\,MeV uncertainty on the value of the charm
mass results in an \amz uncertainty of 0.0005. 
%

%\subsubsection{The Effect of Scale Uncertainties}
%
The experimental results of ZEUS and H1 are consistent, and with more
accurate and the combined data sets a further improved
measurement will be reached.
A rather large theoretical uncertainty of the NLO analysis, however, results
from the choices of the renormalisation scale $\mu_r^2 = m_r \cdot
Q^2$, and of the factorisation scale $\mu_f^2
= m_f \cdot Q^2$ which leads to scale dependent parton distributions.
In the $\overline{MS}$ renormalisation scheme
both scales are set equal to $Q^2$, i.e.
$m_r=m_f=1$. The
effect of both scales on \as was estimated by both ZEUS and H1
through variations of the scale factors
$m_r$ and $m_f$. 

The ZEUS Collaboration \cite{zeusalfas} has varied $m_{r,f}$ by factors
between 0.5 and 2. This caused shifts in \amz of $\simeq \pm 0.004$.
Variations between 0.25 and 4, as are often common, resulted
in unacceptable $\chi^2$ variation effects and were thus not considered.
Similar results were obtained by H1 \cite{h1alfas} as are summarized in
Table~\ref{scales}.
\begin{table}[h]
  \begin{center}
\begin{minipage}{16.5 cm}  
    \caption{ Dependence of  \amz in the H1 fit on the renormalisation
     and factorisation scales $m_f$ and $m_r$, respectively,
     expressed as the difference of \amz obtained for 
     scales different from one and the central value of \amz=0.1150.
    The combination $m_f=4$ and $m_r=0.25$
    is abandoned since the splitting function term $\propto \ln{(m_r/m_f)}^2$ 
    becomes negative at low $Q^2$ which causes a huge
    increase of $\chi^2$.} 
    \label{scales}
\vspace{1.0cm}
\end{minipage}
    \begin{tabular}{|l|l|l|l|l|}
\hline
          &     $m_r = 0.25$      &   $m_r = 1$    &     $m_r =  4$     \\
\hline
 $m_f = 0.25 $     &  $-0.0038 $     & $-0.0001 $    &   $  +0.0043 $   \\
 $m_f = 1 $        &  $-0.0055 $    &  $      --     $    &   $  +0.0047 $   \\
 $m_f = 4 $       &   $ -- $    &  $+0.0005 $    &   $  +0.0063 $   \\
\hline
    \end{tabular}
  \end{center}
\end{table}

The numerical size of the effects depends on the data and $Q^2_{min}$,
the H1 and ZEUS observations are thus found to be consistent.
In agreement with previous studies~\cite{vogtrom} it was observed that the
renormalisation scale causes a much larger uncertainty on \amz than
the factorisation scale. Depending on which set of $m_r$ and $m_f$ is
chosen, the obtained $\chi^2$ typically
differs by a  number much larger than 1. This suggests
that the assumed variations of the scales are too large. It has to be
stressed again that while there is an arbitrariness in the scale choices,
which is inherent to the renormalisation group equation,
the actual prescription of variations by typically $1/4$ to $4$ times
the scale is $ad~hoc$ and needs to be revisited.
% This is especially
%relevant as the experimental accuracy will soon be better, by a factor
%of 5 or so compared to the so-called theoretical uncertainty.
The scale dependence gets diminished  in NNLO \cite{moch},
yet future determinations of \amz from HERA data and possibly 
from new colliders \cite{mkdis07} will push the experimental 
accuracy to a per cent or per mille, respectively.
In order to take advantage of such small experimental uncertainties, 
a new understanding of theoretical
uncertainties--particularly that of the scale uncertainties--will be 
necessary.
%Thus a new accord on how to treat the scale uncertainties is essential.

\clearpage
%\newpage
\section{Jet Measurements}
\label{chjet}
\subsection{Theoretical Considerations}
%The theoretical predictions of jet production are a large subject requiring
%a review in its own right.  Here, a very brief outline of the basic concepts
%will be given.  The readers are referred to the following references for
%further reading~\cite{rkellis_4.1}.

According to the QCD improved parton model, a DIS differential 
cross section, $d\sigma$ can be written as,
\begin{equation}
 d\sigma_{ep} = \sum_{a=q,\bar{q},g} \int 
 dx f_a(x,\mu_F^2;\alpha_s) \cdot
d\hat{\sigma}_a(xP,\alpha_s,\mu_R,\mu_F).  \label{eq_4.1}
\end{equation}
This has the form of a convolution of a partonic, or hard, cross-section
$d\hat{\sigma}$ with the proton PDF $f_a$.  It should be noted
that both $d\hat{\sigma}$ and $f_a$ depend on $\alpha_s$ which depends
on $\mu_R$.

In order to produce theoretical predictions in practice,
semi-analytical programmes such as DISENT~\cite{catani_4.1} 
MEPJET~\cite{mirkes_4.1} or 
DISASTER~\cite{graudenz_4.1} are used.
The partonic cross-section, $d\hat{\sigma}$, above a specified
scale $\mu_R$, is calculated to NLO in $\alpha_s$ by these programs.
The PDF of the proton, $f_a$, is normally taken from the results of
global fits 
such as the MRST~\cite{mrst_4.1} or CTEQ~\cite{cteq_4.1} sets of PDFs.  
These global
sets do not include the HERA jet data in their fits. These PDF fits are made
using a fixed value of $\alpha_s$, but PDFs are, in principle, a function
of $\alpha_s$, as indicated in Equation~\ref{eq_4.1}.  

In order to produce jet cross-sections from the partonic
cross-sections produced by DISENT and DISASTER, 
a jet algorithm\footnote{Jet algorithms produce a small number of ``jets'' from 
a larger number of final state objects.  The four-momenta of jets should correspond
closely to those of hard partons.  Discussions of theory and application of jet 
algorithms at collider experiments may be found in~\cite{chekanov_4.1,blazey_4.1}.}, 
the inclusive $k_T$ algorithm~\cite{catani_4.2,ellis_4.1} in case 
of HERA jet analyses, 
are run on the final partons to produce the ``parton-level'' jet 
predictions to NLO.  The actual measurements, however, are not
made at the parton-level, so additional corrections are necessary.

%In practice, the theoretical predictions are produced 
%in a two step process. The first step is to produce the distributions of
%partons and the final state hadrons using the applicable QCD theory 
%or model.  The second step is to apply the jet finding algorithm. 
%In case the algorithms are applied to the partons before hadronisation,
%the found jets are called ``parton-level'' jets.  In case they are applied
%to the final state hadrons, they are called ``hadron-level'' jets.

Monte Carlo(MC) programs currently available for DIS
event generation, such as LEPTO~\cite{ingelmann_4.1} or
ARIADNE~\cite{loennblad_4.1,loennblad_4.2}, only contain the LO matrix elements.  They
are, however, supplemented by ad-hoc additions of NLO processes
as well as a simulation of higher order processes through 
the Parton-Shower model.  In addition, the partons are ``hadronised''
using the LUND~\cite{andersson_4.1} model.  Thus, after running the jet algorithm 
on the events produced by MC programs, both ``parton-level'' and 
``hadron-level'' MC predictions can be obtained.  

In order to obtain the ``hadron-level'' prediction to order $\alpha_s^2$,
a multiplicative ``hadronisation correction'' is applied to 
the parton-level prediction.  In most cases the hadronisation correction 
is the ratio of hadron-level to parton-level MC predictions.  
While there is no rigorous justification for this procedure, it is a 
reasonable one if a) the parton distributions in the MC are close to those of
the fixed order calculation, b) the MC reproduces the experimental 
measurements reasonably well and 
c) the hadronisation correction is small (5-10\%).  Normally at least
two MC programs are compared and used to evaluate a systematic 
uncertainty due to this procedure.

It is convenient to symbolically write the above procedure to obtain
the jet cross sections in DIS as:
\begin{equation}
 d\sigma^{jet}_{ep} = \sum_{a=q,\bar{q},g} \int 
 dx f_a(x,\mu_F^2;\alpha_s) \cdot
d\hat{\sigma}_a^{jet}(xP,\alpha_s(\mu_R),\mu_R,\mu_F) \cdot
(1+\delta_{had}),            \label{eq_4.2}
\end{equation}
where $d\hat{\sigma}_a^{jet}$ is now the ``jet'', rather than the
partonic, hard cross-section and $\delta_{had}$ is the hadronisation 
correction.

\begin{figure}[tb]
\begin{minipage}[t]{8 cm}
\hspace{20pt}
\epsfig{file=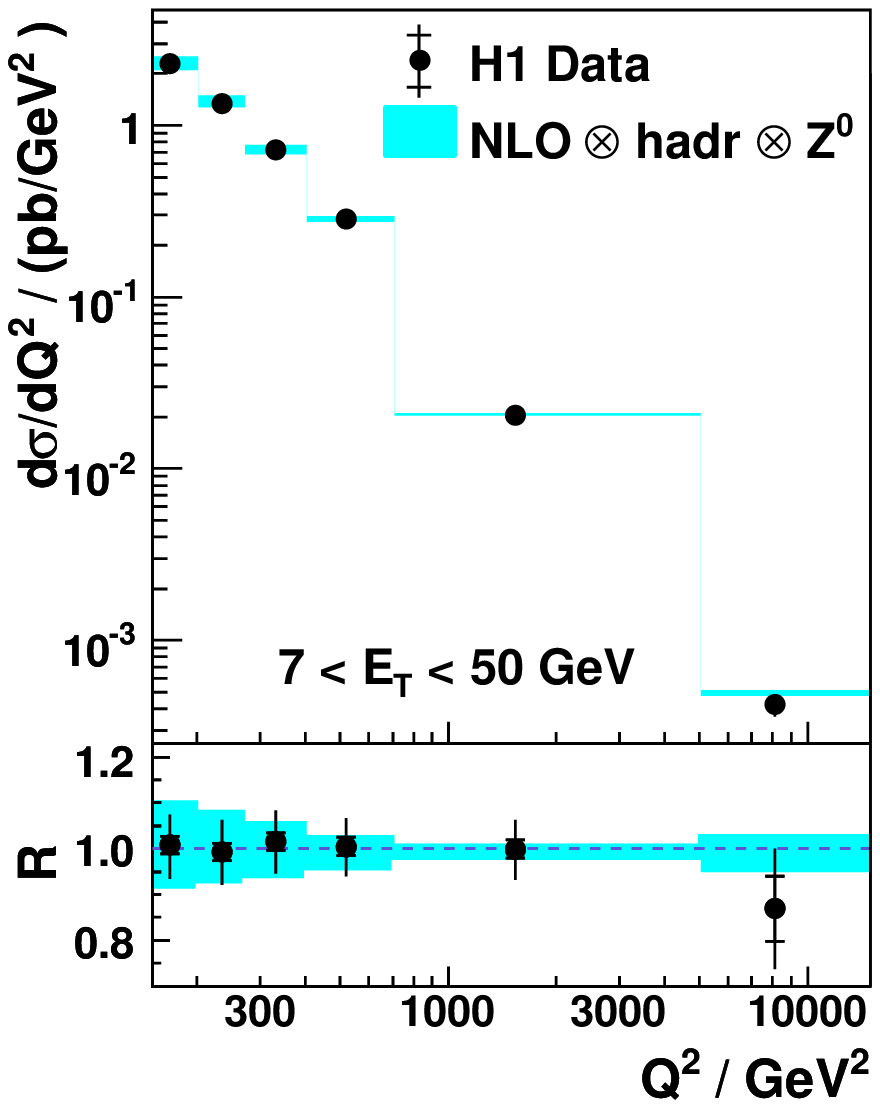,scale=0.75}\epsfig{file=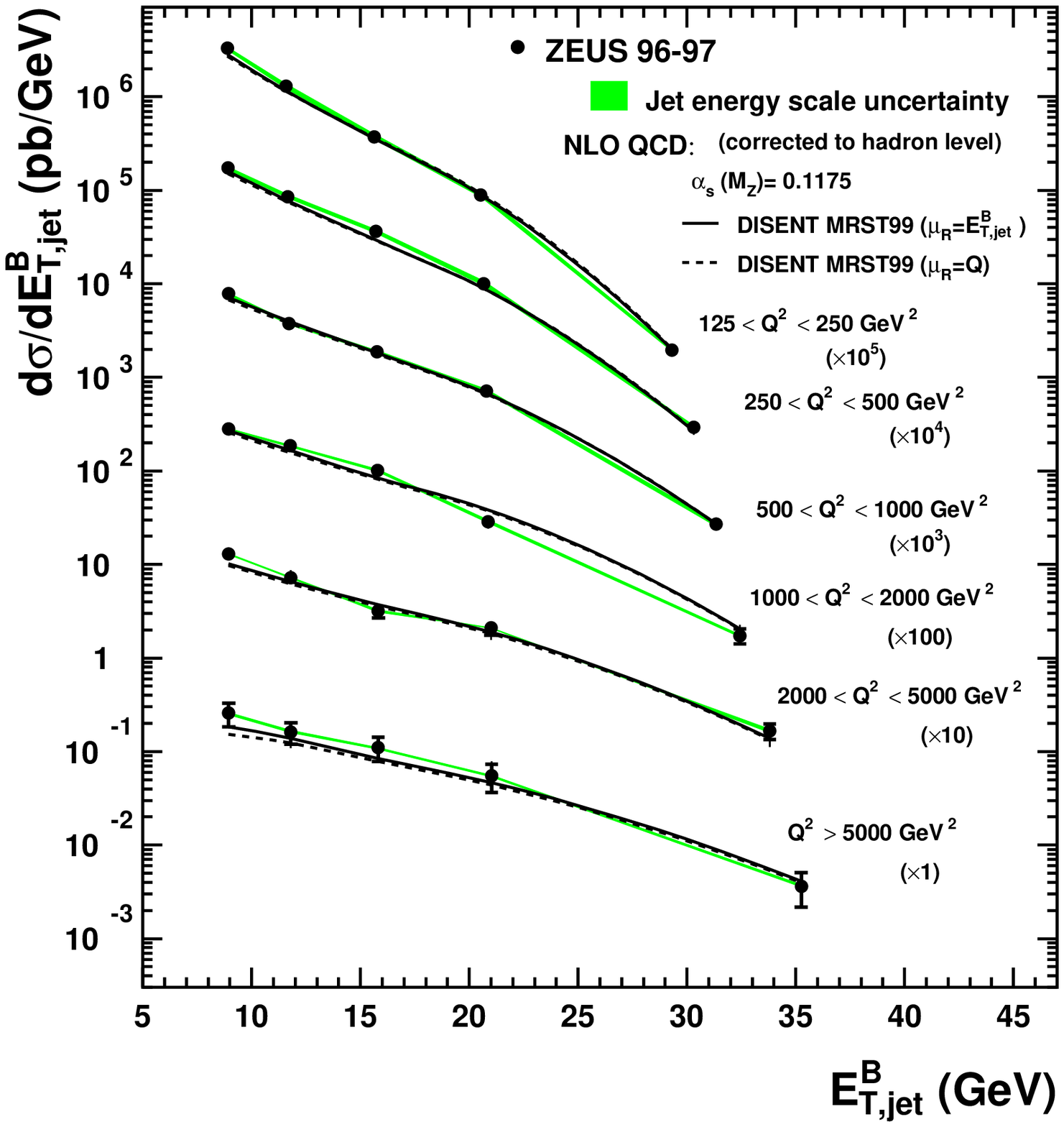,scale=0.5}
\end{minipage}
\vspace{-10pt}
\begin{center}
\begin{minipage}[t]{16.5 cm}
\caption{Measurement of jet cross-sections at HERA. The filled dots
are the data.  The inner error bars represent the statistical uncertainty.
The outer error bars show the statistical and systematic uncertainties
%not associated with the uncertainty in the absolute energy scale of the
%jets.
in quadrature.  
The NLO QCD calculation, corrected for hadronisation effects (and $Z^0$ exchange when
the calculations do not take electroweak effects into account)
is shown.  Left-top) The differential
cross-section $d\sigma/dQ^2$ for the inclusive jet production.  Left-bottom) the
ratio between the measured $d\sigma/dQ^2$ and the NLO QCD calculation.  The
shaded band displays the total theoretical uncertainty.  
%The shaded band
%in Left-bottom) shows the magnitude and uncertainty of the parton-to-hadron
%correction used to correct the NLO QCD predictions. 
Right) The differential
cross-section $d\sigma/dE_{T,jet}^B$ in bins of $Q^2$, where $E_{T,jet}^B$
is the transverse energy of the measured jet in the Breit frame~\cite{Breit}.
\label{fig_4.1}
}
\end{minipage}
\end{center}
\end{figure}

In the case of photoproduction, where 
$Q^2 \approx 0$ (i.e. an almost real photon collides 
with the proton), the photon PDF, as well as the proton PDF is factorized.
Thus, the photoproduction ($\gamma p$) jet cross-section can be written as:
\begin{equation}
 d\sigma^{jet}_{\gamma p} = \sum_{a=q,\bar{q},g}\sum_{b=q,\bar{q},g} \int \int
 dx_p f_a(x_p,\mu_F^2;\alpha_s) \cdot
 dx_\gamma f_b(x_\gamma,\mu_F^2;\alpha_s) \cdot 
d\hat{\sigma}_a^{jet}(x_p,x_\gamma,\alpha_s(\mu_R),\mu_R,\mu_F) \cdot
(1+\delta_{had}),        \label{eq_4.3}
\end{equation}
where $f_b$ is the PDF of the photon, $x_\gamma$ is the Bjorken $x$
of a parton of the photon, and the convolution extends over
partons $a$ in the proton and $b$ in the photon.

At LO (i.e. to zero-th order in  $\alpha_{s}$), the photoproduction
 process can be thought
of as being in two pieces;  the ``direct'' process in which the
 photon couples directly
to the proton ($x_\gamma = 1$), and the ``resolved processes''
 where a parton from
the photon interacts with the proton ($x_\gamma < 1$). 
 While this distinction does not
have a strict meaning at higher orders, it is convenient
 to distinguish these two processes
defined by means of a cut on $x_\gamma$, usually at 0.75 or 0.8
 by convention.  The ``direct'' process
loosely corresponds to the case where $f_b = 1$, and is 
not very sensitive to the hadronic
nature of the photon through its PDFs.

The NLO theoretical prediction for photoproduction is obtained in much the same way as
in the case for DIS.  The NLO analytical programme often used is that of 
Frixione and Ridolfi~\cite{frixione_4.1}.  The most usual LO MC programs are 
PYTHIA~\cite{pythia_4.1} and HERWIG~\cite{herwig_4.1}.

Jet production in the region of transition between photoproduction 
and DIS has also been extensively studied at HERA. The reader is referred to 
the following HERA results for further information on this subject~\cite{h1_zeus_4.1}.

\subsection{Jet Cross-Section Measurements}

The experimental jet  measurements are
made at the ``detector-level''. This means that the energy deposits
in the calorimeters and tracks found in tracking detectors are
subjected to the jet algorithms to determine the jets and their
four momenta.  The jet algorithm used is the same as that used in the
theoretical predictions described in the previous section, i.e. the
inclusive $k_T$ algorithm.

A Monte Carlo simulation is used to correct the measured detector-level
jet cross-sections to the hadron-level cross-sections, which can be
compared to the theoretical predictions.  Events generated by MC generators
such as LEPTO or ARIADNE for DIS, and PYTHIA and HERWIG for photoproduction,
are processed through detector simulations based on GEANT~\cite{geant_4.1} to give
a sample of fully simulated events.  The correction 
factors are calculated using the difference between the detector- and 
hadron-level event rates of the fully simulated event samples.  The
event selection and kinematic cuts are devised in such a way to keep
these corrections relatively small.  Typically the corrections are $< 30\%$.

%The measured jets at the detector-level is then
%corrected, using the ratio of detector-level to hadron-level 
%MC predictions.  
%In this way, measured hadron-level jet cross-sections
%are obtained.

\begin{figure}[tb]
\begin{minipage}[t]{8 cm}
\hspace{25pt}
\epsfig{file=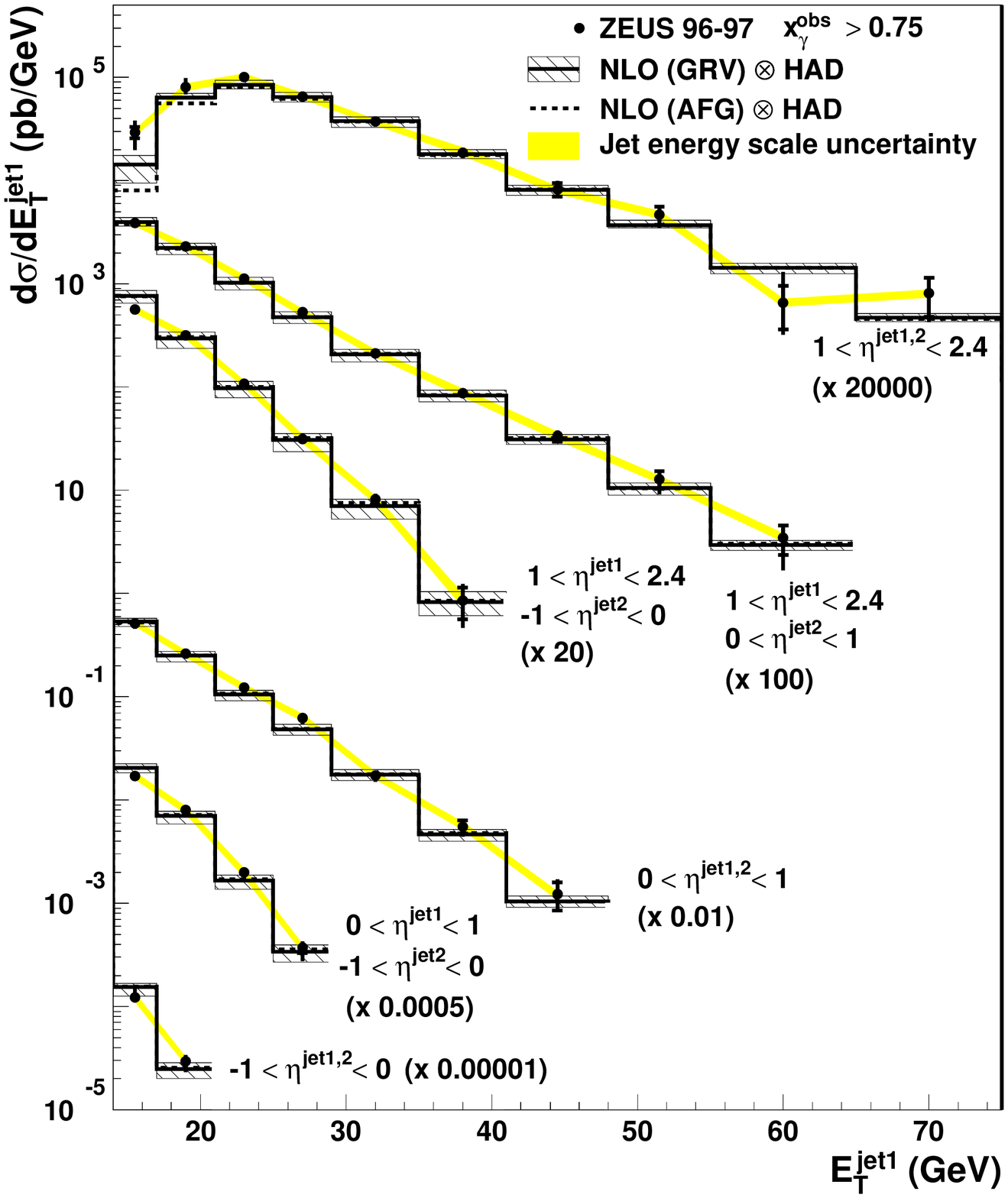,scale=0.4}\epsfig{file=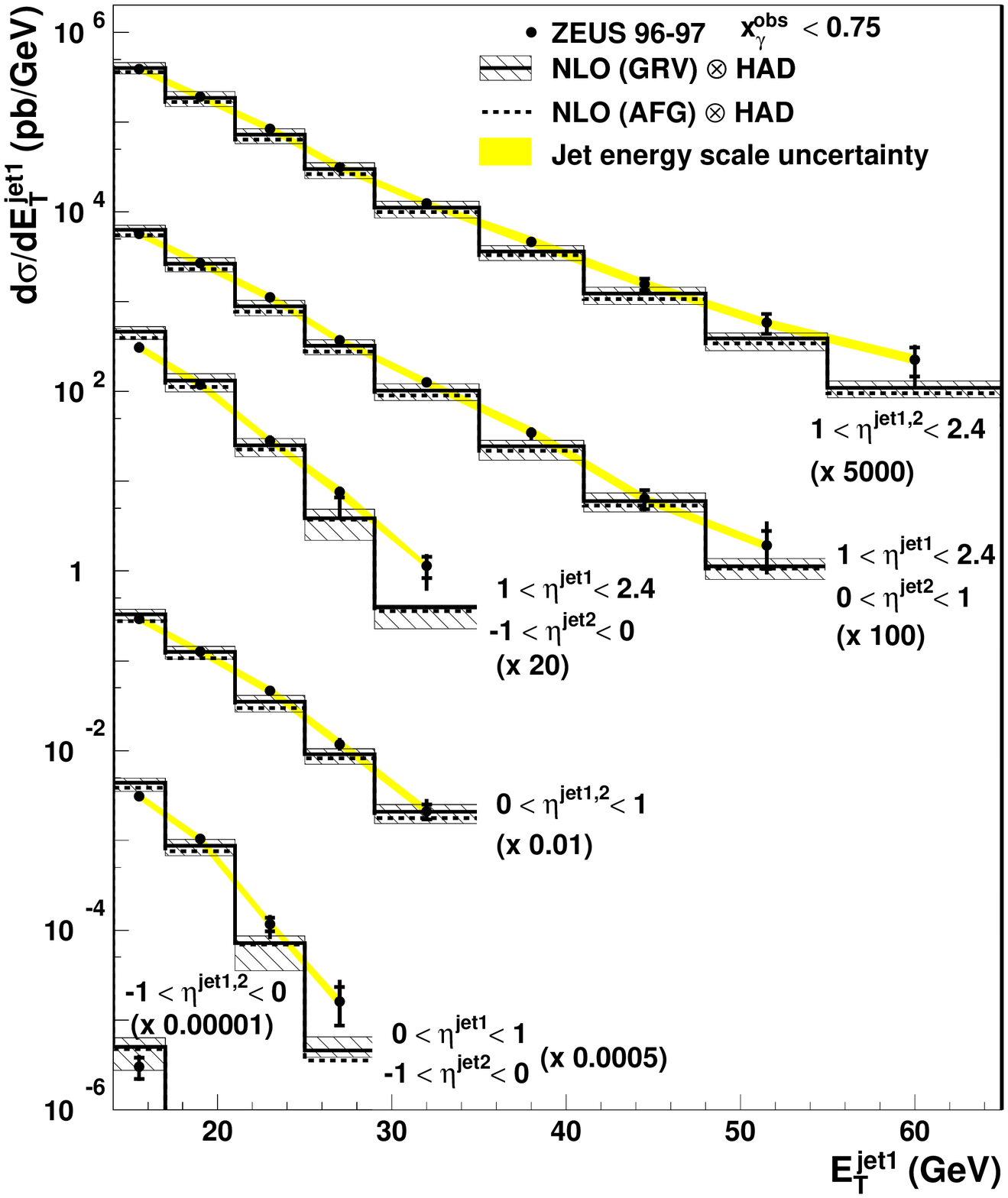,scale=0.4}
\end{minipage}
\begin{center}
\begin{minipage}[t]{16.5 cm}
\caption{The measured photoproduction dijet cross-sections differential
in $E_{T,jet1}$, the transverse energy of the highest energy jet, is shown
separately for $x_\gamma > 0.75$ (left) and $x_\gamma < 0.75$ (right). 
%The presentation
%of the data is the same as that for Figure~\ref{fig_4.1}. 
 The NLO prediction 
corrected for hadronisation effects is shown calculated using the
GRV-HO and CTEQ5M1 PDFs for the photon and the proton, respectively.
The hatched band represents the theoretical uncertainties.  The
prediction using an alternative photon PDF, 
AFG-HO~\cite{aurenche_4.1}, is also shown.
\label{fig_4.2}
}
\end{minipage}
\end{center}
\end{figure}

%The HERA jet measurements, as a rule, is reported at the hadron-level,
%and the comparison to theory is also made at the hadron-level.  This
%is so that future advances in the understanding of hadronisation
%process could be applied to the published measurements without having
%to reproduce the current hadronisation corrections.

Figure~\ref{fig_4.1} shows the measurement of inclusive jet cross-section in 
DIS~\cite{zeus_4.1,h1_4.21}, in this case as a function of $Q^2$.  
The uncertainties of the measurement, except at the highest $Q^2$, where
statistical errors are large, are dominated by the jet energy scale
uncertainties.  The energy uncertainties are 1 to 2\% at HERA experiments which
translates to 5 to 10\% uncertainties in the cross-sections.

\begin{figure}[tb]
\begin{minipage}[t]{8 cm}
\hspace{30pt}
\epsfig{file=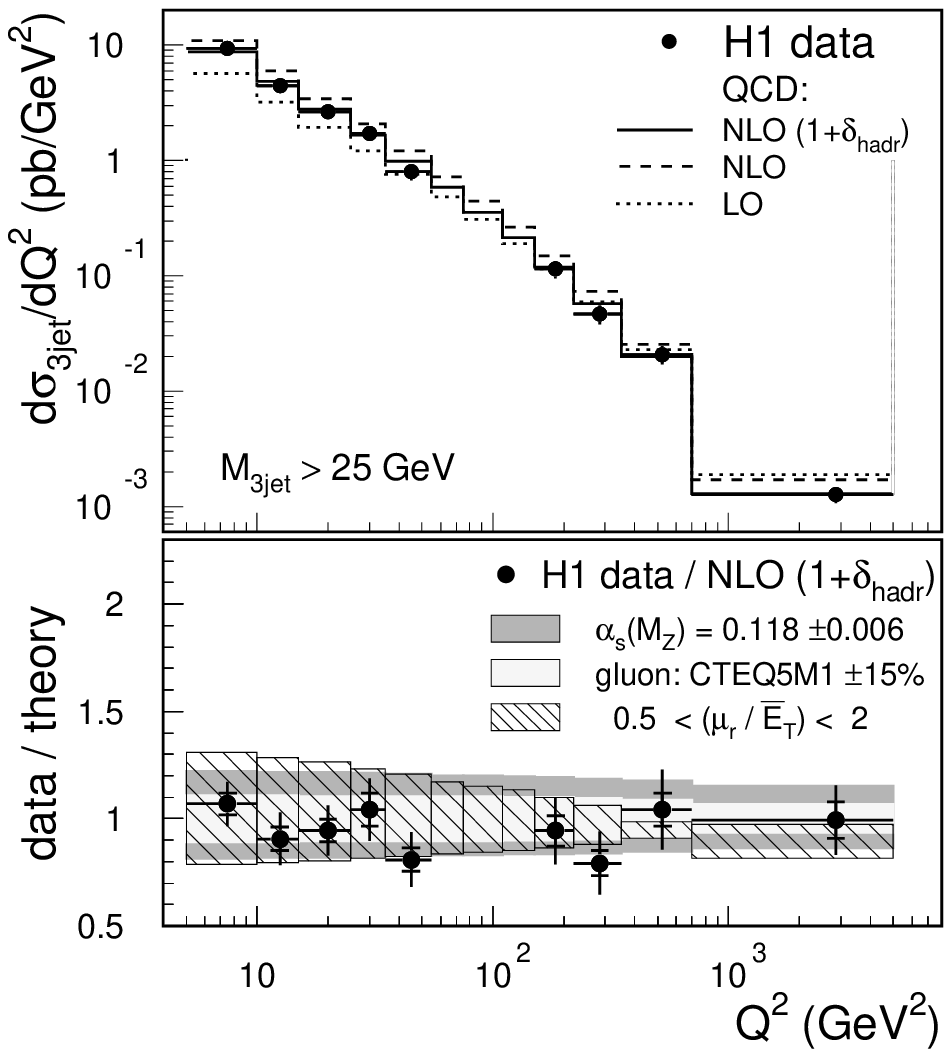,scale=0.74}\epsfig{file=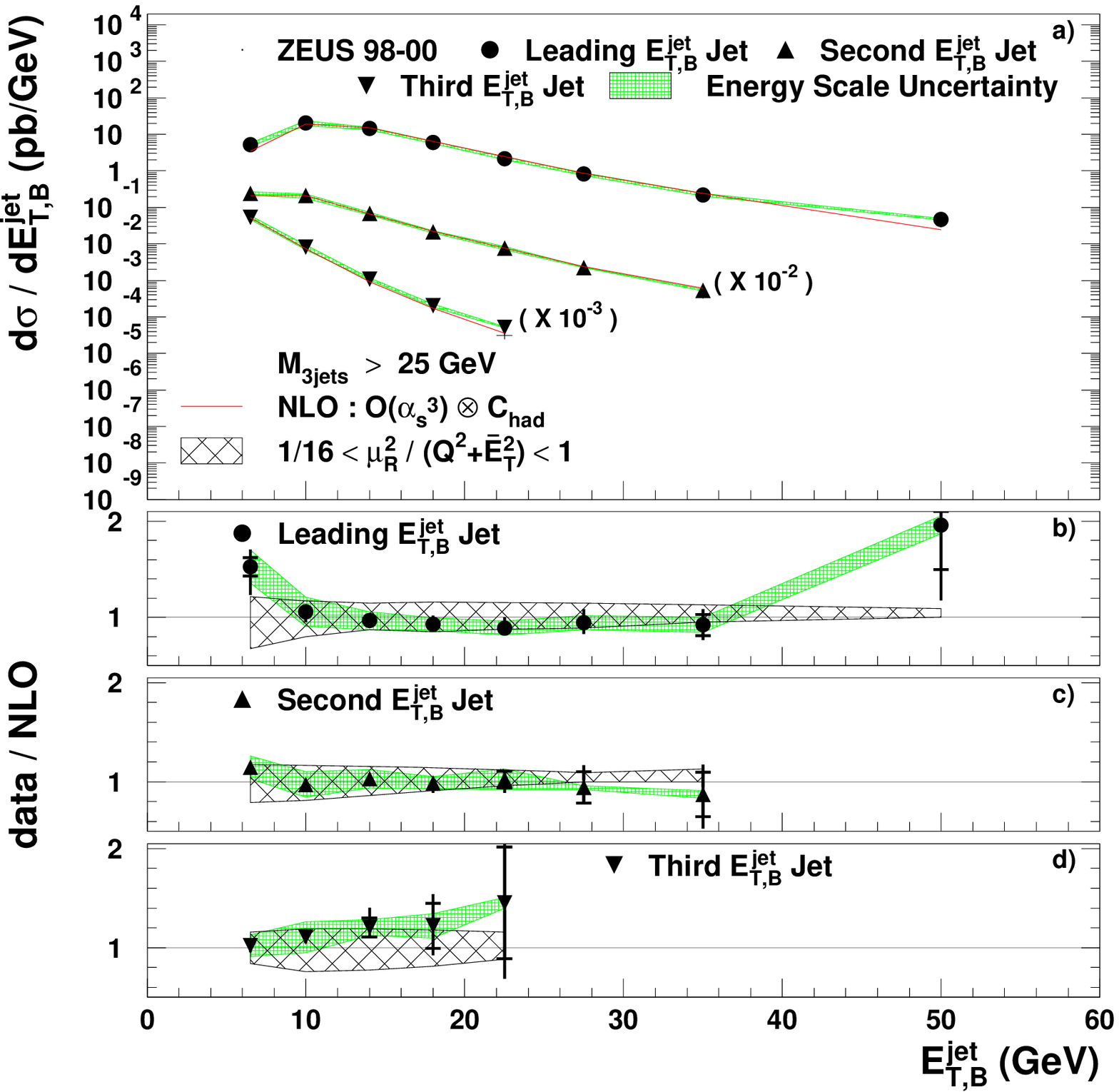,scale=0.45}
\end{minipage}
\begin{center}
\begin{minipage}[t]{16.5 cm}
\caption{Left: The inclusive trijet cross-section measured as a function
of $Q^2$.  The prediction of pQCD in leading order (dotted line) and
in NLO (solid line) and without hadronisation corrections (dashed line)
are compared to the data.  Also shown is the ratio of the measurement and
the theoretical prediction.
Right: a) The inclusive trijet cross-sections as a function
of $E_{T,B}^{jet}$ with the jets ordered in $E_{T,B}^{jet}$.  The cross 
sections of the second and third jet were scaled by factors hown inthe
figure for readability.  The NLO prediction uses CTEQ6 PDFs of the protons.
The other details are the same as in Figure~\ref{fig_4.1}. b) c)
and d) are ratios of the measurements to the NLO predictions. 
\label{fig_4.3}
}
\end{minipage}
\end{center}
\end{figure}

The measured cross-section in Figure~\ref{fig_4.1}(right), for example, is compared to 
the predictions of NLO QCD,
in this case based on the DISENT programme.  The inputs to the calculation
are $\alpha_s$ for the partonic cross-section, chosen to be 0.1175, 
the factorisation and renormalisation scales (set to be 
either $Q$ or $E_T^{jet}$) and the proton PDF, chosen
to be the version of MRST99~\cite{mrst_4.2} extracted with $\alpha_s$ set to 0.1175.  
The bottom plot shows the hadronisation correction,
or parton-to-hadron corrections, which converts the parton-level predictions
of DISENT to the hadron-level predictions to be compared to data.  This
particular correction was determined using ARIADNE and LEPTO MC programmes. 
The comparisons to the theory are generally satisfactory, and the residual
difference can be attributed to the input parameters of theory, namely
$\alpha_s$ and the PDF of the proton. 
%The agreement between data and 
%theoretical predictions are very good, and the remaining difference
%can be attributed to theoretical uncertainties including that 
%of $\alpha_s$.

\begin{figure}[tb]
\begin{center}
\begin{minipage}[t]{18 cm}
\hspace{25pt}
\epsfig{file=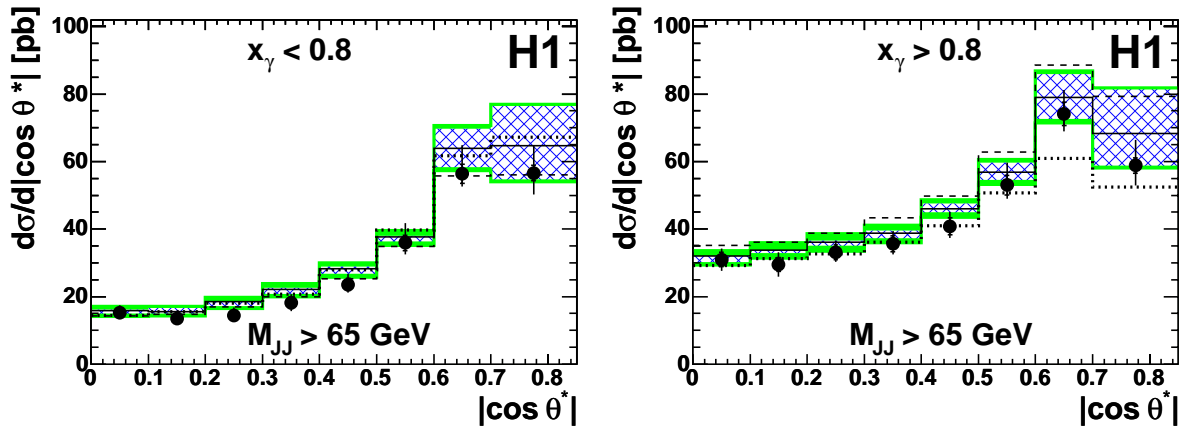,scale=0.8}
\end{minipage}
\begin{minipage}[t]{16.5 cm}
\caption{Photoproduction dijet cross-section as a function of $|cos\theta^*|$
for events with the dijet mass greater than 65 GeV.  
The QCD NLO calculations with (solid line) and
without (dashed line) hadronisation 
corrections, $\delta_{had}$ and for the 
PYTHIA Monte Carlo (dotted ine) scaled by a 
factor 1.2 are shown.  The inner (hatched) band is the scale uncertainty the outer (shaded)
band is the total uncertainty.  The cross-section are shown for the two regions of $x_\gamma$
enhancing the resolved (left) and direct (right) photon contribution. 
\label{fig_4.4}
}
\end{minipage}
\end{center}
\end{figure}

Very similar considerations apply to the study of photoproduced jets
at HERA. One difference is that the PDF of the photon is an additional
input to the theoretical prediction as discussed above. 
Figure~\ref{fig_4.2} shows the comparison of the theory and data~\cite{zeus_4.2} for photoproduced 
dijets for the ``direct'' and
``resolved'' samples.  Many calculations of jet photoproduction at NLO
exist, all of which have been compared to each other and agree to within 
5--10\%.  The calculations shown here are due to Frixione and Ridolfi~\cite{frixione_4.1}.
The photon PDF used in this case is GRV-HO~\cite{glueck_4.1,glueck_4.2}.

\begin{figure}[tbh]
\begin{minipage}[t]{10 cm}
\hspace{5pt}
\epsfig{file=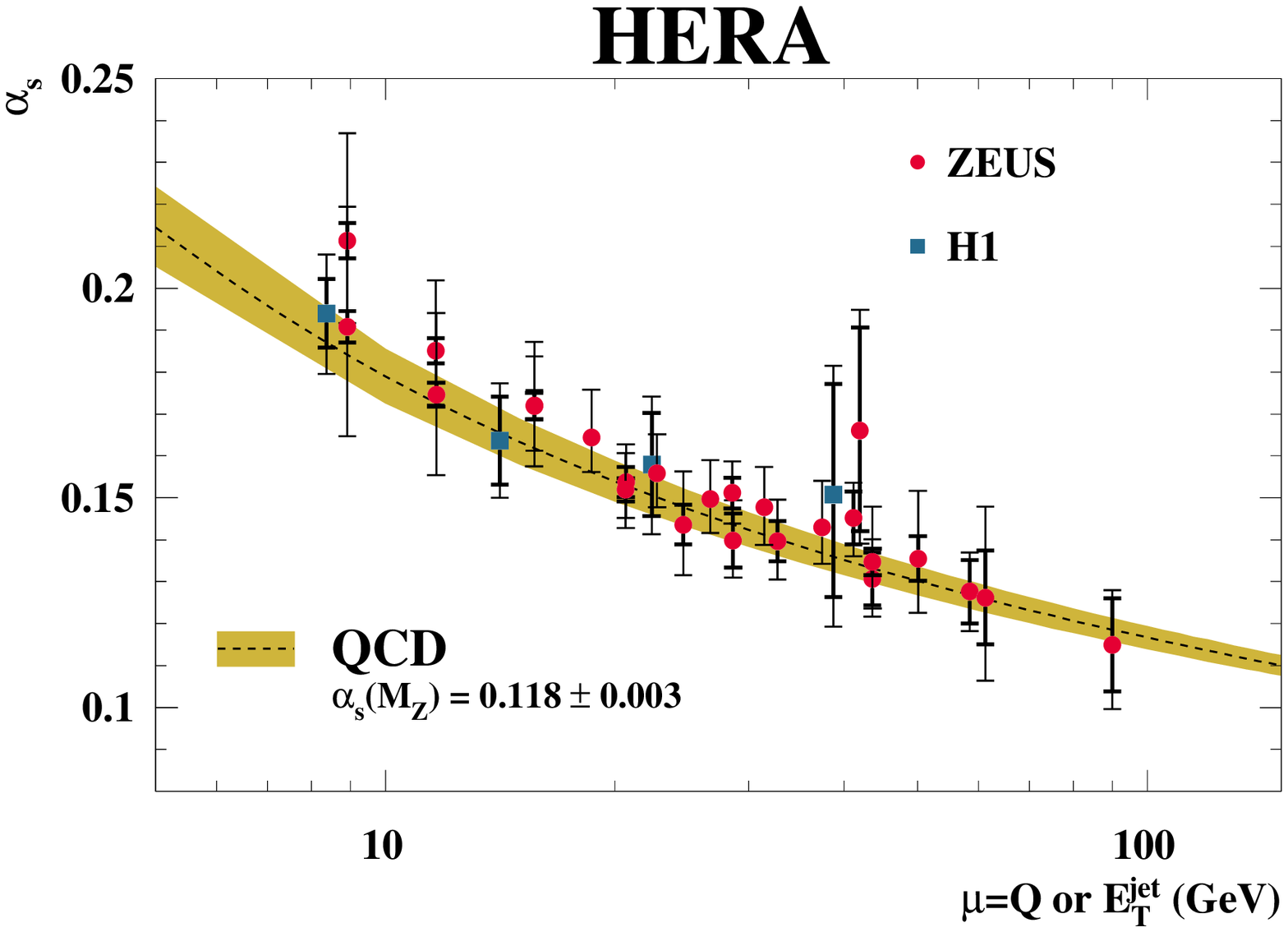,scale=0.45}\epsfig{file=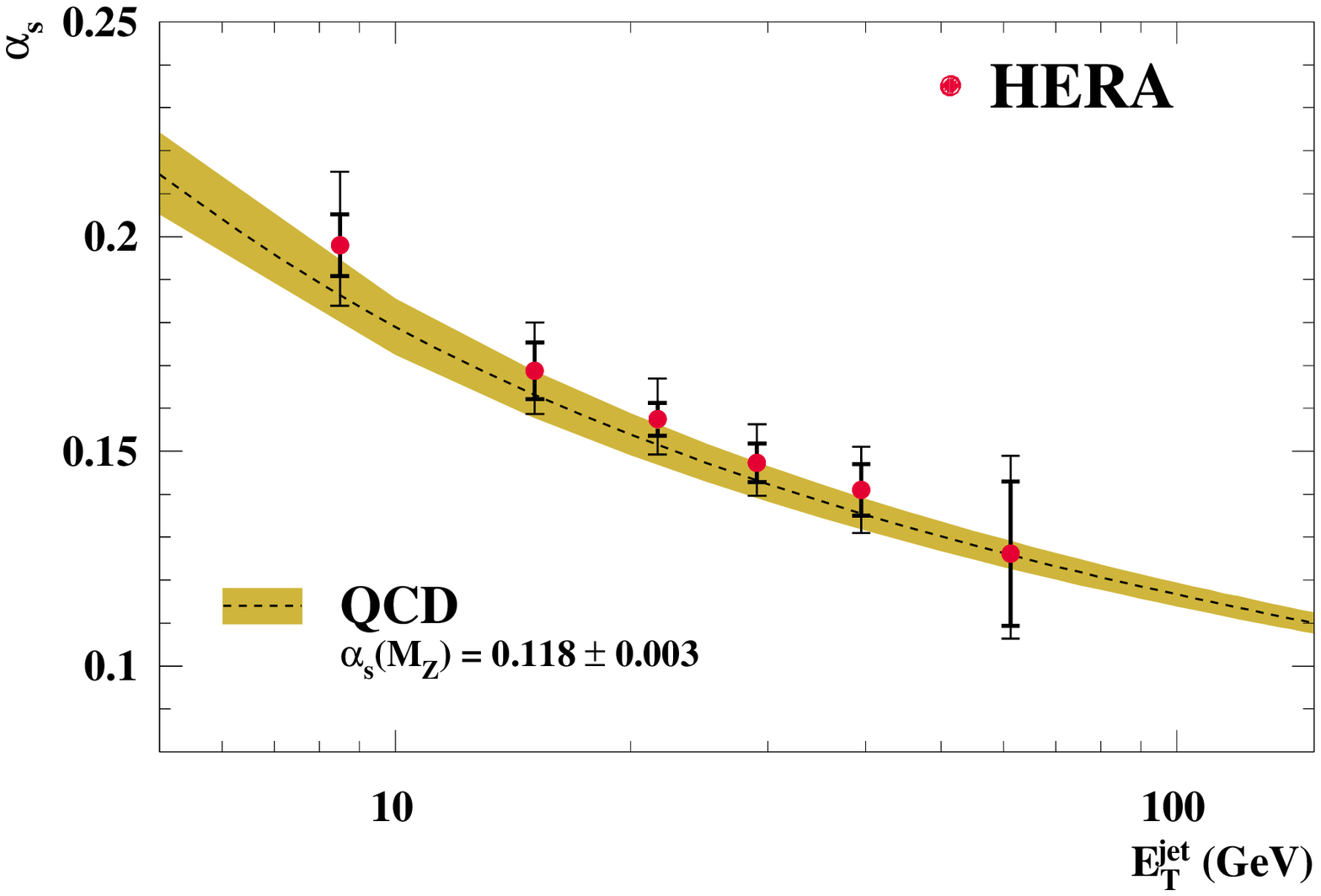,scale=0.45}
\end{minipage}
\vspace{-40pt}
\begin{center}
\begin{minipage}[t]{16.5 cm}
\caption{Left: A compilation of $\alpha_s(E_{T,jet}, Q^2)$ measurements at HERA
determined from the QCD fit of the measured $d\sigma/d(E_{T,jet},Q^2)$ as 
a function of $E_{t,jet}$ or $Q^2$. The band shows the current world average
(excluding HERA data).  Right: The combined HERA measurements from~\cite{glasman_4.1}.
The dashed line and the band shows the result of a global analysis (not using the
HERA data) from \cite{bethke_4.1}.
\label{fig_4.5}
}
\end{minipage}
\end{center}
\end{figure}

The DIS inclusive and photoproduction dijet measurements discussed 
in some detail above are two of the most basic jet measurements at HERA.  The
follow-up measurements to these can be found in \cite{h1_4.2,h1_4.21,zeus_4.21,zeus_4.22}.
An example of a more complex measurement is that of multijet
in DIS which is shown in Figure~\ref{fig_4.3}~\cite{h1_4.1,zeus_4.3}.  
The NLO calculations describe
the data satisfactorily.  There exist
  many more detailed measurements which are
listed here~\cite{zeus_4.31,h1_zeus_4.11,h1_zeus_4.2} for reference.    

In addition to the measurements of cross-sections, there are 
tests of pQCD that can be performed at HERA using jet measurements.  
An example is the angular distribution of photoproduced
dijets.  Figure~\ref{fig_4.4} shows a measurement by the H1 Collaboration~\cite{h1_4.2} 
of the photoproduction dijet cross-section differential in cosine
of the angle of dijets, $|\cos{\theta^*}|$, 
in their centre-of-mass, with respect to the beams.
The difference of steepness of 
the $|\cos{\theta^*}|$ dependence between the cross-sections from resolved and
direct samples is predicted by the dominance of gluon exchange in the
resolved process and is related to the spin of gluons and quarks.

\subsection{Tests of pQCD and Determination of $\alpha_s$}

The strong coupling constant $\alpha_s$ is extracted by
a fit of the theoretical predictions to the data with $\alpha_s$ as
a parameter of the fit.  
%The correlation between the
%$\alpha_s$ in the hard cross-section $\hat{\sigma}$ and the
%PDFs $f$ must be handled consistently.
At HERA, the dependence of PDFs on
$\alpha_s$ may be handled consistently
by a use of sets of PDFs determined at different
$\alpha_s$ such as the MRST99 set.  The best fit is found by simultaneously
varying $\alpha_s$ in the hard cross-section as well as the PDFs. The 
PDFs determined at discrete values of $\alpha_s$ are used to interpolate to
the intermediate values.  In this way, $\alpha_s$ is obtained in
a consistent manner taking the correlations into account. 

The extraction of $\alpha_s$ can be done in bins of $E_T$ (or $Q^2$) to
show its running as a function of the renormalisation scale.  This
is shown in Figure~\ref{fig_4.5}(left)
%for the measurement of inclusive DIS jets from ZEUS.
for a variety of jet measurements at HERA~\cite{h1_zeus_4.2}.
In all measurements of $\alpha_s$ from jets at
HERA, theoretical uncertainties dominate the systematic error, the
largest of these are due to unknown higher order contributions, evaluated by
a variation of the renormalisation scale, $\mu_R$.

%The $\alpha_s$ can be obtained from a variety of jet 
%measurements at HERA~\cite{h1_zeus_4.2}.  
A compilation of the $\alpha_s$ values has been made;
a combined HERA $\alpha_s$ has been derived~\cite{glasman_4.1} and is shown in 
Figure~\ref{fig_4.5} (right).  The overall experimental 
uncertainty of the combined measurement
is 1\%, the theoretical uncertainty, which is dominated by the scale
uncertainty of the NLO theory, is about 5\%.  As was noted in the inclusive
case, there is no firm procedure which would define the theoretical
uncertainty which remains ad-hoc to a large extent.
This combined HERA measurement has been 
used in a determination of the world average of the
 strong coupling constant~\cite{bethke_4.1a}
of 0.1189~$\pm$~0.0010.  Recently, a determination of
 the strong coupling constant
was made using a simultaneous fit of the HERA data from 
\cite{h1_4.21} and \cite{zeus_4.31} to give
a value of 0.1198~$\pm$~0.0019(exp)~$\pm$~0.0026(th) 
\cite{h1_zeus_4.3}. In this analysis only data at large scales  were
used. This halved the theoretical error but enlarged the 
experimental one. Whether theoretical uncertainties can be settled
in a more satisfactory way remains to be seen.
%This uncertainty
%should be compared to that of the world average from~\cite{bethke_4.1}
%(using only measurements for which NNLO thoery is available) which is 
%at the level of 3\% in total.

%systematic
%uncertainties, in particular PDF uncertainties, cancel to a certain
%extent when measuring ratios of jet cross-sections.  
%Figure X shows the ratio of three- to two-jet measurements in NC DIS.  
%
%\begin{figure}[tb]
%%\epsfysize=9.0cm
%\begin{center}
%\begin{minipage}[t]{8 cm}
%\epsfig{file=DESY-05-019_5.eps,scale=0.7}
%\end{minipage}
%\begin{minipage}[t]{16.5 cm}
%\caption{.\label{fig2}}
%\end{minipage}
%\end{center}
%\end{figure}

%\begin{figure}[tb]
%%\epsfysize=9.0cm
%\begin{minipage}[t]{8 cm}
%\epsfig{file=dis_shape237col_dis06.eps,scale=0.4}\epsfig{file=HERA_running_MIIan.eps,scale=0.4}
%\end{minipage}
%\begin{center}
%\begin{minipage}[t]{16.5 cm}
%\caption{a) Compilation of measurements of $\alpha_s(M_Z)$ at HERA compared
%to the world average from other measurements.  b) The combined HERA measurement
%of $\alpha_s$ as a function of $E_T^{jet}$ compared to the world average
%(excluding HERA) and their uncertainty shown as a band.\label{fig5}}
%\end{minipage}
%\end{center}
%\end{figure}

\subsection{Simultaneous Determination of PDF and $\alpha_s$ using Jet Data}

An examination of Equations~\ref{eq_4.1}~and~\ref{eq_4.2} 
shows that the jet cross-sections at HERA
are sensitive to both proton PDFs and $\alpha_s$.  In the previous section,
the dependence on proton PDFs was treated as a systematic uncertainty
in determining $\alpha_s$ from the jet data.  One may also think of
determining the PDFs from jet data using $\alpha_s$ determined 
elsewhere.  Indeed, the simultaneous determination of proton 
PDF and $\alpha_s$ in a combined NLO QCD fit of DIS inclusive 
and jet cross-sections (and possibly other data) is the most consistent 
approach.
%, in principle.  The practical method of doing the latter
%has been achieved only recently.

The inclusion of HERA jet data, or indeed any jet data in an NLO QCD fit, 
such as the one described in Section 4, is not conceptually
difficult; the problem is technical.  While the DGLAP
equation can be rapidly evaluated in a programme such as QCDNUM~\cite{qcdnum},
the NLO jet cross-section calculations are time consuming enough
to make using them in an iterative fitting procedure impractical.
%This means that no NLO QCD fit has been able to 
%make use of jet data rigorously, until recently.
Only recently have technical difficulties been overcome, making this procedure
possible.

In \cite{zeus_4.4}, the hard jet cross-section, $d\hat{\sigma}^{jet}$ in 
Equation~~\ref{eq_4.2} is pre-calculated on a grid of four dimensions, $x$, the proton momentum
fraction, $\alpha_s$, $\mu_R$, the renormalisation scale and $\mu_F$,
the factorisation scale.  By an appropriate choice of the grid points
at which the pre-calculations are carried out, an accuracy of better than
0.5\% with respect to a full calculation can be achieved using interpolations
for all relevant cross-sections.  In this way, the full NLO jet calculations
can be incorporated into a NLO QCD fit on the same rigorous footing as the
DGLAP equations for inclusive cross-sections.

\begin{figure}[tbh]
\begin{minipage}[t]{8 cm}
\hspace{15pt}
\epsfig{file=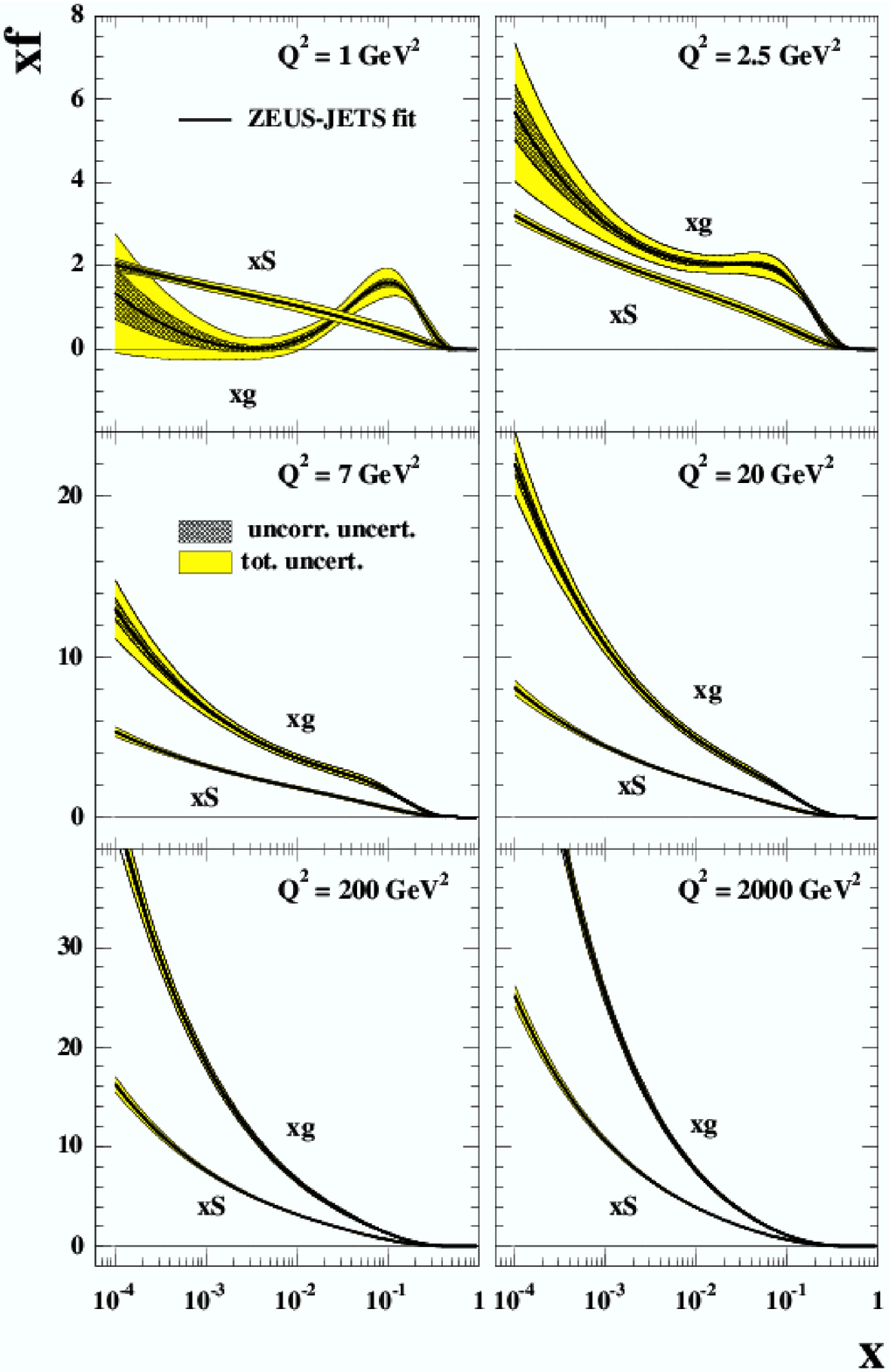,scale=0.40}\epsfig{file=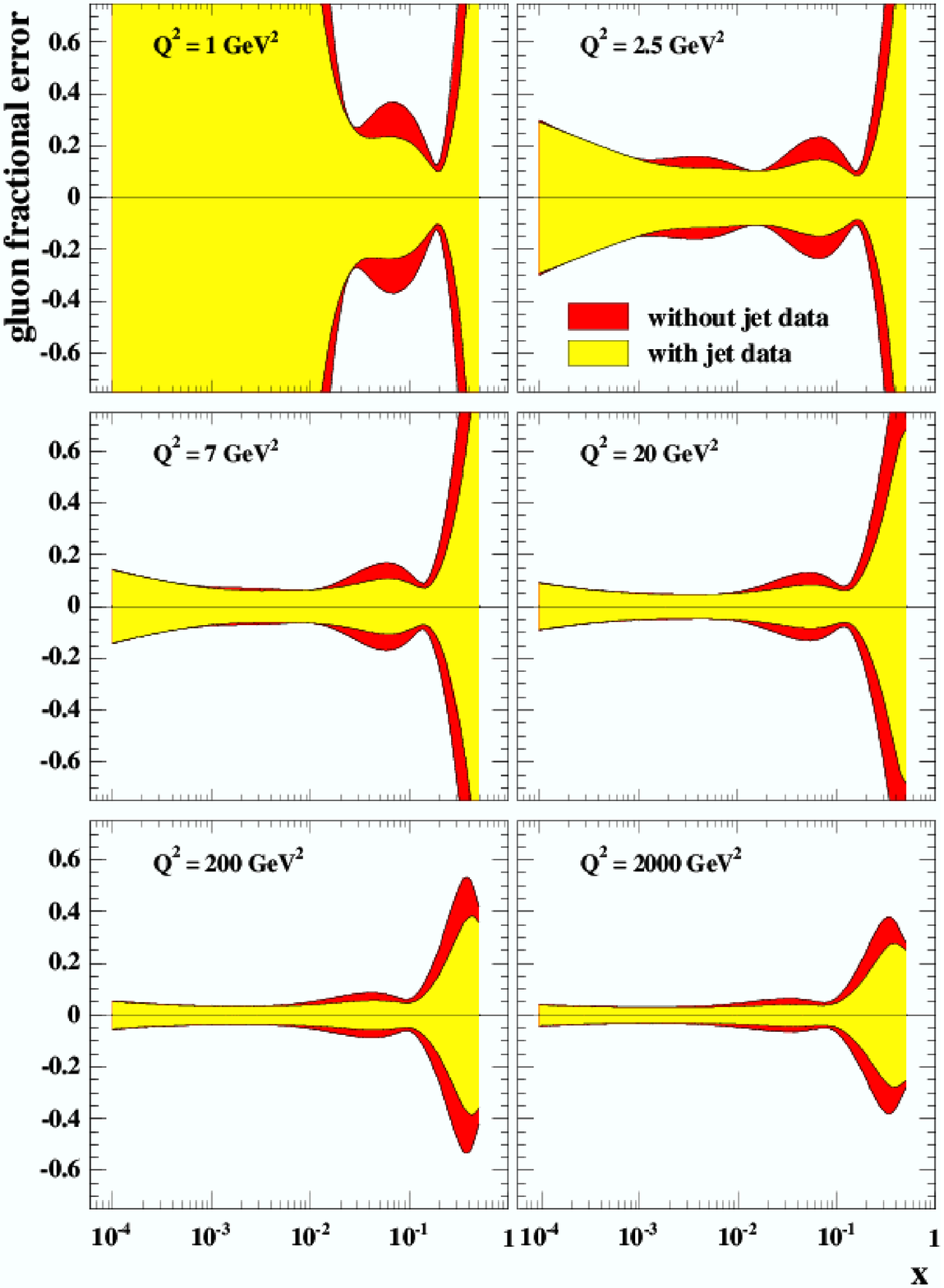,scale=0.40}
\end{minipage}
\begin{center}
\begin{minipage}[t]{16.5 cm}
\caption{Left: The gluon and sea PDFs extracted from the ZEUS-JETS fit.  The inner
cross-hatched error band shows the statistical and uncorrelated systematic uncertainty,
the grey error bands show the total uncertainty including experimental uncorrelated and 
correlated systematic uncertainties, normalisation and model uncertainty.
Right: The total experimental uncertainty on the gluon PDF for the ZEUS-JETS fit
(central error bands) compared to the total experimental uncertainty on
the gluon PDF for a fit not including the jet data (outer error bands).
The uncertainties are shown as fractional differences from the central values
of the fits for various values of $Q^2$. 
\label{fig_4.6}}
\end{minipage}
\end{center}
\end{figure}

The results of the fit using the ZEUS DIS inclusive
cross-sections as well as both DIS
inclusive jet
 cross-sections and the direct photoproduction dijet cross-section are 
shown in Figure~\ref{fig_4.6}.  In this instance, the fit is made
with the value of $\alpha_s$ fixed to the world average value. 
There is a significant improvement to the PDF uncertainties, particularly at values
of $x$ above $10^{-2}$, which is expected since jet production
is dominated by gluon-photon fusion process in this range
of $x$ at HERA.

%\newpage

\begin{figure}[tbh]
\begin{minipage}[t]{8 cm}
\epsfig{file=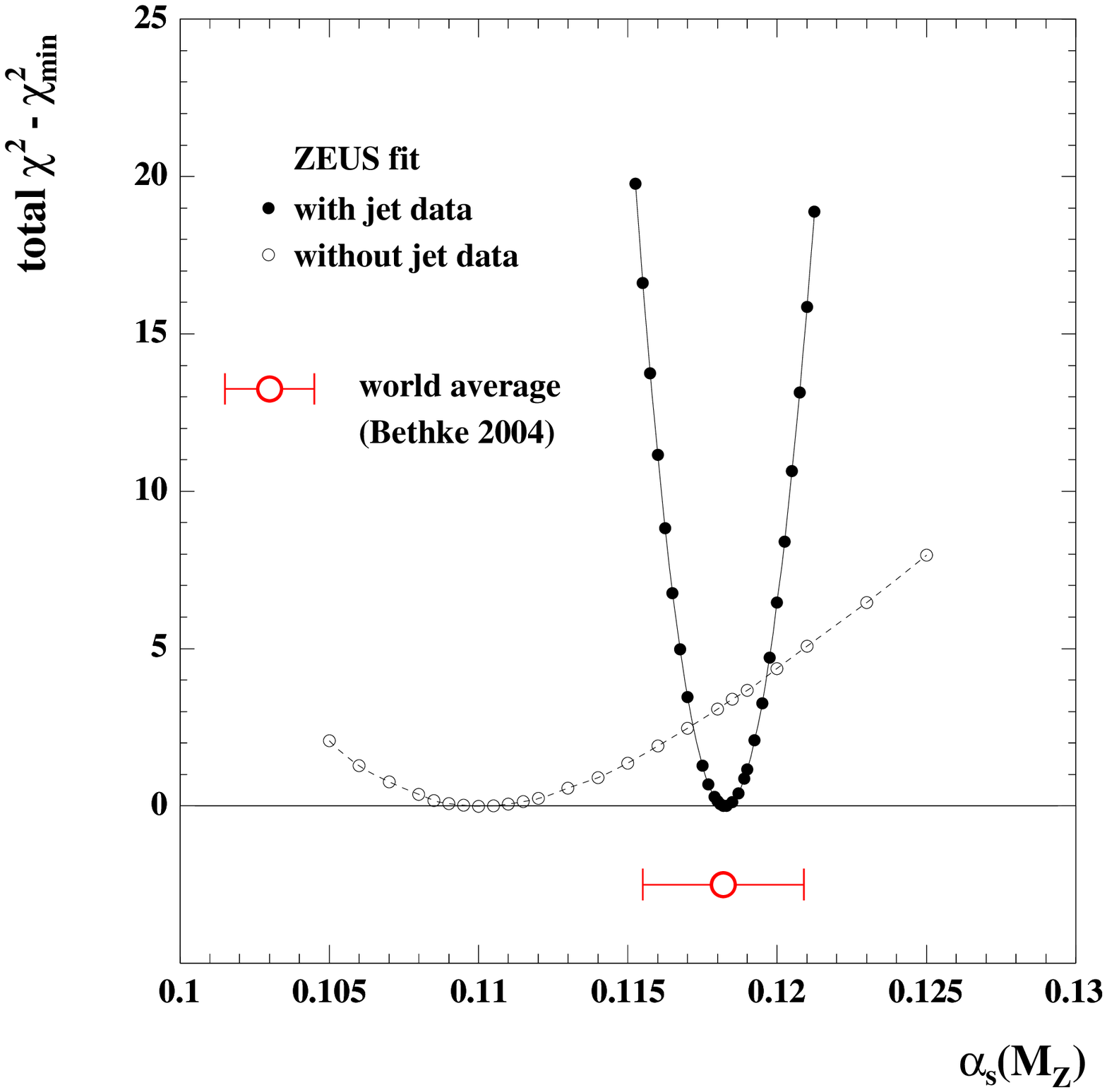,scale=0.5}\epsfig{file=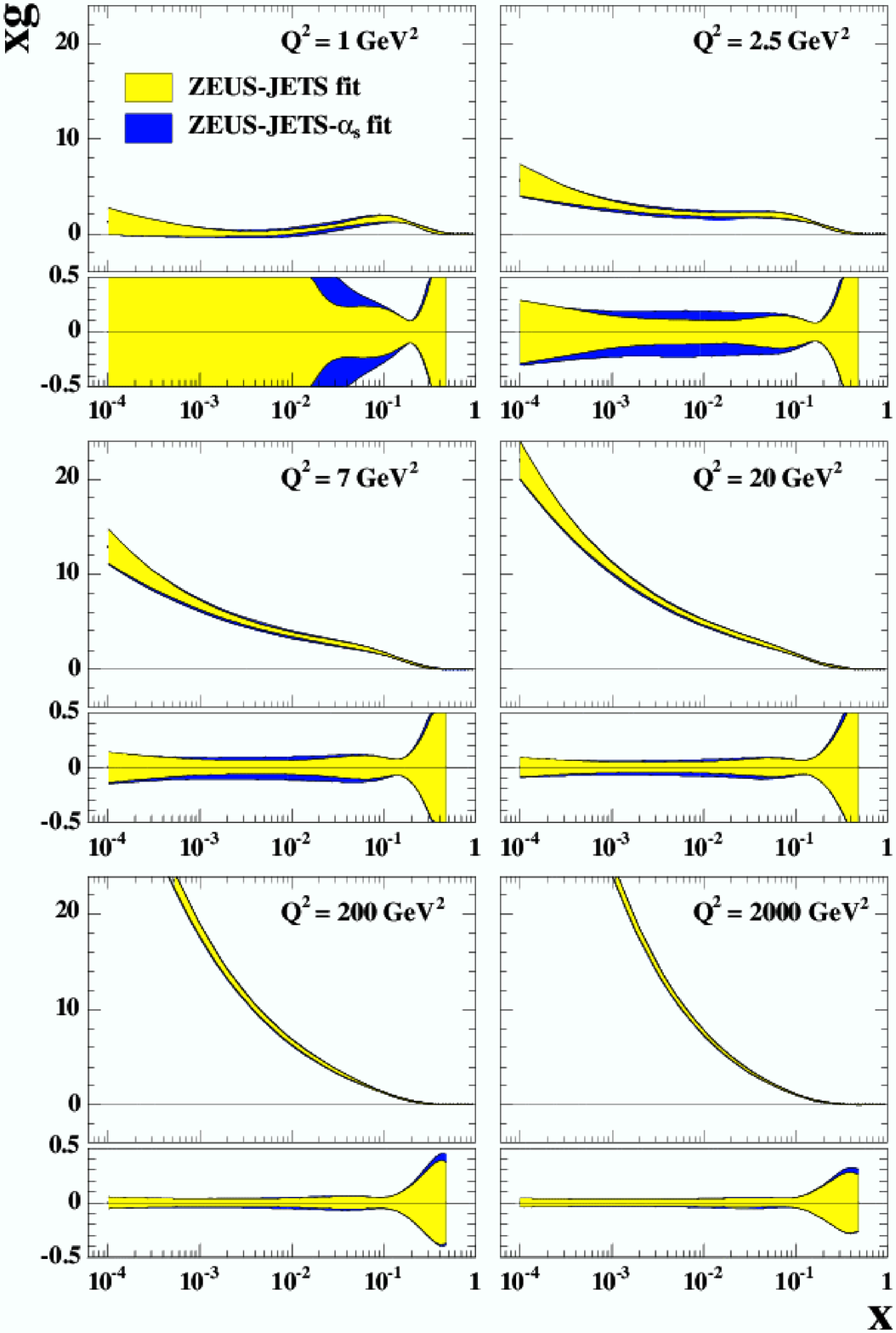,scale=0.3}
\end{minipage}
\begin{center}
\begin{minipage}[t]{16.5 cm}
\caption{Left: The $\chi^2$ profile as a function of $\alpha_s(M_Z)$ for the
fit to ZEUS data with (black dots) and without (clear dots) including the
jet data.  The ordinate is given in terms of the difference between the
total $\chi^2$ and the minimum $\chi^2$ for each fit. Right: 
 The gluon distributions
extracted from the simultaneous fit to ZEUS DIS and jet data.  The uncertainties
on these distribution are shown beneath each distribution as
 fractional differences from
the central value.  The outer error band shows the total 
uncertainty and the inner band
the uncertainty in case $\alpha_s$ is fixed in the fit.
\label{fig_4.7}}
\end{minipage}
\end{center}
\end{figure}

The next step is to simultaneously fit $\alpha_s$ as
 well as the PDFs.  In a QCD fit
using only the structure function data from HERA,
 $\alpha_s$ is so far weakly constrained as 
is shown in Figure~\ref{fig_4.7}.  The same figure shows
 the strong constraint on $\alpha_s$ jet data add to
the fit.  The uncertainties of the PDFs also 
remain stable and relatively small
for the simultaneous fit as shown in the same figure.
 The extracted value
of $\alpha_s(M_Z) = 0.1183\pm 0.0028 {\rm (exp)}\pm 0.0008{\rm (model)}$,
with an additional
uncertainty of $\pm 0.005$ due to unknown higher order
effects to the NLO theoretical
calculations, is very close to the value used in the 
fixed-$\alpha_s$ fit, and thus
does not change the central values of the PDFs significantly.
 The level of the precision
of this determination of $\alpha_s$ is as good as any 
existing single measurement; it is also,
so far, the only accurate
one that is derived from HERA data only, with minimal 
assumptions in the
PDF fit parametrisation that derive from outside data. 
 The other jet measurements
use PDF fit parametrisations derived using world data.

%\newpage

\subsection{Summary and Outlook of Jet Measurements at HERA}

The jet cross-section measurements at HERA have achieved about 5\% precision, with
the main contribution coming from the understanding of the jet-energy
 scale at the level of 1-2\%.
With this precision, a meaningful comparison can be made to the QCD NLO predictions
of jet predictions.  Indeed, in most cases, the precision of the 
experimental measurements
are such that the theoretical part dominates the uncertainties.

Based on the good description of the data, QCD parameters can be extracted.
The strong coupling constant $\alpha_s$ is determined to about 
1\% experimental uncertainty
(5\% theoretical uncertainty) from the current measurements.  The jet measurements
help to constrain the gluon PDF when combined in a simultaneous fit with the DIS inclusive
cross-sections.  The current measurements constrain the gluon to about $\pm 10\%$
at $x$ around 0.05 and $Q^2$ around 200 GeV$^2$, based so far on the ZEUS jet data only. 

All of the results discussed here come from the HERA I data taking phase which comprises about
120~${\rm pb}^{-1}$ per experiment.  While many of the jet measurements are now limited
by the theoretical uncertainties, there are areas, such as the measurements at the
highest $E_T$, which will benefit from the 800~${\rm pb}^{-1}$ of which is available from the
HERA II data taking period.  The projections are that this data should be able to
improve the uncertainty of the gluon PDF in the $x$ range of 0.1 to 0.05 
by about a factor of two~\cite{cooper_4.1}.

%There are many detailed investigation of jets at HERA which is not included here
%due to lack of space.  These include studies of jet substructure [ref], jet radius [ref],
%jet shapes [ref], as well as studies of transition from DIS to photoproduction regime.

%\newpage
\section{Heavy Flavours}
%\begin{figure}[tbh]
%%\epsfysize=9.0cm
%\begin{center}
%\begin{minipage}[t]{10 cm}
%\epsfig{file=DESY-03-115_4.eps,scale=0.6}
%\end{minipage}
%\begin{minipage}[t]{16.5 cm}
%\caption{Differential $D^*$ $ep$ cross-sections as a function of
%a) $Q^2$, b) $x$, c) $p_T(D^*)$ and d) $\eta(D^*)$ compared to NLO
%QCD calculation of HVQDIS.  The data and the error bars are as 
%described in Figure 1. The predictions using ZEUS NLO QCD
%PDFs are shown for $m_c = 1.35 GeV$ (solid line) and its associated
%uncertainty (shaded band).  The prediction of CTEQ5F3 PDF (dash-dotted line)
%and an alternative hadronisation scheme (dotted line) are displayed.
%The ratios of the cross-sections to the central HVQDIS predictions
%are shown beneath each plot. 
%\label{fig_5.1}
%}
%\end{minipage}
%\end{center}
%\end{figure}

\subsection{Introduction}
Processes involving heavy quarks, $Q$, with mass, $M_Q \gg \Lambda_{QCD}$, are, 
in principle, amenable to pQCD calculations.  At HERA, the production of
charm and beauty particles are measured and compared to theoretical predictions.
While $M_Q$ is one QCD scale, there are two other relevant scales in
heavy quark production processes at HERA. These are the virtuality, $Q^2$,
of the exchanged photon, and the transverse energy, $E_T$, of the final
state particles or jets.  The relative
size of these three scales is important in choosing what kind of calculation
is appropriate for comparison to which data.  
For example, in the limit, $Q^2 \gg M_Q^2$, an approximation 
where $M_Q =0$ may be appropriate.   On the other hand,
since there are no quantitative predictions of what 
``much greater than'' may mean, the study of heavy quark production,
as is often the case with QCD, is an iterative one where the appropriate
theory is arrived at gradually 
as better and better data are compared to the theory.

\subsection{Theoretical Calculations of Heavy Quark Production}

%In DIS reactions, where $Q^2 \le M_Q^2$, the heavy quark production
%should be cal

In the conventional co-linear approximation, heavy quark production in DIS
cross-section can be written generically as
\begin{equation}
 d\sigma = \sum_{a} \int 
 dx f_a(x,\mu_F^2;\alpha_s) \cdot
d\hat{\sigma}_{HQ}(xP,\alpha_s(\mu_R),\mu_R,\mu_F,M_Q).
\end{equation}
If the renormalisation scale, $\mu_R$, usually taken to be
the virtuality $Q^2$, is $ \gg M_Q^2$, then the heavy quark
should be treated as a massless quark on the same footing as the other
quarks in the PDF in the framework of $\ln \mu_R^2$ resummation (DGLAP 
approximation).  In this case, $f_a$ is the PDF of the heavy quark, and
the cross section $d\hat{\sigma}$ is the photon- (photoproduction) or
electron- (DIS) quark cross-section for a massless quark.  
This type of calculation is called ``massless''.  

If the scale, $Q^2 \approx M_Q^2$, then it is appropriate to 
treat the heavy quark separately as a massive quark.  In this case,
the sum over $a$ in the above equation will be for all partons $except$ 
the heavy quark.  The cross-section $d\hat{\sigma}$, will be for
a production of a massive quark from an electron colliding
with a (massless) parton at a fixed order in $\alpha_s$.  
This type of calculation is called ``massive''.

There are three types of treatments of heavy quark production
in DIS based on co-linear
factorisation.
\begin{itemize}
 \item{Zero-Mass Variable Flavour Number Scheme (ZM-VFNS), is a 
purely ``massless'' treatment where the heavy quarks are treated 
as massless quarks which become ``active'' 
in the proton PDF at the production thresholds of a $Q\bar{Q}$ pair.  
This is the oldest and still the most common theoretical treatment. As discussed above, 
this is, in principle, incorrect near the production
thresholds.} 
 \item{Fixed Flavour Number Scheme (FFNS) is a purely massive treatment.
This is, in principle, incorrect at high $Q^2$.  The most common treatment
uses the analytic programme HVQDIS~\cite{harris_5.1}.}
 \item{Matching Schemes. Sometimes these are called
Massive Variable Flavour Number Scheme (Massive-VFNS) or
Generalised Mass Variable Flavour Number Scheme (GMVFNS). 
There are several predictions which match the
massless and massive schemes at an intermediate value of $Q^2$. Examples
of these are Roberts-Thorne (RT)~\cite{thorne_5.1,rt_5.1}, ACOT~\cite{acot_5.1}, 
Kniehl et al.~\cite{kniehl_5.1} and ResBos-HQ~\cite{nadolsky_5.1}.
}
\end{itemize}

\begin{figure}[tbh]
\begin{center}
\begin{minipage}[t]{16 cm}
\epsfig{file=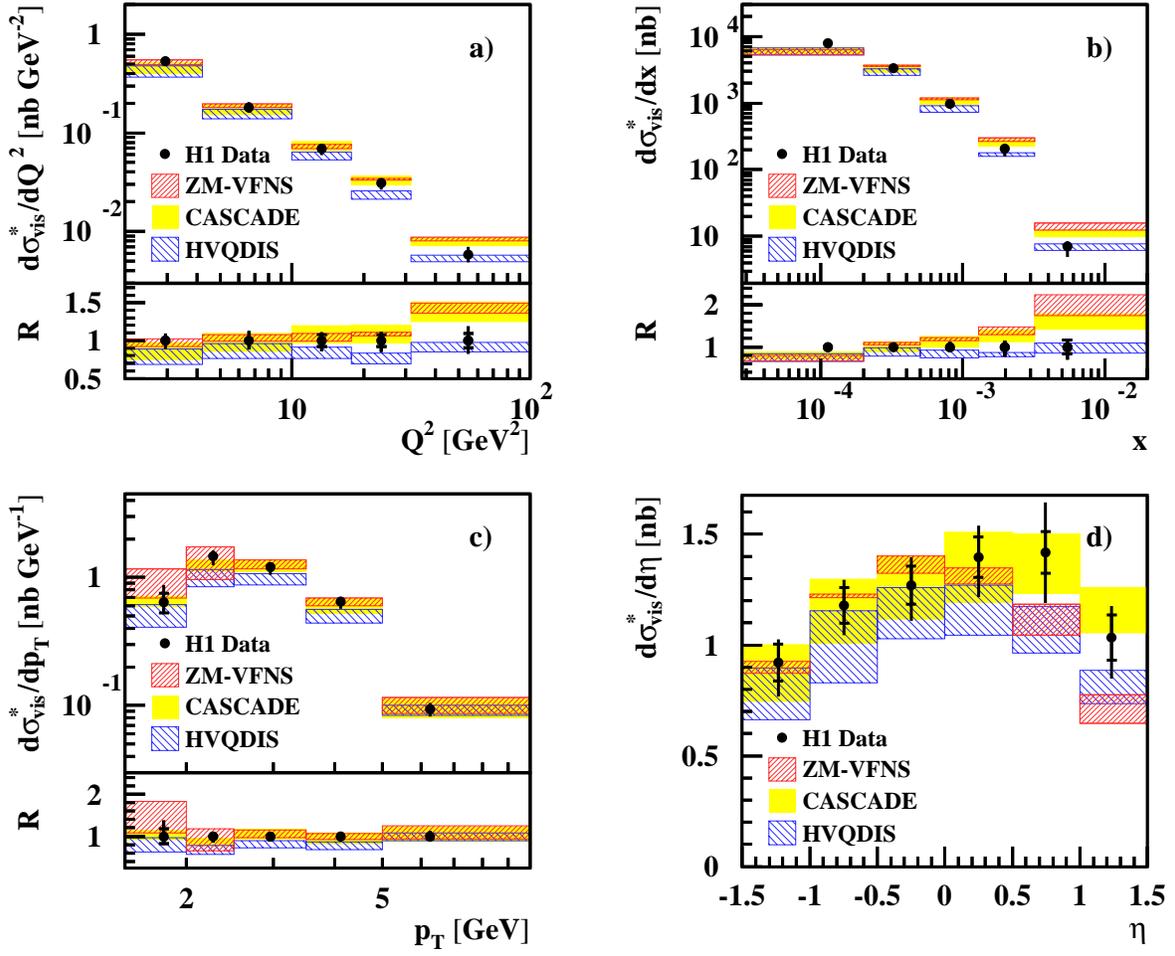,scale=1.0}
\end{minipage}
\begin{minipage}[t]{16.5 cm}
\caption{Differential $D^*$ $ep$ cross-sections as a function of
a) $Q^2$, b) $x$, c) $p_T(D^*)$ and d) $\eta(D^*)$ compared to calculation of HVQDIS
CASCADE and a Massive-VFNS calculation from Kniehl et al.~\cite{kniehl_5.2}.  
The inner error bars indicate the
statistical uncertainty, while the outer show the statistical and 
systematic uncertainties in quadrature.  Figures a), b) and c) also present the
ratios $R=\sigma_{theory}/\sigma_{data}$, taking into account their theoretical
uncertainties.  See \cite{h1_5.0} for details.
\label{fig_5.1}
}
\end{minipage}
\end{center}
\end{figure}

%\newpage
The cross-sections, calculated to order 
$\alpha_s^2$, are available in all three schemes.
Some of the predictions, such as those of HVQDIS and ResBos-HQ, 
are of  semi-analytical type 
similar to those discussed in Chapter 4 for jet production.  In these cases, 
a detailed prediction for comparison to the data of measurements of production
of hadrons containing heavy quarks is possible.  Such predictions 
are made using a fragmentation 
function~\cite{bowler_5.1,kartivelishvili_5.1,peterson_5.1} that give
 the probability of producing 
a particular hadron given an initial heavy quark.  
Also, a cross-section of jets associated with
heavy quark production can be predicted in the same way as for ordinary
jet production discussed in Section 5.  Many predictions give only the $x$ and $Q^2$ 
dependence of the charm cross-section, and thus can only be compared after
some, more or less large, extrapolation of the data
(see also the end of this Section).

The theoretical treatment of the photoproduction of heavy flavours
is similar to that of DIS but complicated by the
second ``hadron'', the real photon.  In this case the photon-proton
cross-section can be represented as;
\begin{equation}
 d\sigma = \sum_{a,b} \int \int 
 dx_p dx_\gamma f_a(x_p,\mu_F^2;\alpha_s) \cdot f_b(x_\gamma,\mu_F^2;\alpha_s) \cdot
d\hat{\sigma}_{HQ}(xP,\alpha_s(\mu_R),\mu_R,\mu_F,M_Q),
\end{equation}
where $f_b$ is the PDF of the photon.  In practice, the measurements
use jets associated with heavy quark production to determine the 
kinematics--in particular the Bjorken variable of the photon, $x_\gamma$.

\begin{figure}[th]
\begin{center}
\begin{minipage}[t]{12 cm}
\epsfig{file=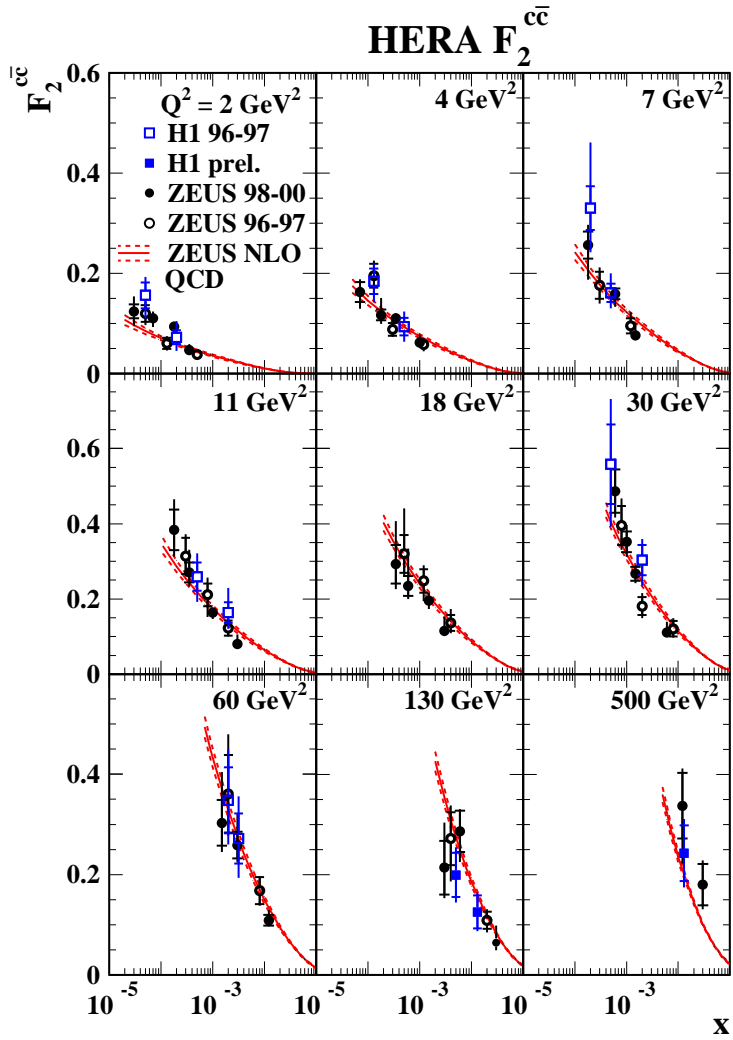,scale=0.6}
\end{minipage}
\begin{minipage}[t]{16.5 cm}
\caption{The measured $F_{2}^{c\bar{c}}$ in bins of
$Q^2$ as a function of $x$.  The data are shown with statistical
uncertainties (inner bars) and statistical and systematic 
uncertainties added in quadrature (outer bars).  The lower and
upper curves show the fit uncertainty propagated from the 
experimental uncertainties of the fitted data. 
\label{fig_5.2}
}
\end{minipage}
\end{center}
\end{figure}
As in the case of DIS, predictions in FFNS, ZM-VFNS and Massive-VFNS
exist for the photoproduction case.  The renormalisation scale 
$\mu_R$ is usually chosen to be some combination of the quark mass
$M_Q$ and the transverse momentum of the produced heavy quarks, $P_T^Q$.

The predictions discussed above are all based on co-linear factorisation.
A different approach is to use $k_t$ factorisation and
 the CCFM equations~\cite{ccfm_5.1}, which,
unlike the DGLAP equations, resum terms proportional to $\ln{1/x}$ as 
well as to $\ln{\mu_R^2}$.  The actual predictions in this approach
are made by the Monte Carlo program 
CASCADE~\cite{cascade_5.1} that uses the off-shell 
photon-gluon fusion process convoluted with 
the gluon distribution unintegrated in $k_t$, 
obtained via a fit to HERA $F_2$ data~\cite{jung_5.1}.  These
predictions are currently available only at order $\alpha_s$. 
Like the predictions of HVQDIS, the
CASCADE programme is capable of providing predictions at the parton level,
and is therefore easily comparable to measurements.

\begin{figure}[tbh]
\begin{minipage}[t]{8 cm}
\hspace{20pt}
\epsfig{file=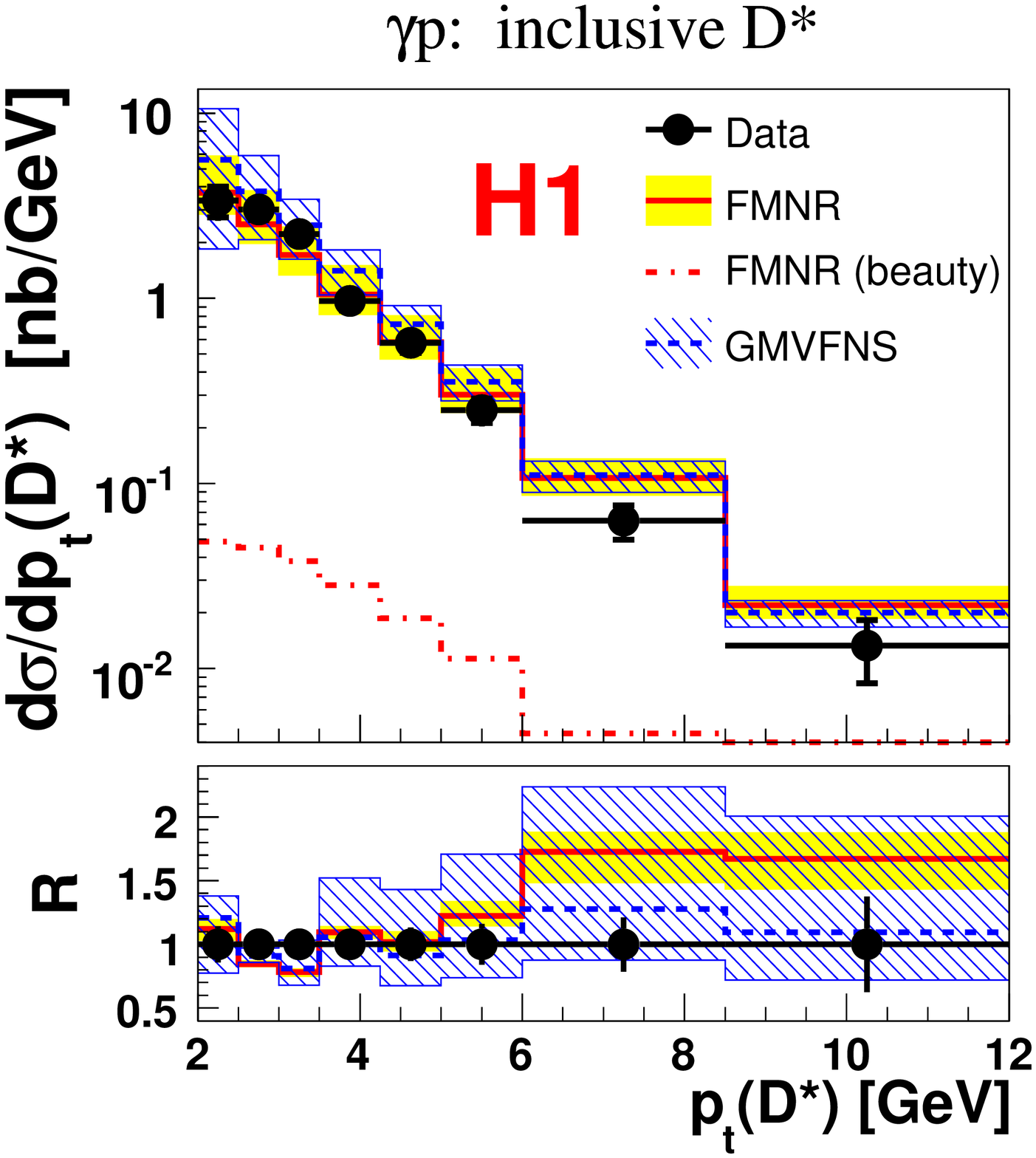,scale=0.4}\epsfig{file=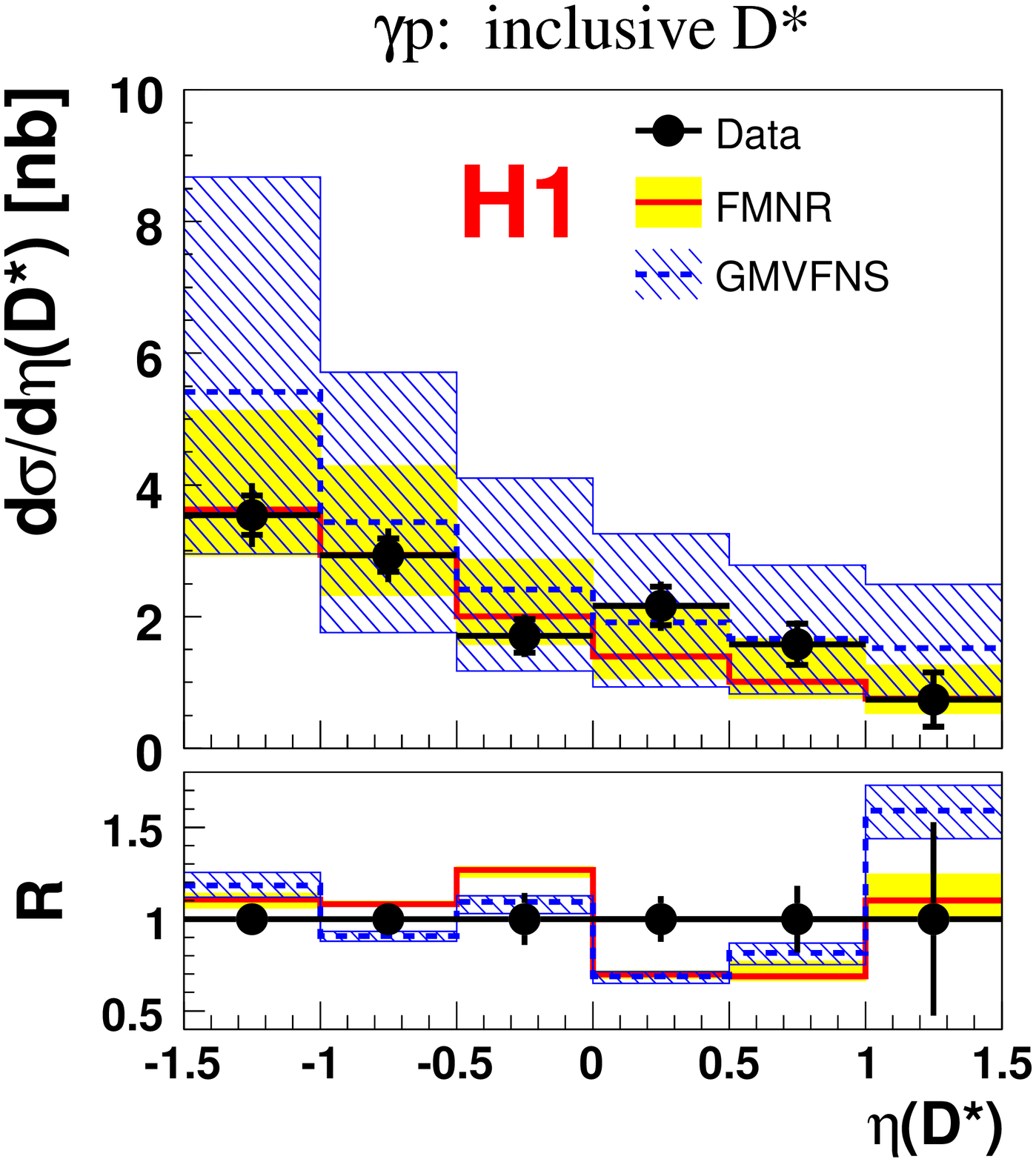,scale=0.4}
\end{minipage}
\begin{center}
\begin{minipage}[t]{16.5 cm}
\caption{Inclusive $D^*$ cross-sections as a function of $p_T(D^*)$
and $\eta(D^*)$ 
compared to the QCD NLO calculations of FMNR and GMVFNS.
For FMNR the beauty contribution is shown 
separately for $p_T(D^*)$. The ratio of the measurement and
the prediction, $R$, is also shown. 
\label{fig_5.3}
}
\end{minipage}
\end{center}
\end{figure}

\begin{figure}[tb]
\begin{center}
\begin{minipage}[tb]{10 cm}
\hspace{30pt}
\epsfig{file=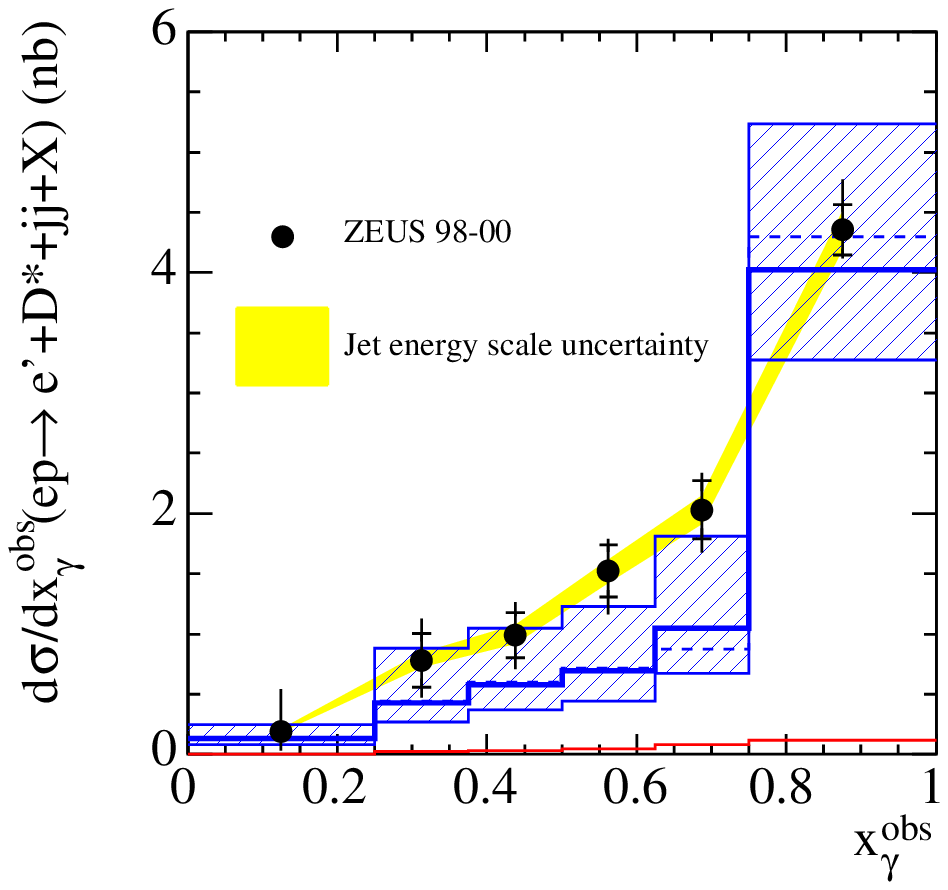,scale=0.7}
\end{minipage}
\begin{minipage}[tb]{16.5 cm}
\caption{Charm dijet cross-section $d\sigma/dx_{\gamma}^{obs}$ 
compared to  QCD predictions in the massive scheme,
with (solid line) and without
(dotted line) hadronisation corrections applied.  The theoretical
uncertainties (hatched band) come from the change in scales simultaneously
with the change in charm mass.  
\label{fig_5.4}}
\end{minipage}
\end{center}
\end{figure}

The reader is referred to \cite{tung_5.1} for a more complete discussion
of the theory of heavy quark production than is possible here.
\subsection{$D^*$ Cross Sections in DIS}

While many different kinds of charmed hadrons have been measured at
HERA, the measurements of the $D^*$ mesons have, by far, the highest
statistical significance.  This is due to the well-known
ease of reconstructing the $D^*$ meson, using tracking
detectors alone, by exploiting the ``slow'' pion from the decay,
$D^* \rightarrow D + \pi_s$.

Figure~\ref{fig_5.1} shows a recent measurement of the $D^*$ cross-sections in DIS~\cite{h1_5.0}.  
The theoretical predictions, which all describe 
the data reasonably well are of the FFNS type, HVQDIS, as well as the CCFM based CASCADE and
a massive VFNS calculation.  The predictions use a range of proton
structure functions and charm quark masses.   It is interesting
to note that the description of FFNS appears to be somewhat better than the
others even at rather high $Q^2$ values near a hundred GeV$^2$
despite the fact that FFNS should be, 
in principle, incorrect when $Q^2 \gg M_Q^2$. Indeed 
this good description is known to hold up
to several hundred GeV$^2$~\cite{zeus_5.1}. 

%The three theoretical predictions, which describe
%the data rather well, are made done
%in the following steps.

%\begin{enumerate}
%\item{A DGLAP fit to the inclusive data are done using a scheme in which
%the three fixed flavours.  The Charm-quark part of the
%cross-sections are generated in these fits using calculations 
%fixed-order, typically NLO, in $\alpha_s$.}
%\item{This three-flavoured PDF is used as input to the calculations of
%the charm production cross-sections differential in $\eta$ and $\P_T$. This
%calculation is done consistently with that used for the total charm
%cross-section in deriving the PDF.}
%\item{A model of fragmentation is used to determine the $D^*$ cross-section 
%from the charm cross-section.  This is usually done by 
%using a fragmentation function, an analytic function whose parameters
%are fitted or derived from experimental data.  The most usual of these
%is the Peterson model.}  
%\end{enumerate}

%It should be noted that, because the charm hadrons are measured only in
%the acceptance of the tracking detectors which is limited in both
%rapidity and $P_T$, the data can only be compared without ambiguity to 
%theoretical predictions of cross-sections differential in rapidity and $P_T$.

%Figure XX shows a measurement of D* mesons with dijets.  D* is in the
%$P_T$ range of 1.5 to 15 GeV and $|\eta| < 1.5$.  The two jets have transverse
%energies of 3 and 4 GeV respectively with -1 $\le \eta \le$ 2.5.  
%The measurements are well described by both HVQDIS and CASCADE.

\subsection{Derivation of the Charm Structure Function $F^{c\bar{c}}_2$}

The measured charm hadron cross-section can be extrapolated 
to the full charm cross-section differential only in $x$ and $Q^2$.
This, in turn, can be interpreted as the charm part of the cross-section,
described essentially by the structure function $F^{c\bar{c}}_2$.  Thus,
\begin{equation}
 \frac{d\sigma^{cc}}{dxdQ^2} = \frac{2\pi\alpha^2}{xQ^4}
\{[1+(1-y)^2]F_2^{c\bar{c}}(x,Q^2)-y^2F_L^{c\bar{c}}(x,Q^2)\}.
\end{equation}
Since the D* cross-section is measured 
in a kinematic range limited in $P_T$ and
$\eta$, a method of extrapolation is needed.  A most straightforward
way to determine this is, for the $D^*$ cross-section $\sigma_i(ep\rightarrow D^*X)$ in
the i-th bin of $x$ and $Q^2$:
\begin{equation}
 F_{2,meas}^{c\bar{c}}(x_i,Q^2_i) = \frac{\sigma_{i,meas}(ep\rightarrow D^*X)}
 {\sigma_{i,theo}(ep\rightarrow D^*X)}F_{2,theo}^{c\bar{c}}(x_i,Q^2_i),
\end{equation}
where $x_i$ and $Q_i^2$ are chosen to be within the measured bin.
The uncertainties in this extrapolation are determined by varying the parameters
of the theoretical models being used.  While these uncertainties, at 10-20\%, are
usually smaller than the uncertainties already present in the $D^*$ cross-section
measurement, it is worth keeping in mind that the extrapolation factors 
range from about 1.5 up to nearly 5.  Thus reliance on the correctness of the
theoretical prediction of the cross-section in $\eta$ and $P_T$ is
relatively large. This is overcome  with impact parameter based measurements
as are discussed below.

Figure~\ref{fig_5.2} shows a compilation of the measurements of $F_2^{c\bar{c}}$ based
on measurements of $D^*$ mesons~\cite{zeus_5.1,h1_5.01}.  
The measurements are compared to a prediction
of a PDF obtained from a fit using a VFNS scheme.  It should be noted that
the comparison of theory to data in the form of $F_2^{c\bar{c}}$ derived 
from the $D^*$ measurements do not contain more information than 
the comparison of the differential cross-sections in the measured kinematic
range to the theory of the kind shown in Figure~\ref{fig_5.1}.

\subsection{$D^*$ Production in Photoproduction}

A recent measurement~\cite{h1_5.1} of charmed meson production in
 photoproduction is shown 
in Figure~\ref{fig_5.3}. 
The theoretical predictions of the massive scheme (FMNR~\cite{fmnr_5.1})
 have a tendency to
have too shallow a slope in $p_T$ of the $D^*$ and be below the data at
forward $\eta$ of the $D^*$. 
The predictions of the variable flavour 
number scheme (GMVFNS) \cite{heinrich_5.1}, while describing the data, have
a very large uncertainty associated with effects of missing higher orders.

In photoproduction, one factorizes the photon structure as well
as the proton structure in order to apply pQCD.
Therefore, in order to compare the predictions to the
data in more detail than above, it is necessary to measure, in addition to
the total $\gamma p$ cms energy, $W$, and the kinematics of the charmed
hadron, quantities that can determine the kinematics of the partons in
the photon.  This can be achieved by measuring the jets produced in
the events with charm hadrons.  The jet finding is done in the same
way as is described in the previous section.  The jets found may 
contain the charmed hadron produced in the event.

\begin{figure}[tb]
\begin{minipage}[t]{8 cm}
\hspace{20pt}
\epsfig{file=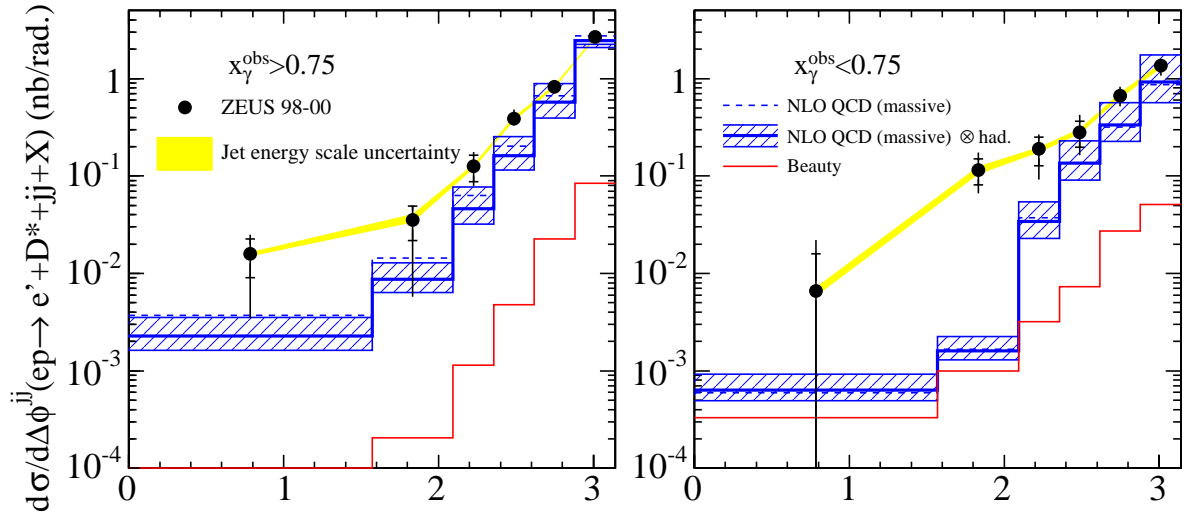,scale=0.8}
\end{minipage}
\begin{center}
\begin{minipage}[b]{16.5 cm}
\caption{Cross-section $d\sigma/d\Delta \phi^{jj}$ separated
into resolved enriched ($x_\gamma^{obs} < 0.75$) and
direct enriched ($x_\gamma^{obs} > 0.75$) region.  The theoretical
uncertainties come from the change in scales simultaneously with the change 
in charm mass.  The beauty component is also shown (lower histogram).
\label{fig_5.5}
}
\end{minipage}
\end{center}
\end{figure}

Figure~\ref{fig_5.4} shows a measurement of dijet photoproduction associated with
a $D^*$~\cite{zeus_5.2}.  The cross-section is plotted
differentially in $x_\gamma$, the
Bjorken variable of the photon.  The theoretical prediction,
a massive calculation, tends to fall below the data in the region
of low $x_\gamma$ indicating that the description in the resolved region
is not adequate.  Figure~\ref{fig_5.5} shows the same data now plotted separately
for high $x_\gamma$ (direct component) and low $x_\gamma$ (resolved component)
in terms of
$\Delta \phi_{jj}$, the azimuthal separation of the two jets.  It is seen
that the theory tends to fall below the data away from $\Delta \phi_{jj}=\pi$ 
(i.e. back-to-back jets) particularly in the resolved region.
 Since $\Delta \phi_{jj}$ is equal to
$\pi$ to LO in $\alpha_s$, NLO is the order at which the first non-trivial
prediction is given.  A failure to describe this distribution may be an indication
of a need for higher order terms in the calculation.

The comparison of the data on photoproduction of charm with the present
theoretical predictions indicate some inadequacies of the theory.  The
indications, both from relatively large renormalisation uncertainties and
from examination of differential distributions, are that the 
theory still requires the inclusion of 
higher order terms in order to give a good description of the available data.

\subsection{Photoproduction of Beauty Particles}
\begin{figure}[tbh]
\begin{center}
\begin{minipage}[tbh]{16 cm}
%\hspace{-200pt}
%\epsfig{file=papfig_extra_etamu_z+h1.eps,scale=0.5}\epsfig{file=plotptb_ICHEP06.eps,scale=0.55}
\epsfig{file=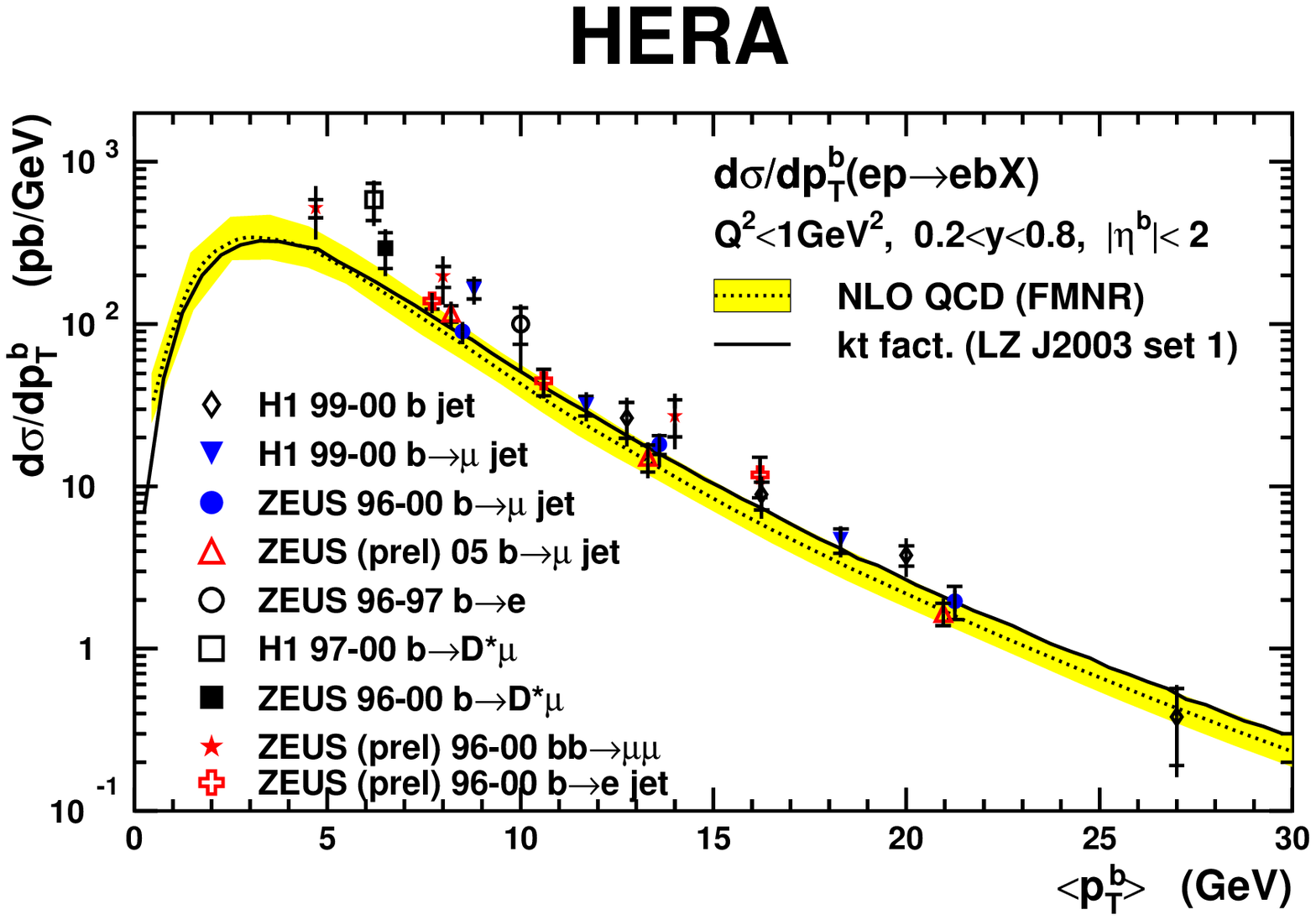,scale=0.8}
\end{minipage}
\begin{minipage}[tbh]{16.5 cm}
\caption{
%a) Cross-section $d\sigma/d\eta^\mu$ for beauty photoproduction at 
%HERA.  The data are compared to predictions of NLOQCD in the massive
%scheme and from Pythia and Herwig Monte Carlo programmes. b) 
A summary of
all available HERA data on beauty photoproduction plotted as the cross-section
$d\sigma/dp_T^b$ and compared to NLO QCD in the massive scheme and a 
$k_T$ factorisation calculation. 
\label{fig_5.6}
}
\end{minipage}
\end{center}
\end{figure}

The measurements of beauty photoproduction are based on 
the semi-leptonic decay of the B-hadron.  Because of the
large mass of the $b$-quark, the leptons from the decay tend
to be produced with large transverse momentum with respect to the
momentum of the rest of the associated jet.  In this case, the
signal is extracted from the distribution of momenta of 
the leptons relative to that of the jets.
Alternatively,
a $D^*$ may be explicitly reconstructed with a decay lepton and
their momentum separations.  In the latter case, which 
exploits the specific decay chain $B\rightarrow D^* \mu \nu$,
further constraints, such as charge of the $D^*$ and the $\mu$,
and the mass of the pair, can be used in the selection.

%Figure~\ref{fig_5.6}a shows two of the measurements~\cite{zeus_5.3,h1_5.2} differential in the
%$\eta$ of the muon compared to a massive calculation; the measurements
%are somewhat above, but consistent with the predictions.  
Figure~\ref{fig_5.6}
is a compilation of all currently available 
measurements~\cite{zeus_5.3,h1_5.2,zeus_5.4,h1_5.3,zeus_5.5,zeus_5.6}
of $b$-photoproduction, 
measured via its decay, converted to a $b$-quark cross-section differential 
in the transverse
momentum of the $b$-quark.  It can be seen that while the massive calculation
gives a prediction which is somewhat lower than the measurements, particularly
at lower ranges of $p_T^b$, the general description is good.  It can also be seen that
the $k_t$ factorisation calculations from \cite{lipatov_5.1} is very similar to that of FMNR.

\subsection{Measurements of Heavy Quark Production using Vertex Separation}

Qualitatively different measurements of heavy quark production at HERA
than discussed above are achieved by the use of the precision
tracking made possible by the use of silicon vertex detectors.  These
devices have been installed as upgrades to the HERA detectors. The  H1 vertex
detector~\cite{h1vtx} was installed in 1996-7 and the ZEUS one~\cite{zeusvtx} in 2001.

The spatial impact parameter 
resolution of the silicon detectors, in the order of tens of microns,
makes possible the use of the long lifetimes of $c$ and $b$ flavoured 
hadrons for selection of these events without explicit reconstruction
of the hadrons.  The events containing heavy quarks are distinguished 
from those containing only light quarks by reconstructing the displacement of
tracks from the primary vertex.  The major advantages of this method compared
to that of reconstructing particular $c$ or $b$ hadrons are that a) there
are little or no uncertainties associated with properties of particular
hadrons, such as its fragmentation function or decay branching ratios, and
b) the severe limitation in the measurable kinematic range in $p_T$ and $\eta$
due to the use of central tracking chambers for the reconstruction of
hadrons is removed, the acceptance for the silicon detectors is
close to 100\%.  The latter point is very important
when extracting the total $c$ and $b$ cross-sections (and therefore
$F_2^{c\bar{c}}$ and $F_2^{b\bar{b}}$) since the extrapolation factors
from previous analyses can be as large as 4 or 5 and, thus rely heavily
on the correctness of the models.

\begin{figure}[tbh]
\begin{minipage}[tbh]{4 cm}
\epsfig{file=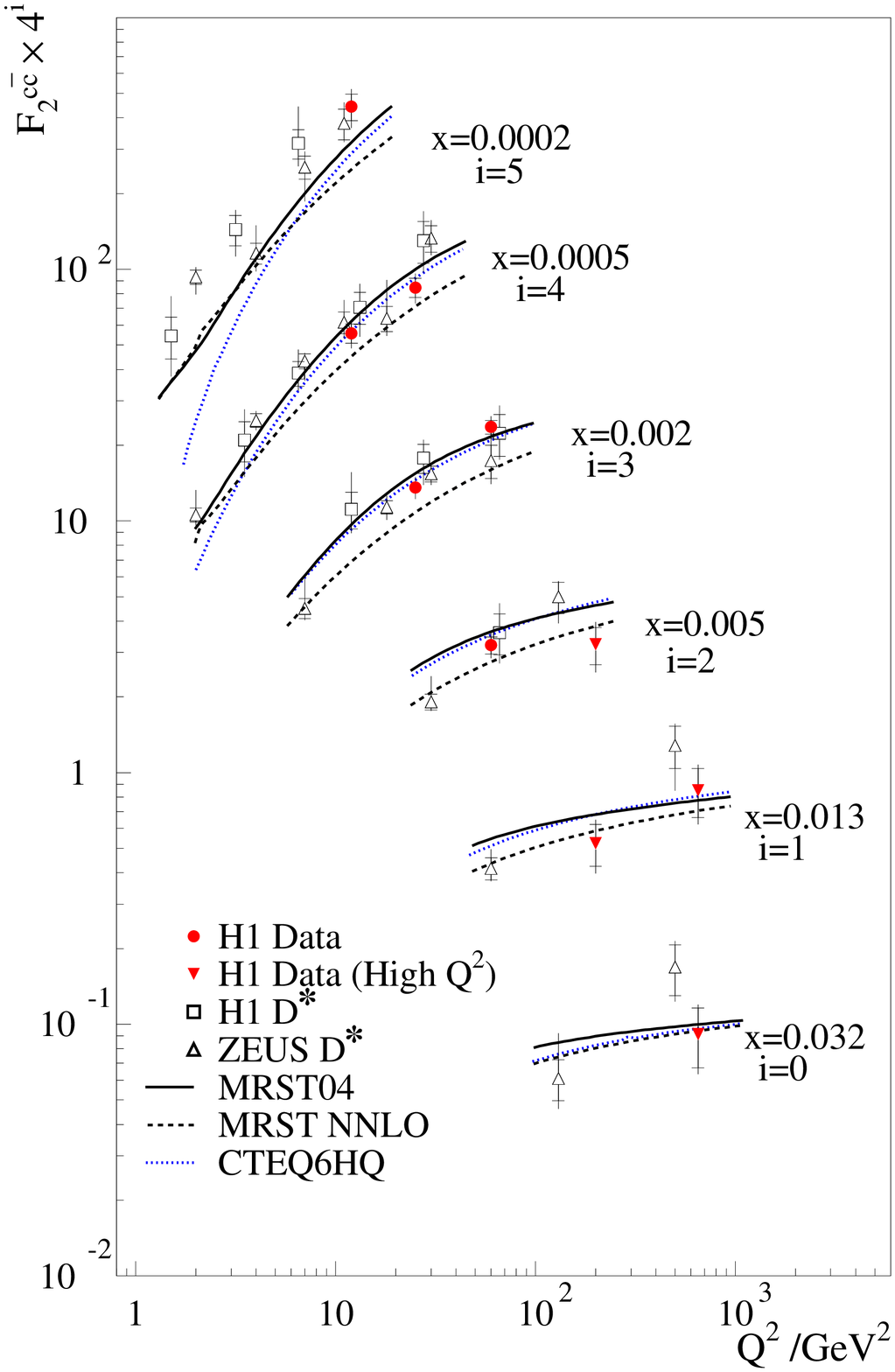,scale=0.45}\epsfig{file=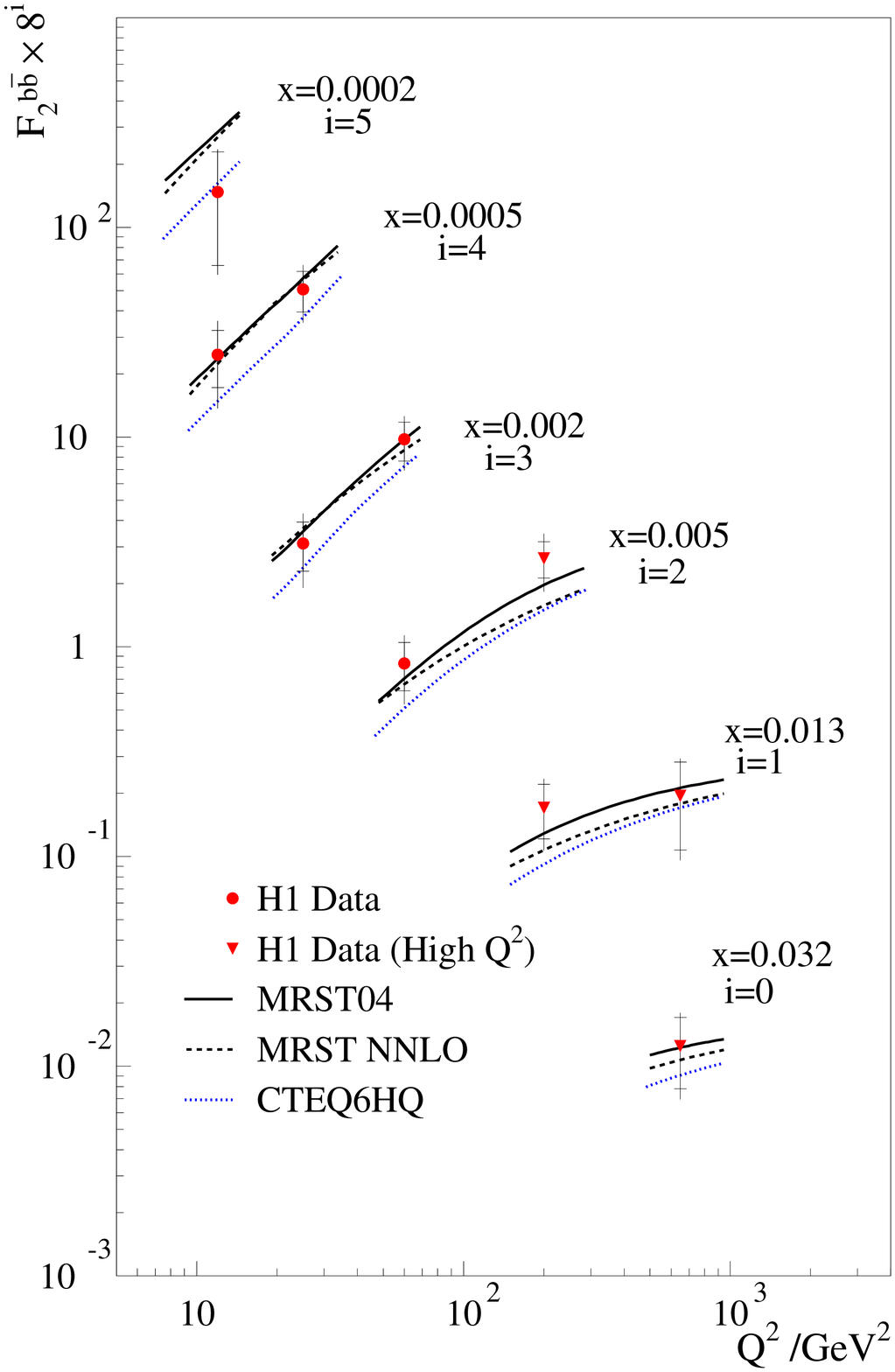,scale=0.45}
\end{minipage}
\begin{center}
\begin{minipage}[tbh]{16.5 cm}
\caption{The measured heavy quark structure functions
 $F_2^{c\bar{c}}$ and $F_2^{b\bar{b}}$
shown as a function of $Q^2$ for various $x$ bins.  The solid points
are the measurements utilising the silicon vertex detector of H1.
The open points on $F_2^{c\bar{c}}$
are older measurements made using the reconstruction
of $D^*$ mesons. 
\label{fig_5.7}
}
\end{minipage}
\end{center}
\end{figure}

The first such type of measurements of
the heavy quark structure functions~\cite{h1_5.4,h1_5.5} 
was performed by H1 and is shown in Figure~\ref{fig_5.7}
together with previous H1 and ZEUS measurements
of $F_2^{c\bar{c}}$ that rely on the reconstruction of $D^*$ mesons. 
It is notable that the $F_2^{c\bar{c}}$ measurements from $D^*$ reconstruction
agree well with the newer results, which is an indication of the reliability
of the extrapolations used in the older results.
While there is an earlier measurement of $\sigma(b\bar{b})$ in DIS
based on the relative transverse momenta of muons~\cite{zeus_5.7} 
this is the first measurement
at HERA of $F_2^{b\bar{b}}$ in a relatively wide kinematic range. 

Both $F_2^{c\bar{c}}$ and $F_2^{b\bar{b}}$ rise steeply as a function
of $Q^2$ at fixed $x$, especially at small $x$, reflecting the large
gluon density there.  The predictions of NLO QCD in the Massive VFN
schemes describe the data well, except at the lowest $x$ for $F_2^{c\bar{c}}$
where the predictions tend to undershoot the data.

\subsection{Summary of Heavy Quark Production at HERA}  

In the deep inelastic scattering regime, the heavy quark production at HERA
is generally well described by NLO QCD.  Massive scheme calculations combined
with a conventional fragmentation function, fitted to
data from $e^+e^-$ colliders, work well in describing $c$-meson
production.  Both the massive VFN and purely massive schemes describe the
$c$ and $b$ production within the current uncertainties.  Within their
uncertainties, $k_t$ factorization calculations are also able to reproduce
the data.

In photoproduction, the descriptions by NLO QCD fall short in some regions.
The predictions tend to undershoot the data in the region of lower
transverse momenta and forward rapidity.  Investigations show that the
discrepancy is largely in the resolved photoproduction regime and 
there are indications that missing higher orders is the cause.

The data covered here are all based on about 120 pb$^{-1}$ each for
H1 and ZEUS in the HERA\,I data taking period.  There is about 400 pb$^{-1}$
of data for each experiment which is yet to be analysed and published
from the HERA II period.  For ZEUS, the HERA II data set is taken with
their newly installed microvertex detector.
 
Due to space limitations, a large part of heavy quark studies at HERA
could not be covered.  These topics include studies of productions
of $c$-hadrons other than the $D^*$ and the study of fragmentation
functions~\cite{h1_zeus_5.1}, the study of the transition region between
DIS and photoproduction~\cite{zeus_5.8} as well as the study of the photoproduction
region using the H1 central silicon tracker~\cite{h1_5.6}. 
%\clearpage
%\newpage
\section{Diffraction}
\subsection{Inclusive Diffraction}
\begin{figure}[t]
\centering
\includegraphics[height=6.3cm]{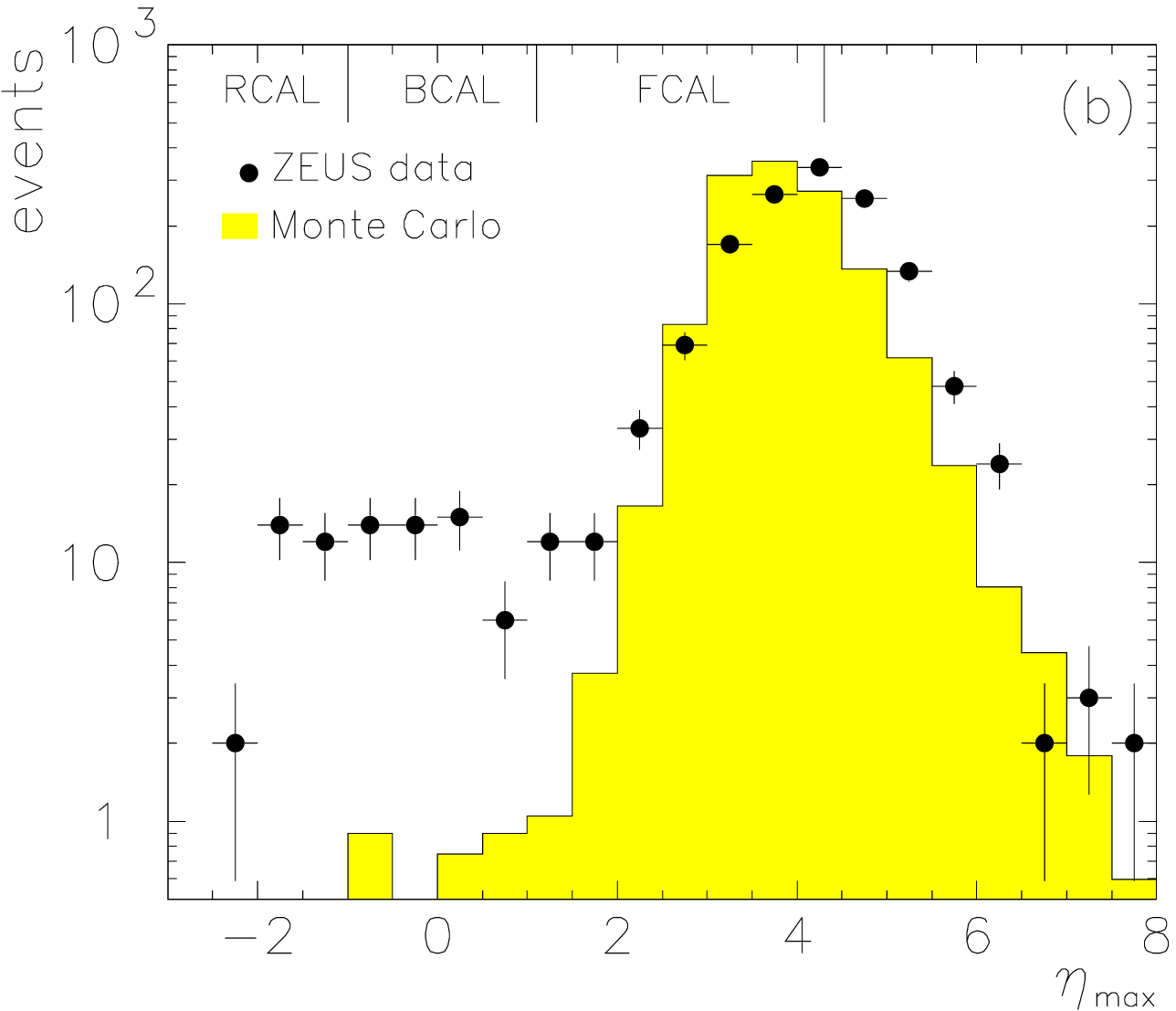}
\includegraphics[height=6.1cm]{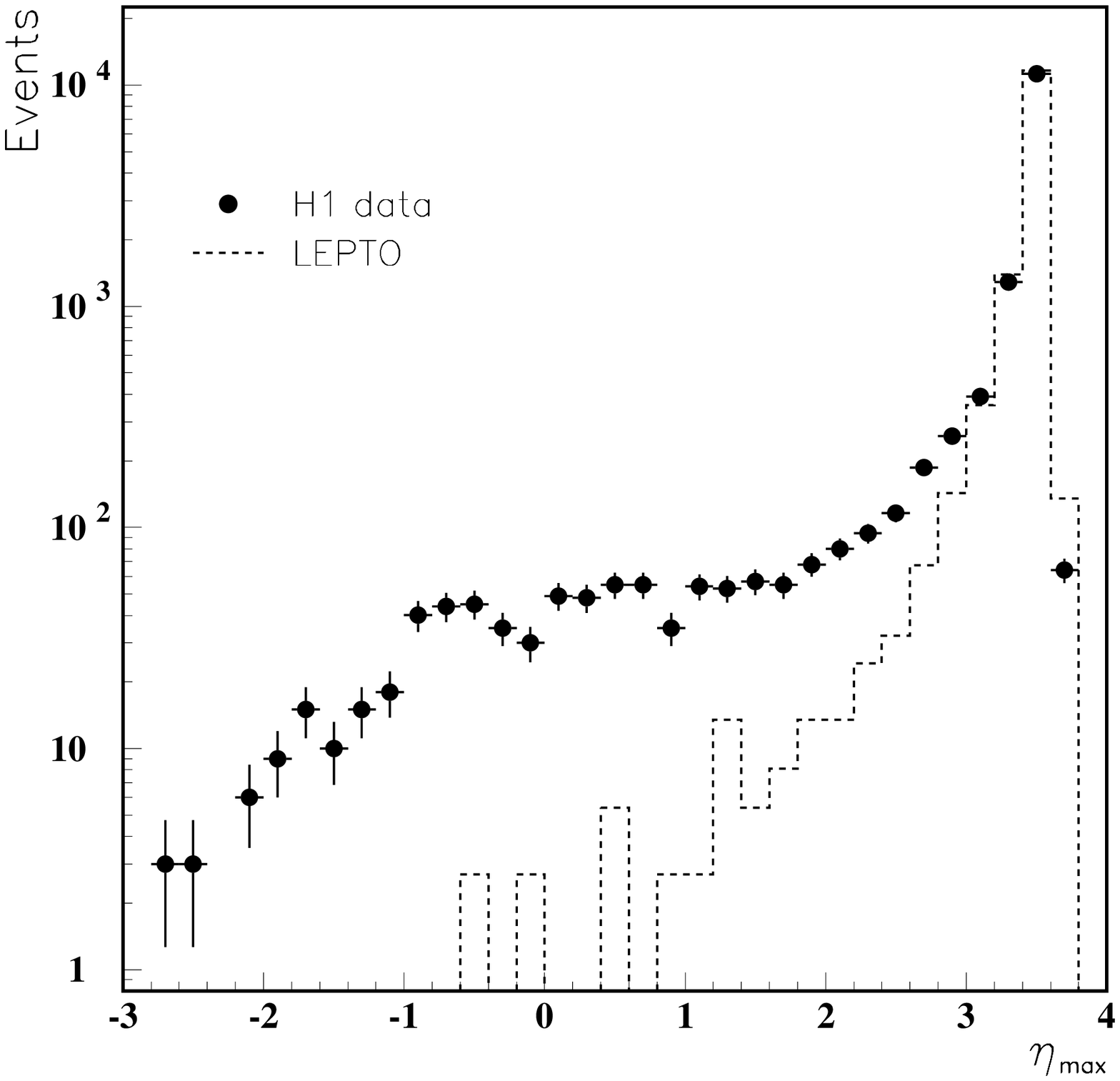}
\begin{minipage}{16.5 cm}
   \caption{The first observations of hard diffraction by ZEUS (left) and 
H1 (right). 
%Left: top: Fraction of DIS events, for which 99\% of the hadronic
%energy lies inside $-3.8 < \eta < \eta_{cut}$, as a function of
%$\eta_{cut}$, for data and a DIS Monte Carlo simulation
%without diffractive component. 
Left:
Distribution  in DIS events of $\eta_{max}$,
the maximum pseudorapidity of a cluster of energy larger than 400 \,MeV,
in the ZEUS calorimeters, the range of which is indicated at the top.
There is a clear excess seen at small $\eta_{max}$ of events in the
data over a conventional DIS Monte Carlo simulation.
Right: Similar observation in the
H1 DIS  $\eta_{max}$ distribution, $E_{min}=400$\,MeV.}
   \label{firstdiff}
\end{minipage}
\end{figure}

\begin{figure}[h]
\centering
\includegraphics[height=6.3cm,angle=90.]{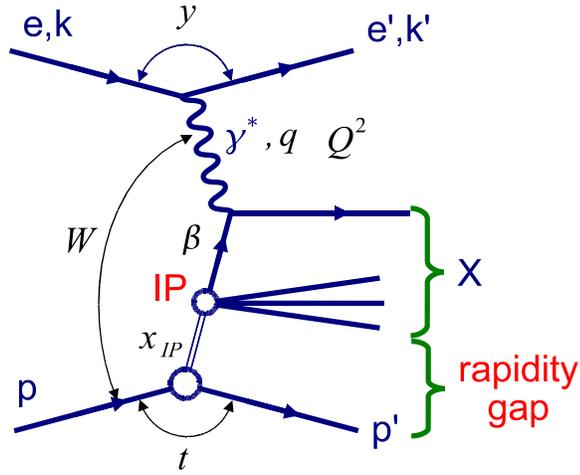}
\begin{minipage}{16.5 cm}
   \caption{Schematic diagram of hard diffractive deep inelastic
scattering. The scattered proton mostly stays intact but also may
dissociate into a system of small mass $M_Y$. The diffractive
exchange carries a fraction $x_{IP}$ of proton's momentum. 
A parton of momentum fraction $\beta=x_{IP}/x$ interacts with
the virtual photon which in DIS has a large virtual mass squared, $Q^2$.
The salient feature of these events is the absence of forward
particle production, near the proton beam pipe, which is measured
as a gap in polar angle, or equivalently rapidity, from the proton
beam axis to the more centrally produced particles which stems from
the struck parton in the diffractive exchange. Despite the violence
of the interaction about a fraction of 10\% (3\%) is observed
at $Q^2$ of 10(1000)\,GeV$^2$. 
%The parton-photon interaction
%can be treated as a hard scattering which obeys the characteristics of
%perturbative QCD. While soft diffractive interactions had indeed
%be expected to be observed at HERA, the occurence of hard diffractive
%DIS came as a surprise.
}
   \label{lrg}
\end{minipage}
\end{figure}

Diffraction as a hard deep inelastic scattering process was discovered at
HERA when ZEUS \cite{zeusdiff1} and H1 \cite{h1diff1} observed an 
unexpected excess of DIS events characterised by the absence of 
activity in the forward direction which is usually occupied by hadrons
emitted from the final state (Figure~\ref{firstdiff}).
In conventional QCD the hadronic final state is produced in the radiation 
from colour charges exchanged in the interaction. The observation of an excess of 
large rapidity gap (LRG) events,
as the absence of hadronic activity at small angles is named, was readily 
understood as being due to a virtual photon-proton interaction without colour exchange.
The basic process is sketched in Figure\,\ref{lrg}.
An exchange of a colour singlet with vacuum quantum numbers is historically called
Pomeron exchange. The early discovery of hard diffraction in DIS at HERA
gave rise to the development of the chromodynamic theory of diffraction,  
based largely on the proof of collinear factorisation\,\cite{diffact} of DDIS
cross sections
at fixed \xpom and $t$. Here \xpom is the fraction of proton momentum
the diffractive exchange carries and $t$ is the 4-momentum transfer at the 
proton vertex. This allowed   diffractive parton densities
to be introduced with the constraint that the proton does not fragment during the violent
$ep$ collision. In fact the calculus of diffractive structure functions can be 
introduced quite analogously to the inclusive case \cite{blumroba}.
%Diffraction has also been studied in the framework of the $k_t$
%factorisation EXPAND.
%

Another way to analyse diffractive DIS is the colour dipole model (CDM) approach
\cite{cdmreview} in which, unlike in the parton model, the proton is studied at rest.
Due to the smallness of $x$, the photon can be considered as disintegrating
at a coherence length $\propto 1/x$ prior to interacting with the 
proton.  This leads to a phenomenological prescription of diffractive
scattering cross sections as the convolution of dipole wave functions
and a dipole-proton cross section, which is to be determined from the
data. 
%The universality of this picture, including inclusive, diffractive and heavy
%flavour production, as well as its simple relation to unitarity, expressed
%as saturation limits
%of the $\gamma^* p$ cross section, has much contributed to a
%remarkable success of the colour dipole approach \cite{cdmreview}.

The accuracy and scope of the diffractive measurements have 
improved dramatically over the past years.  Much more accurate and complete
data are now available based partially on the LRG identification of
diffractive DIS\,\cite{h1diff,zeusdiff}.  The agreement of all data
is approaching an acceptable level, both for the LRG method data and 
also including more recent results applying the so-called $M_X$
selection to diffractive scattering \cite{umoty}. Inclusive
diffractive measurements have been extended to high $Q^2 \leq 2000$\,GeV$^2$.
Diffractive DIS has also been observed in charged current 
scattering. The fraction of hard diffraction to inclusive DIS
is about 10\% at low $Q^2 \sim 20$\,GeV$^2$ and diminishes, both 
in NC and in CC, to a few \% at high $Q^2 \sim 2000$\,GeV$^2$, 
where, however, the mean $x$ is correspondingly larger by
roughly two orders of magnitude at HERA.

Both H1 and ZEUS operated
forward proton spectrometers  in HERA I,  tagging the proton downstream,
near the proton beam line with the 
``FPS'' (H1) and ``LPS'' (ZEUS), at $\simeq + 100$\,m distance from
the interaction point.  In the  HERA II phase, H1 installed
in addition to the FPS a Very Forward Proton Spectrometer,
accessing an \xpom range near to 0.01, while ZEUS had dismantled the LPS.
The FPS and LPS detectors allow a direct comparison of the LRG
based cross sections to the ones obtained with $p$ tagging.
For a diffractive mass range $M_Y < 1.6$\,GeV,  H1
measured  the ratio of the LRG
to the FPS cross sections as $1.23 \pm 0.05$ and to be constant within 
the kinematic range of the measurement,
 $2 < Q^2 < 30$\,GeV$^2$, $0.01 < \beta < 0.7$
and $0.001 < x_{IP} < 0.05$, where $\beta=x/x_{IP}$ is the  momentum
fraction of the diffractive exchange carried by its partons. 
The ZEUS result is $1.23$ with a total error of 0.05 \cite{lukasi}
for the ratio of the LRG to LPS data, in  agreement with the H1
measurement and also constant in the kinematic range of the measurement,
 $2.5 \leq Q^2 \leq 40$\,GeV$^2$, $0.007 \leq \beta \leq 0.816$
and $0.0002 < x_{IP} < 0.02$. Thus the dominant process
in the gap data is indeed diffraction and further processes such as resonance 
or neutron production do not alter the kinematic dependencies beyond the 
point to point uncertainty, of typically 5-10\%,
of the LRG/FPS or LRG/LPS ratio.

It is satisfying to observe
that the tagged differential diffractive cross section
data are consistent between H1 and ZEUS. The tagged data also allow 
the $t$ dependence of the cross sections to be measured which
can be parameterised as $d \sigma /dt \propto e^{-bt}$.
A slope of $7 \pm 0.3$ is obtained by ZEUS \cite{lukasi}.
H1 has parameterised the slope as 
$b = b_{IP} -2 \alpha_{IP} \ln x_{iP}$
corresponding to a trajectory 
$ \alpha_{IP}(t) =  \alpha_{IP}(0) + \alpha^{,}_{IP} t$.
At small \xpom one obtains $b_{IP} = 6.0$ with an error of about 2 and
a small $\alpha^{,}_{IP} = 0.02$ consistent with zero within the 
errors \cite{h1diff}.
A large $\alpha^{,}_{IP}$ would have been an indication for shrinkage
of the diffractive peak.
%, seeFigure~\ref{tagdata}.
%
%\begin{figure}[t]
%\centering
%\includegraphics[height=7.6cm]{herafighztagch6.eps}
%   \caption{Tagged diffractive data of the inclusive cross section
%of H1 (open points) and ZEUS (stars) as obtained in HERA I.}
%   \label{tagdata}
%\end{figure}
%

\subsection{Diffractive Parton Distributions}
The diffractive measurements based on the LRG method, which 
has the widest kinematic coverage, allowed QCD analyses 
to be performed in order to derive the quark and gluon
distributions of the diffractive exchange. Such analyses
are based  on the diffractive cross section $\sigma^{D(3)}$  
\begin{equation}
\frac{{\rm d}^3\sigma^{ep \rightarrow e X Y}} {d x_{IP} d x  d Q^2}
 = \frac{2\pi  \alpha^2} {x Q^4} \cdot Y_+ \cdot \sigma_{r}^{D(3)}(x_{IP},x,Q^2),
\label{sigmard}
\end{equation}
which is integrated over the ranges of $t$ and $M_Y$.
Similarly to inclusive DIS, the reduced $ep$ cross section depends on the 
diffractive structure functions $F_2^{D(3)}$ and $F_L^{D(3)}$ 
in the one-photon exchange approximation according to
\begin{equation}
\sigma_r^{D(3)} = F_2^{D(3)} - \frac{y^2}{Y_+} F_L^{D(3)}.
\label{sfdef}
\end{equation}
For $y$ not too close to unity, 
$\sigma_r^{D(3)} = F_2^{D(3)}$ holds to very good approximation. 

\begin{figure}[t]
\centering
\includegraphics[height=7.cm]{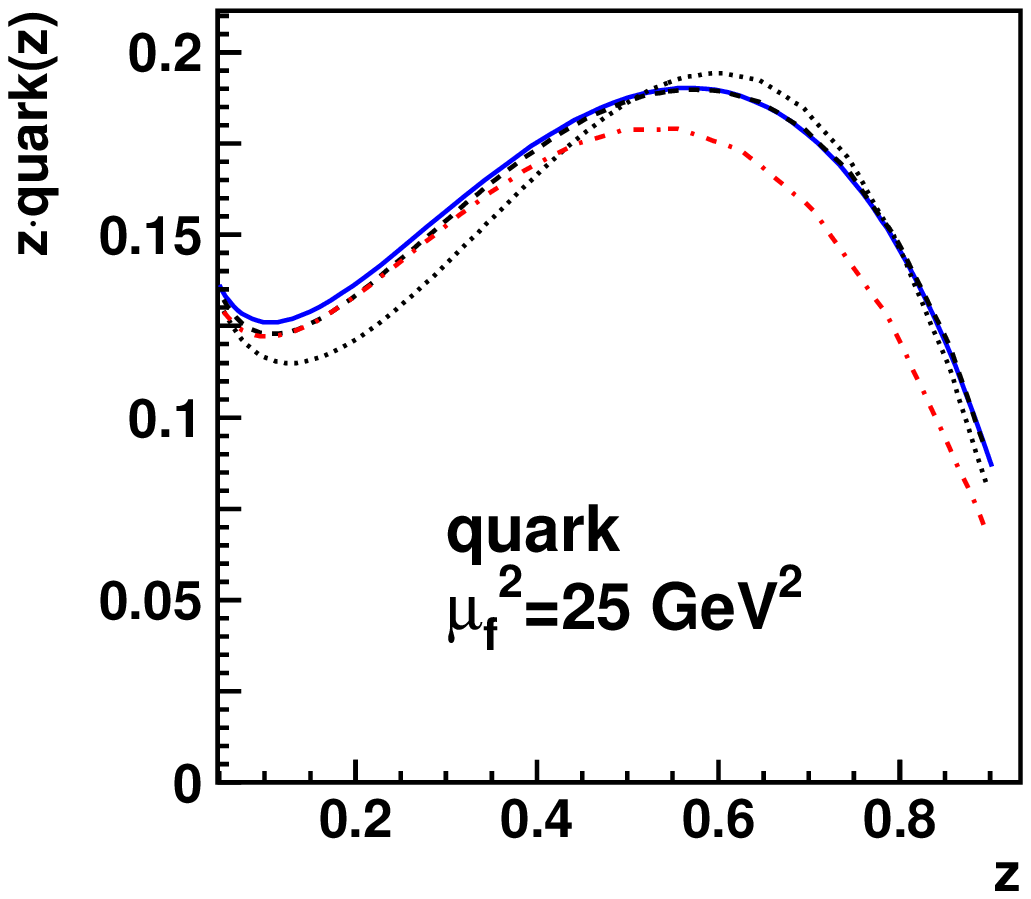}
\includegraphics[height=7.cm]{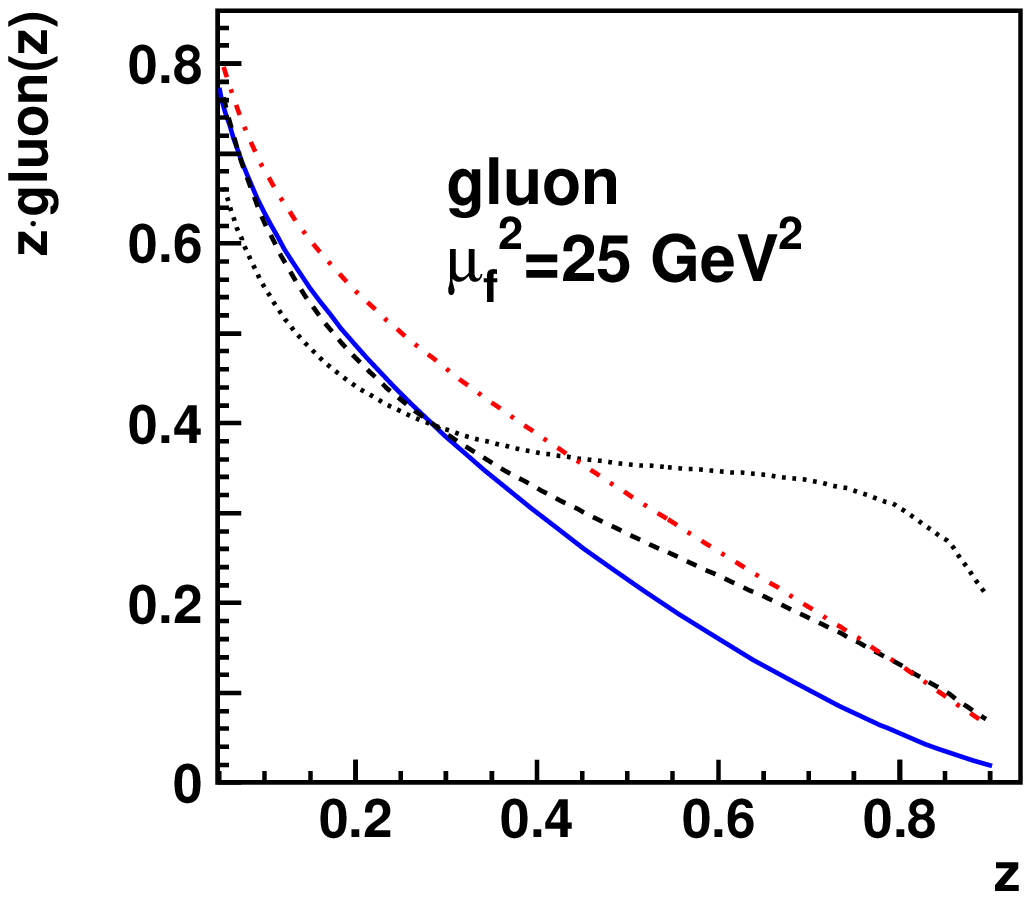}
\begin{minipage}{16.5 cm}
   \caption{Diffractive partons as determined in various NLO QCD fits:
dotted: inclusive H1 fit (A); dashed: inclusive H1 fit (B); solid: H1
fit to inclusive and dijet data; dashed-dotted: from \cite{watts}. Here 
$z$ is the momentum fraction of the diffractive exchange carried 
by the parton. 
}
   \label{fig:difpart}
\end{minipage}
\end{figure}

Recently a detailed QCD analysis was published by H1~\cite{h1diff}.
Since data are used for a range of \xpom values, it is required to
factorise the \xpom dependence out which in \cite{h1diff}
is done with an ansatz inspired by Regge theory. The QCD fit,
similar to truly inclusive scattering as described above, then determines
parton distributions, here a singlet quark distribution $\Sigma(x,Q^2)$,
assuming $u=d=\bu=\bd=s=\bs$, and a gluon distribution $xg(x,Q^2)$,
where $x=\beta x_{IP}$.  The data, as is shown for example
in Figure~\ref{fig:difder} (left) for a medium \xpom value, are well described
in their $x$ and $Q^2$ dependence by this approach.
 In the analysis, rather stable results were obtained
for both distributions, with the exception of the behaviour of the
gluon distribution at large momentum fractions which is rather uncertain.
This can be understood from the consideration of the $ \ln Q^2$ derivative
of the diffractive reduced cross section (see  Figure~\ref{fig:difder} (right)).
As the two QCD fit components reveal, at large $\beta$ there is not much
sensitivity to gluons \footnote{This result is reminiscent of inclusive DIS:
at large $x$ the scaling violations are driven by $\alpha_s$ only
and no direct sensitivity to the gluon distribution appears. This is
the main reason why older fixed target DIS experiments have not
determined $xg$ well. It required a high energy $ep$ collider to 
measure the gluon distribution in the proton. The same holds for 
polarised $ep$ scattering, the determination of the polarised gluon
distribution $G$ cannot reliably be obtained from lower energy data
as these are essentially sensitive to chromodynamic bremsstrahlung only.}.
Thus H1 decided to publish two fits, with very similar $\chi^2$
but different $xg$ at large $\beta$. A distinction between these two possibilities
was made possible with diffractive dijet data \cite{h1ddij}. The diffractive dijet
production leads to improved sensitivity to the gluon distribution in 
the diffractive exchange (see Figure\,\ref{fig:difpart}).
Thus the gluon distribution  uncertainties can be reduced
significantly, similar to the influence of dijet DIS data on the determination
of the gluon distribution in the proton. 

 There are many further similarities
between diffractive and inclusive DIS:
\begin{figure}[h]
\centering
\includegraphics[height=8.8cm]{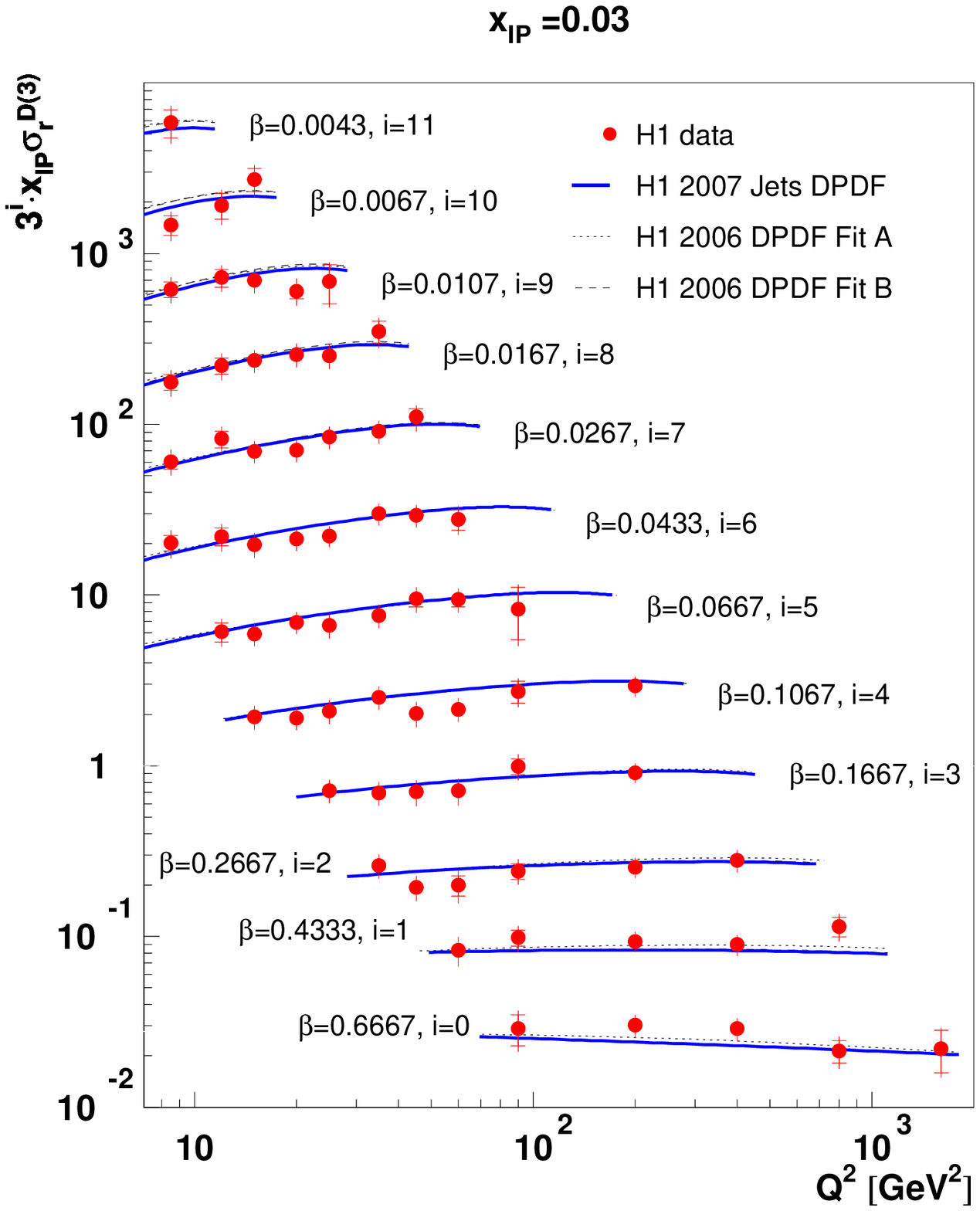}
\includegraphics[height=8.8cm]{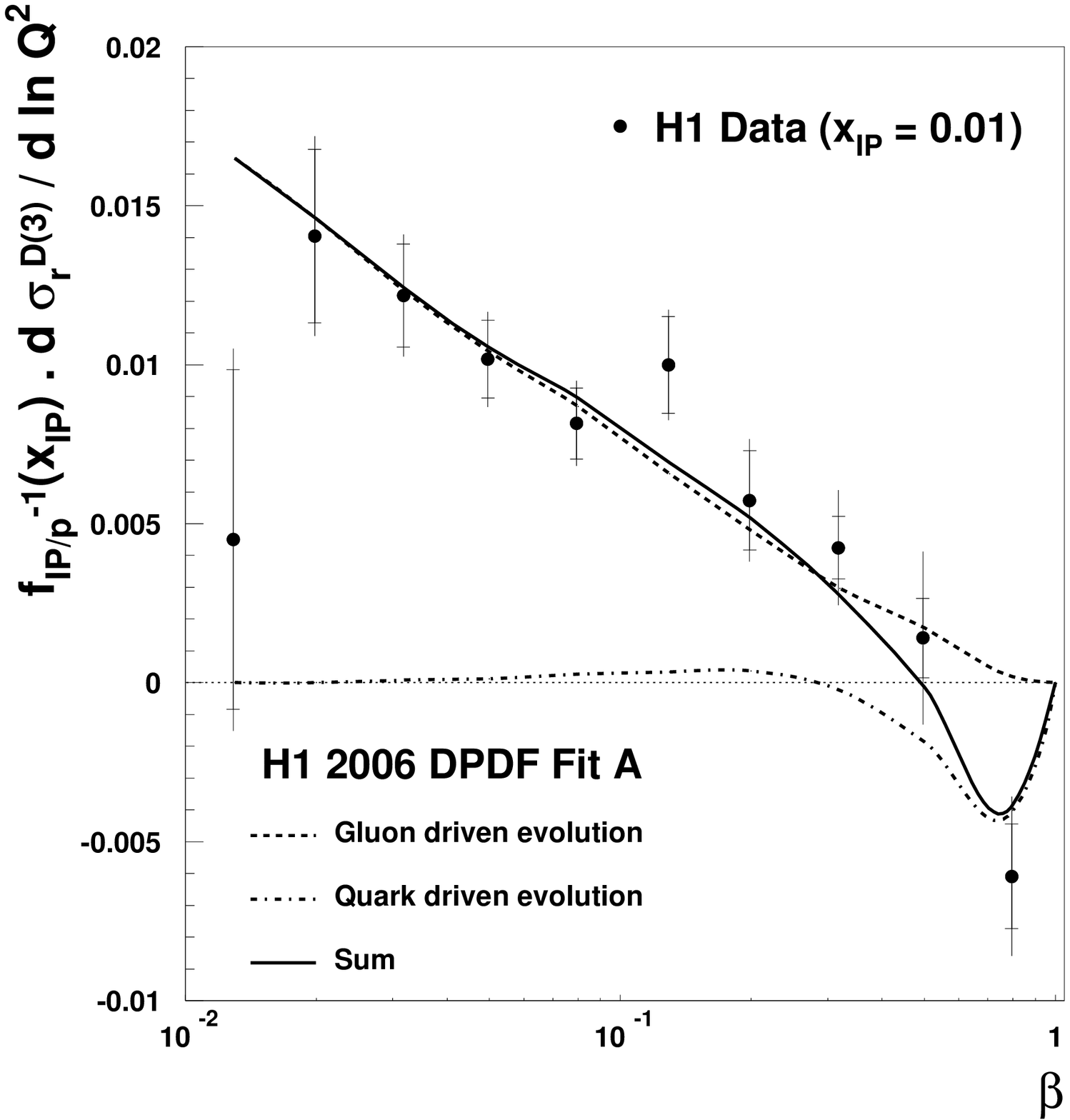}
\begin{minipage}{16.5 cm}
   \caption{Left: Diffractive reduced cross section at \xpom=0.03 as a function
 of $Q^2$ for different $\beta = x/x_{IP}$. The data are H1 data from HERA I.
 The curves are different NLO QCD fits, dashed: fits A and B (see text) 
 to the inclusive data only, solid: fit to inclusive and diffractive dijet data.
 Right: Derivative of the reduced diffractive cross section  at $x_{IP} = 0.01$
 and its decomposition  into the gluon and quark parts in the H1 NLO QCD analysis.}
   \label{fig:difder}
\end{minipage}
\end{figure}
 The logarithmic derivatives
of $\sigma_r^D$ and $\sigma_r$, with respect to $\ln Q^2$, are measured to be the same
for $\beta < 0.6$. Thus low $\beta$ diffraction and low $x$ inclusive scattering seem 
to reflect a common origin, the dynamics of the QCD vacuum.
It is not surprising, then,
that a number of recent observations in diffractive DIS resemble those in inclusive 
DIS; such observations include the rise towards low $\beta$ or $x$ of 
the $\ln Q^2$ derivatives and the fraction of charm of about 20\%, for $Q^2$
away from thresholds,
%~\cite{charmdiff}, 
in both the diffractive cross section and in the inclusive cross section.
Diffraction has evolved from the first surprise of its presence at HERA
to a rather exact and interesting
testing ground of perturbative QCD. Further data from both HERA collaborations
and studies of universality in comparing DIS $ep$ and photoproduction 
predictions
\cite{diffacth1}, as well as of HERA data with Tevatron and later LHC
data, will certainly contribute to this field in the future. 
Recently, diffraction is being considered as a 
means of studying possible SM and SUSY Higgs particle  production at the LHC,
in a probably rare but possibly clean double diffractive scattering
process $pp \rightarrow pHp$ \cite{fp420}. 
\subsection{Colour Dipole Model  Description of Diffraction}
Diffractive parton densities lead to predictions in the same way conventional
parton densities do.
%, it does not directly address the question of the physics
%of this phenomena.  Rather, 
However, the $x_{IP}$ dependence, which is some measure of the
colour singlet formation within the proton, is factored out 
and remains a phenomenological assumption made prior 
to the subsequent QCD analysis.
%and are not considered
%further.

An alternative way to study the phenomenon of diffractive DIS is through the application
of the Colour Dipole Model (CDM).
%, which is formulated in the rest-frame of the proton.
%The virtual photon disintegrates long before interacting with the proton,
%at a coherence length $\propto 1/x$ apart, into a dipole
%of quark and antiquark which hits the proton.  
The cross-section is here given by the convolution
of the quark-antiquark colour dipole wave-function and the dipole-proton cross-section.
This formalism connects the inclusive low $x$ cross-section, discussed in Section 3.4.3,
to the DDIS cross-section in the same way forward elastic hadron scattering is related to
the total cross-section.

The dipole-proton cross-section may be describable in terms of the DGLAP,
or the BFKL formalisms in their respectively applicable kinematic regions. This has not
yet been fully demonstrated.  In the very low $x$
or low $Q^2$ regions, gluon recombination and/or non-perturbative 
effects should
influence the cross-section.  
\begin{figure}[h]
\begin{center}
%\epsfysize=9.0cm
%\begin{minipage}[t]{9 cm}
\epsfig{file=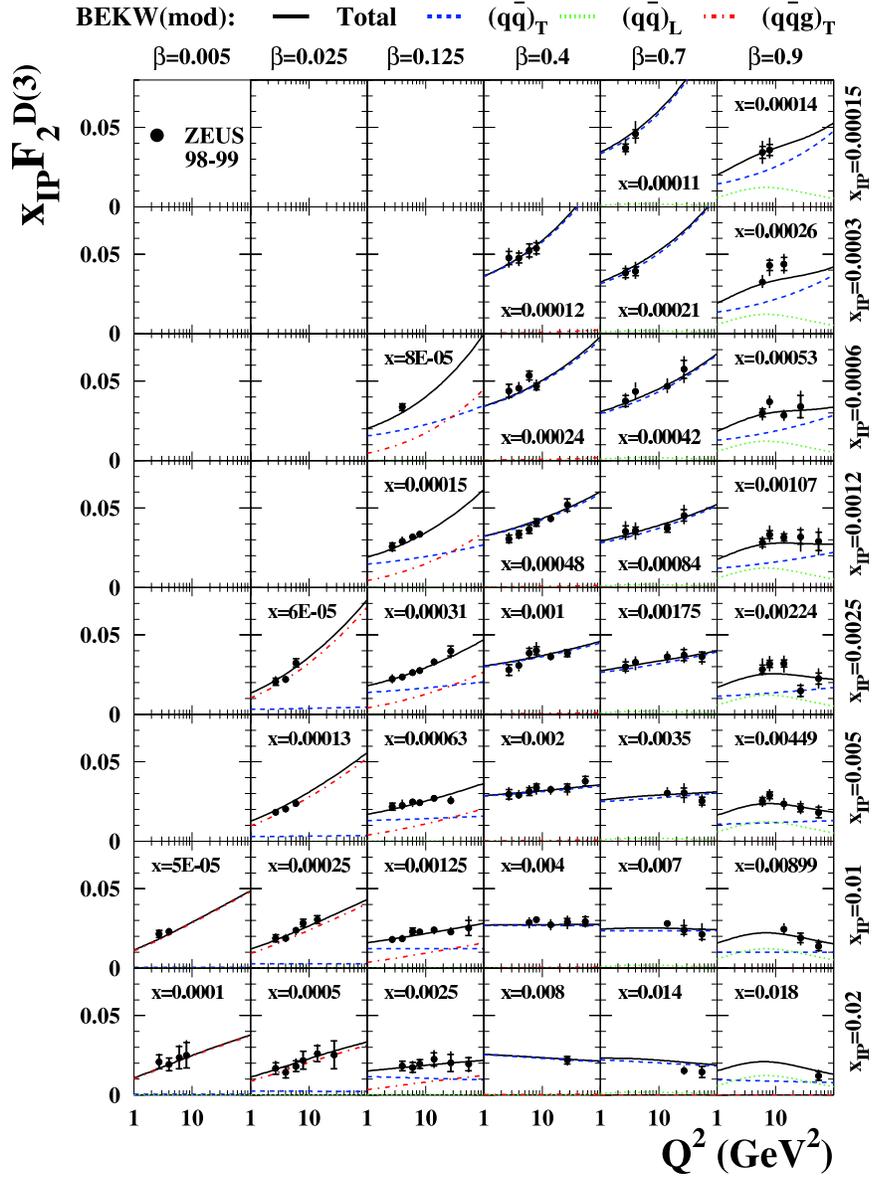,scale=0.6}
%\end{minipage}
\begin{minipage}[t]{16.5 cm}
\caption{DDIS measurements by ZEUS~\cite{umoty}.  The lines are fits to a CDM-type model 
BEKW~\cite{bekw_6.1}.
\label{fig_zeus_diff}
}
\vspace{-0.8cm}
\end{minipage}
\end{center}
\end{figure}

Figure \ref{fig_zeus_diff} shows an example of a 
fit to DDIS cross-section data~\cite{zeus_6.01} 
using a CDM-type model, BEKW~\cite{bekw_6.1}.
The description achieved can be very good, and some models
describe a number of cross-sections, DDIS, inclusive DIS, Heavy Quark 
production, simultaneously. While the phenomenological success is remarkable,
it is fair to say that at this stage, 
there is no theoretical consensus in the interpretation of these models.  The
reader is referred to recent articles and reviews \cite{nachtcdm,cdmreview} for 
further discussions.

\subsection{Elastic Vector Meson production}
The description of exclusive electroproduction of vector mesons, 
$ep \rightarrow e V p$, is closely related to the CDM model described above.
The virtual photon emitted from
the electron fluctuates into a quark-antiquark dipole, which then scatters
elastically off the proton.  Long after the interaction, the $q\bar{q}$
pair forms a vector meson.  At sufficiently
high energies, the three steps,
the $q\bar{q}$ formation, the dipole scattering and the formation of the
vector meson are well separated in time.  The cross-section for the
process, then, is factorised into the $q\bar{q}$ coupling to the
photon, the dipole-proton scattering amplitude, and the final-state
formation~\cite{kopeliovitch_6.1,brodsky_6.1}.  The first and the last steps involve the 
electromagnetic coupling of the $q\bar{q}$ pair and the wave function
of the vector meson, respectively.  In case of light mesons, 
parton-hadron duality is sometimes
invoked for the last step.

If there is a sufficiently large scale involved, the 
dipole-proton cross-section is calculable in terms of two-gluon
exchange.  In principle this cross-section depends on the
generalised parton distribution function (GPD)~\cite{gpd} of the gluon. 
GPDs contain information on the correlation of partons within
the proton.  At low $x$, i.e. where leading $\ln{1/x}$ approximation
is valid (and if $t$ is small), the generalised gluon distribution 
can be approximated by the usual gluon distribution.  
The cross-section then should rise steeply in $1/x$ (or equivalently
in $W$, the virtual photon-proton cms energy, which at low $x$
is given by $\sqrt{s/x}$.) for fixed $Q^2$ reflecting 
the steep rise of the gluon density at low $x$.

For those cases where the interaction is soft, the
cross-section should rise slowly with $W$, similarly to the well-known
Regge energy behaviour of hadron-hadron total cross-sections.
\begin{figure}[tbh]
\begin{minipage}[t]{8 cm}
\hspace{20pt}
\epsfig{file=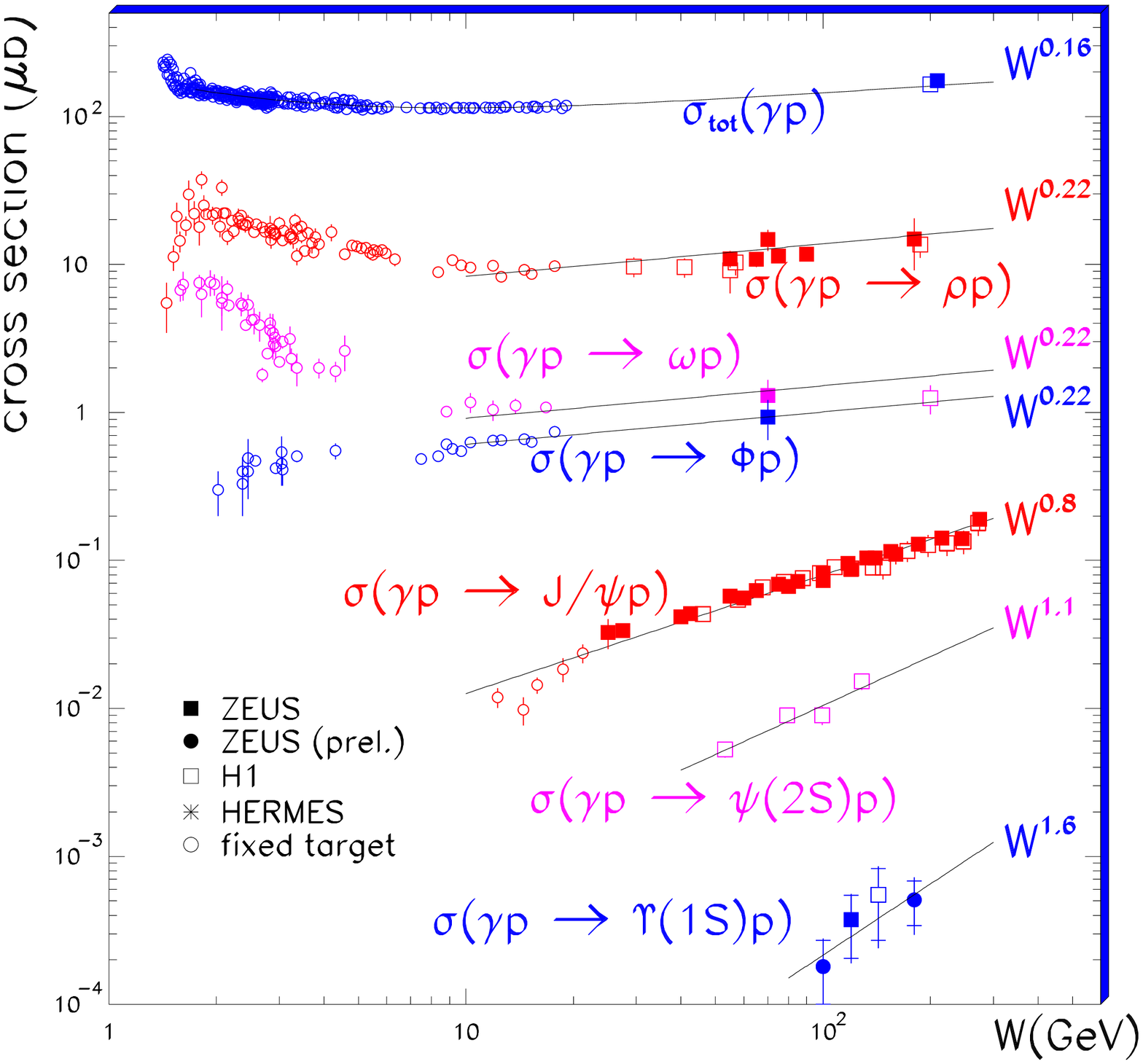,scale=0.36}\epsfig{file=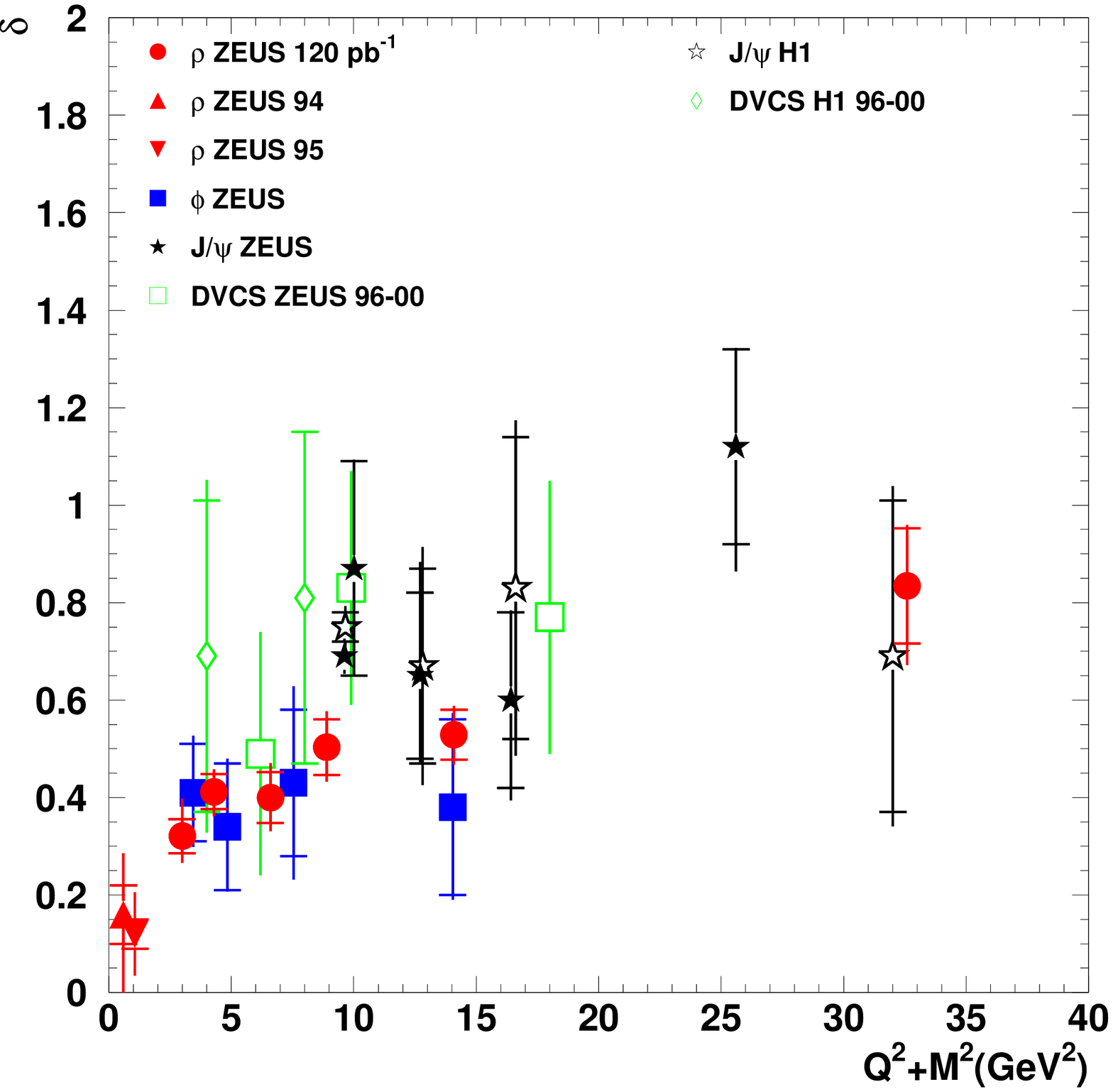,scale=0.38}
\end{minipage}
\begin{center}
\begin{minipage}[t]{16.5 cm}
\caption{
%a) Elastic $J/\Psi$ electroproduction cross-section
%as a function of $W$ for four values of $Q^2$. The results for
%photoproduction ($Q^2 \approx 0$) is also shown.  The data are compared
%to theoretical calculations obtained with diffrent
%parametrisations of the gluon density and normalised to 
%the photoproduction point at $W=90$ GeV.  The inset shows the
%parameter $\delta$ as a function of $Q^2$. The inner error bars are
%statistical uncertainties, and the outer error bars, the statistical
%and systematic uncertainties in quadrature.  
(left) Elastic VM photoproduction cross-section as a function of
$W$ for a variety of species of VMs for HERA and fixed-target
experiments.  The data are fitted to the form $W^\delta$, and the
results are shown in the plot.
(right) The extracted
values of $\delta$ from HERA elastic VM production and DVCS data
as a function of $Q^2+M_{VM}^2$.
\label{fig_6.1}
}
\end{minipage}
\end{center}
\end{figure}

Figure~\ref{fig_6.1}(left) 
%show the measurements of the elastic $J/\Psi$ production~\cite{zeus_6.1}.
%
show the measurements of elastic VM photoproduction~\cite{h1_zeus_6.1} from
HERA as well as the fixed-target experiments.
%% [cite HERMES et al].
The data are fitted to the form $W^\delta$ to quantify the rate of
the rise of the cross-section with energy.  
The $\delta$ increases as the mass of the VM increases. For the $J/\Psi$
one finds a value of
$\delta \approx 0.8$ which is consistent with a gluon density
varying as $x g(x) \propto x^{-0.2}$ close to typical exponents
extracted from the QCD fits to DIS inclusive data at small $Q_0^2 \sim M_{J/\Psi}^2$.  
%The value
%of $\delta$ is consistent with being constant as a function of $Q^2$.
\begin{figure}[bh]
\begin{minipage}[bh]{8 cm}
\hspace{20pt}
\epsfig{file=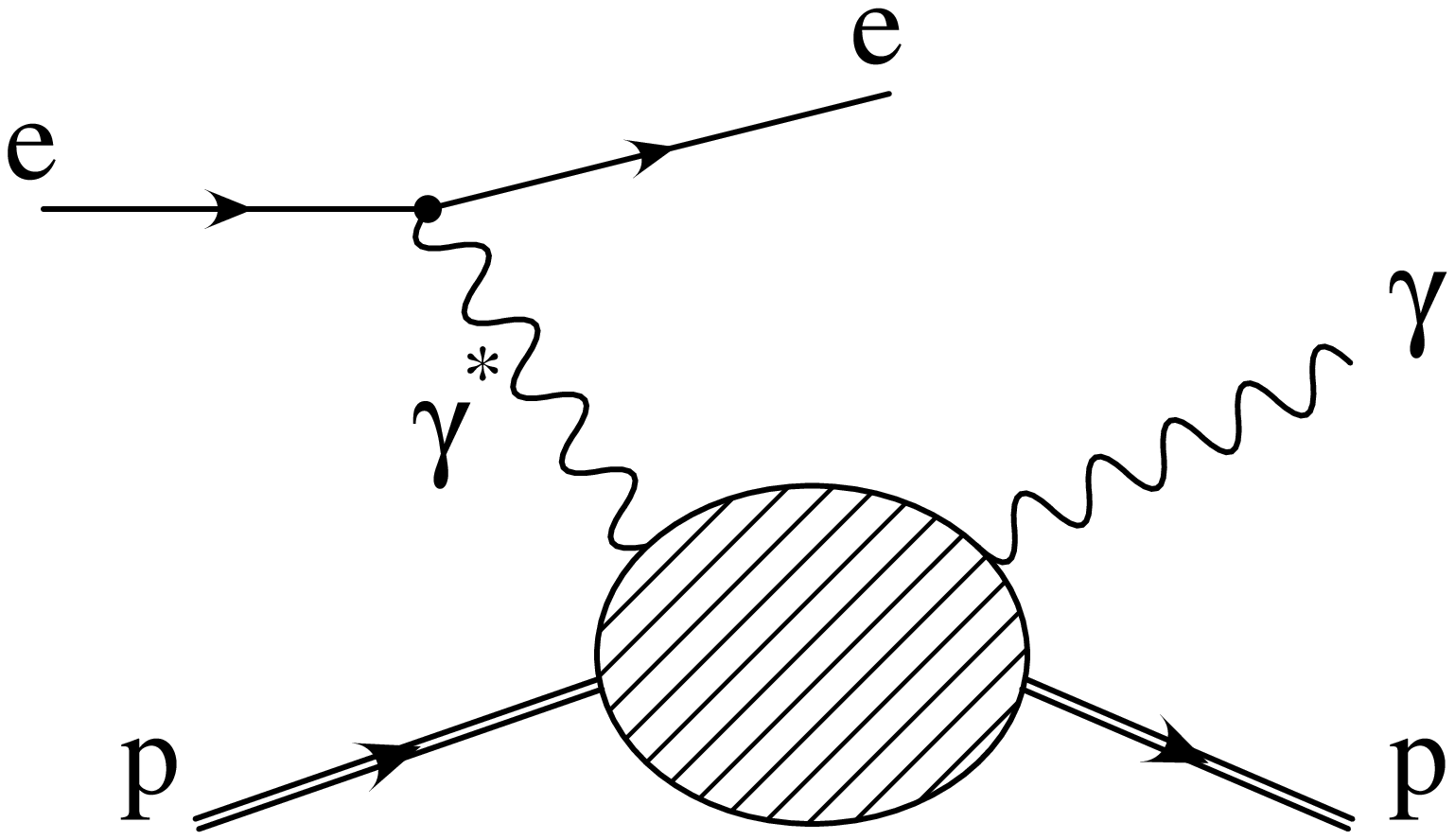,scale=0.44}\epsfig{file=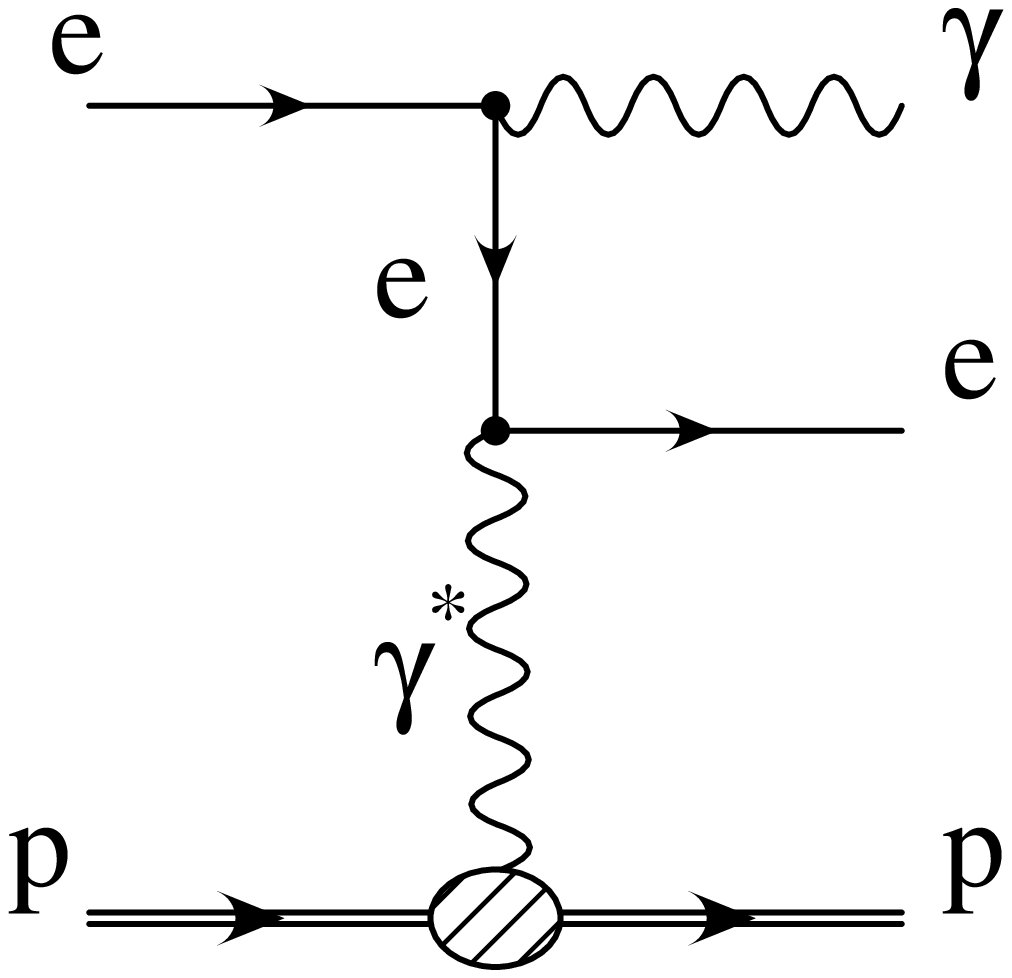,scale=0.44}\epsfig{file=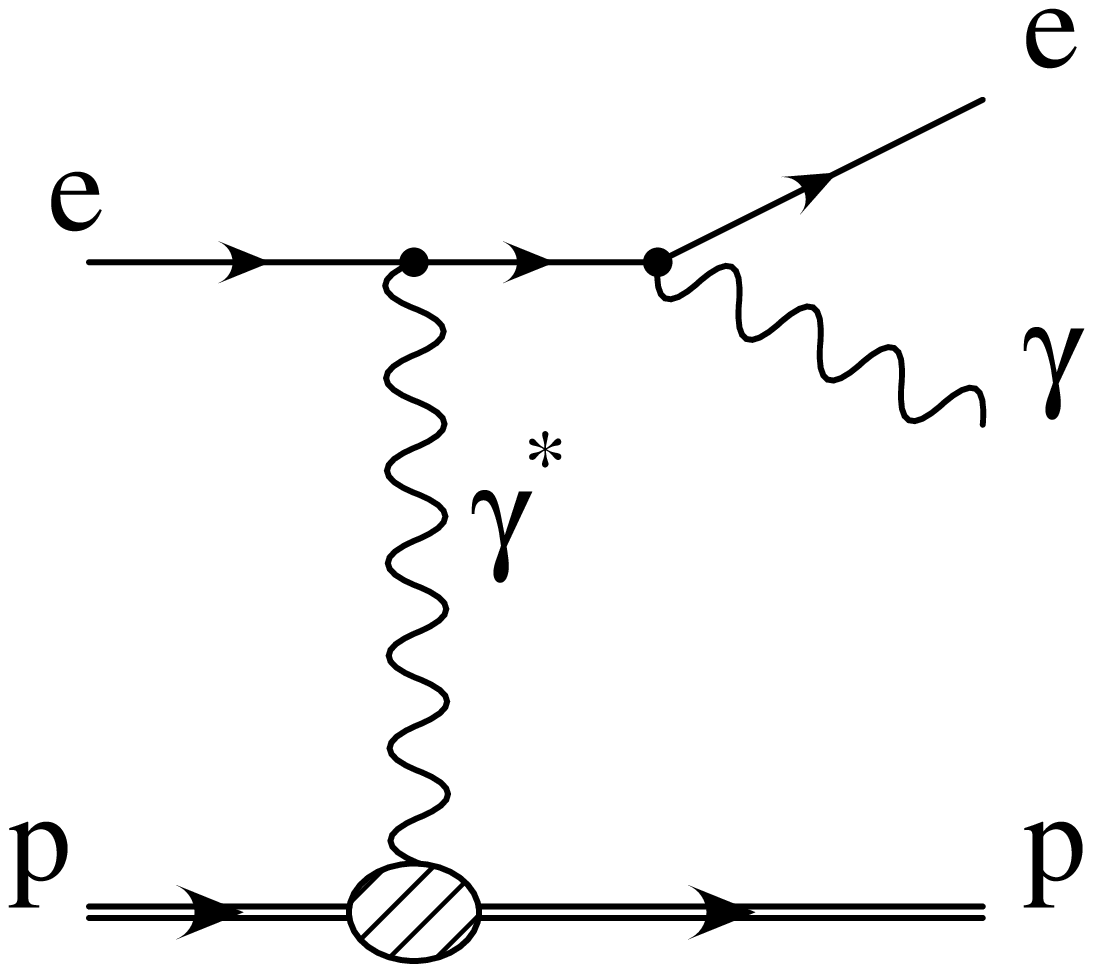,scale=0.44}
\end{minipage}
\begin{center}
\begin{minipage}[bh]{16.5 cm}
\caption{Diagrams illustrating the DVCS (left) and the Bethe-Heitler 
(center and right)
processes.
\label{fig_6.2}
}
\end{minipage}
\end{center}
\end{figure}

Figure~\ref{fig_6.1}(right) shows the quantity $\delta$ extracted from many different
measurements~\cite{zeus_6.1,h1_zeus_6.1} of VM production at HERA, now as a function of
$Q^2 + M_V^2$, where $M_V$ is the mass of the vector meson being
produced.  
While the uncertainties are still large,
at low values of $Q^2 + M_V^2$, $\delta \approx 0.2$
which is the expected value for Regge energy behaviour while at 
higher $Q^2 + M_V^2$ the exponent $\delta$ rises to around 0.8.
The data are consistent with $Q^2 + M_V^2$ being the appropriate hard
scale.
 
%\begin{figure}[h]
%%\epsfysize=9.0cm
%\centering
%\includegraphics[height=8cm]{herafigallVMch6.eps}
%\caption{I THINK WE SHOULD HAVE THIS IN - from ZEUS EPS07 webpage}
%\label{figallvm}
%\end{figure}

\begin{figure}[tb]
\begin{center}
%\begin{minipage}[tb]{15 cm}
\epsfig{file=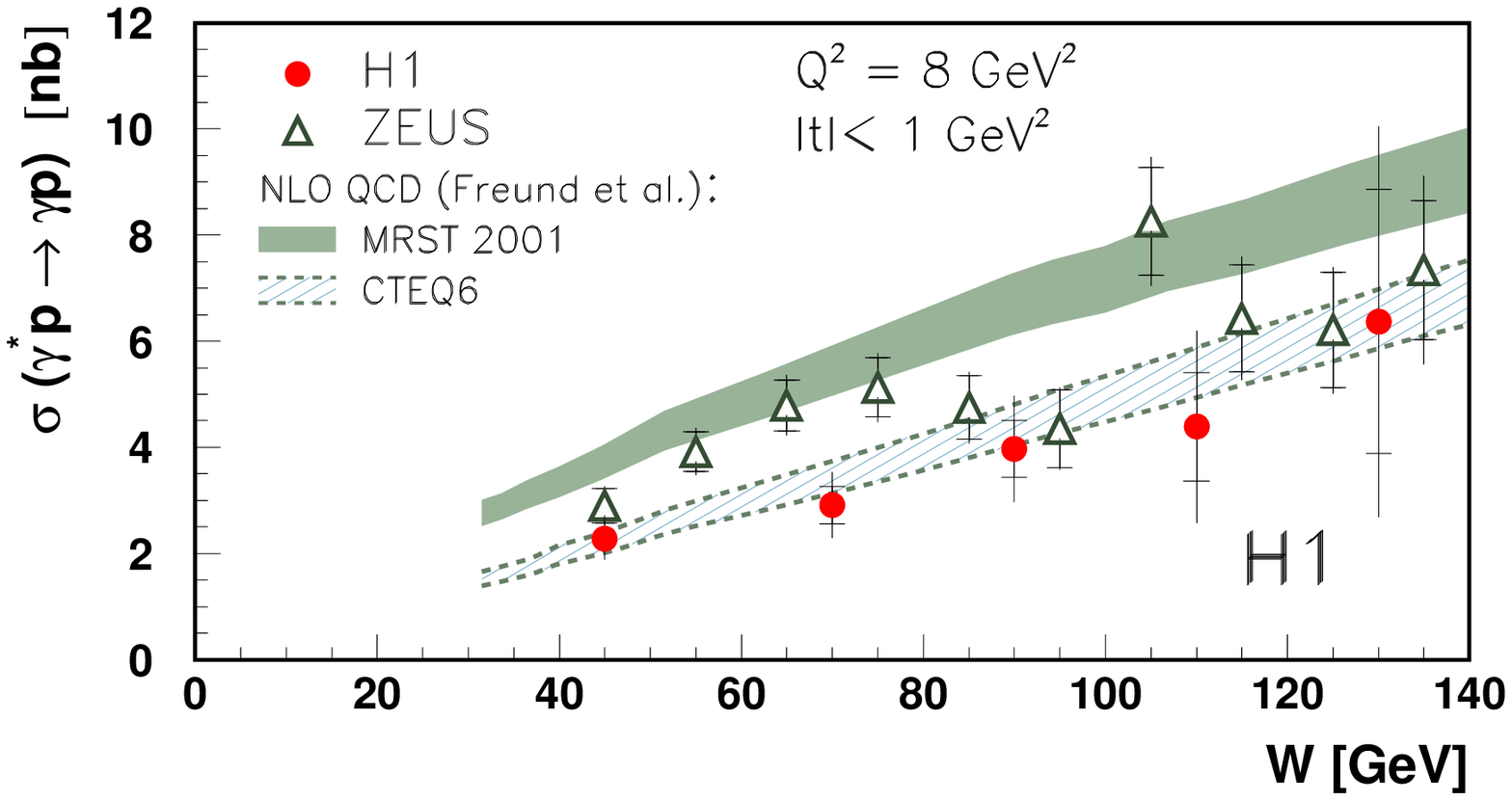,scale=0.8}
%\end{minipage}
\begin{minipage}[tb]{16.5 cm}
\caption{The DVCS cross-section as a function of $W$ for $Q^2 = 8$ GeV$^2$
and $|t| < 1$ GeV$^2$.  The inner error bars are statistical and outer,
statistical and systematic uncertainties taken in quadrature.
\label{fig_6.3}}
\end{minipage}
\end{center}
\end{figure}

\subsection{Deeply Virtual Compton Scattering}

The process $ep \rightarrow e \gamma p$ at the HERA collider has
contributions from both the Bethe-Heitler (BH) and the deeply virtual
compton scattering (DVCS) processes (Figure~\ref{fig_6.2}).   In the kinematic
region of the HERA collider, the interference term between DVCS and BH
processes is small, and the latter, which is precisely calculable,
can be subtracted from the measured cross-section in order to obtain
the DVCS cross-section.

The DVCS process is rather similar to the elastic VM production process,
but has the advantage of having a simple final state with no
need to rely on an understanding of meson wave functions.  
Figure~\ref{fig_6.3} shows 
the measured DVCS cross-section~\cite{zeus_6.2,h1_6.1} as a function of $W$.  The 
rise with $W$ is relatively steep ($\delta \simeq 0.75$,
see Figure \ref{fig_6.1}) demonstrating the relative ``hardness''
of this cross-section.  The theoretical predictions of 
Freund and McDermott~\cite{freund_6.1} use GPDs based on normal 
PDFs (MRST and CTEQ)~\cite{freund_6.2}.

\begin{figure}[h]
\begin{center}
%\begin{minipage}[h]{15 cm}
\epsfig{file=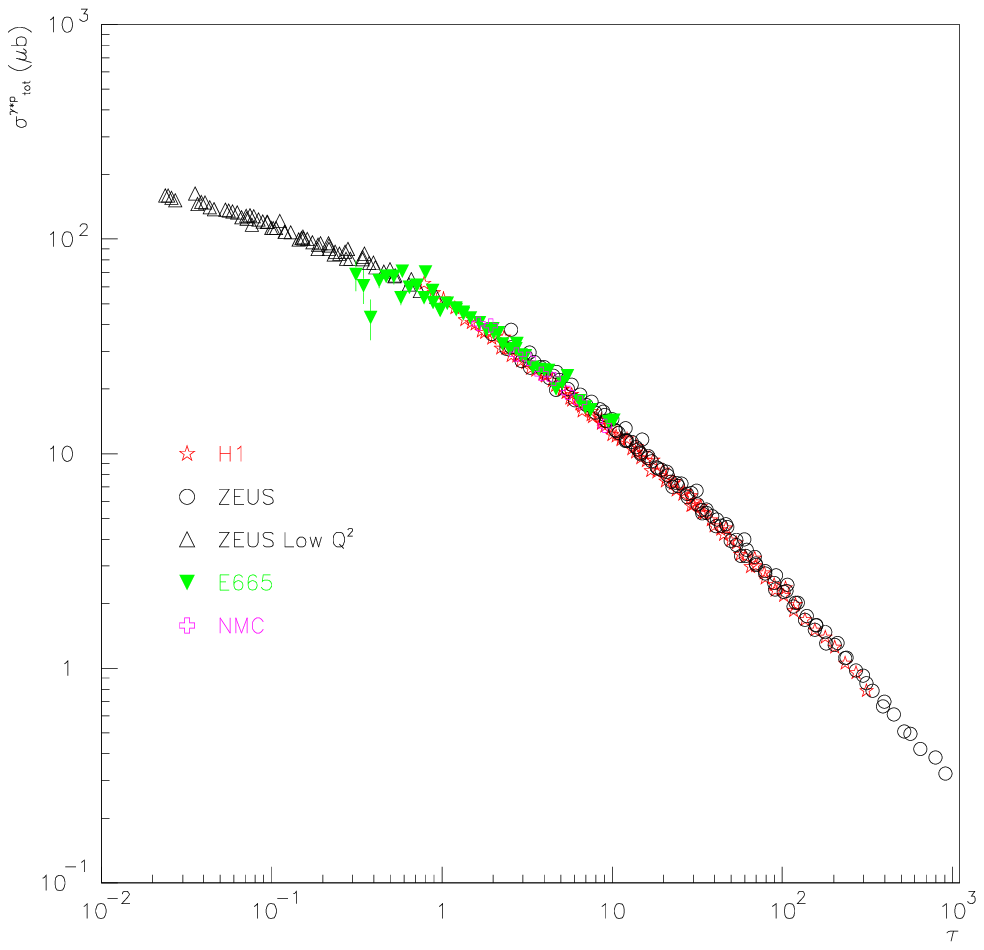,scale=0.8}\epsfig{file=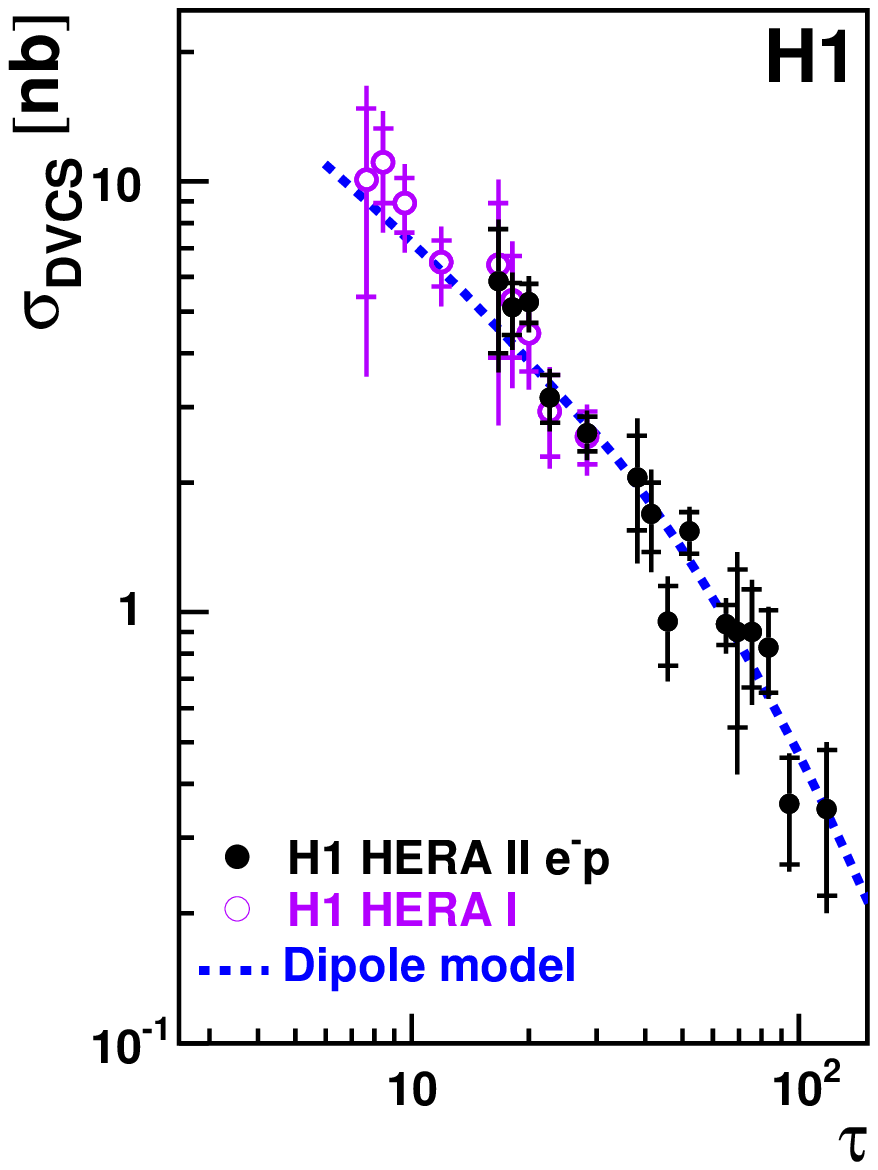,scale=0.63}
%\end{minipage}
\begin{minipage}[h]{16.5 cm}
\caption{Left) Total DIS cross-section (written as the $\gamma^* p$ cross section) as a
function of the saturation scaling variable $\tau$. The figure is taken from~\cite{marquet_6.1}. 
Right) The DVCS cross-section from the H1 collaboration as a function of $\tau$.
\label{fig_6.4}}
\end{minipage}
\end{center}
\end{figure}

In a class of CDM models called the saturation model~\cite{gbw_6.1}, the dipole-proton
cross-section is proportional to the square of the dipole separation $r^2$ at small
$r \propto (1/Q)$.
At sufficiently large $r$, the dipole-proton cross-section saturates (becomes
constant as a function of $r$) to preserve unitarity.  
The transition of a process into the saturation
regime is characterised by the saturation scale $Q_s(x)=Q_0(x_0/x)^{-\lambda/2}$, where
$Q_0$, $x_0$ and $\lambda$ are parameters.
One prediction of such models is that
certain DIS cross-sections can be expressed as a function of the single variable 
$\tau = Q^2 /Q_s^2(x)$.  Figure~\ref{fig_6.4} (left) is the total $ep$ DIS
cross-section data at low-$x$ ($<0.01$) as a function of $\tau$~\cite{marquet_6.1}.  
In this case $Q_0^2=$ 1 GeV$^2$, $x_0 =$ 3.04$\cdot$10$^{-4}$
and $\lambda =$ 0.288, as was determined in~\cite{gbw_6.1}.  Figure~\ref{fig_6.4}\,(right)
show the H1 DVCS data~\cite{h1_6.1} using  slightly different parameters,
 $Q_0^2=$ 1 GeV$^2$, 
$x_0 =$ 2.7$\cdot$10$^{-5}$ and $\lambda =$ 0.25.  It has been shown that such a scaling
holds relatively well also for elastic VM production and DDIS cross-sections.
%\clearpage
%\newpage
\section{Electroweak Measurements}
\subsection{Charged Current Cross-Section}
Rewriting Equation \ref{Rnc}, the charged current DIS cross-section 
$d^2\sigma^{CC}/dxdQ^2$, for the reaction $ep \rightarrow \nu(\bar{\nu})X$,
can be expressed as:
\begin{equation}
 \frac{d^2\sigma^{CC}(e^{\pm})}{dxdQ^2} = \frac{G_F^2 Y_+}{2\pi x}
 [\frac{M_W^2}{M_W^2+Q^2}]^2 \sigma_{r,CC}^\pm,
%\phi^\pm_{CC}(x,Q^2),
\label{eq_7.1}
\end{equation}
where $\sigma_{r,CC}^\pm$
%$\phi^\pm_{CC}(x,Q^2)$ 
contains  three charged current proton
structure functions depending  on the lepton beam charge.
  If the Fermi constant $G_F$, is equated
with that measured in muon decay, then the residual correction to the
above equation from higher order electro-weak corrections amounts to 
only  a few parts per mille~\cite{spiesberger_7.1}.

\begin{figure}[tb]
\begin{minipage}[t]{8 cm}
\hspace{15pt}
\epsfig{file=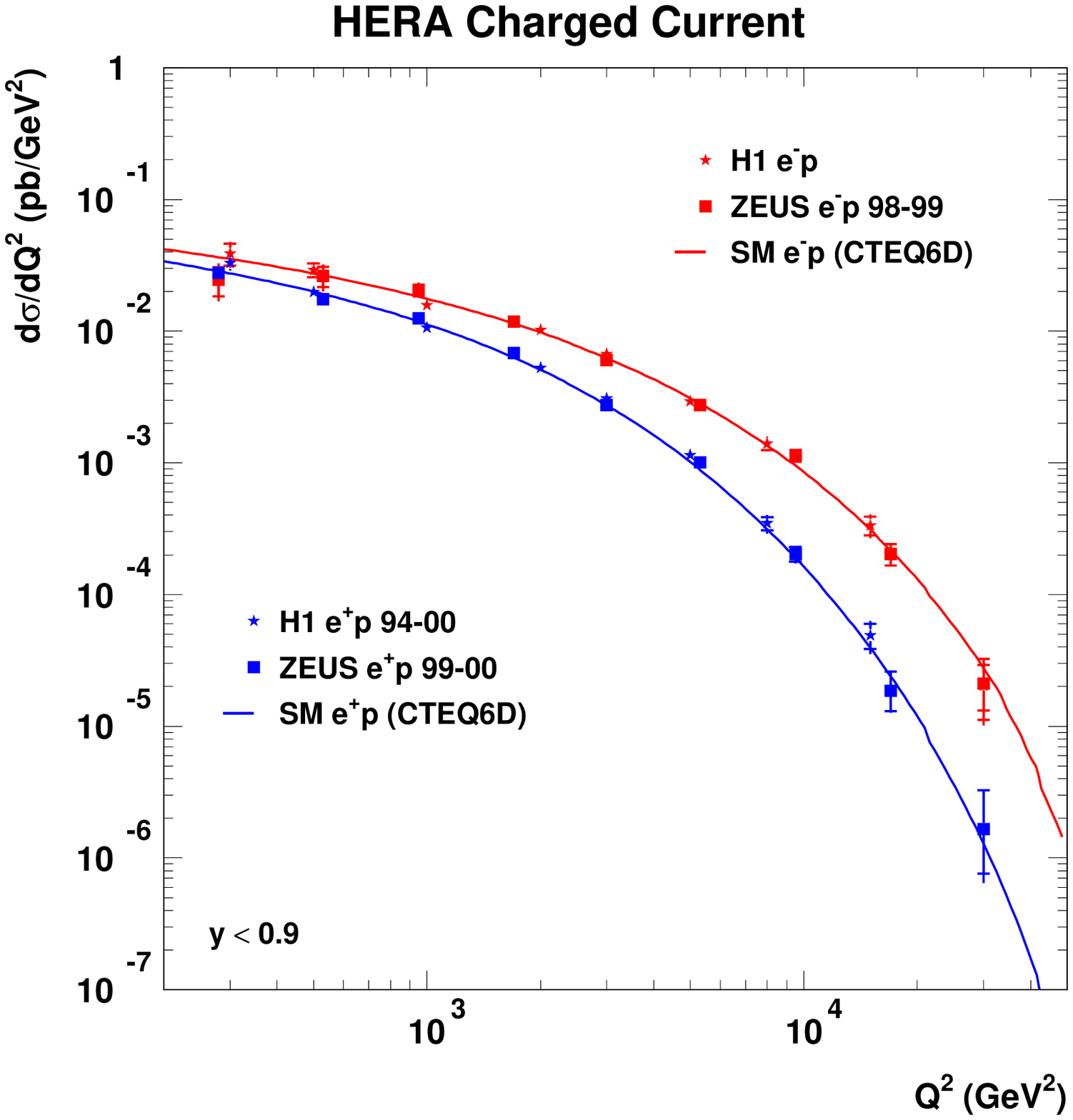,scale=0.44}\epsfig{file=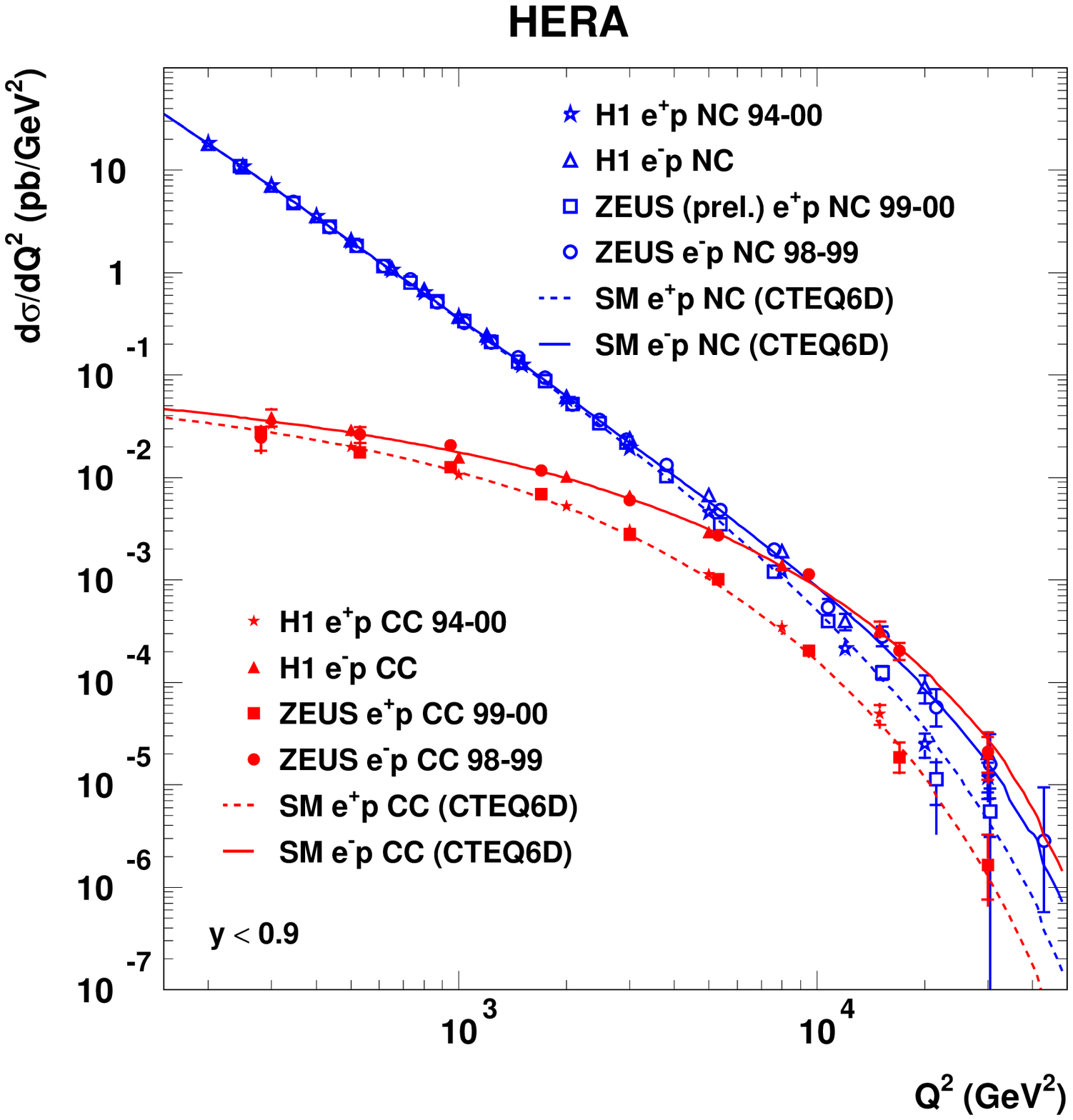,scale=0.44}
\end{minipage}
\begin{center}
\begin{minipage}[t]{16.5 cm}
\caption{Left: The charged current cross-section as a function of $Q^2$
measured at HERA I. Both the $e^+p$ and $e^-p$ cross-sections are
shown.  Right: the charged current cross-sections are compared with
the neutral current cross-sections measured at HERA.
\label{fig_7.1}
}
\end{minipage}
\end{center}
\end{figure}

If the structure functions are known, then the measurement in the
region $Q^2 \ll M_W^2$ is a measurement of $G_F$ independently
of $Q^2$.
%, as can be seen from Equation~\ref{eq_7.1}.  
In the region $Q^2 \approx M_W^2$, the $Q^2$ dependence of the
cross-section is a measurement of the mass of the particle 
exchanged in the $t$ channel, namely the $W$.

Figure~\ref{fig_7.1} shows the charged current cross-section as a function 
of $Q^2$ at HERA for all data~\cite{h1_zeus_7.01} with unpolarised initial leptons. 
The effect of the propagater mass can be seen as $Q^2$ increases
towards $M_W^2$.  The difference between $e^+p$ and $e^-p$ charged
current cross-sections arises mostly from the fact that in the former 
primarily the $d$-valence quark is probed whereas in the latter 
the $u$-valence is probed.

The neutral current cross-section~\cite{h1_zeus_7.01,h1_zeus_7.02} is also compared to that of
the charged current, in Figure~\ref{fig_7.1}.  
It is observed that the two cross-sections indeed become about the
same at the scale of $Q^2 \approx 10^4$ GeV$^2$ giving an explicit 
demonstration of the electro-weak unification as had been anticipated.

A  unique check of the standard model can be made by using the precisely
measured muon decay constant ($1.1639 \cdot 10^{-5}$ GeV$^{-1}$ \cite{pdg})
and determining the propagator mass $M_W$ that 
appears in Equation~\ref{eq_7.1}.
A fit, with $M_W$ as a free parameter gives, for example:
\begin{equation}
 M_W = 78.9\pm 2.0({\rm stat.})\pm 1.8({\rm syst.})^{+2.0}_{-1.8} ({\rm PDF}) {\rm GeV},
\label{eq_7.2}
\end{equation}
for the ZEUS determination using $e^+p$ data from HERA I only. 
The H1 collaboration determined:
\begin{equation}
 M_W = 80.9\pm 3.3({\rm stat.})\pm 1.7({\rm syst.})\pm 3.7 ({\rm theo.}) {\rm GeV},
\label{eq_7.25}
\end{equation}
using 35.6 pb$^{-1}$ of e$^+$p data from HERA I.
These uncertainties are much larger than the direct measurements of $M_W$ 
and the measurements in the time-like region \cite{pdg}.  
However, the HERA measurements demonstrate
a consistency of the space-like measurements with the Standard Model. 

The purely weak nature of the charged current interaction means that
the cross-section for polarised initial leptons is directly related to
the unpolarised cross-sections as:
\begin{equation}
  \frac{d^2\sigma^{CC}_{pol}(e^\pm p)}{dxdQ^2} =
  (1\pm P)\frac{d^2\sigma^{CC}_{unpol}(e^\pm p)}{dxdQ^2}
\label{eq_7.3} 
\end{equation} 
where the longitudinal polarisation of the lepton beam is defined
as 
\begin{equation}
   P = \frac{N_R - N_L}{N_R + N_L},
\label{eq_7.4}
\end{equation}
where $N_R$ and $N_L$ are the numbers of right- and left-handed leptons
in the beam, respectively.

\begin{figure}[tb]
\begin{center}
\begin{minipage}[t]{11 cm}
\epsfig{file=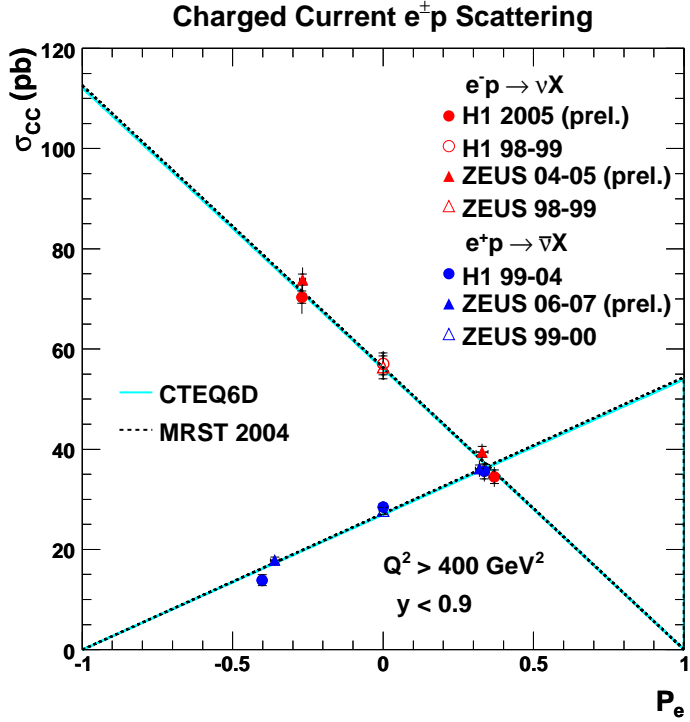,scale=0.5}
\end{minipage}
\begin{minipage}[t]{16.5 cm}
\caption{The total cross-section for $e^\pm p$ CC DIS with $Q^2 > 400$
GeV$^2$ as a function of the longitudinal polarisation of the lepton
beam.  The lines show the prediction of the SM evaluated
using CTEQ and MRST PDFs.
\label{fig_7.2}
}
\end{minipage}
\end{center}
\end{figure}

With the advent of longitudinally polarised lepton beams
in collider mode, in HERA\,II, it
became possible to directly measure the polarisation dependence of 
charged current DIS. 
The result~\cite{h1_zeus_7.01,h1_zeus_7.1,h1_7.1} is shown in 
Figure~\ref{fig_7.2}.

Within the standard model, $G_F$ is a function of the fine structure
constant, $\alpha$, the masses of the gauge bosons $M_W$ and $M_Z$. 
Through radiative corrections, $G_F$ also depends on
%neglecting the masses of the lighter quarks,
 the top mass, $M_t$, as
well as on  the Higgs mass, $M_H$, though only logarithmically.
%Thus it is possible to, instead of using the muon decay constant, take
%the values of $\alpha$ and $M_Z$ from experiments, and determine
%$M_W$ at HERA as a function of $M_t$.  Given the uncertainty of
Rather than taking $G_F$ from the muon decay constant, it is possible to take
values of $\alpha$, $M_Z$ and $M_t$ from other experiments, and determine
$M_W$ at HERA.  Given the uncertainty of
$M_t$ from the Tevatron, which is at the level of 2 GeV \cite{cdf_d0_7.1}, 
it is estimated
that the uncertainty of $M_W$ in the order of 50 MeV will be possible using
the full statistics from HERA~\cite{beyer_7.1}. This uncertainty, which is 
comparable 
to those of current direct measurements~\cite{pdg}, is not enough
to predict the Higgs mass with any certainty; however it is interesting
to see if the trend at the Tevatron for $M_W$ and $M_t$ measurements to
predict a low mass for the standard model Higgs boson is confirmed by this result.
\begin{figure}[tbh]
\begin{center}
\begin{minipage}[t]{11 cm}
\epsfig{file=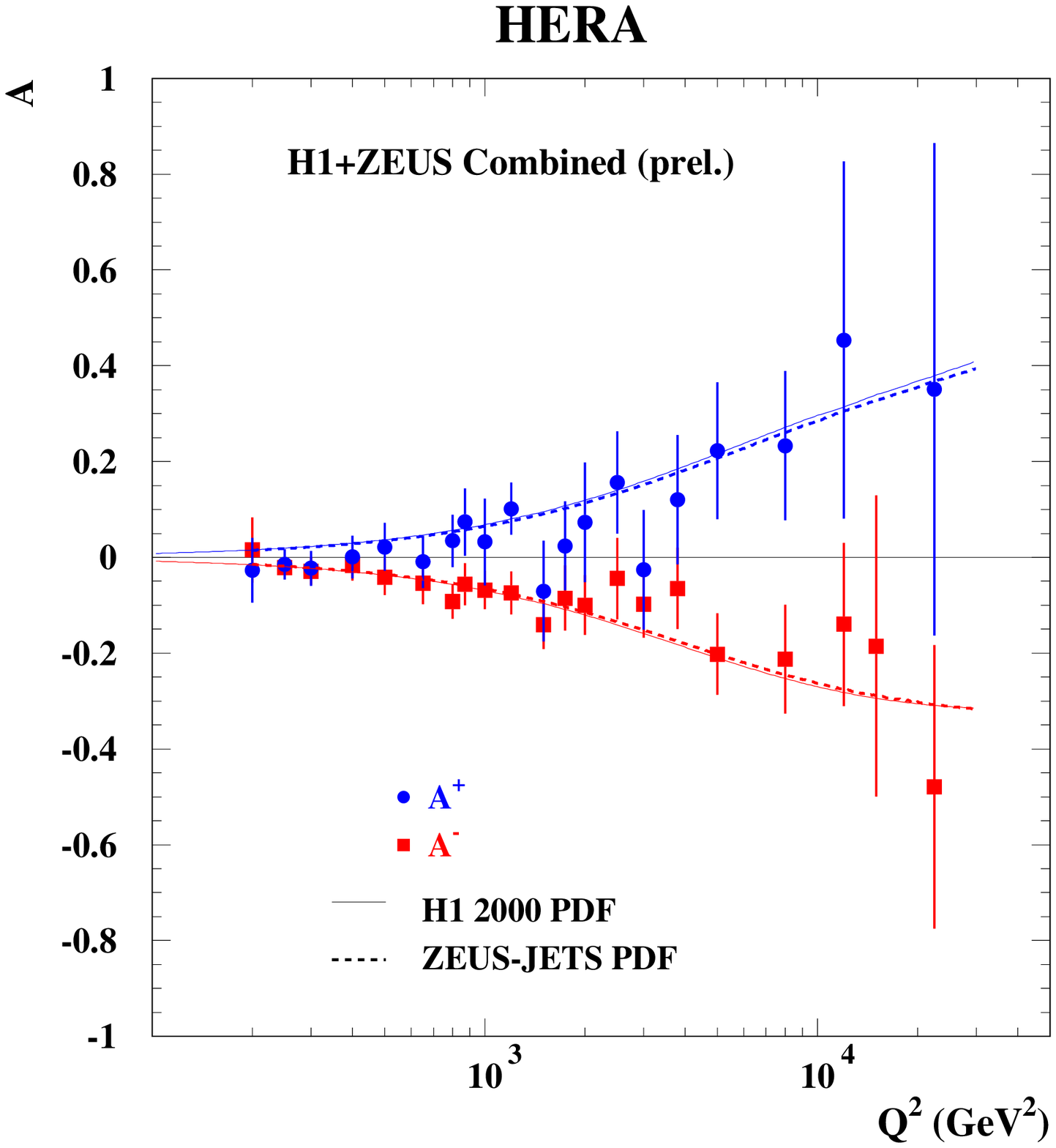,scale=0.5}
\end{minipage}
\begin{minipage}[t]{16.5 cm}
\caption{Measurements of the polarisation asymmetries $A^\pm$ by the
H1 and ZEUS collaborations.  The error bars denote the
total uncertainty.  The curves describe the standard model 
predictions using the PDF sets obtained from unpolarised data.
\label{fig_7.3}}
\end{minipage}
\end{center}
\end{figure}

\subsection{Neutral Current Cross Section}

%The neutral current (NC) $e\pm p$ cross-sections of an $e$ beam with a
%polarisation of $P_e$ can be expressed as:
%\begin{equation}
% \frac{d\sigma^{NC}(e^\pm p)}{dxdQ^2}
%  = \frac{2\pi\alpha^2}{xQ^4} [H^\pm - P_eH_{P_e}^\pm]
%\label{eq_7.5}
%\end{equation} 
%where $H^\pm_{(P_e)}$ contains the unpolarised (polarised) structure
%functions, and has the form,
%\begin{equation}
% H^\pm_{(P_e)} = Y_+F_2^{(P_e)} \mp Y_-xF_3^{(P_e)},
%\label{eq_7.6}
%\end{equation}
%neglecting the longitudinal structure functions and $Y_\pm = 1\pm(1-y)^2$.
%With the electroweak couplings explicitly written out, the unpolarised
%structure functions are,
%\begin{eqnarray}
% F_2 = \sum_q x(q+\bar{q})[e_q^2-2e_qv_qv_e\chi_Z+
%   (v_q^2+a_q^2)(v_e^2+a_e^2)\chi_Z^2], 
%\label{eq_7.7} \\ 
% xF_3 = \sum_q x(q-\bar{q})[-2e_qa_qa_e\chi_Z+4v_qa_qv_ea_e\chi_Z^2]
%\label{eq_7.8}
%\end{eqnarray}
%and the polarised structure functions are:
%\begin{eqnarray}
% F_2^{P_e} = \sum_q x(q+\bar{q})[2e_qv_qa_e\chi_Z-2(v_q^2+a_q^2)
%v_ea_e\chi_Z^2], 
%\label{eq_7.9}\\
% xF_3^{P_e} = \sum_q x(q-\bar{q})[2e_qa_qv_e\chi_Z-4v_qa_q
%(v_e^2+a_e^2)\chi_Z^2],
%\label{eq_7.10}
%\end{eqnarray}
%where $e_f$, $a_f$ and $v_f$, are respectively, the electric charge, the axial coupling
%and the vector couplings of the fermion $f$.  The quantity $\chi_Z$ is:
%\begin{equation}
%  \chi_Z = \frac{1}{\sin{2\theta_W}^2} (\frac{Q^2}{M_Z^2+Q^2}).
%\label{eq_7.11}
%\end{equation}

\begin{figure}[tb]
\begin{minipage}[t]{8 cm}
\hspace{15pt}
\epsfig{file=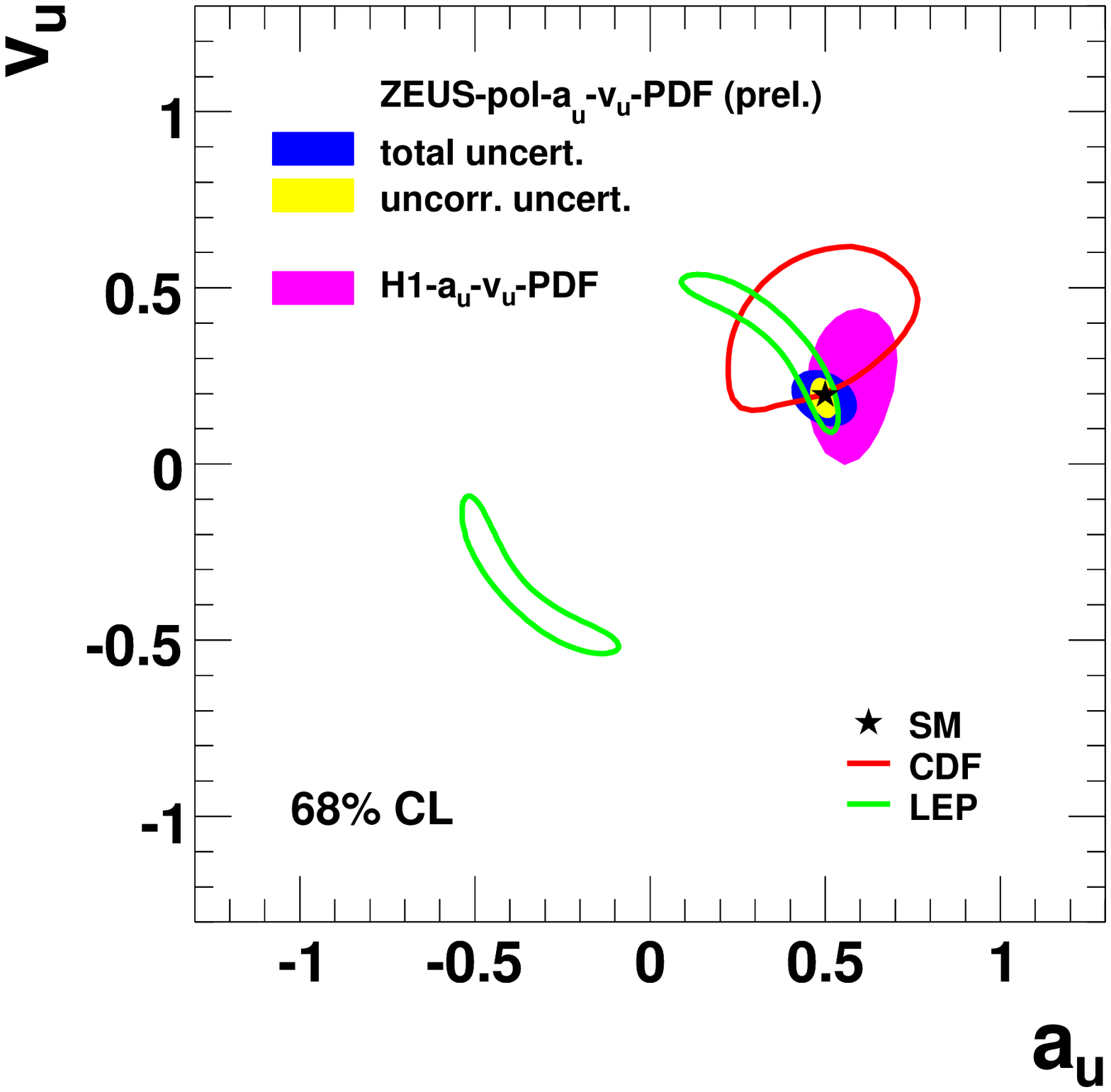,scale=0.44}\epsfig{file=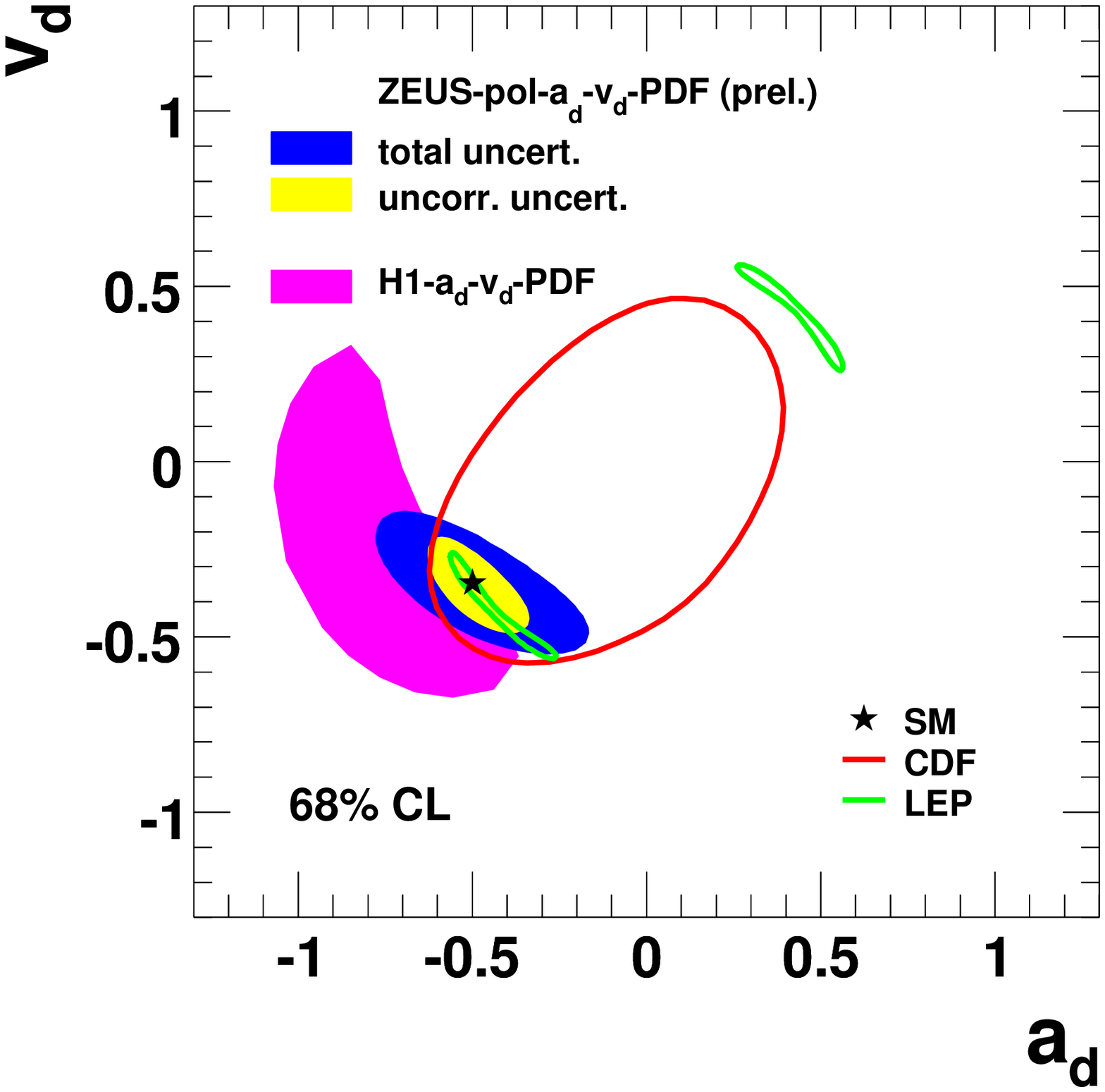,scale=0.44}
\end{minipage}
\begin{center}
\begin{minipage}[t]{16.5 cm}
\caption{The 68\% confidence level contours of the electroweak
parameters $a_u$ vs $v_u$ and $a_d$ vs $v_d$ from the fits
to H1 (polarised only) and preliminary 
ZEUS (polarised and unpolarised data)
compared to the results from LEP and Tevatron.
\label{fig_7.4}
}
\end{minipage}
\end{center}
\end{figure}

%With polarised lepton beam, the cross-section asymmetry in polarisation
%can be defined as:
%\begin{equation}
%   A^\pm = \frac{2}{P_R-P_L}\cdot
%   \frac {\sigma^{P_R}(e^\pm p)-\sigma^{P_L}(e^\pm p)}
%        {\sigma^{P_R}(e^\pm p)+\sigma^{P_L}(e^\pm p)}
%\label{eq_7.12}
%\end{equation}
%which is to a rather good approximation,
%\begin{equation}
%   A^\pm  \simeq \mp \chi_Z \frac{a_e}{e_q} v_q
%\label{eq_7.13}
%\end{equation}
%and thus a direct measurement of electroweak parity violation.  

A polarisation asymmetry measurement of the NC scattering, 
for helicities $P_R$ and $P_L$,
according to  Equations\,\ref{ncsi} and \ref{strf}, determines a
combination of $F_2^{\gamma Z}$ and $xF_3^{\gamma Z}$
\begin{equation} \label{pasy}
\frac{ \sigma_{r,NC}^{\pm}(P_R) - \sigma_{r,NC}^{\pm}(P_L)}{P_R -P_L} 
 = \kappa_Z    [ \mp a_e F_2^{\gamma Z} +  \frac{Y_-}{Y_+} v_e xF_3^{\gamma Z}]
% \simeq  \mp \kappa_Z  a_e G_2 
\end{equation}
neglecting the pure $Z$ exchange terms, which are small at HERA. The second term is 
a small correction since the vector coupling is small, as is the factor
$Y_-=1-(1-y)^2$, in most of the kinematic range at HERA. 
The product $a_e F_2^{\gamma Z}$ is proportional to 
combinations $a_e v_q$ and is thus
a direct measure of parity violation at very small distances,
$\sim 10^{-18}$\,m, as they are probed with electroweak cross section
measurements at HERA.

The polarisation NC cross-section asymmetry 
\begin{equation} \label{apm}
A^\pm = \frac{2}{P_R-P_L} \cdot \frac{\sigma_{NC}^{\pm}(P_R) -\sigma_{NC}^{\pm}(P_L)}
                          {\sigma_{NC}^{\pm}(P_R) +\sigma_{NC}^{\pm}(P_L)}
\end{equation}
to a very good approximation measures the structure function ratio  
\begin{equation}  \label{fgf}                                 
        A^\pm      
%        \simeq  \mp \kappa   a_e \frac{F_2^{\gamma Z}}{(F_2 + \kappa
%         a_e Y_- xF_3^{\gamma Z}/Y_+)} 
         \simeq \mp  \kappa_Z a_e \frac{F_2^{\gamma Z}}{F_2}.
\end{equation}
Thus $A^+$ is expected to be positive and about equal to $-A^-$.
 At large $x$  these asymmetries measure 
the $d/u$ ratio of the valence quark distributions according to
\begin{equation} \label{doveru} 
           A^\pm   \simeq  \pm  k \frac{1+d_v/u_v}{4+d_v/u_v}.
\end{equation}
The
preliminary measurements of $A^\pm$ from HERA, Figure~\ref{fig_7.3}, based on the 
combined data~\cite{h1_zeus_7.1} from H1 and ZEUS show a significant
polarisation effect and the asymmetries $A^\pm$ to be of opposite sign
as predicted.  The lines, which describe the data well, are the
predictions of the standard model as obtained from the H1 and ZEUS QCD fits.

%It is clear from Equations~\ref{eq_7.7} to \ref{eq_7.10} that 
It is clear from Equations~\ref{ncsi} to \ref{va} that 
neutral current cross-sections
at HERA are sensitive to the vector and axial-vector couplings of the
quarks. The sensitivity of the cross-sections   
to $a_q$ through $\rm \bf xF_3$ is enhanced by polarisation.
%while polarised cross-sections
%have enhanced sensitivity to $a_q$ through $xF_3$ and $xF_3^{P_e}$.
Unpolarised cross-sections are sensitive to $v_q$ mainly via $\rm \bf F_2 $ 
whereas polarised cross-sections have further sensitivity through $\rm \bf xF_3 $. 
Figure~\ref{fig_7.4} shows the published measurement of $a_q$ and $v_q$ by the
H1 collaboration compared to determinations from LEP and Tevatron.
The figure also shows
the preliminary determinations~\cite{h1_zeus_7.03} of ZEUS using
the polarised data which shows the expected improvement in precision.

%From Equations~\ref{eq_7.5} and \ref{eq_7.6}, it can be 
From Equations~\ref{ncsi} and \ref{strf}, it can be 
seen that the difference between the
unpolarised $e^+ p$ and $e^- p$ cross-sections can be defined to be
${ \bf xF_3}$.  Terms proportional to $v_e$, which is small (0.036), can be
neglected  so that, to leading order in pQCD, one has
\begin{equation}
% xF_3 = -2x a_e  [e_u a_u (U-\bar{U}) + e_d a_d (D-\bar{D})] 
%\frac{1}{\sin{2\theta_W}^2} (\frac{Q^2}{M_Z^2+Q^2})
{ \bf xF_3^\pm } \simeq \pm a_e \kappa_Z xF_3^{\gamma Z} 
\end{equation}
to a good approximation, with
\begin{equation}
xF_3^{\gamma Z} =2 x [e_u a_u (U-\bar{U}) + e_d a_d (D-\bar{D})]   
\label{eq_7.14}
\end{equation} 
and $U = u+c$ and $D=d+s$ for four flavours.
Assuming that there are no anomalous differences between sea quark and anti-quark
distributions, $xF_3^{\gamma _Z}$ depends on the valence quark distributions
of the proton only, and on the  axial vector couplings of
the $u$ and $d$ quarks.  

\begin{figure}[tb]
\begin{center}
%\begin{minipage}[t]{8 cm}
\epsfig{file=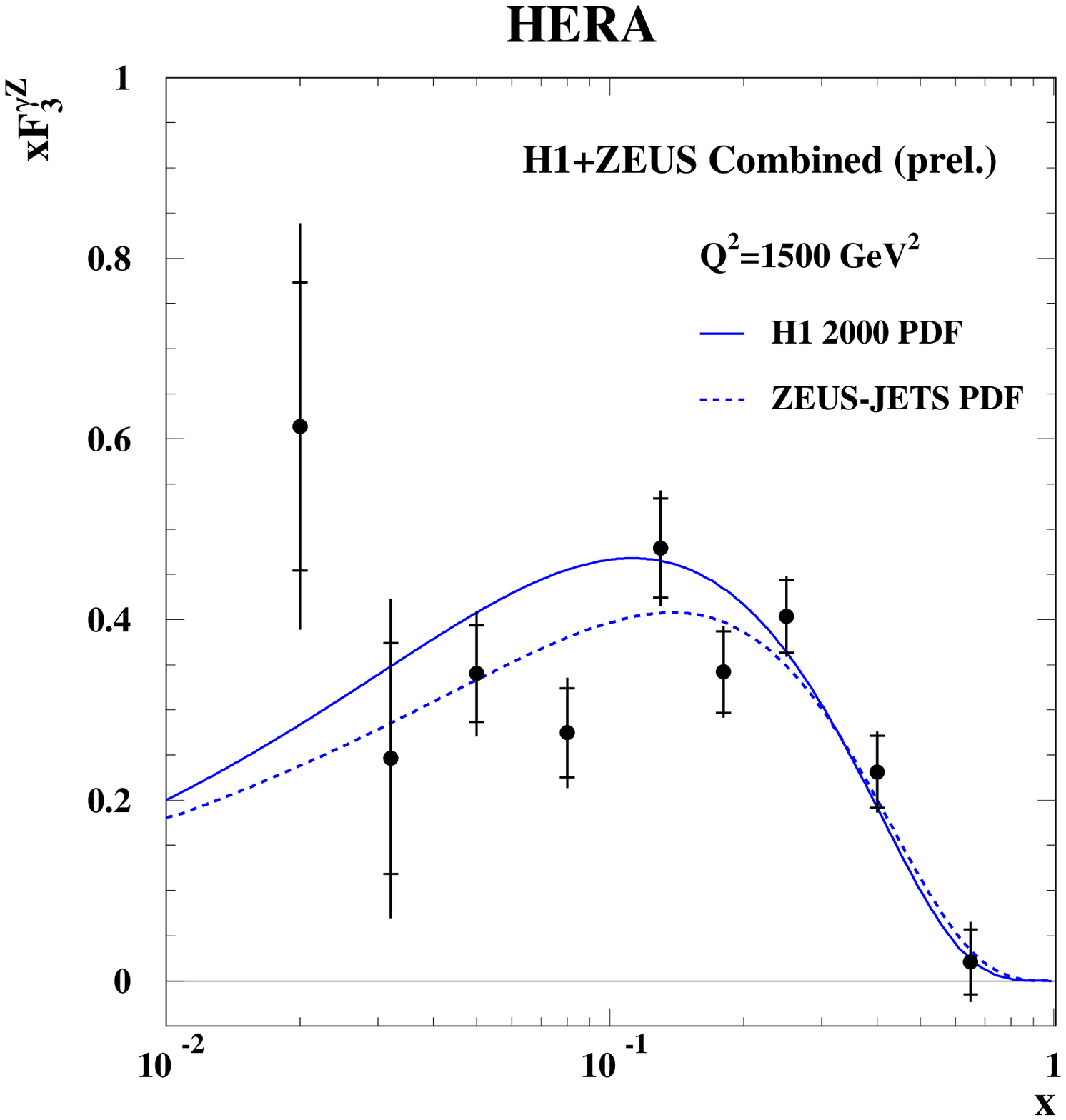,scale=0.5}
%\end{minipage}
\begin{minipage}[t]{16.5 cm}
\caption{The measurement of the structure function $xF_3^{\gamma Z}$ by
the H1 and ZEUS collaborations.  The inner error bars denote the
statistical uncertainty while the full error bars comprise the statistical
and systematic uncertainty added in quadrature. The data are from HERA\,I
but include also part of the polarised HERA\,II data, in which, however,
effectively polarisation effects average out leading to a small
correction only, see \cite{h1_zeus_7.1}. The curves describe the
Standard Model predictions as obtained in the H1 and ZEUS NLO QCD fits.
% to data not including these measurements directly.
\label{fig_7.5}
}
\end{minipage}
\end{center}
\end{figure}

%Since the above expression is an interference term between the
%photon and $Z$ exchanges, and pure $Z$ exchange term can be neglected, 
%$xF_3^{\gamma Z} = xF_3\frac{1}{2\chi_Z}$ can be defined. 
%\begin{equation}\label{xf3gamz}
%xF_3^{\gamma Z} = xF_3 \cdot 2\sin{2\theta_W}^2 (\frac{M_Z^2+Q^2}{Q^2}).
%\end{equation}
The preliminary measurement~\cite{h1_zeus_7.1} by the 
ZEUS and H1 collaborations 
of this quantity as a function of $x$ at $Q^2$ of
1500 GeV$^2$ is shown in Figure~\ref{fig_7.5}.   The measurements are well
described by the standard model predictions using PDFs as obtained
from the HERA data. 

Substituting the standard model values for the couplings, one finds
\begin{equation}
  xF_3^{\gamma Z}= \frac{x}{3}(2u_v + d_v)
\label{eq_7.15}
\end{equation}
where $u_v$ and $d_v$ are the valence up and down quarks
distributions.  To leading
order, the above equation leads to a sum rule~\cite{rizvi_7.1}
\begin{equation}
  \int_0^1 xF_3^{\gamma Z} \frac{dx}{x} =
  \frac{1}{3}\int_0^1 (2u_v+d_v)dx = \frac{5}{3}.
\label{eq_7.16}
\end{equation}
In the range of the data, the integral of $F_3^{\gamma Z}$ is measured to be
\begin{equation}
  \int_{0.02}^{0.65} F_3^{\gamma Z} dx = 1.21 \pm 0.09(stat)\pm 0.08(syst)
\label{eq_7.17}
\end{equation}
which is in agreement with predictions of the SM for this $x$ range, using,
for example, H1 and ZEUS-Jets PDFs.

%\clearpage
%\newpage
\section{Searches}
Searches for ``Beyond the Standard Model'' (BSM) phenomena at HERA 
is a large topic, reviewed at some length in~\cite{kuze_8.1} relatively 
recently.  Here only a few selected topics are covered based on updated results.
\begin{figure}[h]
\begin{center}
%\begin{minipage}[t]{13cm}
\epsfig{file=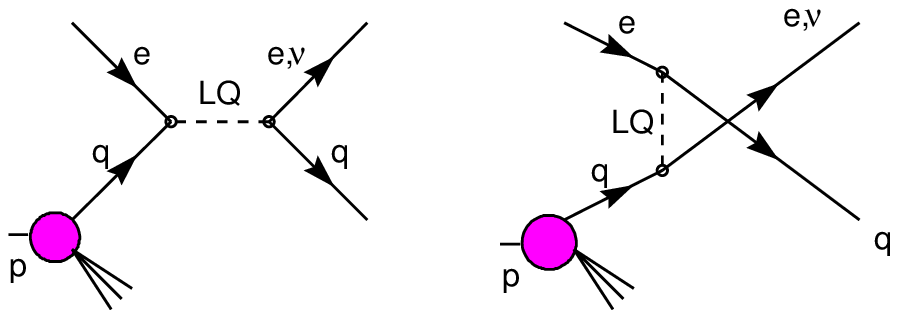,scale=1.5}
%\end{minipage}
\begin{minipage}[t]{16.5 cm}
\caption{Diagrams for $s$-channel leptoquark production and $u$-channel
LQ exchange.  Here $e$ stands for both electron and positron, and $\nu$ for
both neutrino and anti-neutrino.
\label{fig_8.1}
}
\end{minipage}
\end{center}
\end{figure}

Unlike LEP and the Tevatron, HERA does not have
a particle and its anti-particle in the colliding beams.  Therefore,
pair production cross-sections of particles are rather small
but new particles may favourably be produced singly. 
%.  Searches
%at HERA, then, tend to concentrate on production of single particles.  
As a
result, the limits obtained at HERA are generally presented as a function
of the couplings of the new particle as well as its mass.
\begin{figure}[bth]
\epsfysize=9.0cm
\begin{minipage}[bth]{14 cm}
\epsfig{file=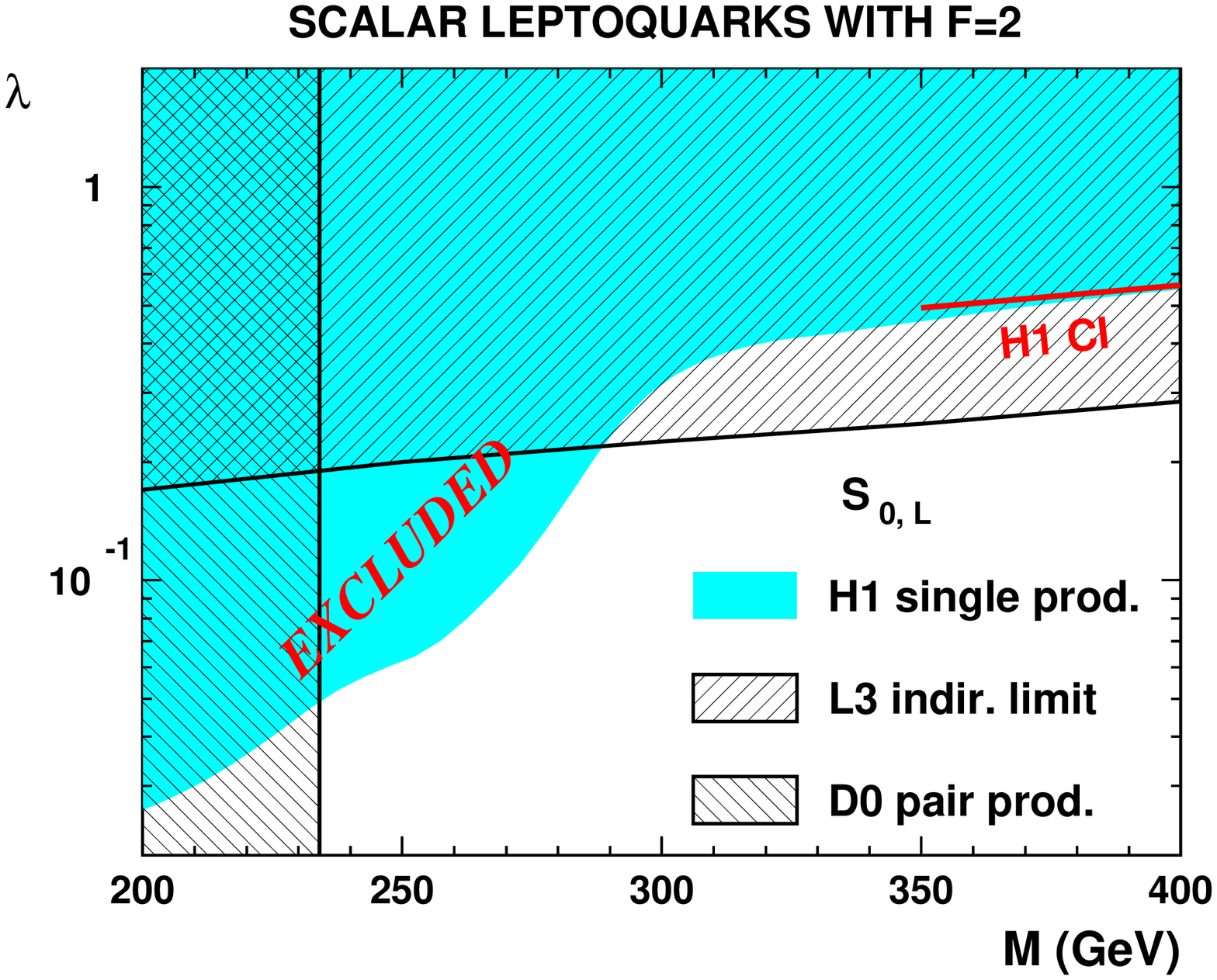,scale=0.45}\epsfig{file=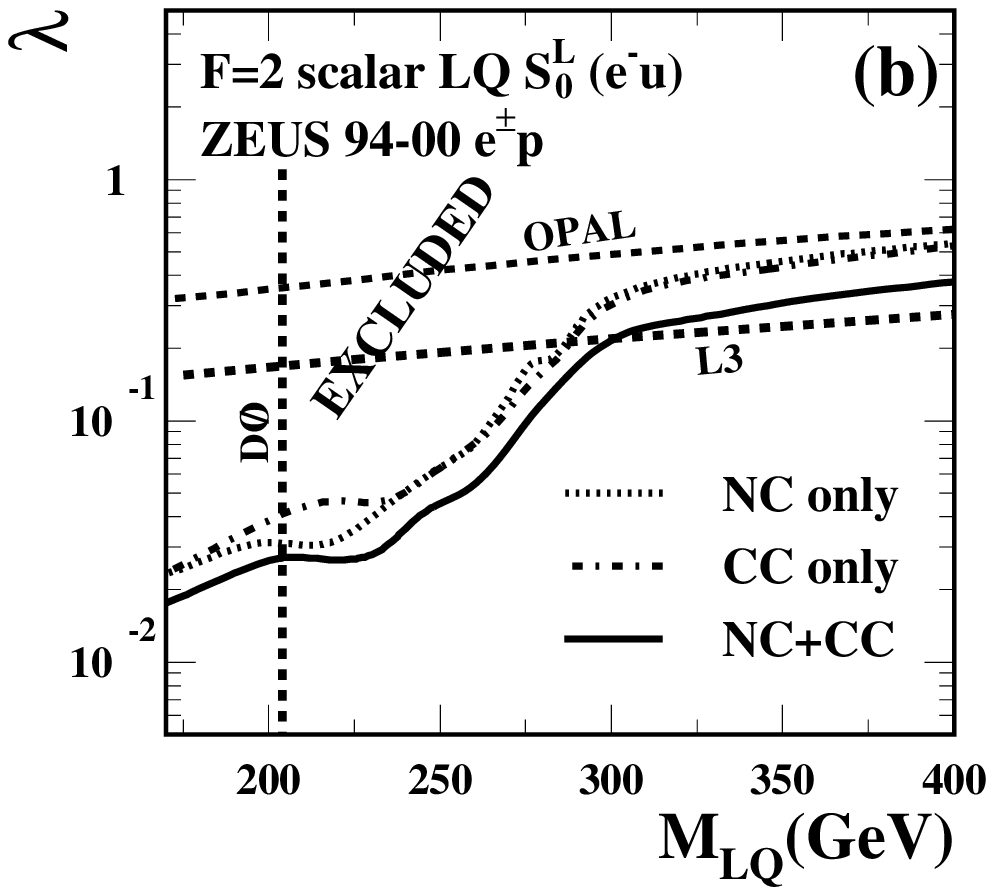,scale=0.8}
\end{minipage}
\begin{center}
\begin{minipage}[bth]{16.5 cm}
\caption{Exclusion limits at 95\% confidence level on the coupling $\lambda$
as a function fot leptoquark mass for $S_{0,L}$ in the framework
of the BRW model from H1 and ZEUS collaborations.  The limits from
L3, OPAL  and D0 collaborations \cite{l3_8.1,opal_8.1,d0_8.1} are also shown.  The D0 limit in the
ZEUS plot is an older one \cite{d0_8.2} current at the time of the publication of the
ZEUS result.  
\label{fig_8.2}
}
\vspace{-0.3cm}
\end{minipage}
\end{center}
\end{figure}

\subsection{Leptoquarks}

In general, measurements for BSM searches may be interpreted, or analysed,
in the context of a particular model (``model-dependent analysis'').
On the other hand, the data, very often the very same data used in the
model-dependent analysis, can be interpreted simply as possible deviations from
the Standard Model.  A first  example of the former is
the search for Leptoquarks.

Leptoquarks (LQs)~\cite{buchmueller_8.1}, colour triplet bosons carrying both baryon and lepton
numbers are postulated in many extensions of the Standard Model. An
example are squarks in R-parity violating supersymmetry~\cite{butterworth_8.1}.
In $ep$ collisions at HERA, LQ states may be produced directly though 
$e$-quark fusion, or in the $s$-channel, with subsequent decay 
into $eq$ or $\nu q$ yielding resonant peaks at the LQ mass.  LQs may
also be exchanged in the $u$-channel, in which case characteristic
deviations from the SM cross-sections are expected at high $Q^2$.

Both H1 and ZEUS have
published the results of searches~\cite{zeus_8.1,h1_8.1} for LQs from the
HERA\,I period  using 120-130 pb$^{-1}$ of data each.
In the H1 search for LQs, the reconstructed CC and NC events 
are searched for deviations from the SM.  In the $s$-channel production of LQs,
an enhancement is expected at the $x$ corresponding to the LQ mass of 
$M_{LQ} = \sqrt{xs_{eq}}$.  In the ZEUS search, the invariant mass 
of the final-state $e$-jet and $\nu$-jet system is reconstructed directly; in 
this case $M_{LQ} = M_{e(\nu)-jet}$. The two methods are rather similar 
in the large $Q^2$ region which corresponds to large LQ masses.

Neither search finds evidence for LQs and limits are set.  At HERA, it is 
usual to set limits based on the phenomenological model proposed by 
Buchm\"{u}ller, R\"{u}ckl and Wyler (BRW)~\cite{buchmueller_8.1} which describes fourteen 
types of LQs.  It is out of the scope of this review to give the limits
for each of these.  As an example, Figure~\ref{fig_8.2} shows the limits obtained by
the experiments for one of the 14 BRW LQs compared to the limits set
at LEP and the Tevatron.   At LEP, the searches are sensitive mostly
to the coupling, $\lambda$, as is the case at HERA above the kinematic
limit of $M = \sqrt{s}$ since both rely on the virtual exchange of the LQs.
The Tevatron searches, on the other hand, are mainly sensitive to the mass since
the LQs would be pair produced via quark-anti-quark annihilation.

There are many ways to state the limits, and the reader is referred 
to the original papers for a fuller discussion. 
Also limits for LQ couplings to higher generation quarks have been set 
in searches for lepton-flavour violation by both ZEUS and H1 collaborations~\cite{zeus_8.2,h1_8.2}.

\subsection{Excesses Beyond the SM}

Rather than doing a model-dependent search for phenomena beyond the Standard
Model, one may study processes with a low cross-section in the SM in order
to observe possible deviations from expectations.

In 2002, the H1 collaboration reported an excess of events with high energy
isolated electrons or muon accompanied by missing
 transverse momentum~\cite{h1_8.3}.
In approximately 100 pb$^{-1}$ of data, H1 observed 10 events with 
transverse energy of the hadronics system, P$^X_T$, greater than 25 GeV.
Such events are expected in the SM from $W$ boson production with 
subsequent leptonic decay. The prediction from the SM and 
background from misidentified CC and NC events was 2.9$\pm$0.5.
In a similar search, the ZEUS collaboration found no excess
over the SM~\cite{zeus_8.3}.

\begin{figure}[t]
\epsfysize=9.0cm
\begin{minipage}[t]{14 cm}
\epsfig{file=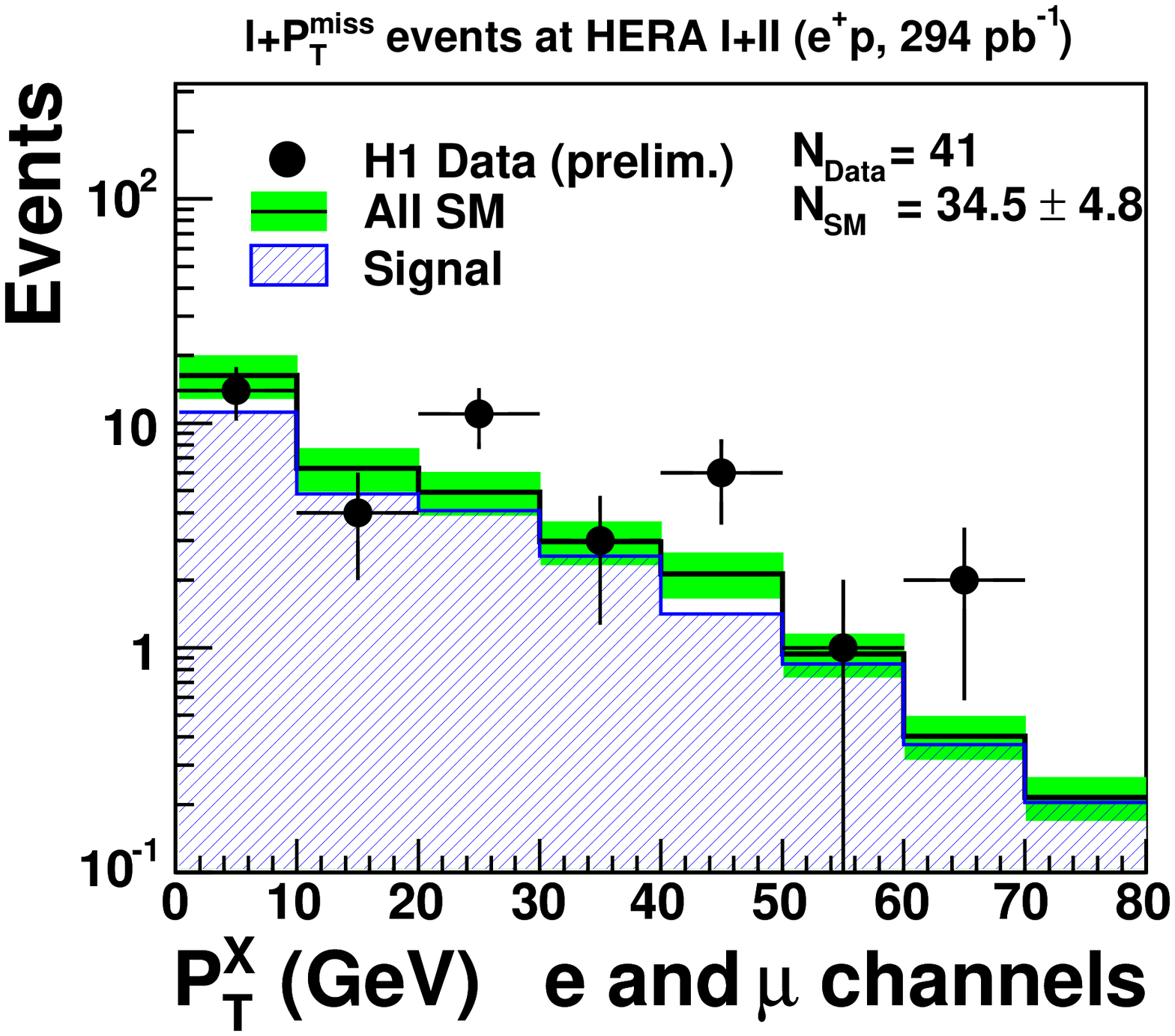,scale=0.45}\epsfig{file=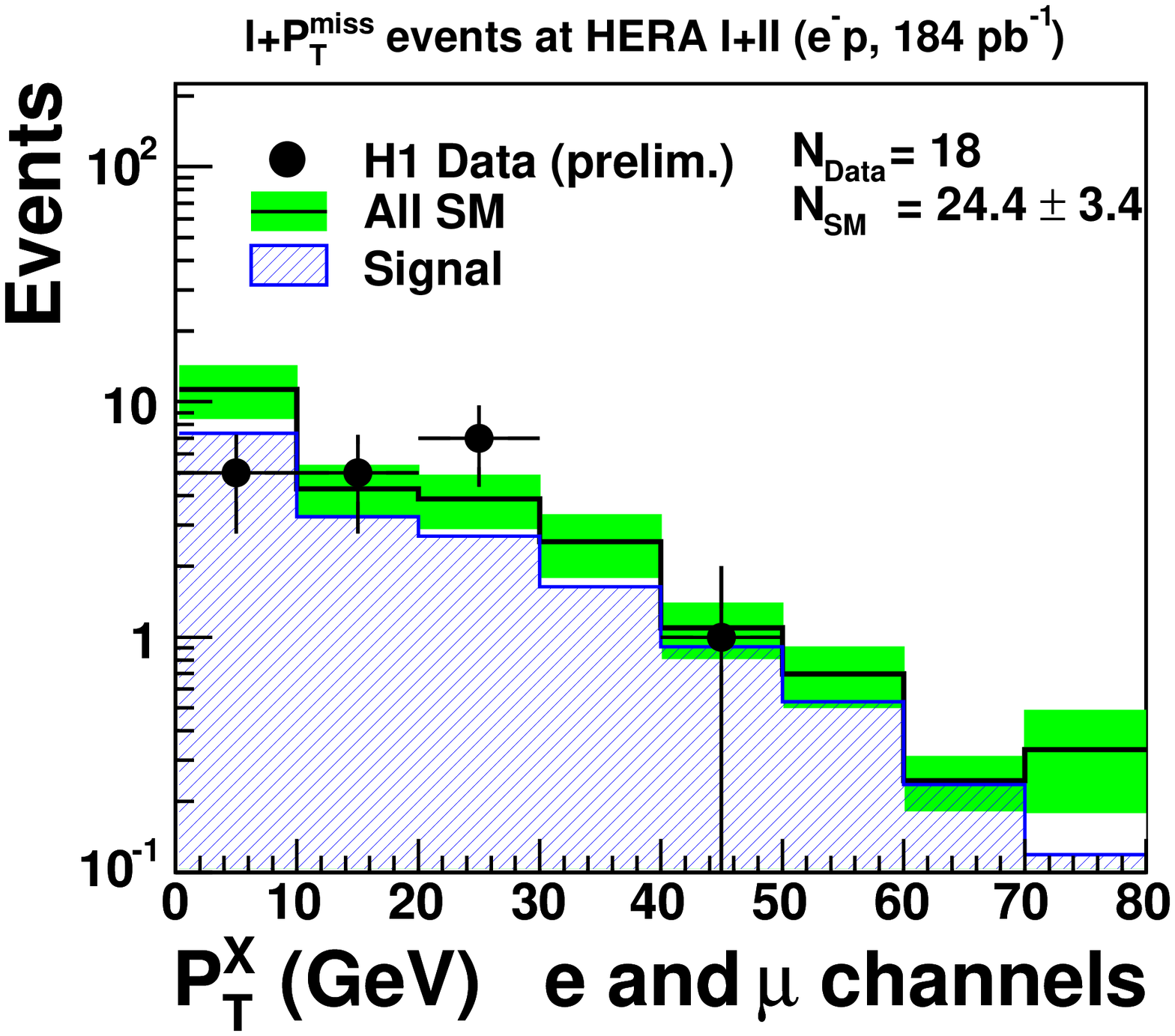,scale=0.45}
\end{minipage}
\begin{center}
\begin{minipage}[t]{16.5 cm}
\caption{The distribution of transverse hadronic energy, $P_T^X$, for events
with isolated leptons as measured by the H1 collaboration.  
The $e^+p$ (left) and $e^-p$ (right) data are compared to the SM
expectations (histogram labeled ``All SM''). The component due
to true SM signal, dominated by $W$ production, is given by the
histogram labeled ``Signal''.  N$_{data}$ is the total
number of data events observed, N$_{SM}$ is the total
SM expectation.  The total uncertainty of the SM expectation
is given by the shaded band.
\label{fig_8.3}
}
\end{minipage}
\end{center}
\end{figure}

Both collaborations have recently made preliminary
 updates~\cite{h1_zeus_8.1} including 
the data from HERA II which amounts to a five fold increase in 
integrated luminosity. Figure~\ref{fig_8.3} shows the distributions of 
these events as a function of P$^X_T$ as reported by the H1 collaboration.

The results from both collaborations in the region of excess observed 
by H1 originally is summarised in Table~\ref{tab_8.1}.  Overall, the data show 
a good agreement with SM expectations.  In detail, there is an 
excess observed by H1 in the $e^+p$ sample for 
both electrons and muons.  The excess is
at the level of 3 $\sigma$ if these two H1 samples are summed.  
The ZEUS data in the same channel, on the other hand, is in good agreement with the SM. 
In a combined analysis~\cite{h1_zeus_8.15}, where a common phase space was defined, 29 isolated electrons and muons
were observed at $P^X_T > 25$ GeV, in 0.97 fb$^{-1}$ of data, where 25.3$\pm$3.2 events are expected from SM.
Both H1 and ZEUS collaborations have also looked in the $\tau$ channel, and have found no significant 
excess beyond the SM~\cite{zeus_8.4,h1_8.4}.

\begin{table}
\begin{center}
\begin{minipage}[hb]{16.5 cm}
\caption{The summary of HERA preliminary results for events with isolated
electrons or muons and missing transverse momentum with $P_T^X > 25$ GeV.
The number of observed events is compared to the SM prediction.  The signal
component of the SM expectation, dominated by $W$ production, is given as a 
percentage in the parenthesis.  The selections for H1 and ZEUS results are 
similar but not identical.  See \cite{h1_zeus_8.1} for details.
\label{tab_8.1}}
\end{minipage}
\vspace{0.1in}

\begin{minipage}[hb]{14 cm}
\begin{tabular}{|c|c|c|l}
\cline{1-3}
{\small Isolated $e$ candidates  } & {\small  H1 Prelim.  } & {\small  ZEUS Prelim. } & \\
\cline{1-3}
{\small $e^{-}p$ (H1: 184 pb$^{-1}$, ZEUS: 204 pb$^{-1}$)} &  {\small 3$/3.8 \pm 0.6$ (61\%)} &  {\small 5$/3.8 \pm 0.6$ (55\%)}\\
\cline{1-3}
{\small $e^{+}p$ (H1: 294 pb$^{-1}$, ZEUS: 228 pb$^{-1}$)} &  {\small 11$/4.7 \pm 0.9$ (75\%) }& {\small 1$/3.2 \pm 0.4$ (75\%) }& \\
\cline{1-3}
{\small $e^{\pm}p$ (H1: 478 pb$^{-1}$, ZEUS: 432 pb$^{-1}$)} & {\small 14$/8.5 \pm 1.5$ (68\%)} & {\small 6$/7.0 \pm 0.7$ (64\%)} & \\
\cline{1-3}
\end{tabular}
\end{minipage}
\end{center}

\begin{center}
\begin{minipage}[hb]{14 cm}
\begin{tabular}{|c|c|c|l}
\cline{1-3}
{\small Isolated $\mu$ candidates  } & {\small  H1 Prelim.  } & {\small  ZEUS Prelim. } & \\
\cline{1-3}
{\small $e^{-}p$ (H1: 184 pb$^{-1}$, ZEUS: 204 pb$^{-1}$)} &  {\small 0$/3.1 \pm 0.5$ (74\%)} & {\small 2$/2.2 \pm 0.3$ (86\%)}\\
\cline{1-3}
{\small $e^{+}p$ (H1: 294 pb$^{-1}$, ZEUS: 228 pb$^{-1}$)} & {\small 10$/4.2 \pm 0.7$ (85\%)} & {\small 3$/3.1 \pm 0.5$ (80\%) }& \\
\cline{1-3}
{\small $e^{\pm}p$ (H1: 478 pb$^{-1}$, ZEUS: 432 pb$^{-1}$)} & {\small 10$/7.3 \pm 1.2$ (79\%)} &  {\small 5$/5.3 \pm 0.6$ (82\%)} & \\
\cline{1-3}
\end{tabular}
\end{minipage}
\vspace{-0.5cm}
\end{center}
\end{table}

Having done a model-independent search, one can use the data to set 
limits on particular models. One interpretation of this type of excess 
would be an anomalous coupling  of the u-quark to the t-quark.  
Such a coupling in the SM is negligibly
small.  Indeed the same type of analysis has been used to set limits on
a single top production at HERA, and thus on the anomalous $tu\gamma$ and
$tuZ$ couplings~\cite{zeus_8.2,h1_8.5}.

\vspace{0.1in}
The H1 collaboration had observed events with high $P_T$ di- or tri-electrons 
in excess of those expected by the SM via higher order electroweak
events~\cite{h1_8.6}; there were three di-electron events and three tri-electron
events, with the mass of a pair of electrons in excess of 100 GeV, 
in a data sample corresponding to the integrated luminosity
of 115 pb$^{-1}$ where the expectations were 0.30$\pm$0.04 and
0.23$\pm$0.04 events respectively. 

In recent updates, both H1 and ZEUS collaborations have released 
preliminary results on multi-electron production using over 450 pb$^{-1}$
each~\cite{h1_zeus_8.2}.  Neither collaboration finds
 a significant excess beyond the
SM in the larger data sample.  H1 collaboration searched for multi-lepton
events, where leptons are either electrons or muons, and
found four events with a scalar
sum of lepton transverse momenta greater than 100 GeV, whereas 
1.9$\pm$0.4 events are expected from SM.  ZEUS searched for di-
and tri-electrons with the mass of a pair of electrons in excess
of 100 GeV, and found 6 events where 5.4$\pm$1.1 events are expected
from SM.  In a combined analysis\cite{h1_zeus_8.3},
done in a common phase space,
H1 and ZEUS  together observed 6 di- and tri-electron events 
with a scalar sum of electron transverse momenta greater than 100 GeV,
in 0.94 fb$^{-1}$ of data, where 3.0$\pm$0.3 events are expected from SM.

%\clearpage
\section{Summary}
HERA was the first electron-proton collider ever built. It allowed 
the investigation of physics of deep inelastic lepton hadron scattering to be
extended by two orders of magnitude into a new kinematic domain
in terms of four-momentum transfer squared, $Q^2$, and Bjorken $x$.
Unlike previous fixed target experiments, which scattered either
charged leptons (electron and muons) or neutrinos off nucleon
targets, the very large $Q^2$ region at HERA, at which weak interactions 
become manifest, allowed the simultaneous investigation of 
neutral current ($ep \rightarrow eX$) and charged current
($ep \rightarrow \nu X$) interactions. Furthermore almost real
photon-proton interactions -- the
photoproduction processes -- could be
investigated in which particles or jets at high masses are
produced. 

The accelerator, comprising separate $e^{\pm}$ and $p$ sources, 
pre-accelerators and rings, was a new challenge for machine physics.
The challenges included the problem of polarised $ep$ collisions,
which was mastered successfully and will provide the a knowledge base
for future $ep$ machines such as that being considered in conjunction
with the LHC $p$ beam. The collider detectors, H1 and ZEUS,
were developed independently and used very different
detector techniques; for example, calorimeter
techniques were based on liquid argon and uranium at H1 and ZEUS,
respectively. 
The physics results obtained turned out to be of similar quality and coverage
and in good agreement. Occasional discrepancies between
the data or methods used led to constructive and sometimes
challenging interactions between the two collaborations.

The intra-H1-ZEUS collaboration
is currently being moved to a new level, in which H1 and
ZEUS will combine their final data sets where appropriate.
An example is the combination and  common QCD analysis
of the inclusive NC and CC cross section data. Such an
analysis can be used to, in essence, cross-calibrate the
results, taking advantage of the different detector
techniques and different kinematic reconstruction
methods. The combination is expected to lead
to results of a new quality, exceeding what is to be expected
in accuracy from a simple statistical average of the data.
HERA will, in this way, provide a base for predictions of
various processes to be studied at the LHC;
and QCD will have to be used to extrapolate HERA
measurements to the LHC kinematic regime. While
the final results of HERA are being obtained, the first
observations at the LHC are expected to be made.
This will lead to a fruitful link between $ep$ and
$pp$ physics, between HERA and the LHC.
 
The main results of HERA as had been presented in this overview
may be summarised as follows:
\begin{itemize}
\item
HERA opened the field of low $x$ physics having  
reached values of $x = Q^2/sy$ of the order of $10^{-4}$
at $Q^2$ larger than a few GeV$^2$, owing to the unprecedented
large energy $s$. In this newly accessed range, parton densities
were discovered to be rising towards lower $x$ at fixed $Q^2$
which is the consequence of gluon dominance at low $x$. This is in
contrast to valence quark dominance at high $x$.
Thus, a new phase of matter was discovered in which
the densities are high but the coupling constant is small
compared to unity. This discovery has immediate consequences
for the understanding of quark-gluon dynamics in QCD as well as for
high density states which are being investigated,
for example, in nucleus-nucleus scattering or in super-high
energy neutrino interactions.
\item
A further discovery of HERA was the observation of hard
diffractive scattering comprising a significant fraction
(order 10\%) of the cross-section. This, attributed to
 the exchange of vacuum
quantum numbers, is a process subject to perturbative
QCD calculations due to the large momentum transfer squared 
involved. These colourless exchange reactions may lead to a
firmer understanding of confinement, and, at the
LHC, to the measurement of quantum numbers of the
Higgs boson in a supersymmetric extension of the
theory,
in a rather clean experimental environment.
% due to the
%absence of proton dissociative particles. 
\item
As a machine of cleanest resolution of the structure of matter,
HERA has set a new limit of $7 \cdot 10^{-19}$\,m to a possible
substructure of quarks, about $10^3$ times smaller than the
proton radius. While one could have hoped to find a preonic
substructure, this result constitutes an important
milestone and signals that higher energy experiments will have to 
be performed in order to look even deeper into the structure of matter.
\item
The results of H1 and ZEUS have led to an unexpectedly rich harvest
of QCD related results which are the basis of developing
a further understanding of 
the theory of strong interactions. This refers,
for example, to the findings on multi-jet production in DIS,
the measurement of the strong coupling constant and 
the determination of a rather
complete set of parton distributions, 
the analysis of heavy quarks as dynamically produced
in boson-gluon fusion, and the first measurements on deeply virtual
Compton scattering at low $x$ which access parton correlations.
\item
The recent results of HERA have a precision which is becoming
competetive for certain measurements of electroweak theory
parameters at highest energies; an example is the determination
of the light quark weak neutral current couplings.
\end{itemize}
There have been many searches performed for signals of new 
physics, new particles, new symmetries, $eq$ resonances
and QCD phenomena. At the present stage of the data analyses, 
a few years prior to the final results, there have been no
signals observed of SUSY phenomena, 
R-parity  violating single production of SUSY particles,
leptoquarks or 
high density instanton states. The limits set on new particles
are competitive and complementary to the limits set
in the crossed amplitude reactions at similar fermion energies,
as have been studied in $e^+e^-$ reactions at LEP and
in $p \overline{p}$ reactions at the Tevatron. 
If there is physics beyond the Standard Model,
near the accelerator energy frontier, it is thus most 
likely to appear at TeV energies rather than at
energies of a few hundred GeV, which characterise the
Fermi scale as has been thoroughly studied in the last 
two decades of particle physics.

The operation of HERA ended in 2007. The analysis of data
taken by the collider experiments ZEUS and H1,
at the time of this publication, is expected to still take
a few years. More precise results and also new results
will still emerge.
The remaining puzzles, such as that on the existence of
penta-quark states, is expected to be resolved. The physics of 
deep inelastic $ep$ scattering has proven to be a vital
part of particle physics, and it may possibly be continued
at even higher energies than HERA.

%\newpage
%\vspace{1cm} 
%{\bf Acknowledgement} 
%\vspace{0.5cm}
\section*{Acknowledgements}

The overview presented here is a brief summary of the work
of perhaps a thousand technicians, engineers and physicists
for more than two decades. The authors had the great privilege
to be part of a most exciting period of particle physics which
in a few years will come to an end. It is our great pleasure to thank many 
colleagues, too numerous to be named here, for the possibility
to share their wisdom and expertise and for most fruitful
years of collaboration. As spokesmen of H1 and ZEUS during the
times where HERA's upgrade was in some difficulty, it is particularly
pleasant to acknowledge the success of the HERA crew and
to thank them for their impressive efforts without which 
the collider experiments could not have been operated so well.
We acknowledge the strong support of the H1 and
ZEUS experiments by the theoretical particle physics
community.  DESY had been a generous host
to us and many of our collaborators from
many countries.  It is 
not forgotten, that the wisdom and dedication of Volker Soergel, 
Bjoern Wiik and others laid the foundations so that the following
generation (to which we belong) was able to
work on a most exciting machine and on some of the most
fundamental problems in elementary particle physics.

We would like to thank M. Derrick, S. Magill and
P. Nadolsky for careful reading of parts of this manuscript. Any
remaining inconsistencies or errors are, of course, our own.
We would like to thank the H1 and ZEUS collaborations for 
permission to show some of their results, presented so far
at conferences only, also in this review.

RY's work at Argonne National Laboratory was supported by the U.S.
Department of Energy, Office of Science, under
contract DE-AC02-06CH11357.

\end{document}